\documentclass[11pt]{article}
\global\arraycolsep=1pt
\usepackage[T1]{fontenc}
\usepackage{hyphenat}
\usepackage{amsmath,amssymb,graphicx}
\usepackage{amsfonts}
\usepackage{mathrsfs}
\usepackage{mathdots}
\usepackage{hyperref}
\usepackage{multicol,color,longtable}
\definecolor{darkred}{rgb}{0.65,0.15,0}
\hypersetup{pdfborder={0 0 0},colorlinks=true,urlcolor=darkred,citecolor=blue,linkcolor=darkred,linktocpage=true}

\usepackage{array}
\newcolumntype{P}[1]{>{\centering\arraybackslash}p{#1}}
\usepackage{lscape}

\usepackage{empheq}

\newcommand{\lsharp}{\text{\raisebox{-0.45ex}{\Large\guilsinglleft}}}
\newcommand{\rsharp}{\text{\raisebox{-0.45ex}{\Large\guilsinglright}}}

\newcommand{\doi}[2]{\href{http://dx.doi.org/#2}{#1}}
\newcommand{\eprint}[1]{{\href{http://arxiv.org/abs/#1}{\texttt{[#1}]}}}
\newcommand{\eprintN}[1]{{\href{http://arxiv.org/abs/#1}{\texttt{#1 [hep-th]}}}}

\newcommand{\scr}{\mathscr}
\newcommand{\jp}{\Omega}

\usepackage{cite}

\newcommand{\mf}[1]{{\mathfrak{#1}}}

\newcommand{\lb}{\left[}
\newcommand{\rb}{\right]}

\newcommand{\ord}[1]{{\scriptscriptstyle (#1)}}

\newcommand{\Scal}[1]{\Bigl ({#1} \Bigr )}
\newcommand{\scal}[1]{\bigl ({#1} \bigr )}

\makeatletter

\@addtoreset{equation}{section} \makeatother

\newcommand{\CR}{\nonumber \\*}
\newcommand{\nn}{\nonumber}

\newcommand{\ie}{{\textit{i.e.},}\ }
\newcommand{\eg}{{\textit{e.g.},}\ }

\newcommand{\West}{{\scalebox{0.7}{$E_{11}$}}}

\newcommand{\cN}{\mathcal{N}}

\newcommand{\gl}{\mathfrak{gl}}
\newcommand{\de}{\delta}

\def\cF{\mathcal{F}}
\def\cG{\mathcal{G}}
\def\cH{\mathcal{H}}

\newcommand{\be}{\begin{equation}}
\newcommand{\ee}{\end{equation}}
\newcommand{\bea}{\setlength\arraycolsep{2pt} \begin{eqnarray}}
\newcommand{\eea}{\end{eqnarray}}
\newcommand{\eq}[1]{(\ref{#1})}

\newcommand{\w}[1]{\\[0.#1cm]}

\newcommand{\del}{\partial}
\newcommand{\ft}[2]{\tfrac{#1}{#2}}

\newcommand{\mfr}{\mathfrak r}

\begin{document}

\begin{flushright}
%\today 
\hfill arXiv:1703.01305v2\\
%e11tha\underline{ }v2 \hfill 
CPHT-RR-007-032017\\  
MI-TH-1745
\end{flushright}
\vspace{10mm}

\begin{center}

{\LARGE {\bf Beyond $E_{11}$}} \\[5mm]

\vspace{8mm}
\normalsize
{\large  Guillaume Bossard${}^{1}$, Axel Kleinschmidt${}^{2,3}$, Jakob Palmkvist${}^4$,\\ Christopher N. Pope${}^4$ and Ergin Sezgin${}^4$}

\vspace{10mm}
${}^1${\it Centre de Physique Th\'eorique, Ecole Polytechnique, CNRS\\
Universit\'e Paris-Saclay 91128 Palaiseau cedex, France}
\vskip 1 em
${}^2${\it Max-Planck-Institut f\"{u}r Gravitationsphysik (Albert-Einstein-Institut)\\
Am M\"{u}hlenberg 1, DE-14476 Potsdam, Germany}
\vskip 1 em
${}^3${\it International Solvay Institutes\\
ULB-Campus Plaine CP231, BE-1050 Brussels, Belgium}
\vskip 1 em
${}^4${\it Mitchell Institute for Fundamental Physics and Astronomy\\ Texas A\&M University
College Station, TX 77843, USA}

%\vspace{20mm}
\vspace{15mm}

\hrule

\vspace{8mm}

\begin{tabular}{p{14cm}}
{\small
 We study the non-linear realisation of $E_{11}$ originally proposed by West with particular emphasis on the issue of linearised gauge invariance. Our analysis shows even at low levels that the conjectured equations can only be invariant under local gauge transformations if a certain section condition that has appeared in a different context in the $E_{11}$ literature is satisfied.  This section condition also generalises the one known from exceptional field theory. Even with the section condition, the $E_{11}$ duality equation for gravity is known to miss the trace component of the spin connection. We propose an extended scheme based on an infinite-dimensional Lie superalgebra, called the tensor hierarchy algebra, that incorporates the section condition and resolves the above issue. The tensor hierarchy algebra defines a generalised differential complex, which provides a systematic  description of gauge invariance and Bianchi identities. It furthermore provides an $E_{11}$ representation for the field strengths, for which we define a twisted first order self-duality equation underlying the dynamics. 
}
\end{tabular}
\vspace{7mm}
\hrule
\end{center}

\newpage

\setcounter{tocdepth}{2}
\tableofcontents

%%%%%%%%%%%%%%%%%%%%%%%%%%%%%%%%%%%%%%%%

\section{Introduction}

%%%%%%%%%%%%%%%%%%%%%%%%%%%%%%%%%%%%%%%%

In an attempt to find the structure underlying M-theory, West has proposed to study non-linear realisations based on the Lorentzian Kac--Moody group $E_{11}$~\cite{West:2001as,West:2011mm,West:2014eza} and this proposal has been developed further in~\cite{Tumanov:2015iea,Tumanov:2016abm,Tumanov:2016dxc}.\footnote{A conceptually different approach based on the hyperbolic Kac--Moody group $E_{10}$ can be found in~\cite{Damour:2002cu}.} One of the reasons for considering $E_{11}$ is that it contains the covariance group $GL(11)$ of eleven-dimensional supergravity as well as the Cremmer--Julia sequence of split $E_d$ symmetry groups of maximal supergravity~\cite{West:2001as,Cremmer:1979up,Cremmer:1997ct,Cremmer:1998px}. A convenient way of organising the infinitely many generators of the corresponding Lie algebra $\mathfrak{e}_{11}$ is by decomposing its adjoint representation under $\mathfrak{gl}(11)$ and this immediately reveals a possible connection to eleven-dimensional supergravity. One finds as the first generators in this so-called level decomposition the adjoint of $\mathfrak{gl}(11)$ (that is associated with the vielbein), an antisymmetric three-form (that is associated with the three-form gauge field), an antisymmetric six-form (that is associated with the magnetic dual of the three-form) and a mixed symmetry generator with index structure $(8,1)$ (that is associated with the (linearised) magnetic dual of the vielbein)~\cite{West:2002jj,Nicolai:2003fw,Kleinschmidt:2003mf}. These are but the first of an infinity of generators contained in $\mathfrak{e}_{11}$.

\begin{figure}[t!]
\centering
\begin{picture}(300,50)
\thicklines
\multiput(10,10)(30,0){10}{\circle*{10}}
\put(10,10){\line(1,0){270}}
\put(220,40){\circle*{10}}
\put(220,10){\line(0,1){30}}
\put(7,-5){$1$}
\put(37,-5){$2$}
\put(67,-5){$3$}
\put(97,-5){$4$}
\put(127,-5){$5$}
\put(157,-5){$6$}
\put(187,-5){$7$}
\put(217,-5){$8$}
\put(247,-5){$9$}
\put(277,-5){$10$}
\put(225,36){$11$}
\end{picture}
\caption{ 
\small \textit{Dynkin diagram of $E_{11}$ with labelling of nodes used in the text.}}\label{fig:e11dynk}
\end{figure}
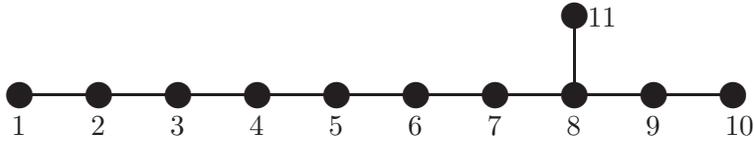

In order to construct a theory with $E_{11}$ symmetry one has to consider also an extended (infinite-dimensional) space-time as well as a local symmetry that is associated with a maximal subgroup of $E_{11}$ that we will call $K(E_{11})$ and that plays the role of a generalised $R$-symmetry group.\footnote{In the literature one often finds the notation $I_C(E_{11})$ since it is defined as the fixed point set of a Cartan involution. In order to obtain Lorentz symmetry $SO(1,10)\subset K(E_{11})$ one has to also allow for multiple time signatures~\cite{Englert:2003py,Keurentjes:2004bv}.} The infinite-dimensional space-time is associated with an infinite-dimensional lowest weight representation of $\mathfrak{e}_{11}$ that is called the $\ell_1$ representation in the literature~\cite{West:2003fc,Kleinschmidt:2003jf}, in accordance with the labelling of the nodes in the $\mathfrak{e}_{11}$ Dynkin diagram shown in Figure~\ref{fig:e11dynk}. Thus, the Dynkin labels that we associate with the lowest weight representation $\ell_1$ are $(1,0,\ldots,0)$ with $1$ at the first node, and $0$ at all other nodes. Decomposed under the $\mathfrak{gl}(11)\subset \mathfrak{e}_{11}$ subalgebra the $\ell_1$ representation comprises standard translation generators as well as generators that are associated with the two-form and five-form central charges of the $D=11$ supersymmetry algebra~\cite{West:2003fc}.\footnote{A coordinate $y_{mn}$ for the membrane central charge was already discussed in~\cite{Duff:1989tf,Duff:1990hn}.} In the $E_{11}$ framework there is a coordinate $z^M$ for every basis element $P_M$ of the $\ell_1$ representation and all fields depend on all these coordinates. 
A set of first-order equations of motion and a set of gauge transformations have been proposed in~\cite{West:2011mm,Tumanov:2016abm,West:2014eza} to describe an $E_{11}$ invariant extension of eleven-dimensional supergravity. This far-reaching proposal has a number of points related to the dynamics and gauge invariance that deserve further study. In this paper, we investigate these points and we make a proposal for an extended framework which may overcome some difficulties that we encounter in the original scheme.

More precisely, the non-linear realisation of $E_{11}$ on a space-time based on the $\ell_1$ representation leads to objects that transform in the tensor product of the coset representation of $K(E_{11})$ and the $\ell_1$ representation (viewed as a $K(E_{11})$ representation). A construction of dynamics that respects the $E_{11}$ symmetry then could be based on requiring that the projection of the general Maurer--Cartan coset velocity to certain invariant subspaces of this tensor product has to vanish. The equations obtained in this way will be a set of $K(E_{11})$ covariant first order equations that are similar to the (twisted) duality equations introduced in~\cite{Cremmer:1998px}. Since the decomposition of the tensor product of $\ell_1$ and the coset representation under $K(E_{11})$ is not known, the construction of such subspaces can only be probed in a pedestrian way in a level decomposition, starting for example from known duality equations such as the one between the four-form field strength and its dual seven-form in $D=11$ supergravity. The multiplet should then also involve first order equations for gravity. This is the approach followed in~\cite{West:2011mm}. The level decomposition does not allow, however, to prove the existence of a suitable $K(\mf{e}_{11})$ invariant subspace, and one will eventually need to introduce more sophisticated methods to define the theory. Note that the construction does not assume these first order equations to be invariant under generalised gauge transformations. In fact, it is expected from the point of view of unfolded field equations of higher spin gauge fields (starting from gravity) that these first order equations are not gauge invariant~\cite{Riccioni:2006az,Riccioni:2007hm,Boulanger:2012df,Boulanger:2015mka}.

First order duality equations imply second order field equations by integrability. Given the $K(\mf{e}_{11})$ multiplet of first order duality equations one can in principle construct a $K(\mf{e}_{11})$ multiplet of second order field equations in this way. The construction of a $K(\mf{e}_{11})$ multiplet of second order equations has been initiated in~\cite{Tumanov:2016abm} and continued to higher derivative orders in~\cite{Tumanov:2016dxc}. There are two important aspects to this construction that have not been addressed in detail in the literature. First, one forms a compatible system of equations, in the sense that the $K(\mf{e}_{11})$-multiplet of second order equations is automatically solved by the solutions to the $K(\mf{e}_{11})$-multiplet of first order equations by integrability. This requires in particular the first and second order equations to transform consistently with respect to $K(\mf{e}_{11})$. The second aspect concerns gauge invariance of the second order field equations. The dynamics must be gauge invariant and so one may hope that these second order field equations are invariant under the generalised gauge transformations acting on the fields of the theory in much the same way that the Einstein equation and matter equations are gauge invariant. However, it was explained in \cite{Tumanov:2016dxc} that the order of the differential equations that can possibly be gauge invariant increases linearly with the $\mf{gl}(11)$ level of the associated gauge fields, more precisely the number of columns of the associated Young tableau. These gauge invariant equations of high differential order can be integrated to lower order differential equations at the price of introducing undetermined total derivatives. It is proposed in \cite{Tumanov:2016dxc} that these ambiguous total derivatives can be interpreted as certain (yet to be determined) gauge transformations of the theory. As $E_{11}$ contains fields with an arbitrarily high number of columns, seeking a full set of gauge and $K(\mf{e}_{11})$ invariant differential equations of finite order in derivatives, one has to introduce additional fields and this is the approach we will pursue in this article.

Independently of this additional difficulty, we argue in this paper that, as is visible already at low levels, any kind of integrability condition or gauge invariance can only be realised upon imposition of a \textit{section constraint}. This section constraint is of the type that has also featured prominently in recent efforts devoted to defining exceptional field theory for finite-dimensional symmetry groups $E_d$ with
$d\leq 8$~\cite{Hohm:2013vpa,Hohm:2013uia,Hohm:2014fxa} using also earlier ideas on exceptional generalised geometry~\cite{Hull:2007zu,Coimbra:2011nw,Coimbra:2011ky,Berman:2012vc,Godazgar:2013dma,Godazgar:2014sla} and double field theory~\cite{Siegel:1993xq,Hull:2009mi,Hohm:2010jy,Bergshoeff:2016ncb,Bergshoeff:2016gub}. In the context of $E_{11}$,
the section constraint has been discussed in relation to  generalised BPS conditions in~\cite{West:2012qm}, but it has been also
argued for example in~\cite{Tumanov:2015iea} that the section constraint
is not necessary for the consistency of the full non-linear realisation of $E_{11}$. 

In exceptional field theory all fields depend on an extended space-time that is determined by the finite-dimensional analogue of the $\ell_1$ representation mentioned above. However, consistency of the gauge algebra and the theory requires that all fields in the theory satisfy the (strong) section constraint, which effectively limits the dependence to that on the coordinates of ordinary space-time,
by requiring that certain combinations of two derivatives vanish on any field, or on any product of fields (where the derivatives act separately on one field
each).
In group theoretic terms, the section constraint 
says that the product of 
two derivatives $\partial_M\otimes \partial_N$
has to vanish when projected to a certain subrepresentation of
the tensor product $\ell_1 \otimes \ell_1$. Since most of our analysis is at the linear order in the fields, we will only
encounter the weak version of the section constraint here, where both derivatives act on the same field, 
and thus only the symmetric part of the tensor product
is relevant. The section constraint then relies on the decomposition
\begin{align}
\label{eq:SC}
(\ell_1\otimes \ell_1)_{\text{sym}} = (2\ell_{1} )\oplus \left[\ell_{10} \oplus\cdots\right]\ ,
\end{align}
where $\ell_{10}$ denotes the $\mf{e}_{11}$ representation with Dynkin labels
$(0,0,\ldots 0,1,0)$,
and $(2\ell_1)$ denotes the representation with Dynkin labels $(2,0,\ldots,0)$. 
The part projected out by the section constraint is the 
complement of the 
$(2\ell_1)$ representation 
that is shown in square brackets.
In the analogous discussion for the finite-dimensional Lie algebras $\mf{e}_d$ with $d\leq 7$, the analogue of $\ell_{10}$ (\ie~$\ell_{d-1}$) is in fact the only other irreducible representation, besides $(2\ell_1)$, in the symmetric part of the tensor product
$\ell_1\otimes
\ell_1$.\footnote{In the last finite-dimensional case $d=8$ the symmetric product contains in addition an $\mathfrak{e}_8$ singlet.} In these cases, one could therefore alternatively write $(\partial_M\otimes\partial_N)_{\text{sym}}|_{\ell_{d-1}}=0$. A discussion of section constraints for arbitrary groups was initiated in~\cite{StricklandConstable:2013xta}.

In this article, we will present a new scheme that is based on an extension of $\mathfrak{e}_{11}$
to a Lie superalgebra which is the $d=11$ analogue of the tensor hierarchy algebras extending $\mathfrak{e}_d$ for $d\leq 8$ \cite{Palmkvist:2013vya,Greitz:2013pua,Howe:2015hpa}. This tensor hierarchy algebra provides a framework for constructing gauge invariant objects by furnishing a differential complex of functions satisfying the section constraint. The tensor hierarchy algebra also provides new generators in addition to the ones of $\mf{e}_{11}$ and the associated fields allow a consistent description of the dualisation of linearised gravity~\cite[Sec.~4]{West:2002jj}.\footnote{There is no obvious relation between our new fields and the section constrained forms that appear in exceptional field theory~\cite{Hohm:2014fxa} and that are not part of $E_{11}$ either. The fields of~\cite{Hohm:2014fxa} are relevant for the gauging of the trombone symmetry and the field strengths defined in the present article do not accommodate these gaugings.} We will also explain how this algebraic structure could provide a (linearised) two-derivative Lagrangian whose equations of motion, together with a duality relation, reduce to the standard $D=11$ supergravity field equations upon choosing the standard $D=11$ solution of the section constraint that only retains the eleven-dimensional coordinates. 

The tensor hierarchy algebra has two features that we find particularly remarkable. The first is that it extends in a controlled way the adjoint representation of $\mathfrak{e}_{11}$. The resulting representation contains the adjoint of $\mathfrak{e}_{11}$ as a
subrepresentation but is not fully reducible.
In particular, the tensor hierarchy algebra introduces new generators starting from $\mf{gl}(11)$ level three, the first of which has nine antisymmetric indices. It combines with the irreducible $(8,1)$ hook structure of the $\mathfrak{e}_{11}$ dual graviton generator to produce the correct dual gravity equation with the correct gauge transformations. Understanding this has been a long-standing puzzle. This point is explained in more detail in Sections~\ref{sec:THA} and \ref{sec:THA2}. 

The second remarkable feature of the tensor hierarchy algebra is that it includes an $E_{11}$ module that allows to define natural field strengths in the theory. This module is equipped with an invariant symplectic form, that descends from a non-degenerate bilinear form with $\mathbb{Z}_2$-graded symmetry on the whole tensor hierarchy algebra. The symplectic form together with an appropriate $K(\mf{e}_{11})$ invariant bilinear form on the field strength representation can be used to write down a first order duality equation. This equation is not gauge invariant (in generalised space-time) but corresponds exactly to the duality equation of $D=11$ supergravity. However, it is compatible with the gauge-invariant second order field equations that we also construct.

As another new result we present the decomposition of all equations in a language adapted to type IIB supergravity. This is relevant since $E_{11}$ is known to relate to type IIB supergravity as well~\cite{Schnakenburg:2001he} and the section constraint~\eqref{eq:SC} has type IIB as another maximal vector space solution~\cite{Blair:2013gqa,Baguet:2015xha,Bossard:2015foa}.

The structure of this article is as follows. In Section~\ref{sec:NLR}, we review the construction of the non-linear realisation of $E_{11}$ and identify the building blocks for constructing field equations respecting $E_{11}$ symmetry. In Section~\ref{sec:West1}, we discuss potential paths to constructing first order field equations and identify a particular candidate multiplet of first order duality equations. In Section~\ref{sec:West2}, we investigate second order field equations that can be derived from the candidate multiplet of first order duality equations and study their consistency with $K(E_{11})$. Moreover, we study linearised gauge transformations of the second order equations and find that their gauge invariance requires as a novel feature the section constraint. Our results in Sections~\ref{sec:NLR} and~\ref{sec:West1} extend the analysis in~\cite{West:2011mm,Tumanov:2016abm} by including higher level fields and by noticing the necessity of working modulo a section constraint. 

In Section~\ref{sec:THA}, we introduce our new scheme based on the tensor hierarchy algebra, providing a construction of an $E_{11}$ multiplet of gauge invariant field strengths (modulo the section constraint). We also introduce a Lagrangian based on this construction in Section~\ref{sec:THA2} and show that its Euler--Lagrange second order field equations are gauge invariant and agree with those of $D=11$ supergravity. Furthermore we discuss the existence of a natural set of first order duality equations compatible with the field equations. We also connect our construction to non-geometric fluxes and the unfolding construction. In Section~\ref{sec:typeIIB}, we rediscuss our analysis of the preceding sections in a language where everything is written in terms of type IIB variables rather than $D=11$. This will bring out more clearly the difference between our scheme based on the tensor hierarchy algebra and the original $E_{11}$ formulation. In Section~\ref{sec:NL}, we offer some comments on non-linear extensions of our theory. Section~\ref{sec:concl} contains some concluding comments. In two appendices we collect more technical details on some of the arguments and calculations used in the body of this article.

%%%%%%%%%%%%%%%%%%%%%%%%%%%%%%%%%%%%%%%%%%%%%%%%%%%%%%%%%%%%%%%%%%%%%%%%%%%%%%%%%%%%%%%%
\section{Non-linear realisation of \texorpdfstring{$E_{11}$}{E11} and \texorpdfstring{$D=11$}{D=11} supergravity}
\label{sec:NLR}
%%%%%%%%%%%%%%%%%%%%%%%%%%%%%%%%%%%%%%%%%%%%%%%%%%%%%%%%%%%%%%%%%%%%%%%%%%%%%%%%%%%%%%%%

After reviewing first the non-linear realisation and the $\mf{gl}(11)$ level decomposition of $\mf{e}_{11}$, we discuss the construction of dynamics associated with it following the $E_{11}$ proposal~\cite{West:2001as,Tumanov:2016abm}. In most of the paper we will be dealing with Lie algebras that we write in fraktur font. For Kac--Moody Lie algebras like $\mf{e}_{11}$ the definition of the corresponding groups is more subtle than just taking the exponential map due to the existence of imaginary roots. One can define an associated group by considering only the real roots and the associated one-parameter subgroups. The Kac--Moody group is generated from these one-parameter groups, see~\cite{KP,Kumar,DeMedts} for detailed discussions.

%%%%%%%%%%%%%%%%%%%%%%%%%%%%%%%%%%%%%%
\subsection{Non-linear realisation}
\label{NLreal}
%%%%%%%%%%%%%%%%%%%%%%%%%%%%%%%%%%%%%%

The fields of the theory parametrise the coset $E_{11}/K(E_{11})$, and are functions on the $E_{11}$ module  $\ell_1$. To define the action of  $E_{11}$ on the module  $\ell_1$, it is convenient to define the semidirect sum $\mathfrak{e}_{11}\oplus \ell_1$.  We introduce the following abstract notation for the generators of the various representations. The generators of the adjoint of $\mathfrak{e}_{11}$
are called $t^\alpha$ with commutation relations %
\begin{align}
[t^\alpha, t^\beta] = C^{\alpha\beta}{}_\gamma t^\gamma\ .
\end{align}
The generators of the $\ell_1$ representation are called $P_M$. They transform in a representation of $\mf{e}_{11}$ according to{\footnote{In Sections~\ref{sec:THA} and~\ref{sec:THA2}, we shall instead use the indices $\alpha_0$ and $M_0$ for the adjoint and $\ell_1$ representations of $\mf{e}_{11}$ in order to distinguish them from additional representations that arise in the tensor hierarchy algebra. No confusion should arise, given the context in which the formulas are given.}
\begin{align}\label{FundE11}
\lb t^\alpha, P_M \rb = -D^{\alpha N}{}_M P_N
\end{align}
and are abelian, $\lb P_M,P_N\rb=0$. 

One parametrises an element $z$ of the module $\ell_1$ as
\be z = z^M P_M \ , \ee
which parametrises an \textit{a priori} infinite-dimensional extended space-time. $g_0 \in E_{11}$ acts linearly on these coordinates through the action \eqref{FundE11}
\be z\rightarrow g_0 z g_0^{\; -1} \ . \ee
The $E_{11}$ group element $g(z)$ depends on these coordinates. On $g(z)$ we define the action of global $E_{11}$ and local $K(E_{11})$ as
\begin{align}
\label{eq:trms}
g(z) \to g_0 g(g_0 z g_0^{\; -1}) k(z) \ ,  \end{align}
for $k(z)\in K(E_{11})$. Here `local' means that $k(z)$ depends on the extended space-time. In practice one represents the coset $E_{11}/K(E_{11})$ through a  representative $g(z)$ satisfying a specific gauge condition (which is possible almost everywhere). Then $k(z)$ becomes an induced compensating transformation function of $g_0$ and $g(g_0 z g_0^{\; -1})$.

The first building block for the dynamics comes from the Maurer--Cartan form 
\begin{align}
{\cal V}(z)\equiv g(z)^{-1} d g(z) \ ,
\end{align}
where the differential
\begin{align}
d= dz^M \frac{\partial}{\partial z^M}
\end{align}
corresponds to taking derivatives with respect to all coordinates $z^M$ of the $\ell_1$ module. 

As a form it is valued in the adjoint of $\mf{e}_{11}$ and transforms as
\begin{align}
{\cal V}(z) \to k(z)^{-1} {\cal V}(g_0 z g_0^{\; -1}) k(z) + k(z)^{-1} dk(z)
\end{align} 
under~\eqref{eq:trms}. The global $E_{11}$ transformation only acts on the argument of ${\cal V}$. The second inhomogeneous term on the right is a connection term valued in the Lie algebra $K(\mf{e}_{11})$ of $K(E_{11})$. Under this subalgebra, $\mf{e}_{11}$ decomposes %generally
as
\begin{align}\label{CartanInvo} 
\mf{e}_{11} = \mf{p} \oplus K(\mf{e}_{11})\ ,
\end{align}
where $\mf{p}$ is 
a $K(\mf{e}_{11})$-module, which we shall refer to as the coset representation. It is not known whether it is irreducible or not, even in the affine case.

If one splits the Maurer--Cartan one-form ${\cal V}$ according to the decomposition \eqref{CartanInvo} as
\begin{align}
\label{eq:MCdec}
{\cal V}(z) = {\cal P}(z) + {\cal K}(z)\ ,
\end{align}
then the `coset component' ${\cal P}(z)$ transforms as a linear $K(E_{11})$ representation,
\be {\cal P}(z) \to k(z)^{-1} {\cal P}(g_0 z g_0^{\; -1})   k(z)\ ,  \ee
and the `connection part' ${\cal K}$ as 
\be {\cal K}(z)\to k(z)^{-1} {\cal K}(g_0 z g_0^{\; -1})  k(z) + k(z)^{-1} dk(z)\ .  \ee
It is convenient to define the vielbein basis\footnote{Note that unlike the original papers~\cite{West:2003fc,Kleinschmidt:2003jf,West:2011mm}, we do not include a factor $e^{z^M P_M}$ in the group element entering the non-linear realisation. The only purpose that it serves there is to obtain the vielbein $E(z)_M{}^A$ from the non-linear realisation. Here, we obtain this simply as the representative of the $E_{11}$ group element $g(z)$ in the $\ell_1$ representation.} 
\be g(z)^{-1} dz g(z) = E(z)^A P_A  = E(z)_M{}^A dz^M P_A \ , \ee
where the `vielbein' $E(z)_M{}^A$ is the matrix representation of the coset representative $g(z)$ written in the $\ell_1$ representation where $M$ is a `curved index' transforming under $E_{11}$ and $A$ is a flat index transforming under local $K(E_{11})$. When one expands out the one-form ${\cal P}$ in this basis, one obtains
\begin{align}
\label{eq:CMcos}
{\cal P}(z) = {\cal P}_M(z) dz^M = {\cal P}_A(z) E(z)^A \ .
\end{align}
The remaining tangent space components ${\cal P}_A(z)$ then transform under $K(E_{11})$ both on the $A$ index (in the $\ell_1$ representation branched to $K(E_{11})$) and in the coset representation of $K(E_{11})$. The ${\cal P}_A$ are the basic dynamical variables of the non-linear realisation of $E_{11}$ with the group element $g(z)$ depending on variables $z$ in the $\ell_1$ representation.

%%%%%%%%%%%%%%%%%%%%%%%%%%%%%%%%%%%%%%%%%%%%%%%%%%%%%%%%%%%%%%%%%%%%%%%%%%%%%%%%%%
\subsection{\texorpdfstring{$GL(11)$}{GL(11)} level decomposition of \texorpdfstring{$E_{11}$}{E11} and its \texorpdfstring{$\ell_1$}{l1} representation}
\label{sec:GL11E11}
%%%%%%%%%%%%%%%%%%%%%%%%%%%%%%%%%%%%%%%%%%%%%%%%%%%%%%%%%%%%%%%%%%%%%%%%%%%%%%%%%%

\renewcommand{\arraystretch}{1.5}
\begin{table}[t!]
\centering
\begin{tabular}{|c|c|c|c|}
\hline
Level $\ell=q$ & $\mf{sl}(11)$ representation & {Generator}&Potential  \\
\hline
$0$ & $\begin{matrix}(1,0,0,0,0,0,0,0,0,1) \\ (0,0,0,0,0,0,0,0,0,0)\end{matrix}$ & $K^m{}_n$ & $h_m{}^n$\\
\hline
1 & $(0,0,0,0,0,0,0,1,0,0)$ & $E^{n_1n_2n_3}$ & $A_{n_1n_2n_3}$\\
\hline
2 & $(0,0,0,0,1,0,0,0,0,0)$ & $E^{n_1\cdots n_6}$ & $A_{n_1\cdots n_6}$\\
\hline
3 & $(0,0,1,0,0,0,0,0,0,1)$ & $E^{n_1\cdots n_8,m}$ & $h_{n_1\cdots n_8,m}$\\
\hline
4 & 
$\begin{matrix}
(0,1,0,0,0,0,0,1,0,0)\\
(1,0,0,0,0,0,0,0,0,2)\\
(0,0,0,0,0,0,0,0,0,1)
\end{matrix}$
&
$\begin{matrix}
E^{n_1\cdots n_9,p_1p_2p_3}\\
E^{n_1\cdots n_{10},p,q}\\
E^{n_1\cdots n_{11},m}
\end{matrix}$
&
$\begin{matrix}
A_{n_1\cdots n_9,p_1p_2p_3}\\
B_{n_1\cdots n_{10},p,q}\\
C_{n_1\cdots n_{11},m}
\end{matrix}$\\
\hline
5
& 
$\begin{matrix}
(0,1,0,0,1,0,0,0,0,0)\\
(1,0,0,0,0,0,1,0,0,1)\\
(0,0,0,0,0,0,0,1,0,1)\\
(0,0,0,0,0,0,1,0,0,0)
\end{matrix}$
&
$\begin{matrix}
E^{n_1\cdots n_9,p_1\cdots p_6}\\
E^{n_1\cdots n_{10},p_1\cdots p_4,q}\\
E^{n_1\cdots n_{11},p_1p_2p_3,q}\\
E^{n_1\cdots n_{11},p_1\cdots p_4}
\end{matrix}$
&
$\begin{matrix}
A_{n_1\cdots n_9,p_1\cdots p_6}\\
B_{n_1\cdots n_{10},p_1\cdots p_4,q}\\
C_{n_1\cdots n_{11},p_1p_2p_3,q}\\
C_{n_1\cdots n_{11},p_1\cdots p_4}
\end{matrix}$\\
\hline
\end{tabular}
\caption{\label{tab:e11adj} \small\textit{Level decomposition of $\mf{e}_{11}$ under its $\mf{gl}(11)$ 
subalgebra obtained by deleting node $11$ from the Dynkin diagram in Figure~\ref{fig:e11dynk}, up to level $\ell=5$. The level $\ell$ is the eigenvalue of the generator $\frac13K^m{}_m$. The degree $q$ is defined in~\eqref{eq:qdegree} and for the adjoint of $\mf{e}_{11}$ equals the level $\ell$.}}
\end{table}

We will require a more explicit parametrisation of $\mf{e}_{11}$ and its $\ell_1$ representation and use a decomposition into $\mf{gl}(11)$ representations for this.
As is visible from the Dynkin diagram in Figure~\ref{fig:e11dynk},
the Lie algebra $\mf{e}_{11}$ contains a $\mf{gl}(11)$ subalgebra,
since the Dynkin diagram of $\mf{sl}(11)$ is obtained by deleting node $11$,
and the Cartan generator associated to the deleted node extends $\mf{sl}(11)$ to $\mf{gl}(11)$.
The generators $K^m{}_n$ of this subalgebra satisfy the commutation relations
\begin{align}
\lb K^m{}_n , K^p{}_q \rb = \delta^p_n K^m{}_q -\delta^m_q K^p{}_n
\end{align}
with $\mf{gl}(11)$ tensor indices $m,n,\ldots = 0,1,\ldots, 10$.
Any representation of 
$\mf{e}_{11}$ can then be decomposed into representations of $\mf{gl}(11)$.
In the cases we consider here, these are finite-dimensional representations that can be specified by $\mf{sl}(11)$ Dynkin labels together with a level $\ell$, which is the eigenvalue of $\tfrac13 K$, where $K=K^m{}_m$ is the trace of the $\mf{gl}(11)$ generators. We use the convention that is common in the context of hyperbolic and Lorentzian Kac--Moody algebras~\cite{Damour:2002cu}, namely to use as Dynkin labels the negative of the lowest weight. We give more details on our conventions for the $\mf{gl}(11)$ representations and the translation to tensors in Appendix \ref{app:conv}.

Table~\ref{tab:e11adj} lists the result of the $\mf{gl}(11)$ level decomposition for the adjoint
of $\mf{e}_{11}$ at levels $0\leq \ell \leq 5$~\cite{West:2002jj}. The generators  $E^{n_1n_2n_3}$ and $E^{n_1\ldots n_6}$ are
completely antisymmetric, while the level $\ell=3$ generator $E^{n_1\ldots n_8,m}$ transforms in an $(8,1)$ hook tableau\footnote{Projectors on tensors with hook symmetry are discussed more generally in Appendix~\ref{app:conv}, see~\eqref{hookproj}.} of $\mf{gl}(11)$:
\begin{align}
\label{eq:l3hook}
E^{n_1 \cdots n_8,m}=E^{[n_1 \cdots n_8],m}\ ,
\quad E^{[n_1 \cdots n_8,m]}=0\ .
\end{align}
We will always use the notation that comma-separated sets of indices belong to an irreducible tensor whereas a semi-colon denotes a reducible tensor. 
Conjugate to the positive level generators one has negative level generators down to level $\ell\geq -3$ consisting of
\begin{align}
F_{n_1\cdots n_8,m},\  F_{n_1\cdots n_6},\  F_{n_1n_2n_3}
\end{align}
with analogous symmetry properties. Together they constitute all $t^\alpha$ of the adjoint of $E_{11}$ for $|\ell|\leq 3$. Their complete commutations relations are given in Appendix~\ref{app:conv}. We note that our conventions for the commutators differ slightly from the ones used in~\cite{Tumanov:2016abm}. As an example, we have
\begin{align}
\lb E^{n_1n_2n_3}, E^{n_4n_5n_6}\rb = E^{n_1\cdots n_6}\ .
\end{align}
This is the reason for some differences in coefficients of our expressions below compared to the literature.

\begin{table}[t!]
\centering
\begin{tabular}{|c|c|c|c|c|c|}
\hline
$\ell$ & $q=\ell-\tfrac32$ & $\mf{sl}(11)$ representation & Generator & Coordinate& Parameter\\
\hline
$\tfrac32$ & $0$
& $(1,0,0,0,0,0,0,0,0,0)$ & $P_m$ & $x^m$& $\xi^m$\\
\hline
$\tfrac52$ & $1$
& $(0,0,0,0,0,0,0,0,1,0)$ & $Z^{mn}$ & $y_{mn}$ &$\lambda_{mn}$\\  
\hline
$\tfrac72$ & $2$
& $(0,0,0,0,0,1,0,0,0,0)$ & $Z^{n_1\cdots n_5}$ & $y_{n_1\cdots n_5}$&
$\lambda_{n_1\cdots n_5}$\\  
\hline
$\tfrac92$ & $3$
& $\begin{matrix} (0,0,0,1,0,0,0,0,0,1)\\ (0,0,1,0,0,0,0,0,0,0)\end{matrix}$
&  $\begin{matrix} P^{n_1\cdots n_7,m}\\P^{n_1\cdots n_8}\end{matrix}$
& $\begin{matrix}x_{n_1\cdots n_7,m}\\x_{n_1\cdots n_8}\end{matrix}$ & $\begin{matrix}\xi_{n_1\cdots n_7,m}\\
\lambda_{n_1\cdots n_8}\end{matrix}$\\
\hline
$\tfrac{11}2$ & $4$
& $\begin{matrix} (0,0,0,0,0,0,0,0,0,0)\\ (1,0,0,0,0,0,0,0,0,1)\\(1,0,0,0,0,0,0,0,0,1)\\(0,1,0,0,0,0,0,0,1,0) \\ (0,1,0,0,0,0,0,0,0,2)\\ (0,0,1,0,0,0,0,1,0,0)\end{matrix}$ 
&$\begin{matrix}P^{n_1\ldots n_{11}} \\ P^{n_1\ldots n_{10},m}\\\tilde{P}^{n_1\ldots n_{10},m}\\P^{n_1\ldots n_{9},m_1m_2}\\P^{n_1\ldots n_{9},m,p}\\P^{n_1\ldots n_{8},m_1m_2m_3}
\end{matrix}$
& $\begin{matrix}y_{n_1\ldots n_{11}} \\ y_{n_1\ldots n_{10},m}\\\tilde{y}_{n_1\ldots n_{10},m}\\y_{n_1\ldots n_{9},m_1m_2}\\y_{n_1\ldots n_{9},m,p}\\y_{n_1\ldots n_{8},m_1m_2m_3}
\end{matrix}$
& $\begin{matrix}\lambda_{n_1\ldots n_{11}} \\ \lambda_{n_1\ldots n_{10},m}\\\tilde{\lambda}_{n_1\ldots n_{10},m}\\\lambda_{n_1\ldots n_{9},m_1m_2}\\\lambda_{n_1\ldots n_{9},m,p}\\\lambda_{n_1\ldots n_{8},m_1m_2m_3}
\end{matrix}$\\\hline
\end{tabular}
\caption{\label{tab:e11l1} \small\textit{Level decomposition of the $\ell_1$ representation 
of $\mf{e}_{11}$ under $\mf{gl}(11)$, up to level $\ell=11/2$. This is a lowest weight representation and therefore the top entry is annihilated by all lowering generators. The names of the generators already anticipate their roles as translation and central charge type coordinates in a $D=11$ interpretation.  The degree $q$ in this case differs from the $\mf{gl}(11)$ level $\ell$ in the way indicated in the table and in \eqref{eq:qdegree}. }}
\end{table}

The coordinate representation $\ell_1$ is a lowest weight representation of $\mf{e}_{11}$ with the following low-lying generators in $\mathfrak{gl}(11)$ basis~\cite{West:2003fc}
\begin{align}
P_M = \left\{ P_m,\ Z^{mn},\ Z^{n_1\cdots n_5},\ P^{n_1\cdots n_8},\ P^{n_1\cdots n_7,m},\ldots \right\} \ .
\label{gl1}
\end{align}
The last two that are displayed here appear on the same $\mf{gl}(11)$ level
$\ell=9/2$ and this information is also summarised in Table~\ref{tab:e11l1}. The action of $\mf{e}_{11}$ on the representation in this decomposition is  given in Appendix~\ref{app:conv}. We stress that the objects in~\eqref{gl1} are not tensors of $\mf{gl}(11)$ but tensor \textit{densities}. Under the $\mf{gl}(11)$ generators $K^m{}_n$ one has for example that
\begin{align}
\label{eq:KPc}
\lb K^m{}_n , P_k \rb = -\delta^m_k P_n + \frac12 \delta^m_n P_k\ .
\end{align}
This is the reason that we introduce an additional degree $q$ that uses as an offset the $\mf{gl}(11)$ level $\ell$ of the lowest weight component in $\ell_1$ with respect to $\mf{gl}(11)$. This degree $q$ is not the eigenvalue of any semisimple operator of $\mf{e}_{11}$ but very useful to keep track of the number of steps one has taken from the lowest component. Thus we have
\begin{align}
\label{eq:qdegree}
\begin{array}{ll} q=\ell & \hspace{10mm}\textrm{for the adjoint of $\mf{e}_{11}$,}\\
q=\ell-\frac32 & \hspace{10mm}\textrm{for the $\ell_1$ representation of $\mf{e}_{11}$.}
\end{array}
\end{align}

Using this more explicit parametrisation of $\mf{e}_{11}$ and its $\ell_1$ representation in the $\mf{gl}(11)$ {\it maximal parabolic gauge}, we can write the group element $g(z)$ and its argument $z$ more precisely as
\begin{align}
\label{eq:GE}
g &= \cdots e^{\frac{1}{8!} h_{n_1\cdots n_8,m} E^{n_1\cdots n_8,m}}\,
e^{\frac{1}{6!} A_{n_1\cdots n_6} E^{n_1 \cdots n_6}}\,
e^{\frac{1}{3!} A_{n_1n_2n_3} E^{n_1n_2n_3}}\,
e^{\varphi_n{}^m K^n{}_m}\ \nn\\
&= 1 + \sum_{\alpha\textrm{ with } \ell\geq 0} A_\alpha t^\alpha+ \cdots\ ,\\ 
z &= x^m P_m+\frac{1}{2!}y_{mn}Z^{mn}+\frac{1}{5!}y_{n_1 \cdots 
n_5}  Z^{n_1\cdots n_5}+\frac1{7!}x_{n_1\cdots n_7,m} P^{n_1\cdots n_7,m} + \frac1{7!}x_{n_1\cdots n_8} P^{n_1\ldots n_8}+ \cdots \ .\nn
\end{align}
The local $K(E_{11})$ invariance has been used to take a coset representative solely in terms of non-negative levels.\footnote{This is not always possible but we restrict ourselves here to work on a patch of $E_{11}$ where it is. This difficulty is due to the non-compact involution defining $K(E_{11})$ with `Lorentz signature'. There is a second difficulty with the parametrisation above that is due to the fact that some of the generators are associated with imaginary roots and therefore not locally nilpotent such that the exponential map is not {\it a priori}  well-defined~\cite{Kleinschmidt:2014uwa}.} At this point $\varphi_n{}^m$ at level $\ell=0$ is not constrained, and so it is a general $(11\times 11)$-matrix. This means that we have not completely fixed the local $K(E_{11})$ invariance but are left with a local Lorentz invariance coming from $SO(1,10)\subset GL(11)$ at level $\ell=0$. This type of $K(E_{11})$ gauge is referred to as a maximal parabolic gauge. We have also used different letters for the coordinates and fields according to whether they are part of the gravity or of the matter sector of the theory. Note the prefactor $1/7!$ in front of $x_{n_1\cdots n_8} P^{n_1\cdots n_8}$,
which turns out to be more convenient than $1/8!$.\footnote{This is related to the fact that
$P^{n_1\cdots n_8}$ appears at the same level as $P^{n_1\cdots n_7,m}$, and one can combine them into $\frac{1}{7!} x_{n_1\dots n_7;m} P^{n_1\dots n_7;m}$.}

In the explicit parametrisation of fields~\eqref{eq:GE} one can then construct the ${\cal P}_A$ of equation~\eqref{eq:CMcos}. Working at the
linearised level one obtains the following components: 
\begin{align}
\label{eq:BB}
\partial_a &h_{bc}, &\, \partial_a &A_{b_1b_2b_3}, &\, \partial_a &A_{b_1\cdots b_6},\,& \partial_a &h_{b_1\cdots b_8,c},&\,& \ldots\nn\\
\partial^{a_1a_2} &h_{bc}\,,& \partial^{a_1a_2} &A_{b_1b_2b_3}, &\, \partial^{a_1a_2}& A_{b_1\cdots b_6},\, &\partial^{a_1a_2}& h_{b_1\cdots b_8,c},&\, &\ldots\nn\\
&\vdots&\,&\vdots&\,&\vdots&\,&\vdots&\,&\ddots
\end{align}
Latin indices from the beginning of the alphabet are tangent space indices. We note that the components in ${\cal P}_A$ depending on the vielbein fluctuation $\varphi_m{}^n$ only depend on the derivative of the metric fluctuation 
\be 
h_{ab} = \varphi_{ab} + \varphi_{ba} \ , \label{gg}
\ee
so we shall use the symmetric tensor $h_{ab}$ instead of the generic tensor $\varphi_{ab}$. Note that this does not mean that we have gauge fixed the local Lorentz invariance. Sometimes in the literature the complete Maurer--Cartan form ${\cal V}_A=\mathcal{P}_A + \mathcal{K}_A$ is used rather than only ${\cal P}_A$. In this case the full $\varphi_{ab}$ appears. For completeness we have checked that our computations lead automatically to the condition that only ${\cal P}_A$ is involved in the first order equation without assuming it to start with.

First order field equations will be constructed out of the objects~\eqref{eq:BB} and in order to maintain $E_{11}$ symmetry the resulting equations will have to form a $K(E_{11})$ multiplet since the induced $K(E_{11})$ action is all that remains when working with $\mathcal{P}_A$.  The action of $K(E_{11})$ on the various quantities above have been worked out for example in~\cite{West:2011mm} (see also~\cite{Damour:2007dt,Damour:2009ww}) and we give them here in our conventions. For defining the action of rigid $K(\mf{e}_{11})$ it suffices to give the action of the `level one' generator 
\be
\Lambda = 
\frac1{3!}\Lambda_{a_1a_2a_3} \left( E^{a_1a_2a_3} - F_{a_1a_2a_3}\right)\,
\ee
since we are working in a manifestly Lorentz covariant formalism and all other $K(\mf{e}_{11})$ generators can be obtained from this by multiple commutation. 

Using the commutators of Appendix~\ref{app:conv}, the coset potentials transform under the linearised action of this $K(\mf{e}_{11})$ generator as\footnote{These transformations correspond to the symmetric gauge for the potentials (in which $\varphi_{ab} = \frac12 h_{ab}$), but hold for the components of $\mathcal{P}_A$ in $\partial_M A_\alpha$ for any gauge, and in particular for the parabolic gauge we consider.}
\begin{align}
\label{eq:KE11pot}
\delta_\Lambda h_{ab} &= \Lambda^{c_1c_2}{}_{(a} A_{b)c_1c_2} - \frac{1}{9} \eta_{ab} \Lambda^{c_1c_2c_3} A_{c_1c_2c_3} \,,\nn\\
\delta_\Lambda A_{a_1a_2a_3} &= - 3\Lambda_{b[a_1a_2} h_{a_3]}{}^{b}+\frac{1}{6} \Lambda^{b_1b_2b_3} A_{a_1a_2a_3b_1b_2b_3}  \,,\nn\\
\delta_\Lambda A_{a_1a_2a_3a_4a_5a_6} &= 20 \Lambda_{[a_1a_2a_3} A_{a_4a_5a_6]} +\frac{1}{2} \Lambda^{b_1b_2c} h_{a_1a_2a_3a_4a_5a_6b_1b_2,c}  \,,\nn\\
\delta_\Lambda h_{a_1a_2a_3a_4a_5a_6a_7a_8,b} &= 56 \Lambda_{\lsharp a_1a_2a_3} A_{a_4a_5a_6a_7a_8,b \rsharp} + \cdots ,
\end{align}
These transformations were obtained in~\cite{Tumanov:2016abm} in a different normalisation of the fields and without the symmetric gauge choice~\eqref{gg}.
The angle brackets \raisebox{.4ex}{$\lsharp\,\rsharp$} denote projection on the $(8,1)$ hook representation (see Appendix~\ref{app:conv}).
The derivatives transform as
\begin{align}
\label{eq:KE11der}
\delta_\Lambda \partial_a &= \frac{1}{2} \Lambda_{ab_1b_2} \partial^{b_1b_2} \,,\nn\\
\delta_\Lambda \partial^{a_1a_2}  &= -\Lambda^{a_1a_2b}\partial_b + \frac{1}{6} \Lambda_{b_1b_2b_3} \partial^{a_1a_2b_1b_2b_3} \,,\nn\\
\delta_\Lambda \partial^{a_1a_2a_3a_4 a_5}  &= -10 \Lambda^{[a_1a_2a_3} \partial^{a_4a_5]} + \frac{1}{2} \Lambda_{b_1b_2c} \partial^{a_1a_2a_3a_4 a_5b_1b_2,c} + \frac{1}{6} \Lambda_{b_1b_2b_3} \partial^{a_1 a_2a_3a_4 a_5b_1b_2b_3}  \,,\nn\\
\delta_\Lambda \partial^{a_1 a_2a_3a_4a_5a_6 a_7,b}  &= -\frac{105}{8} \left(  \Lambda^{[ a_1a_2a_3} \partial^{a_4a_5a_6a_7]b} + \Lambda^{b[ a_1a_2} \partial^{a_3a_4a_5a_6a_7]} \right) +\cdots  \,,\nn\\
\delta_\Lambda \partial^{a_1a_2a_3a_4a_5a_6a_7a_8}  &= 7  \Lambda^{[ a_1a_2a_3} \partial^{a_4a_5a_6a_7a_8]} +\cdots .
\end{align}
In the last two equations the ellipses indicate terms involving derivatives of $\mf{gl}(11)$ level $\ell \le - \frac{11}{2}$.

%%%%%%%%%%%%%%%%%%%%%%%%%%%%%%%%%%%%%%%%%%%%%%%%%%%%%%%%%%%%%%%%%%%%%%%%%%%%%%
\subsection{Gauge transformations and \texorpdfstring{$E_{11}$}{E11}}
%%%%%%%%%%%%%%%%%%%%%%%%%%%%%%%%%%%%%%%%%%%%%%%%%%%%%%%%%%%%%%%%%%%%%%%%%%%%%%

The local gauge transformations of the above non-linear realisation of $E_{11}$ are just the local $K(E_{11})$ transformation in~\eqref{eq:trms}. In order to obtain (generalised) diffeomorphisms, one must introduce additional gauge transformations as was discussed in~\cite{West:2014eza}. For this one introduces gauge parameters $\Xi^M$  that transform in the $\ell_1$ representation. In the present basis, this means one has
\begin{align}
\Xi^M = \left\{\xi^m, \, \lambda_{mn}, \, \lambda_{n_1\cdots n_5}\,, \xi_{n_1\cdots n_7,m}\,, \lambda_{n_1\cdots n_8}\,,\ldots\right\}\,.
\end{align}
The reason for using different letters in the decomposition is that the $\xi$ are associated with the gravity sector (diffeomorphisms and dual diffeomorphisms) whereas the $\lambda$ are thought of as associated with the matter sector in this decomposition. One exception to this labelling occurs for the antisymmetric parameter
$\lambda_{n_1\cdots n_8}$ that is also associated with dual diffeomorphisms. The reason it is denoted with $\lambda$ rather than $\xi$ is to reduce the risk of confusion with the mixed symmetry parameter $\xi_{n_1\cdots n_7,m}$ with the same number of indices.

The \textit{linearised} gauge transformations for fields $A_\alpha$ parametrising the coset $E_{11}/K(E_{11})$ as in~\eqref{eq:GE} can be defined by using the fact that the adjoint appears in the tensor product of the translation representation $\ell_1$ and its dual \eqref{gaugeNL}:
\begin{align}
\label{eq:GT}
\delta_\Xi A_\alpha = \kappa_{\alpha\beta} D^{\beta M}{}_N \partial_M \Xi^N + b_\alpha\ .
\end{align}
Here $\kappa_{\alpha\beta}$ is the inverse of the symmetric invariant bilinear form on $\mf{e}_{11}$
(see Appendix \ref{app:conv}) and $D^{\beta M}{}_N$ are the structure constants in the $\ell_1$ representation. The fields $A_\alpha$ are the components of an element in the maximal parabolic subalgebra of $\mf{e}_{11}$ (see \eqref{eq:GE}) and the compensator $b_\alpha \in K(\mf{e}_{11})$ is defined such as to remove any component of negative level generated this way. For \textit{non-linear} gauge transformations one must also introduce an appropriate connection in a (gauge) covariant derivative $\nabla$ replacing the partial derivative above,
\begin{align}
\delta_\Xi g(z) = \bigl( \kappa_{\alpha\beta} D^{\beta N}{}_M \nabla_N \Xi^M(x) t^\alpha \bigr) g(z) + g(z) b(g,\nabla \Xi) \ , \label{gaugeNL} 
\end{align}
where we have written out the local $K(\mf{e}_{11})$ transformation $b(g,\nabla \Xi)$ that restores the gauge fixing.  The covariant derivative $\nabla_M$ is not {\it a priori} determined directly from a group theory construction~\cite{West:2014eza}. Its definition is an open problem that we shall not address in this paper since we shall almost always work at the linearised level.

The linearised gauge transformations~\eqref{eq:GT} for our fields are then found, using the commutators provided in  Appendix~\ref{app:conv}, to be
\begin{align}
\label{eq:GTM}
\delta_\Xi h_{ab} &=2\,\partial_{(a} \xi_{b)} -2 \partial_{(a}{}^c \lambda_{b)c} - \frac{2}{4!} \partial_{(a}{}^{c_1\cdots c_4} \lambda_{b)c_1\cdots c_4} - \frac{2}{6!} \partial_{(a}{}^{c_1\cdots c_6,d} \xi_{b)c_1\cdots c_6,d}  
\nn\\
&
\quad - \frac{2}{7!} \partial^{c_1\cdots c_7,}{}_{(a} \xi_{|c_1\cdots c_7|,b)} - \frac{16}{7!} \partial_{(a}{}^{c_1\cdots c_7} \lambda_{b)c_1\cdots c_7} 
\nn\\
&\quad + \frac{1}{3} \eta_{ab} \left( \partial^{c_1c_2} \lambda_{c_1c_2} + \frac{4}{5!} \partial^{c_1\cdots c_5} \lambda_{c_1\cdots c_5} + \frac{6}{7!} \partial^{c_1\cdots c_7,d} \xi_{c_1\cdots c_7,d} + \frac{6}{7!} \partial^{c_1\cdots c_8}
\lambda_{c_1\cdots c_8}\right) + \ldots \ ,
\nn\\
\delta_\Xi A_{a_1a_2a_3} &= 3 \,\partial_{[a_1} \lambda_{a_2a_3]} + \frac{1}{2} \partial^{b_1b_2} \lambda_{a_1a_2a_3b_1b_2} + \frac{1}{4!} \partial^{b_1\cdots b_5} \xi_{a_1a_2a_3b_1\cdots b_4,b_5}- \frac{1}{5!} \partial^{b_1\cdots b_5} \lambda_{a_1a_2a_3b_1\cdots b_5}
\nn\\
&\quad +3 \,\partial_{[a_1a_2} \xi_{a_3]} + \frac{1}{2} \partial_{a_1a_2a_3}{}^{b_1b_2} \lambda_{b_1b_2} + \frac{1}{4!} \partial_{a_1a_2a_3}{}^{b_1\cdots b_4,c} \lambda_{b_1\cdots b_4c} \nn\\
&\quad -  \frac{1}{5!} \partial_{a_1a_2a_3}{}^{b_1\cdots b_5} \lambda_{b_1\cdots b_5}  + \ldots\,,\nn\\
\delta_\Xi A_{a_1\cdots a_6} &= 6 \, \partial_{[a_1} \lambda_{a_2\cdots a_6]} - \partial^{bc} \xi_{a_1\cdots a_6b,c} +  \partial^{b_1b_2} \lambda_{a_1\cdots a_6b_1b_2} 
\nn\\
&\quad - 6 \,\partial_{[a_1\cdots a_5} \xi_{a_6]} - \partial_{a_1\cdots a_6}{}^{b,c} \lambda_{bc} +\partial_{a_1\cdots a_6}{}^{b_1b_2}   \lambda_{b_1b_2} + \ldots \ ,
\nn\\
\delta_\Xi h_{a_1\cdots a_8,b} &= 8 \,\partial_{[a_1} \xi_{a_2\cdots a_8],b}+\frac{8}{3}\left( \partial_{[a_1} \lambda_{a_2\cdots a_8]b}-\partial_{b} \lambda_{a_1\cdots a_8}\right) -8 \,\partial_{[a_1\cdots a_7|,b|} \xi_{a_8]}
\nn\\
&\quad -\frac{8}{3}\left( \partial_{a_1\cdots a_8} \xi_{b}- \partial_{b[a_1\cdots a_7} \xi_{a_8]}\right) + \ldots\ .
\end{align}
The ellipses indicate terms involving derivatives or gauge parameters of $\mf{gl}(11)$ level $|\ell| \ge \frac{11}{2}$.\footnote{We note that these expressions, as similar ones below for the tensor hierarchy algebra, are formally infinite sums and therefore not fully well-defined algebraically. A discussion of this point in the context of affine Kac--Moody algebras can be found in~\cite{ChariPressley}.}} These transformations extend the ones given in the original paper~\cite{West:2014eza}. Indices have been raised and lowered with the flat background metric. Alternatively, the coefficients in all the transformations above can be fixed by the requirement that they commute with the $K(E_{11})$ transformations. We will use these gauge transformations later to check gauge invariance of the field equations that we construct.

%%%%%%%%%%%%%%%%%%%%%%%%%%%%%%%%%%%%%%%%%%%%%%%%%%%%%%%%%%%%%%%%%%%%
\subsection{\texorpdfstring{$D=11$}{D=11} supergravity and its first order duality relations}
%%%%%%%%%%%%%%%%%%%%%%%%%%%%%%%%%%%%%%%%%%%%%%%%%%%%%%%%%%%%%%%%%%%%

We will consider $D=11$ supergravity~\cite{Cremmer:1978km} in conventions 
such that the bosonic second order field equations are  
given (in tangent space indices) by
\begin{subequations}
\begin{align}
\label{eq:11EE}
R_{ab}&= \frac1{12} F_{ac_1c_2c_3} F_b{}^{c_1c_2c_3} - \frac1{144} \eta_{ab} F_{c_1\ldots c_4} F^{c_1\ldots c_4}\ ,
\\
\label{eq:11FF}
D_c F^{ca_1a_2a_3} &= -\frac1{1152} \varepsilon^{a_1a_2a_3b_1\ldots b_8 } F_{b_1\ldots b_4} F_{b_5\ldots b_8}\ ,
\end{align}
\end{subequations}
where $D_a = e_a{}^m ( \partial_m + \omega_m )$ is the tangent frame covariant derivative with the torsion free spin connection $\omega_m$. The field strength is given by $F_{a_1a_2a_3a_4} = 4 D_{[a_1} A_{a_2a_3a_4]}$. The flat indices have ranges $a,b,\ldots=0,1,\ldots,10$, with $0$ indicating the time direction and $\eta_{ab}=(-+\cdots +)$ the flat Minkoswki metric. 

As is well-known, the non-linear matter equation of motion~\eqref{eq:11FF} can be recast in a first order form by pulling a covariant derivative out of the Chern--Simons contribution on the right-hand side, leading to 
\begin{align}
D_c\left( F^{ca_1a_2a_3} - \frac{1}{144} \varepsilon^{ca_1a_2a_3b_1\ldots b_7} A_{b_1b_2b_3} F_{b_4\ldots b_7}\right) =0\ ,
\end{align}
and the existence of a six-form potential $A_{a_1\ldots a_6}$ satisfying
\begin{align}
\label{eq:DA}
F^{a_1\ldots a_4} - \frac1{144} \varepsilon^{a_1\ldots a_4 b_1\ldots b_7}A_{b_1b_2b_3} F_{b_4\ldots b_7} =- \frac1{6!} \varepsilon^{a_1\ldots a_4 b_1\ldots b_7} \underbrace{D_{[b_1} A_{b_2\ldots b_7]}}_{=: \frac17 F_{b_1\ldots b_7}}\ .
\end{align}
By contrast, the non-linear Einstein equation is not amenable to a similar treatment~\cite{Bekaert:2002uh,Bergshoeff:2008vc}. However, once one linearises the theory one can obtain a dual graviton field and write the linearised Einstein equation in first order form~\cite{Curtright:1980yk,Nieto:1999pn,Hull:2000zn,West:2001as,Hull:2001iu,West:2002jj}; the matter contribution disappears in this approximation. We will perform the dualisation from the linearised (vacuum) equation of motion $R_{ab}-\frac12\eta_{ab}R=0$. Expanding the vielbein around flat space, $e_m{}^a = \delta_m^a + \varphi_m{}^a$, the Ricci tensor and scalar become at linear order
\begin{align}
R_{ab} = \partial_a \omega_{cb}{}^c - \partial_c\omega_{ab}{}^c \,,\quad  R= 2 \partial^d \omega_{cd}{}^c\,,\quad\textrm{where}\quad
\omega_{abc} = -\partial_{[b} \varphi_{c]a} -\partial_{[b} \varphi_{|a|c]}+\partial_a \varphi_{[bc]}  \ . 
\end{align}
Note that local Lorentz invariance has not been fixed here and the linearised vielbein $\varphi_{ab}$ contains an antisymmetric part. Introducing a mixed symmetry field $C_{a_1\ldots a_8;b}$ with
\begin{align}
\label{eq:DG1}
\omega_{a b_1b_2} - 2 \omega_{c [b_1}{}^c  \eta_{b_2]a} =- \frac1{8!} \varepsilon_{b_1b_2}{}^{c_1\ldots c_9} \partial_{c_1} C_{c_2\ldots c_9;a}\ ,
\end{align}
one finds that the integrability of this equation (taking $\partial^{b_1}$ on both sides) implies
\begin{align}
R_{ab}-\frac12\eta_{ab} R =0 \ , 
\end{align}
and therefore~\eqref{eq:DG1} is equivalent to the linearised Einstein equation.

It is important to note that the field $C_{a_1\ldots a_8;b}$ that one calls the dual graviton does not satisfy $C_{[a_1\ldots a_8;b]}=0$ from~\eqref{eq:DG1}. This is indicated by the notation with the semi-colon. Recall that we will always use the notation that a comma on a set of indices denotes an irreducible Young tableau as in~\eqref{eq:l3hook}. Indeed, taking the trace $\eta^{ab_2}$ of that equation leads to
\begin{align}
\omega_{ca}{}^c  =2\partial_{[c} \varphi_{a]}{}^c =
-\frac1{9!} \varepsilon_{a}{}^{c_1\ldots c_{10}}  \partial_{c_1} C_{c_2\ldots c_9;c_{10}}\,,
\end{align}
so that the vanishing of the completely antisymmetric part would mean that the spin connection has to be traceless, whereas it is not in general. Following \cite{West:2002jj,Boulanger:2008nd}, one defines the local Lorentz transformations at the linearised level as
\be 
\delta \varphi_{ab}  = \Lambda_{ab} \;, \qquad \delta C_{a_1\dots a_8;b} = - \frac{1}{2} \varepsilon_{a_1\dots a_8bc_1c_2} \Lambda^{c_1c_2} \ , 
\ee
such that one can fix the gauge by setting $C_{[a_1\ldots a_8;b]}= 0$, if one allows for an antisymmetric component of $\varphi_{ab}$ with the constraint $\partial_{[c} \varphi_{a]}{}^c=0$. Note that it is not possible, however, to use Lorentz invariance to set $\varphi_{[ab]}=0$ and $C_{[a_1\ldots a_8;b]}= 0$ at the same time \cite{Boulanger:2003vs}.

Alternatively, we can write linearised gravity in terms of the metric $g_{mn} = \eta_{mn} + h_{mn} $ (with $h_{mn}$ symmetric) by defining
\begin{align}  \label{Omega21sugra}
\Omega_{n_1n_2}{}^m \equiv 2 g^{mp} \partial_{[n_1} g_{n_2]p} \ ,
\end{align}
such that the linearised equations of motion are equivalently written in terms of the Ricci tensor $R_{ab} = \partial^c \Omega_{cab} - \partial_{b} \Omega_{ac}{}^c$ and the duality equation~\eqref{eq:DG1} takes the form 
\begin{align}
\label{eq:DG2}
\Omega_{n_1n_2}{}^{m} + 2 \delta^m_{[n_1} \Omega_{n_2]p}{}^p  = \frac2{8!}  \varepsilon_{n_1n_2}{}^{q_1\ldots q_9} \partial_{q_1} C_{q_2\ldots q_9;}{}^m\ .
\end{align}
The two definitions \eqref{eq:DG1} and  \eqref{eq:DG2} are identical for $\varphi_{ab} = \frac12 h_{ab}$. However, in this case there is no freedom to set the antisymmetric component of $C_{n_1\ldots n_8;m}$ to zero by a Lorentz transformation since $C_{n_1\ldots n_8;m}$ is inert at the linearised level. This second formulation in terms of $g_{mn}$ is closer to the $E_{11}$ formulation to be developed below.

In the following it will be convenient to decompose the dual graviton into a field $h_{n_1\ldots n_8,m}$  with vanishing antisymmetric component and a nine-form field $X_{n_1\ldots n_9}$ as follows\footnote{This is not a complete decomposition into Lorentz irreducible representations since $h_{n_1\dots n_8,m}$ still decomposes into a traceless and a trace component $h_{n_1\dots n_7m,}{}^m$.} 
\be
2 C_{n_1\ldots n_8;m} =  h_{n_1\ldots n_8,m} +  X_{n_1\ldots n_8 m}\,.
\ee
Then the duality equation~\eqref{eq:DG2} splits into its trace and its traceless component as
\begin{subequations}
\begin{align}
\Omega_{np}{}^p &= \frac1{9!} \varepsilon_{n}{}^{p_1\ldots p_{10}} \partial_{p_1} X_{p_2\ldots p_{10}}\; ,\\
\Omega_{n_1n_2}{}^m + \frac{1}{5} \delta^m_{[n_1} \Omega_{n_2]p}{}^p &= \frac1{8!}\varepsilon_{n_1n_2}{}^{p_1\ldots p_9} \eta^{mq} \bigl( \partial_{p_1} h_{p_2\ldots p_9,q} +\tfrac{1}{10} \partial_{n_1} X_{n_2\ldots n_9q} +\tfrac{1}{10} \partial_{q} X_{p_1\ldots p_9} \bigr) \; .
\end{align}
\end{subequations}
If one considers the gauge fixing for linearised diffeomorphisms 
\be \partial^{b} h_{ab} - \partial_a h_b{}^b = 0  \qquad \Rightarrow  \qquad 10 \partial_{[n_1} X_{n_2\dots n_{10}]} = 0 \ , \label{DiffGauge} 
\ee
$X_9$ is then pure gauge and there is an appropriate gauge for dual diffeomorphisms such that $X_9=0$.

Finally, we can also linearise the duality equation~\eqref{eq:DA} to obtain
\begin{align}
\label{eq:FDl}
F_{a_1\ldots a_7} = \frac1{4!} \varepsilon_{a_1\ldots a_7}{}^{b_1\ldots b_4} F_{b_1\ldots b_4}\,.
\end{align}
These first order duality equations are the ones we will now try to reproduce from a first order dynamical system based on $E_{11}$. The occurrence of the fields $h_{ab}$, $A_{a_1a_2a_3}$, $A_{a_1\ldots a_6}$ and $h_{a_1\ldots a_8,b}$ is not surprising from the perspective of $E_{11}$ in view of the low level generators of Table~\ref{tab:e11adj}. What is seemingly missing from $E_{11}$ is the component $X_{a_1\ldots a_9}$ as was already noted in~\cite{West:2001as,West:2002jj}. Although $X_9$ can a priori be set to zero in an appropriate gauge, we shall see, {\it e.g.} in Section~\ref{sec:trace}, that its presence is important for the $K(\mf{e}_{11})$ invariance of the first order equations.

%%%%%%%%%%%%%%%%%%%%%%%%%%%%%%%%%%%%%%%%%%%%%%%%%%%%%%%%%%%%%%%%%%%%%%%%%%%%%%%%%%%
\section{Dynamics for \texorpdfstring{$E_{11}$}{E11} and the section constraint}
\label{sec:West1}
%%%%%%%%%%%%%%%%%%%%%%%%%%%%%%%%%%%%%%%%%%%%%%%%%%%%%%%%%%%%%%%%%%%%%%%%%%%%%%%%%%%

In this section we investigate possible first order dynamics that respect $E_{11}$ symmetry. We begin with some general analysis that will lead to the conclusion that at present no general prescription exists that would yield unique dynamics. Then we probe a construction `by hand' that is built from the $D=11$ equations above. In doing so, we shall extend the results in~\cite{West:2011mm,Tumanov:2016abm} by including higher level derivatives and fields.  We shall then discuss in some detail the important shortcoming of the formalism in that it gives traceless Lorentz spin connection. We finally proceed with the two-derivative field equations, in which case we find that gauge invariance of the equations of motion require section constraints. 

%%%%%%%%%%%%%%%%%%%%%%%%%%%%%%%%%%%%%%%%%%%%%%%%%%%%%%%%
\subsection{First order dynamics: General remarks}
%%%%%%%%%%%%%%%%%%%%%%%%%%%%%%%%%%%%%%%%%%%%%%%%%%%%%%%%
\label{sec:GM}

Having established the Maurer--Cartan form as the starting point of the non-linear realisation, the next question to address is how to define $E_{11}$ invariant dynamics from it. We are aiming for a set of first order differential equations. Using as building blocks the components ${\cal P}_A$ of the Maurer--Cartan form~\eqref{eq:CMcos}, we have at our disposal $E_{11}$ invariant quantities that transform in the tensor product representation $\mf{p}\otimes \ell_1$ of $K(\mf{e}_{11})$ where $\mf{p}$ denotes the coset representation and $\ell_1$ is viewed as a representation of $K(E_{11})\subset E_{11}$ (associated with the $A$ index). If a $K(E_{11})$ gauge is fixed, then $E_{11}$ acts on $P_M$ by the induced compensating $K(E_{11})$ transformation in the coset representation of $K(E_{11})$. After conversion to tangent indices ${\cal P}_A=E_A{}^M {\cal P}_M$ therefore transforms in the tensor product of the coset representation of $K(E_{11})$ with the $\ell_1$ representation viewed as a $K(E_{11})$ representation. Given a decomposition 
\begin{align}
\label{eq:tpdec}
\mf{p} \otimes \ell_1 = \bigoplus_{i\in I} V_i 
\end{align}
of the tensor product into $K(E_{11})$ invariant and indecomposable subspaces $V_i$ (labelled by some index set $I$), setting 
\begin{align}
\label{eq:dyn}
{\cal P}_A |_{\bigoplus_{j\in J}V_j} = 0
\end{align}
for any subset $J\subset I$ would clearly constitute a set of $E_{11}$ invariant first order differential equations that potentially define some `dynamics'. Obviously, setting all of ${\cal P}_A$ equal to zero ($J=I$) is a too strong choice since it would trivialise the whole dynamics whereas the other extreme $J=\varnothing$ does not put any constraints on the dynamics. We also note that for $|I|>1$ the first-order dynamics of the non-linear realisation is not unique and the question remains how to pick the right set $J$ of equations.

For the case at hand, we are actually faced with the problem that no non-trivial decomposition of the type~\eqref{eq:tpdec} is known, where we stress that the decomposition can be into invariant subspaces and not necessarily irreducible representations of $K(\mf{e}_{11})$.\footnote{Incidentally, it is not even known whether $\mf{p}$ and $\ell_1$ themselves have invariant subspaces, not even in the affine case when $\mf{e}_{11}$ is replaced by the affine $\mf{e}_9$ and $\ell_1$ by the basic representation of $\mf{e}_9$.} This can be traced back to the fact that the Lie algebra $K(\mf{e}_{11})$ is \textit{not} a Kac--Moody algebra with a triangular decomposition into raising, lowering and Cartan generators, see~\cite{Kleinschmidt:2005bq,Damour:2006xu,Kleinschmidt:2006dy} for a more detailed discussion of this point. In the absence of such a decomposition one can try to construct an invariant subspace in a `level by level' fashion using supergravity as a guiding principle and aiming for a \textit{small} invariant subspace $\bigoplus_{j\in J} V_J$ in order not to overconstrain the system. This is the approach we will follow below for $E_{11}$, using linearised $D=11$ supergravity in first order form as presented in the preceding section.

%%%%%%%%%%%%%%%%%%%%%%%%%%%%%%%%%%%%%%%%%%%%%%%%%%%%%%%%%%%%%%%%%%%%%%%%%%%%%%%%

\subsection{First order duality relations for \texorpdfstring{$E_{11}$}{E11}}
\label{PeterThing}
%%%%%%%%%%%%%%%%%%%%%%%%%%%%%%%%%%%%%%%%%%%%%%%%%%%%%%%%%%%%%%%%%%%%%%%%%%%%%%%%

We now proceed to construct a tentative invariant subspace in the sense of~\eqref{eq:dyn} using $D=11$ supergravity as a guiding principle as done originally in~\cite{West:2011mm}. For determining  a $K(E_{11})$ invariant subspace it is sufficient to use the linearised building blocks~\eqref{eq:BB}.

The starting point of the construction is equation~\eqref{eq:FDl} involving the field strength $F_{a_1a_2a_3a_4}$ of the three-form potential $A_{a_1a_2a_3}$. In the Maurer--Cartan form we have at our disposal $\partial_a A_{b_1b_2b_3}$, which is generally not completely antisymmetric. Therefore projecting to the antisymmetric part could correspond to the requirement above that one uses only a true subspace of the general tensor product~\eqref{eq:tpdec}.\footnote{A different interpretation was pursued in~\cite{Godazgar:2013dma} for finite-dimensional $E_n$ where the mixed symmetry part of $\partial_a A_{b_1b_2b_3}$ was interpreted as part of an exceptional connection.}  The starting ansatz for the construction is thus a four-form and we begin with terms not involving any epsilon tensors as this generalises $F_{a_1a_2a_3a_4}$. 
Indeed, the construction of all field strengths will not involve the epsilon tensor since it is not produced by the action of  $K(E_{11})$. We consider the most general expression that involves all fields up to level $\ell=3$ and also derivatives up to the same level (relative to the highest level derivative):\footnote{Since $\varphi_{ab}$ is projected to its symmetric component $h_{ab}$ in the coset there can be no terms of type $\partial_{[a_1a_2} \varphi_{a_3a_4]}$ in the ansatz.}
\begin{align}
\label{eq:G4ans}
\mathcal{G}_{a_1a_2a_3a_4} &= 4 \partial_{[a_1} A_{a_2a_3a_4]} - \frac{1}{2} \alpha_1 \partial^{b_1b_2} A_{a_1\dots a_4b_1b_2} - 2 \alpha_2 \partial_{[a_1a_2a_3}{}^{b_1b_2} A_{a_4]b_1b_2} \nn\\
&\quad\, + \frac{\alpha_3}{6} \partial_{[a_1}{}^{b_1\dots b_4} h_{a_2\dots a_4]b_1\dots b_4c,}{}^c  + \frac{\alpha_5}{4} \partial_{[a_1a_2}{}^{b_1\dots b_4c,}{}_c A_{a_3a_4]b_1\dots b_4}
\\
&\quad\,  - \frac{\beta_1}{24}\partial^{b_1\dots b_5} h_{[a_1\dots a_4]b_1\dots b_4,b_5} + \frac{\beta_2}{6} \partial_{[a_1a_2a_3}{}^{b_1\dots b_4,c} A_{a_4]b_1\dots b_4c}  
\nn\\
&\quad\, - \frac{\beta_3}{30} \partial_{[a_1a_2a_3}{}^{b_1\dots b_5} A_{a_4]b_1\dots b_5} + {\cal O}(4,4) \ .
\nn
\end{align}
We are employing a notation for the maximum order of terms in an expression that works as follows.  For derivatives, we define a degree $n_d=-\ell -\ft32$ (equal to $q$ for the corresponding coordinate given in Table \ref{tab:e11adj}), and for 
potentials, we set $n_p=\ell=q$, so that 
\bea
(\del_a, \del_{a_1 a_2}, \del_{a_1\cdots a_5}, \del_{a_1...a_7,b}, \del_{a_1...a_8}, \ldots ) \quad &{\rm have\ degree}&\quad  n_d=(0,1,2,3,3,\ldots)\ \quad \text{and}
\nn\\
(h_{a}{}^b,A_{a_1 a_2 a_3},A_{a_1\cdots a_6}, h_{a_1\cdots a_8, b},\ldots ) \quad
&{\rm have\ degree}& \quad n_p=(0,1,2,3,\ldots)\ . 
\label{order}
\eea
Note that $\del^{a_1\cdots a_7,b}$ and $\del^{a_1\cdots a_8}$ both have $n_d= 3$.  
The notation ${\cal O}(N_d,N_p)$ then indicates that we are presenting all terms which have $n_d<N_d$ and $n_p<N_p$. On rare occasions, we do not present all possible terms that may arise at order ${\cal O}(N_d,N_p)$, in which case we will use the notation ${\cal O} (N_d,N_p, N_t)$, signifying that only the terms that satisfy the additional condition $ n_d+n_p <N_t$ are kept.

We now consider the $K(\mf{e}_{11})$ variation of this `field strength' using~\eqref{eq:KE11pot} and~\eqref{eq:KE11der} while attempting to keep the result as small as possible, meaning that we try not to generate too large Lorentz representations in the process. This is in line with the general discussion in Section~\ref{sec:GM}. This will constrain some of the parameters in the ansatz. It is useful to consider terms in $\delta_\Lambda \mathcal{G}_{a_1a_2a_3a_4}$ structure by structure. On the grounds of comparison with supergravity we would expect a transformation into things related to the seven-form field strength, the spin connection and possibly into the field strength of the dual graviton. This last term, however, cannot be computed reliably in the present truncation.

Here and in the following we will often make use of the shorthand for indicating tensorial derivatives of tensors where we only list the numbers of antisymmetric indices (lengths of columns in Young tableau) separated by commas (see Appendix \ref{app:conv}). In this notation a term $\partial_1 A_3$ represents a generic structure of type $\partial_a A_{b_1b_2b_3}$ whereas $\partial^2 h_{8,1}$ would be any structure involving $\partial^{a_1a_2} h_{b_1\cdots b_8,c}$. 

We now consider $\delta_\Lambda \mathcal{G}_{a_1a_2a_3a_4}$, beginning with terms that vary into $\partial_1 A_6$:
\begin{align}
\delta_\Lambda \mathcal{G}_{a_1a_2a_3a_4}|_{\partial_1 A_6} = \frac23 \Lambda^{b_1b_2b_3} \partial_{[a_1}  A_{a_2a_3a_4]b_1b_2b_3} + \frac12\alpha_1 \Lambda^{b_1b_2b_3} \partial_{b_1} A_{b_2b_3a_1\ldots a_4}\,.
\end{align}
In order for the terms on the right-hand side to combine into a seven-form (which would be the smallest possible representation one can have), one needs to fix $\alpha_1=1$ and then gets
\begin{align}
\delta_\Lambda \mathcal{G}_{a_1a_2a_3a_4}|_{\partial_1 A_6} = \frac{7}{3!} \Lambda^{b_1b_2b_3} \partial_{[a_1}  A_{a_2a_3a_4b_1b_2b_3]}\,.
\end{align}

Next we consider terms of the form $\partial_2 A_3$ where we obtain
\begin{align}
\label{eq:dG4p2A3}
\delta_\Lambda \mathcal{G}_{a_1a_2a_3a_4}|_{\partial_2 A_3} &= - 6 \Lambda_{c[a_1a_2} \left( \partial^{cb} A_{a_3a_4]b} +2\alpha_2 \partial_{a_3}{}^b A_{a_4]}{}^c{}_{b} + \frac{1-\alpha_2}{3} \delta_{a_3}^c \partial^{b_1b_2} A_{a_4]b_1b_2}\right)\nn\\
&\quad  +6  \alpha_2\Lambda^{b_1b_2}{}_{[a_1} \partial_{a_2a_3} A_{a_4]b_1b_2}\ .
\end{align}
Demanding that the last term be absent (since it would correspond to the 
generic five-index tensor in $\partial^2 A_3$) leads to the constraint 
$\alpha_2=0$. The remaining terms are then only in the (not traceless) 
representation of type $(2,1)$ which is identical to that of the spin 
connection.  We note that in~\cite{West:2011mm}, 
a term of the form $\del_{[a_1 a_2} \varphi_{a_3 a_4]}$ was added to  (\ref{eq:G4ans}), in order to constrain the $\del^2 A_3$ terms in the first line of~\eqref{eq:dG4p2A3} to be even more restricted and to be
antisymmetric in their three free indices. However, this is not needed for it
to belong to the representation of the spin connection. In fact,
although we have presented our calculation in terms of the coset component $\mathcal{P}$ involving the symmetric
$h_{ab}$ only,  we have also  
checked using the ansatz in terms of $\mathcal{V}$  that the next term in 
$\del^5 \varphi_1{}^1$ arising from the variation of $\partial_{[a_1a_2} \varphi_{a_3a_4]}$ in $\mathcal{G}_{a_1a_2a_3a_4}$ cannot be of the correct 
structure. 
(This was not yet apparent at the level of truncation considered in 
\cite{West:2011mm}.)  

Continuing to terms of the type $\partial_5 \varphi_2$ we find the same constraint $\alpha_2=0$. If one next analyses the terms of type $\partial^5 A_6$ and demands that the terms in $\Lambda_{a_1a_2a_3}$ combine into a 7-form and the terms in $\Lambda_{a_1b_1b_2}$ combine at least into a 5-form instead of a generic tensor, all remaining coefficients  are fixed to $\beta_1=0,\,  \alpha_3=1,\, \alpha_5=1,\, \beta_2=1,\, \beta_3=1$ such that the final fixed version of~\eqref{eq:G4ans} is found to be
\bea
\mathcal{G}_{a_1a_2a_3a_4} &=& 4 \partial_{[a_1} A_{a_2a_3a_4]} - \frac{1}{2} \partial^{b_1b_2} A_{a_1\dots a_4b_1b_2}  \CR
&& \quad+ \frac{1}{6} \partial_{[a_1}{}^{b_1\dots b_4} h_{a_2\dots a_4]b_1\dots b_4c,}{}^c  + \frac{1}{4} \partial_{[a_1a_2}{}^{b_1\dots b_4c,}{}_c A_{a_3a_4]b_1\dots b_4}\CR
&& \quad + \frac{1}{6} \partial_{[a_1a_2a_3}{}^{b_1\dots b_4,c} A_{a_4]b_1\dots b_4c}- \frac{1}{30} \partial_{[a_1a_2a_3}{}^{b_1\dots b_5} A_{a_4]b_1\dots b_5}  +{\cal O}(4,4)\ .
\label{G4WestCorr}
\eea
This result extends the previous expressions in the literature. We stress that the ansatz~\eqref{eq:G4ans} that was the starting point of this analysis included the most general terms up to this order. Thus there is no definite degree structure (as defined in~\eqref{order}) that governs the resulting expression. This implies in particular that one cannot prove that the full expression does not involve $\partial_a$ derivatives of higher $\mf{gl}(11)$ level fields, which might make the interpretation of this field strength in eleven-dimensional supergravity problematic. The $K(\mf{e}_{11})$ variation of this expression is given by
\begin{align}
\delta_\Lambda \mathcal{G}_{a_1a_2a_3a_4}  = \frac{1}{6} \Lambda^{b_1b_2b_3} \mathcal{G}_{a_1\dots a_4b_1b_2b_3} - 6 \Lambda_{b[a_1a_2} \Omega_{a_3a_4]}{}^b +{\cal O}(3,3)\ ,
\label{dg4}
\end{align}
where to this level of truncation 
\bea
\label{eq:NFS}
\Omega_{a_1a_2}{}^b &=& 2 \partial_{[a_1} h_{a_2]}{}^b + \partial^{bc} A_{a_1a_2c} + \frac{1}{4!} \partial^{bc_1\dots c_4} A_{a_1a_2c_1\dots c_4}
\nn\\
&&\quad \quad\quad+ \frac{1}{3}\delta^b_{[a_1} \left( \partial^{c_1c_2} A_{a_2]c_1c_2} + \frac{4}{5!} \partial^{c_1\cdots c_5} A_{a_2]c_1\cdots c_5}\right) +{\cal O}(3,3)\ ,
\eea
and
\be
\mathcal{G}_{a_1\dots a_7} = 7 \partial_{[a_1} A_{a_2\dots a_7]} - \frac{35}{2} \partial_{[a_1a_2a_3}{}^{b_1b_2} A_{a_4\dots a_7]b_1b_2} + {\cal O}(3,3)\ .
\ee
These are the highest terms that can be trusted, given the order to which we
have presented our ansatz for ${\mathcal G}_{a_1 a_2 a_3 a_4}$, because after a 
$K(\mf{e}_{11})$ variation any higher term in the varied expression has 
the possibility of contributions from level $\ell=4$ potentials in the original 
field strengths that could vary into the same structure. 

One can, however, `improve' for example $\mathcal{G}_{a_1\cdots a_7}$ by terms up to $\partial^{7,1}$, $\partial^8$ and $h_{8,1}$ and run through the same logic as for $\mathcal{G}_{a_1a_2a_3a_4}$. This means that one computes the variation of the improved ansatz and demands that the variation produce only small representations. In this way one finds the following improved expression for $\mathcal{G}_{a_1\cdots a_7}$:
\begin{align}
\mathcal{G}_{a_1\dots a_7} &= 7 \partial_{[a_1} A_{a_2\dots a_7]} - \frac{35}{2} \partial_{[a_1a_2a_3}{}^{b_1b_2} A_{a_4\dots a_7]b_1b_2} + 7 \partial_{[a_1}{}^b h_{a_2\dots a_7]bc,}{}^c 
\CR
&\quad\quad + 7 \partial_{[a_1\dots a_6}{}^{b,c} A_{a_7]bc} + 21 \partial_{[a_1\dots a_5}{}^{bc,}{}_c A_{a_6a_7]b} - 7 \partial_{[a_1\dots a_6}{}^{b_1b_2} A_{a_7]b_1b_2} +{\cal O}(4,4,6) \ .
\label{G7E11}
\end{align}
Note that here we have not included terms of the form $\del_8 h_{8,1}$, which would have total level $n_d+n_p=6$, and hence the order ${\cal O}(4,4,6)$, as explained below \eq{order}. This is the $E_{11}$ generalisation of the field strength $F_{a_1\ldots a_7}$ of the six-form $A_{a_1\cdots a_6}$ that also appears in~\eqref{eq:FDl}. Its variation under $K(\mf{e}_{11})$ is given by
\begin{align}
\delta_\Lambda \mathcal{G}_{a_1\cdots a_7} = - 35 \Lambda_{[a_1a_2a_3} \mathcal{G}_{a_4a_5a_6a_7]}+ \frac{1}{2} \Lambda^{b_1b_2c} \Omega_{a_1\cdots a_7b_1b_2,c} 
+{\cal O}(3,3)\ ,
\end{align}
where
\begin{align}
\Omega_{a_1\cdots a_9,b} &= 252 \left( \partial_{[a_1a_2} A_{a_3\cdots a_8} + 2 \partial_{[a_1\dots a_5} A_{a_6a_7a_8}\right) \eta_{a_9],b} +{\cal O}(3,3)\ . 
\end{align}

For our purposes here, it suffices to vary $\Omega_{a_1a_2}{}^b$ without adding any improvement terms.  Thus, under $K(\mf{e}_{11})$ the quantity $\Omega_{a_1a_2}{}^b$ defined in \eq{eq:NFS} varies into
\begin{align}
\label{eq:GTsc}
\delta_\Lambda \Omega_{a_1a_2}{}^b &= \frac{1}{2} \Lambda^{bc_1c_2} \mathcal{G}_{a_1a_2c_1c_2} + \frac{1}{9} \Lambda^{c_1c_2c_3} \delta^b_{[a_1} \mathcal{G}_{a_2]c_1c_2c_3} \nn\\
&\quad+ \Lambda_{c_1c_2[a_1} \mathcal{H}_{a_2]}{}^{bc_1c_2} - \frac{1}{9}  \Lambda_{c_1c_2c_3} \delta^b_{[a_1} \mathcal{H}_{a_2]}{}^{c_1c_2c_3} + \Lambda_{ca_1a_2} \Theta^{b,c} 
+ {\cal O}(2,2) \ ,
\end{align}
where
\begin{align}
\mathcal{H}_a{}^{b_1b_2b_3} &= - \partial_a A^{b_1b_2b_3} + 3 \partial^{[b_1b_2} h_a{}^{b_3]} + \frac{3}{2} \delta_b^{[a_1} \partial^{a_2|c} h_c{}^{a_3]}
+ {\cal O}(2,2)\ ,
\label{eq:Hs}\\
\Theta^{a,b} &= \partial^{c(a} h_c{}^{b)} + {\cal O}(2,2) \ .
\label{eq:hab}
\end{align}
As for $\Omega_{a_1\cdots a_9,b}$, we proceed by improving the ansatz for it as
\bea
\label{eq:O91}
\Omega_{a_1\cdots a_9,b} &=&
252 \left( \partial_{[a_1a_2} A_{a_3\dots a_8} + 2 \partial_{[a_1\cdots a_5} A_{a_6a_7a_8}\right) \eta_{a_9],b}
\label{345}\\
&&\quad +9 \gamma_1 \partial_{[a_1} h_{a_2\cdots a_9],b}  
+ \eta_{b[a_1} \left( \gamma_2 \partial^c   h_{a_2\dots a_9],c} +\gamma_3  \partial_{a_2}   h_{a_3\cdots a_9]c,}{}^{c} \right)
+{\cal O}(3,4,5)\ 
\nn
\eea
and its $K(\mf{e}_{11})$ variation yields
\begin{align} 
\delta_\Lambda  \Omega_{a_1\cdots a_9,}{}^b 
&= - 168\gamma_1 \left( \Lambda^b{}_{[a_1a_2} \partial_{a_3} A_{a_4\cdots a_9]} 
+ \Lambda_{[a_1a_2a_3} \partial_{a_4} A_{a_5\dots a_9]}{}^b \right)
+ \delta^b_{[a_1} ( \cdots ) 
\nn\\
&= - \gamma_1 \left( 24\Lambda^b{}_{[a_1a_2} \mathcal{G}_{a_3\cdots a_9]} 
   - 28\Lambda_{[a_1a_2a_3}\mathcal{G}_{a_4\cdots a_9]}{}^b 
   + 28\Lambda_{[a_1a_2a_3} {\mathcal H}^b{}_{a_4\cdots a_9]}\right)\nn\\
&\quad   + \delta^b_{[a_1} ( \cdots ) +{\cal O}(2,3)\ ,
\end{align}
up to the trace components, where
\begin{align} 
\mathcal{H}_a{}^{b_1\dots b_6}  &= \partial_a A^{b_1\dots b_6}+ {\cal O}(2,3)\ .
\label{eq:h16}
\end{align} 
The last entry ($N_t=5$) in the order displayed in \eq{345} is due to the fact that terms of the form $\del_5 h_{8,1}$ are not included.

One might be worried at first sight at seeing non-antisymmetrised derivatives of the potentials in \eqref{eq:Hs}, \eqref{eq:hab} and \eqref{eq:h16}. This is not a problem if one considers that $\mf{e}_{11}$ includes at level $\ell=4$ potentials of the type $A_{9,3}$, $B_{10,1,1}$ and $C_{11,1}$ whose field strengths include $\mathcal{H}_{10,3}$ and $\Theta_{11,1,1}$, leading to the conclusion that there should be well defined first order equations between $\mathcal{H}_{1,3}$ and $\mathcal{H}_{10,3}$, and between $\Theta^{1,1}$ and $\Theta_{11,1,1}$. We do indeed find part of such duality equations in the $K(\mf{e}_{11})$ variation of the first order gravity equations. Proceeding to include higher level contributions to these equations we expect them to take the form 
\begin{align}
\mathcal{T}_a{}^{b_1b_2b_3}  &\equiv \mathcal{H}_a{}^{b_1b_2b_3} -\frac{1}{10!} \varepsilon_a{}^{c_1\dots c_{10}} \mathcal{H}_{c_1\dots c_{10},}{}^{b_1b_2b_3} = 0\ ,
\label{t13}\\
\mathcal{T}^{a,b} &\equiv  \Theta^{a,b} -\frac{1}{11!} \varepsilon^{c_1\dots c_{11}} \Theta_{c_1\dots c_{11},}{}^{a,b} = 0\ ,
\label{t11}\\
\mathcal{T}_a{}^{b_1\dots b_6} & \equiv \mathcal{H}_a{}^{b_1\dots b_6} - 
\frac{1}{10!} \varepsilon_a{}^{c_1\dots c_{10}} 
\mathcal{H}_{c_1\dots c_{10},}{}^{b_1\dots b_6}=0 \ . 
\label{t16}
\end{align}
These would correspond to duality equations that appear in the unfolding 
approach \cite{Boulanger:2015mka}, but in a first-order form.

We now postulate the $E_{11}$ version of the duality equation~\eqref{eq:FDl} to be 
\be
\label{eq:E11S1}
\mathcal{S}_{a_1\ldots a_4} \equiv \mathcal{G}_{a_1\ldots a_4} +\frac1{7!} 
\varepsilon_{a_1\ldots a_4b_1\ldots b_7} \mathcal{G}^{b_1\dots b_7} =0\ .
\ee
Putting the results above together we obtain under $K(\mf{e}_{11})$
\begin{align}
\delta_\Lambda \mathcal{S}_{a_1\ldots a_4} &= -\frac1{3!\cdot 4!} \Lambda^{b_1b_
2b_3} \varepsilon_{a_1\ldots a_4 b_1b_2b_3 c_1\ldots c_4} \mathcal{S}^{c_1\ldots
 c_4} - 6 \Lambda_{b[a_1a_2} S_{a_3a_4]}{}^b\ ,
\end{align}
where 
\bea
\label{eq:E11S2}
\mathcal{S}_{a_1a_2}{}^b \equiv \Omega_{a_1a_2}{}^b - \frac{1}{9!} \varepsilon_{a_1a_2}{}^{c_1\dots c_9} \Omega_{c_1\dots c_9,}{}^b =0\ .
\eea
Requiring that the $K(\mf{e}_{11})$ variation of $S_{a_1a_2}{}^b$ gives back equation $S_{a_1a_2a_3a_4}$ fixes $\gamma_1=1$ and $\gamma_2=\gamma_3=0$ in \eqref{eq:O91}. One obtains then
\bea 
\delta_\Lambda S_{a_1a_2}{}^b &=& \frac{1}{2} \Lambda^{bc_1c_2} \mathcal{S}_{a_1a_2c_1c_2} + \frac{1}{9} \Lambda^{c_1c_2c_3} \delta^b_{[a_1} \mathcal{S}_{a_2]c_1c_2c_3} - \frac{28}{9} \Lambda^{c_1c_2c_3}  \varepsilon_{a_1a_2c_1\dots c_9} \mathcal{T}^{b,c_4\dots c_9} \nn\\
&&\quad+ \Lambda_{c_1c_2[a_1} \mathcal{T}_{a_2]}{}^{bc_1c_2} - 
\frac{1}{9}  \Lambda_{c_1c_2c_3} \delta^b_{[a_1} 
\mathcal{T}_{a_2]}{}^{c_1c_2c_3} + \Lambda_{ca_1a_2} 
\mathcal{T}^{b,c} + \mathcal{O}(2,2)\ ,
\eea
where the  $\mathcal{T}$ tensors are equal to the corresponding
$\Theta$ and $\mathcal{H}$ tensors \eqref{eq:Hs}, \eqref{eq:hab} and \eqref{eq:h16}, at this level of truncation.

The duality equations \eq{t13}, \eq{t11} and \eq{t16} are not
invariant under the gauge transformations~\eqref{eq:GTM}. 
However, taking the derivative of, for example, the duality equation \eq{t13} in the following fashion,
\be
\partial^{[a_1} {\mathcal S}_b{}^{a_2a_3a_4]} =
\partial_b F^{a_1a_2a_3a_4} - \frac1{10!}\, 
\varepsilon_b{}^{c_1 \cdots c_{10}}\, 
  R_{c_1\cdots c_{10}}{}^{a_1 a_2 a_3 a_4} =0\,,\label{unfold}
\ee
%%%%%
where
%%%%%
\be
R_{c_1\cdots c_{10}}{}^{a_1 a_2 a_3 a_4} \equiv
40\, \partial_{[c_1} \partial^{[a_1} A_{c_2\dots c_{10}],}{}^{a_2a_3a_3]}\,, 
\ee
gives an equation that is gauge invariant, and is known to appear in 
the unfolding approach 
\cite{Boulanger:2015mka}.\footnote{The integrability condition of the
gradient $\del_b F^{a_1\cdots a_4}$ then gives, from (\ref{unfold}),
the condition
\begin{align}
\del_{[b_1}\del_{b_2]} F^{a_1\cdots a_4} = \frac1{2\cdot 9!}\,
\varepsilon_{b_1 b_2}{}^{c_1\cdots c_9} \, 
\del^{c_{10}}\, R_{c_1\cdots c_{10},}{}^{a_1 a_2a_3a_4}=0\, , 
\nn
\end{align}
which is an equation of motion satisfied by $A_{9,3}$.}
Just as for gravity,  we can expect only the second order equations to be fully gauge invariant (for two-column fields).

In this section we have revisited the proposal of \cite{West:2011mm} to construct  a $K(\mf{e}_{11})$ multiplet of first order duality equations starting from the duality equation ${\cal G}_7 =  \star {\cal G}_4$~\eqref{eq:FDl} in supergravity, with the requirement that the total set of first order constraints is small enough to allow for dynamical equations. We confirmed that one obtains in this way a duality equation for the gravitational field  $\Omega_{2,1} = \star \Omega_{9,1}$ that enforces, however, the additional constraint that the spin connection be traceless, incidentally violating general covariance.  Pushing the program to higher levels, we see the premises of an infinite chain of unfolding duality equations advocated in \cite{Boulanger:2015mka} that relate level $\ell$ to level $\ell+3$ fields 
\begin{align}
\label{de1}
{\cal H}_{1,3} = \star {\cal H}_{10,3}\ ,\quad
{\cal H}_{1,6} = \star {\cal H}_{10,6}\ , \quad
{\cal H}_{1,8,1} = \star {\cal H}_{10,8,1}\ , \quad
{\cal H}_{1,9,3} =  \star {\cal H}_{10,9,3}\ , \ \dots
\end{align}
Here, the field strengths on the left-hand side are derivatives of a potential $A_{R}$ at level $\ell$ with $R$ being given by some Young tableau not containing any column with ten or eleven indices, and the field strength on the right-hand side is the curl of the next dual potential $A_{9,R}$ at level $\ell+3$. The terminology of ``unfolding'' refers to the fact that there is a field $A_{9^n,R}$ dual to each gradient $(\partial_1)^n A_{R}$ of a given field $A_{R}$,\footnote{The derivative $\partial_1  A_{R}$ is dual to curl $d A_{9,R}$, the derivative $\partial_1 A_{9,R}$ is dual to $d A_{9,9,R}$, and so on, such that by recurrence $\partial^n_1 A_{R}$ can be reduced to $ (\star d)^n A_{9^n,R}$.} such that all the degrees of freedom of the fields are unfolded into infinitely many potentials in one-to-one correspondence with the solutions to the wave equation.\footnote{We note also the alternative formulation in terms of `Ogievetsky generators' given in~\cite{Riccioni:2009hi}.}

We identify also the appearence of gauge non-invariant non-dynamical dualities that relate field strengths involving the exceptional $\partial^2$ derivative of potentials of level $\ell$ in some representations to the curl of level $\ell+4$ fields that carry a column of $10$ antisymmetrised indices
\begin{align}
\Theta_{1,1} = \star \Theta_{11,1,1}\ ,\quad 
\Theta_{4,1} = \star \Theta_{11,4,1}\ ,\quad 
\Theta_{7,1} = \star \Theta_{11,7,1}\ ,\quad 
\Theta_{6,2} = \star \Theta_{11,6,2}\ ,\quad 
\Theta_8 = \star \Theta_{11,8}\ ,\, \dots
\end{align}
and more generally the exceptional derivative of level $-3/2-n$ of potentials of level $\ell$ to the curl of level $\ell+3+n$ fields  that carry a column of ten antisymmetrised indices. They are non-dynamical because the left-hand side does not include ordinary derivatives so that they vanish identically when interpreted in eleven-dimensional supergravity. The duality relation then implies that the fields including a column of ten antisymmetrised indices have a vanishing curl, and are therefore non-dynamical, as expected from the standard free field analysis. We shall argue in Section \ref{NonGeoFluxes} that backgrounds with such field strengths turned on are non-geometrical, and that the latter can be identified with components of the embedding tensor in gauged supergravity.

As was emphasized above, the field strengths that appear in this construction of the field equations are not governed by any grading; the only requirement that is imposed on the terms is that they have the correct Lorentz tensor structure. Therefore there is no argument to rule out the contribution of standard derivatives $\partial_1$ of potentials of arbitrarily high level contributing to for instance $\mathcal{G}_{a_1a_2a_3a_4}$. Simple examples would come from ordinary derivatives of the form $\partial_1 A_{9^{2n},3}$ where one simply contracts all $2n$ of the columns of nine indices in a pairwise fashion. Therefore it is not clear whether one can safely interpret the equations restricted to eleven-dimensional supergravity in a given level truncation. Similar terms were discussed in~\cite{West:2011mm}.

%%%%%%%%%%%%%%%%%%%%%%%%%%%%%%%%%%%%%%%%%%%%%%%%%%%

\subsection{On the trace of the spin connection}

%%%%%%%%%%%%%%%%%%%%%%%%%%%%%%%%%%%%%%%%%%%%%%%%%%%
\label{sec:trace}

As already emphasised, in order for the first order dual gravity equation to be formulated in a gauge invariant formulation, one needs a nine-form potential, which does not appear in $\mf{e}_{11}$ at $\mf{gl}(11)$ level $\ell=3$. If one were to give up gauge invariance, one would need to find a consistent $K(\mf{e}_{11})$-multiplet of gauge-fixing conditions that would represent an additional $K(\mf{e}_{11})$-multiplet of first order constraints.
However, one has to make sure that the gauge-fixing conditions are not too strong, \eg, they should not contain $\partial_a A_{b_1b_2b_3}=0$ for arbitrary indices. If one starts with a Lorentz-covariant ansatz for a metric gauge of the form $\partial^c h_{ac} - \alpha_1 \partial_a h_c{}^c + \alpha_2 \partial^{b_1b_2} A_{ab_1b_2}+ \dots =0$ and demands that its $K(\mf{e}_{11})$ variation only includes the derivative of the three-form gauge field through the Lorenz gauge term $\partial^b A_{a_1a_2b}$, this fixes the two coefficients with the result
\bea
&&  \delta \Bigl(  \partial_b h_{a}{}^{b} - \tfrac12 \partial_a h_b{}^b + \tfrac12 \partial^{b_1b_2} A_{ab_1b_2} + \tfrac{1}{5!} \partial^{b_1\dots b_5} A_{ab_1\dots b_5} + \dots \Bigr) \CR
&= &\frac12 \Lambda_{ab_1b_2} \Bigl( \partial_c A^{b_1b_2c} + 2 \partial^{cb_1} h_c{}^{b_2} - \tfrac12 \partial^{b_1b_2} h_c{}^c + \tfrac16 \partial^{b_1b_2c_1c_2c_3} A_{c_1c_2c_3} + \dots   \Bigr) \ . 
\eea
Varying it again under $K(\mf{e}_{11})$ this then gives consistently the original condition on the metric together with a Lorenz gauge for the six-form potential 
\bea
&& \delta \Bigl(  \partial_c A^{a_1a_2c} + 2 \partial^{c[a_1} h_c{}^{a_2]}- \tfrac12 \partial^{a_1a_2} h_c{}^c+  \tfrac16 \partial^{a_1a_2b_1b_2b_3} A_{b_1b_2b_3} \Bigr)  \CR
&=& - \Lambda^{ba_1a_2} \Bigl(  \partial_{c} h_{b}{}^c- \tfrac12 \partial_b h_c{}^c  +  \tfrac12 \partial^{c_1c_2} A_{bc_1c_2}+ \dots  \Bigr)  \CR
&& \qquad - \frac{1}{6} \Lambda^{b_1b_2b_3} \bigl(   \partial^c A_{a_1a_2b_1b_2b_3c} -10 \partial_{[a_1a_2} A_{b_1b_2b_3]} + \dots \bigr)   \ .
\eea
One concludes that an appropriate $K(\mf{e}_{11})$-multiplet of first order gauge-fixing conditions must involve the harmonic gauge $\partial_b h_a{}^b-\tfrac12 \partial_a h_b{}^b=0$ for the graviton rather than the one in \eqref{DiffGauge}. In the harmonic gauge for the metric, it is not consistent to gauge fix the nine-form component of the dual graviton to zero. We conclude that there is no $K(\mf{e}_{11})$-multiplet of first order gauge-fixing conditions that is consistent with the condition that the nine-form vanishes.

We will therefore discuss whether such a nine-form may possibly arise in the theory. Inspecting the Table~\ref{tab:e11adj} or the tables of~\cite{Nicolai:2003fw}, the lowest level field that includes a nine-form in its $\mf{so}(1,10)$ decomposition is the $\ell=5$ field $C_{11,3,1}$. A suitable triple trace of this field yields a nine-form potential, $C_9$.  However, the local gauge transformation of this potential will not agree with the required gauge transformation, $\delta_\Xi X_9= d\lambda_8$, since we have
\begin{align}
\label{eq:GT1131}
\delta_\Xi C_{a_1\cdots a_9 b_1 b_2,}{}^{b_1 b_2 b_3,}{}_{b_3} = 9\partial_{[a_1}
 \xi_{a_2\cdots a_9]b_1 b_2,}{}^{b_1 b_2 b_3,}{}_{b_3}
 -2 \partial_{b_1} \xi_{a_1\cdots a_9 b_2,}{}^{b_1 b_2 b_3,}{}_{b_3}+ \cdots \,.
\end{align}
The first term is of the correct form, but the second is not. We expect a similar phenomenon for higher level fields whose trace may yield a nine-form. For this we note that the analysis of~\cite{Henneaux:2011mm} shows that for any generator of a given Young tableau, the $\ell_1$ representation contains corresponding gauge parameters of the form where a single box is removed from the Young tableau in all possible admissible ways. In the example above with a potential $C_{11,3,1}$ this means that there are parameters of the form $\lambda_{10,3,1}$, $\lambda_{11,2,1}$ and $\lambda_{11,3}$. These can be paired with ordinary derivatives in the gauge transformation~\eq{eq:GT} to yield the linearised gauge transformations of $C_{11,3,1}$. In equation~\eqref{eq:GT1131} above, we have only displayed the transformation under $\lambda_{11,2,1}$ that suffices to make our argument as all gauge parameters are independent. A similar calculation to~\eqref{eq:GT1131} will then show for any higher level gauge potential of mixed symmetry type that even if its
$\mf{so}(1,10)$ decomposition contains a nine-form, the gauge transformation of that nine-form will not be of the standard type that is needed in the dual gravity equation. The fact that there are no pure nine-forms contained in the adjoint representation of
$\mf{e}_{11}$ follows from the arguments in~\cite{Nicolai:2003fw}. From this discussion we conclude that one cannot reconcile the standard gauge transformations required for the trace of the spin connection with the gauge symmetries present in $\mf{e}_{11}$. 

An additional problem in the analysis of the duality equations arises as follows. Their construction as described in the previous subsection is such that the terms in the equations are determined up to order ${\cal O}(n_d , n_p, n_t)$ (see below \eq{eq:G4ans} for the definition of this notation).  The problem with this procedure is that the ordinary derivatives of a potential will arise at arbitrarily high levels. Therefore, even if we encounter a potential or its trace playing the role of $X_9$ at some level that is needed for the dual graviton equation, the latter could be spoiled by a term that could  arise at  some higher level which would involve the ordinary derivative. For example, the level $\ell=8$ field
$B_{10,9,3,2}$ could spoil the dual graviton equation $S_{a_1a_2}{}^b=0$ by a contribution of the form
\be
\varepsilon_{a_1a_2}{}^{a_3\cdots a_{11}} \varepsilon^{c_1\cdots c_{11}}\partial_{c_1} B_{c_2\cdots c_{11},a_3\cdots a_{11},}{}^{b_1b_2b}{}_{, b_1b_2}\ .
\ee
Turning to the problem associated with the incorrect gauge transformation rule for the nine-form potential that might arise at higher levels, assuming that 
the set of first order equations must reproduce the corresponding equations in the bosonic sector of eleven-dimensional supergravity, the system described above will not 
give the trace part of the dual graviton equation. A related discussion of this issue without reference to gauge transformations can be found in the appendix of~\cite{West:2011mm}. 

As we shall see in section~\ref{sec:THA}, the difficulties associated with the trace of the spin connection are circumvented in the extension of the theory that we propose in this article.

%%%%%%%%%%%%%%%%%%%%%%%%%%%%%%%%%%%%%%%%%%%%%%%%%%%%%%%%%%%%%%%%%%%%%%%%%%%%%%%%%%%%%%%%%%%%%%%%%%
\subsection{Second order \texorpdfstring{$E_{11}$}{E11} field equations and the section constraint}
\label{sec:West2}
%%%%%%%%%%%%%%%%%%%%%%%%%%%%%%%%%%%%%%%%%%%%%%%%%%%%%%%%%%%%%%%%%%%%%%%%%%%%%%%%%%%%%%%%%%%%%%%%%%

So far we have not considered the behaviour of the first order equations under gauge transformations. One reason for this is  that 
we already know from ordinary gravity theory that the first order duality equation is not gauge invariant unless one introduces a 
St\"uckelberg field \cite{Boulanger:2008nd}. The duality equation for gradients of the physical fields are not gauge invariant either. 
However, gauge invariant second order duality equations do exist without introducing any St\"uckelberg fields. Therefore we demand gauge invariance of the second order field equations of the theory based on $E_{11}$.  In general, for potentials with more than two columns one might expect gauge invariant equations only involving as many derivatives as there are columns~\cite{Bekaert:2002dt,Bekaert:2003az} and indeed this is what appears in the recent work~\cite{Tumanov:2016dxc}. We note that this approach entails equations of arbitrarily high derivative order and there is no closed $K(\mf{e}_{11})$ multiplet of gauge invariant equations (as $K(\mf{e}_{11})$ does not change the number of derivatives).

Given the first order duality equations one can deduce second order equations as compatibility relations for them. 
Starting from $\mathcal{S}_{a_1a_2a_3a_4}$  in~\eqref{eq:E11S1} we can form, for example,
\begin{align}
\partial^{b} \mathcal{S}_{ba_1a_2a_3} =  \partial^b \mathcal{G}_{ba_1a_2a_3} -\frac1{7!} \varepsilon_{a_1a_2a_3b_1\cdots b_8} \partial^{b_1}\mathcal{G}^{b_2\cdots b_8} = 0\ .
\end{align}
In usual supergravity this would be the field equation for $A_{a_1a_2a_3}$ and its validity would be ensured by the seven-form field strength being closed and it would be gauge invariant. However, it is easy to check that $\partial^{b} \mathcal{S}_{ba_1a_2a_3}$ is not gauge invariant with ${\cal G}_{a_1\cdots a_4}$ and ${\cal G}_{a_1\cdots a_7}$ given in \eqref{eq:E11S1} and \eqref{G7E11}. Since the gauge and  $K(\mf{e}_{11})$ variations do not produce a Levi-Civita tensor, there is no loss of generality in considering second order field equations in terms of the field strengths without their Hodge duals. In fact, such terms should vanish as a consequence of generalised Bianchi identities for the consistency with the first order duality equations. Thus, we start from the ansatz
\begin{align}
\label{eq:Ans1}
\mathcal{E}_{a_1a_2a_3} = \partial^{b}\mathcal{G}_{ba_1a_2a_3} + \alpha_1 \partial_{b[a_1} \Omega_{a_2a_3]}{}^b + \alpha_2 \partial_{[a_1a_2} \Omega_{a_3]b}{}^b +{\cal O}(2,2)\ ,
\end{align}
with ${\mathcal G}_{a_1a_2a_3a_4}$ and $\Omega_{a_1 a_2}{}^b$ from \eq{G4WestCorr} and \eq{eq:NFS}, respectively. 
Under the gauge transformation~\eqref{eq:GTM} we for example obtain the following terms in the 
variation of $\mathcal{G}_{a_1a_2a_3a_4}$ and $\Omega_{a_1a_2}{}^b$:
\begin{align}
\delta_\Xi \mathcal{G}_{a_1\cdots a_4} &= \partial^{b_1b_2} \partial_{b_1} \lambda_{a_1\cdots a_4b_2} + 12 \partial_{[a_1} \partial_{a_2a_3} \xi_{a_4]} + \cdots\CR
\delta_\Xi \Omega_{a_1a_2}{}^b &=2\partial^b \partial_{[a_1}  \xi_{a_2]} + \cdots \,,
\end{align}
so the field strengths are clearly not gauge invariant by themselves.  Under these variations,  \eqref{eq:Ans1} transforms as
\bea
\delta_\Xi{\cal E}_{a_1a_2a_3} &=& \partial^b\left(\partial^{cd}\partial_d \lambda_{a_1a_2a_3bc}\right) +(\alpha_2-3)\left( \partial_{[a_1a_2}\partial_{a_3]} \partial^b\xi_b - \partial^b\partial_b \partial_{[a_1a_2} \xi_{a_3]}\right)
\nn\\
&& +2(\alpha_1+3) \partial^b\partial_{b[\alpha_1}\partial_{a_2} \xi_{a_3]}\ . 
\label{var1}
\eea
Nothing in~\eqref{eq:Ans1}, including its higher level extensions, can compensate for the first term above, as well as the terms proportional to $(\alpha_2-3)$. 
Therefore we conclude that 
\be
\alpha_2=3\ ,
\ee
and the combination 
\begin{align}
\label{eq:WS1}
\partial^{b_1b_2} \partial_{b_1} \lambda_{a_1\dots a_4b_2} =0
\end{align}
has to vanish. This is indeed what happens if we impose the $d=11$ analogue of the section constraint
encountered in exceptional field theories for finite-dimensional $\mf{e}_d$~\cite{Hohm:2013vpa,Hohm:2013uia,Hohm:2014fxa},
which at the lowest level in the $\mf{gl}(d)$ decomposition implies $\partial^{ab} \partial_{b}=0$. The same condition ensures the vanishing of the last term in  \eq{var1} without determining the value of $\alpha_1$. Thus the gauge invariance of 
\eqref{eq:Ans1} is established to the level we are working for any $\alpha_1$ up to the section constraint.

The coefficient $\alpha_1$ can be fixed by considering gauge invariance at the next level. However, it is more convenient to fix it by demanding 
that the $K(\mf{e}_{11})$ transformation of the equation~\eqref{eq:Ans1} leads to a sensible second order equation for the graviton. 
Upon $K(\mf{e}_{11})$ variation of \eqref{eq:Ans1} one finds 
\be
\delta_\Lambda \mathcal{E}_{a_1a_2a_3} =    - 3 \Lambda_{c[a_1a_2} 
{\cal E}^c{}_{a_3]}  
-(\alpha_2-3)\Lambda_{c[a_1 a_2}\, \del^c\,\Omega_{a_3] b}{}^b + (\alpha_1-3)\, \Lambda_{cb [a_1}\, \del^b \, \Omega_{a_2 a_3]}{}^c + {\cal O}(1,1)\ ,
\label{e1a}
\ee
%%%%%
where 
\be
{\cal E}_{ab} = \partial_{a} \Omega_{bc}{}^c -\partial^{c} \Omega_{bc,a} 
+ {\cal O}(1,1)
=\partial_a \partial_b h_c{}^c -2\partial^c\partial_{(a} h_{b)c} + \partial^2 h_{ab}\ ,
\label{e2}
\ee
is the linearised Ricci tensor and we have used $h_{ab}=h_{(ab)}$. Since the last term is the linearised Riemann tensor, we need to impose 
\be
\alpha_1=3 \ .
\ee
Note also that the field equation ${\cal E}_{a_1...a_6}= \partial^b {\cal G}_{ba_1...a_6}$ has the minimum order ${\cal O}(1,2)$, and that is why it does not appear in \eq{e1a}, which holds to order ${\cal O}(1,1)$. 

We conclude that the second order field equations that are built out of the first order duality equations constructed above and in~\cite{West:2011mm,Tumanov:2016abm} is only gauge invariant if one imposes the section constraint. The need for this condition was not seen in \cite{Tumanov:2016abm}, since the invariance under gauge transformations was not investigated there. From the point of view of $E_{11}$, the section constraint is part of an infinite multiplet that contains as leading contribution the lowest weight representation $\ell_{10}$, and is the complement of $2\ell_1$ in the decomposition \eqref{eq:SC}. This representation is analysed in more detail in Appendix~\ref{app:SCmult}.\footnote{Checking similar equations for the other fields, different components of the section constraint are generated. Instead of providing the details here, we will present a more systematic construction based on the tensor hierarchy algebra in the following section.} Note, however, that it was emphasized in \cite{Tumanov:2016dxc} that there were no gauge invariant second order field equations for the $\mf{gl}(11)$ level $4$ fields, even modulo the section constraint. It follows that one should find obstructions in the construction of such gauge invariant second order equations modulo the section constraint when continuing the construction to $\mf{gl}(11)$ levels beyond those considered here.

Nonetheless, one expects that in the $E_{11}$ formalism gauge invariance is satisfied up to a certain level. In order to increase the level at which the equations are gauge invariant, one apparently needs to also increase the order of the field equations. Our analysis exhibits that demanding generalised gauge invariance of the field equations at a given truncation level necessarily requires the fields to satisfy the section constraint. This is the case, for example, for the first order duality equation $\cG_7 =\star \cG_4$, and of the second order Einstein equation. It seems that this pattern should extend to higher level fields, such that generalised gauge invariance of the third order equation for the fields $B_{10,1,1}$ considered in \cite{Tumanov:2016dxc} might also require the section constraint to be satisfied.\footnote{Checking gauge invariance only for terms involving the ordinary derivative $\partial_m$ will not reveal the necessity to impose the section constraint because one is effectively working on a solution of the section constraint.}
Moreover, the compatibility of the second order equations displayed in this section with the first order duality equations discussed in the previous section also requires the section constraint to be satisfied. We conclude that demanding any kind of generalised gauge invariance in the $E_{11}$ framework requires constraining the fields to satisfy the section constraint.

%%%%%%%%%%%%%%%%%%%%%%%%%%%%%%%%%%%%%%%%%%%%%%%%%%%%%%%%%%%%%%%%%%%%%%%%%%%%%%%
\section{Tensor hierarchy algebra and gauge invariant field strengths}
\label{sec:THA}
%%%%%%%%%%%%%%%%%%%%%%%%%%%%%%%%%%%%%%%%%%%%%%%%%%%%%%%%%%%%%%%%%%%%%%%%%%%%%%%

We will now change gears and present a different construction based on the tensor hierarchy algebra that provides a definition of the field strengths in a representation of $\mf{e}_{11}$. At the same time this construction will automatically remedy the issue with the trace of the spin connection encountered above.

%%%%%%%%%%%%%%%%%%%%%%%%%%%%%%%%%%%%%%%%%
\subsection{The tensor hierarchy algebra}
%%%%%%%%%%%%%%%%%%%%%%%%%%%%%%%%%%%%%%%%%
\label{sec:tha1}

For $4 \leq d \leq 8$, the finite-dimensional Lie algebra $\mf{e}_d$
was extended in \cite{Palmkvist:2013vya} to an infinite-dimensional Lie superalgebra.
It was called the tensor hierarchy algebra, since its level decomposition
into $\mf{e}_d$ representations $R_p$ for all integers $p$ gives exactly the tensor hierarchy that appears in gauged maximal supergravity in $D=11-d$ dimensions for
$p\geq 1$~\cite{deWit:2005hv,deWit:2008ta}. Moreover, $R_{-1}$ is the 
representation in which the embedding tensor transforms, and by considering it as an element in this subspace of the algebra, the 
approach in \cite{Cremmer:1998px} to
$D$-dimensional maximal supergravity can be extended to the gauged theory \cite{Greitz:2013pua,Howe:2015hpa}. The possibility to interpret  
the embedding tensor as an element in $R_{-1}$ is the crucial difference between
the tensor hierarchy algebras for $\mf{e}_d$ are the similar Lie superalgebras of Borcherds type that have also been considered in the context of maximal supergravity and exceptional geometry \cite{HenryLabordere:2002dk,Greitz:2011da,Kleinschmidt:2013em,Palmkvist:2015dea},
and in relation 
to $\mf{e}_{11}$ \cite{Henneaux:2010ys,Palmkvist:2011vz}.

In a further level decomposition with respect to $\mathfrak{gl}(d)$, the $\mf{e}_d$ representation $R_{-1}$ contains a four-form as well as 
a seven-form (for $d=7,8$). This observation suggests that
the field strengths of eleven-dimensional supergravity should transform in an
$\mf{e}_{11}$ representation that
would be $R_{-1}$ in a tensor hierarchy algebra analogously defined for $d=11$.
Although the construction in \cite{Palmkvist:2013vya} is not applicable to the cases $d\geq9$, where the Lie algebras $\mf{e}_d$ are infinite-dimensional,
we show in Appendix \ref{app:THA} that there exists such an extension of $\mf{e}_{11}$. 
We shall in the following describe some of its features,
and in the next subsection argue that
it indeed gives the right representation for the field strengths in the present set-up.

\begin{landscape}
\begin{table}
\setlength{\arraycolsep}{4pt}
\renewcommand{\arraystretch}{1.38}
\begin{align*}
\begin{array}{c|c|ccc|ccc|cc|c|c|cc|ccc|ccc|c}
p&\cdots&
\multicolumn{3}{c|}{q=-5}&
\multicolumn{3}{c|}{q=-4}&
\multicolumn{2}{c|}{q=-3}&q=-2 & q=-1 & \multicolumn{2}{c|}{q=0} &  \multicolumn{3}{c|}{q=1} & \multicolumn{3}{c|}{q=2}&
\cdots
\\
\hline
\vdots &\ddots
\begin{picture}(0,0)(-16.25,0)
\thinlines
\put(150.7,-87.7){\line(1,0){193}}
\put(150.7,-87.7){\line(0,1){102}}
\put(150.7,14.3){\line(-1,0){160}}
\put(343.6,-87.7){\line(0,-1){122}}
\put(343.6,-209.7){\line(-1,0){352.9}}
\put(-9.3,-209.7){\line(0,1){224}}
\put(238.5,-101){$\times$}
\end{picture}
&\ddots&\vdots&\vdots&\vdots&\vdots& \vdots &\vdots&\vdots&\vdots&&&&&&&& &&\\ 
3
&\cdots&\cdots&F^4{}_{4,1}&F^5{}_6&F^3{}_1&F^4{}_{1,1}&F^5{}_3& L_6& L_{5,1}& L_3& L&&&&&&  &  & & \\ 
2
&\cdots&\cdots&F^3{}_{4,1}&F^4{}_6&F^2{}_1&F^3{}_{1,1}&F^4{}_3& L_7& L_{6,1}& L_4& L_1&&&&&&  &  & &\\ 
1
&\cdots&\cdots&F^2{}_{4,1}&F^3{}_6&F^1{}_1&F^2{}_{1,1}&F^3{}_3& 
P_8& P_{7,1}
& Z_5& Z_2 & \multicolumn{2}{c|}{P^1} &&& & & & &  \\ 
0 &\cdots&\cdots&F^1{}_{4,1}&F^2{}_6&F_1&F^1{}_{1,1}& F^2{}_3 & 
F_9 & F_{8,1}
& F_6& F_3 & 
\multicolumn{2}{c|}{K^1{}_1}
& E^3 &  &
& E^6 &  & &\cdots\\ 
-1
&\cdots&&F_{4,1}&F^1{}_6&&F_{1,1}&F^1{}_3&\multicolumn{2}{c|}{ K^2{}_1}& K^4
& K^7 & K^{9,1}&K^{10}&
E_1{}^3 & E^{1,1}&& E_1{}^6 & E^{4,1}
&  &\cdots \\ 
-2
&\cdots&&&\widetilde F_6&&&\widetilde F_3 &  \multicolumn{2}{c|}{ \widetilde K^1{}_1}    & \widetilde E^3& 
\widetilde E^6 &\widetilde E^{8,1} & \widetilde E^9  & 
E_2{}^3 & E_1{}^{1,1}  
& E^1
& E_2{}^6&E_1{}^{4,1}&\cdots&\cdots\\ 
-3
&&&&&&&&\multicolumn{2}{c|}{P_1} & Z^2 & Z^5 & P^{7,1} & P^8 &  E_3{}^3  &E_2{}^{1,1}  & 
E_1{}^1 
& E_3{}^6
& E_2{}^{4,1}&\cdots&\cdots\\ 
-4
&&&&&&&&& & L^1 & L^4 & L^{6,1} & L^7 &  E_4{}^3  &E_3{}^{1,1}  & 
E_2{}^1 
& E_4{}^6
& E_3{}^{4,1}&\cdots&\cdots\\
-5
&&&&&&&&& & L & L^3 & L^{5,1} & L^6 &  E_5{}^3  &E_4{}^{1,1}  & 
E_3{}^1 
& E_5{}^6
& E_4{}^{4,1}&\cdots&\cdots\\
\vdots &&&&&  & && & & &\vdots &\vdots &\vdots &\vdots&\vdots&\vdots   & \vdots&\vdots&\ddots&\ddots
\end{array}
\end{align*}
\caption{\label{tab:THA}\small \it 
Part of the tensor hierarchy algebra $\mathscr T$, decomposed under $\mathfrak{gl}(11)$.
This decomposition is a $(\mathbb{Z} \times \mathbb{Z})$-grading, where we denote the
vertical and horizontal $\mathbb{Z}$-degrees by $p$ and $q$, respectively.
The (possibly reducible) $\gl(11)$ representation at any bi-degree $(p,q)$ is given by 
the index structure of one or more tensor densities,
using our shorthand notation explained in Appendix \ref{app:conv}, up to the 
eigenvalue $3 \ell$ of the generator $K^m{}_m$, which is given by the linear combination $\ell=q-\frac32 p$ of $p$ and $q$.
The subalgebra at $q=0$ is the Cartan
superalgebra $W(11)$ (see Appendix~\ref{app:THA}), and the subalgebra at $p=0$ is the extension of $\mathfrak{e}_{11}$ that we denote by $\mathfrak{t}_{11}$. The only new generator in this extension explicitly shown in the table is $F_9$ at $q=-3$, but there are also other new generators coming from the traces of the tensor densities at $q=-4$ and $q=-5$. It follows by the grading that
the components of fixed $p$ (the rows) are in $\mf{e}_{11}$ representations, whereas the components of fixed $q$ (the columns) are in representations of $W(11)$. The gauge potentials lie at degree $p=-2$, the field strengths at degree $p=-1$ and the derivatives at degree $p=1$.
The cross marks the fixed point of the reflection symmetry
explained in the text, mapping any $\mathfrak{gl}(11)$ representation to its conjugate; more precisely mapping the bi-degree $(p,q)$ to $(-2-p,-q-3)$.
This symmetry is most evident when one of the two entries mapped to
each other is dualised using the epsilon tensor (after a decomposition into irreducible parts). In the L-shaped area we have performed such a dualisation. See Table \ref{tab:THA3} for part of the tensor hierarchy algebra without performing such a dualisation.
}
\end{table}
\end{landscape}

We denote the tensor hierarchy algebra for $d=11$ defined in Appendix \ref{app:THA} by $\mathscr T$.
As described above for $d\leq 8$, it decomposes into a direct sum of $\mf{e}_{11}$ representations $R_p$ for all integers
$p$.\footnote{Our embedding of $\mf{e}_{11}$ into $\scr T$ is different from (in fact, conjugate to) the embedding of $\mf{e}_{d}$ into the tensor hierarchy algebras defined for $d\leq 8$
in \cite{Palmkvist:2013vya,Greitz:2013pua,Howe:2015hpa}. As a result, our representations
$R_1,R_2,\ldots$ for $d\leq 8$ are conjugate to those appearing in the tensor hierarchy. In particular, $R_1$ is here the conjugate of $\ell_1$ for $d\leq 8$ (and contains the conjugate of $\ell_1$ for $d=11$).}
This is a $\mathbb{Z}$-grading, $[R_i,R_j]\subset R_{i+j}$, which is consistent with the $\mathbb{Z}_2$-grading
that $\scr T$ has as a superalgebra (in the sense that $R_p$ is an odd subspace if $p$ is odd, and an even subspace if $p$ is even). We sometimes write commutators for ease of notation even though $\scr T$ is a Lie superalgebra, and $[R_i, R_j]$ then denotes a graded commutator, \ie either a commutator or an anti-commutator depending on the parity of the
product $ij$.
It follows from the $\mathbb{Z}$-grading that $R_0$ is a subalgebra of $\scr T$,
and we shall denote it by ${\mathfrak t}_{11}$.
Any representation $R_p$ of $\mathfrak{e}_{11} \subset \mf{t}_{11}$ can be further decomposed into representations of $\mf{gl}(11)$, and
this decomposition corresponds to
another $\mathbb{Z}$-grading of $\scr T$, which is not consistent in the sense above.
We choose this
other $\mathbb{Z}$-grading such that the degree $q$ is not equal to the
$\mf{gl}(11)$ level $\ell$, but related to it by
\begin{align}
\ell=q-\frac32 p\,.
\end{align}
Thus we have two different $\mathbb{Z}$-gradings of $\scr T$, with degrees $p$ and $q$.
To distinguish them from each other, 
we call them {\it vertical} and {\it horizontal}, respectively. This is in accordance with
Table~\ref{tab:THA}, where we show the decomposition of $R_p$ for vertical degree
$-3 \leq p \leq 3$ into representations of $\gl(11)$ for horizontal degree
$-5 \leq q \leq 2$.

A feature that the tensor hierarchy algebra $\scr T$ has in common with its analogues for $d\leq 8$ (up to a singlet at $p=-1$ for $d=8$) 
is the fact that it is 
conjugated to itself through the action of a (vector space) involution
such that for any vertical degree $p$, the representations $R_p$ and $R_{9-d-p}$ are conjugate to each other, $\bar R_p = R_{9-d-p}$.
This involution is related to the usual Hodge duality of the $(p+1)$-form field strengths of maximal supergravity in $D=11-d$ dimensions. In the case $d=11$ it maps $R_{-1}$ to itself, and as we will see, it can be used to generalise the duality relation for the
four- and seven-form field strengths in eleven-dimensional supergravity to a self-duality relation valid for field strengths living in the whole
of $R_{-1}$. In the further decomposition of the $\mf{e}_{d}$ representations $R_p$ into
$\gl(d)$ representations labelled by the horizontal degree $q$, the `reflection symmetry' of the algebra
(up to conjugation of the representations)
$p \leftrightarrow 9-d-p$ is refined to $(p,q)\leftrightarrow (9-d-p,-q-3)$, which for
$d=11$ means $(p,q)\leftrightarrow (-p-2,-q-3)$
as can be seen in Table \ref{tab:THA}.

As we will see in the next subsection, an important difference compared to the cases
$d\leq 8$ is that the representation
$R_0$ is \textit{not} the adjoint of $\mf{e}_{11}$.
It contains the adjoint as an
irreducible subrepresentation, but is not fully reducible. In other words, the Lie algebra 
${\mathfrak t}_{11}$
contains ${\mathfrak e}_{11}$ as a subalgebra, but is not semisimple; 
it is the semidirect sum of ${\mathfrak e}_{11}$ and an additional subspace. An example of a basis element in this additional 
subspace of $\mf{t}_{11}$ occurs at $(p,q)=(0,-3)$, where, 
in addition to the ${\mathfrak e}_{11}$ generator $F_{n_1\cdots n_8,m}$ with irreducible $(8,1)$ index structure,
$\mf{t}_{11}$ contains also an extra 9-index totally antisymmetric
generator $F_{n_1\cdots n_9}$.
This additional generator  $F_{n_1\cdots n_9}$ vanishes when the range of indices is restricted 
to $d\leq 8$, and it is a scalar density under $\gl(9)$ for $d=9$.\footnote{The extension of $\mf{e}_9$ with this additional generator, which is the Virasoro raising generator $L_1$, has been applied 
to gauged supergravity in two dimensions in~\cite{Samtleben:2007an}.} 
For $q<-3$
there will be further additional generators. However, the generators at $q\ge -2$
coincide with those of ${\mathfrak e}_{11}$, as we explain in Appendix \ref{app:THA}. Thus the Cartan involution of ${\mathfrak e}_{11}$ does not extend to
the whole of $\mf{t}_{11}$. 

In what follows it will be useful to introduce the generators of the subspaces $R_p$. Schematically they can be grouped as follows\footnote{Below and in the rest of the paper the indices $\alpha$ and $M$ refer to level $p=0$ and level $p=1$ generators of the tensor hierarchy algebra, respectively, and they contain generators in addition to those of $E_{11}$ and its $\ell_1$ representation described in section~\ref{sec:NLR}.} 
\begin{align}
\label{gens}
&\vdots &\vdots  
\nn\\
p &= 2 &P^{MN} & =\Pi^{MN}{}_{,\Xi} \; P^\Xi 
\nn\\
p &= 1 &P^{M}&
\nn\\
p&= 0 &t^\alpha &
\nn\\
p&= -1 &t^\alpha{}_M &= \Pi^{\alpha}{}_{M,I}\; t^I 
\nn\\
p&= -2 &t^\alpha{}_{MN} &= \Pi^{\alpha}{}_{MN,}{}^\beta \;{\widetilde t}_\beta 
\nn\\
p&= -3 &t^\alpha{}_{MNP} &= \Pi^{\alpha}{}_{MNP,}{}^Q \;P_Q 
\\
&\vdots  &\vdots \nn
\end{align}
where the $\Pi$ tensors are suitable linear homomorphisms. The (anti-)commutation rules that will be needed below in the construction of the theory are
\begin{align} 
\{P^M,P^N\} &= 2\Pi^{MN}{}_{,\Xi}\,  P^\Xi \ , \quad & [P^M , t^\alpha]  &=- D^{\alpha\,M}{}_N P^N \ , 
\nn\w2
\{  P^M , t^I \} &= f^M{}_{\alpha,J} \Omega^{JI} t^\alpha \  , \quad &[P^M,{\widetilde t}_\alpha] &= f^M{}_{\alpha,I}\,  t^I  \ , 
\nn\w2
\{P^M,P_N \} &= D^{\alpha\,M}{}_N\, {\widetilde t}_\alpha  \ , \quad &  [P^\Xi , P_M] 
&= f^\Xi{}_{M,I}\, t^I\ , 
\end{align}
where $D^{\alpha \, M}{}_{N}$ are the representation matrices of the Lie algebra 
$\mf{t}_{11}$ on $R_1$, and $\Omega^{IJ}$ is the inverse of the $R_0$ symplectic form on $R_{-1}$, such that the quadratic Casimir is
\be C_2 = \Omega_{IJ} t^I t^J +\{ \widetilde{t}_\alpha , t^\alpha\}   + [P_M , P^M ]  + \dots \ . 
\ee
The existence of this quadratic Casimir is proved in Appendix \ref{app:THA}. Note that it has weight $p=-2$ and corresponds to the ``reflection symmetry'' $(p,q)\leftrightarrow (-p-2,-q-3)$ discussed above.\footnote{Note that only the $\mathfrak{gl}(11)$ level $\ell = q-\frac{3}{2}p$ is defined by the action of an element of the superalgebra $\mathscr{T}$, and is therefore preserved by the Casimir operators.} 

An important point of the construction of the tensor hierarchy algebra is that it defines, along the vertical $\mathbb{Z}$-grading,
a differential complex of functions depending on coordinates $x^M$,
where the differential is defined through the adjoint action of the basis elements
$P^M$ in $R_1$ as 
\begin{align}
d = ({\rm ad}\,P^M)\,\partial_M\ . \label{Dext}
\end{align}

The requirement that this differential squares to zero,
\begin{align}
d^2 = ({\rm ad}\,P^M)\,({\rm ad}\,P^N) \,\partial_M \partial_N = \Pi^{MN}{}_{,\Xi}\, ({\rm ad}\,P^\Xi) \,\partial_M \partial_N = 0\ ,
\end{align}
is equivalent to the condition that all fields in the complex satisfy the weak section constraint (at the linearised level, the issue of a strong section constraint does not arise):
\be 
\Pi^{MN}{}_{,\Xi} \ \partial_M  \partial_N \Phi(x) = 0 \ . 
\ee
Note that one can equivalently define the standard de Rham complex from the graded abelian superalgebra freely generated by anticommuting variables $\theta^m$ of degree $1$ and commuting variables $x^m$ of degree $0$, such that the differential complex is the module of superfields $\omega(x,\theta)$ and $d = \theta^m \frac{\partial \, }{\partial x^m}$. A differential complex can still be defined for a non-abelian superalgebra, provided one enforces a section constraint ensuring that $d$ is nilpotent. The differential complex defined above will serve as a basis for the construction of the field equations in the next section, such that the degree $p=-3$ supports the gauge parameters, $p=-2$ the potentials, $p=-1$ the field strengths, and $p=0$ the Bianchi identities, as one can anticipate by looking at Table~\ref{tab:THA}. It might seem  counter-intuitive that the potentials do not belong to the degree $p=0$ component $\mf{t}_{11}$, which is a subalgebra of the tensor hierarchy algebra $\mathscr{T}$, but instead belong to a module in the co-adjoint representation of
$\mf{t}_{11}$. However, this definition is determined by the property that the exterior derivative \eqref{Dext} increases the degree $p$ by one unit, and the fact that gauge parameters are defined in the $\ell_1\subset R_{-3}$ module and the potentials in $\mf{e}_{11}\subset R_{-2}$. Note moreover that the functions in $R_{-2}$ are valued in the full representation without restriction whereas the physical potentials parametrise a coset, and are defined modulo $K(\mf{e}_{11})$ in the linearised approximation. To avoid confusion between the fields valued in $R_{-2}$ discussed in this section and the physical potentials $A^\alpha$, we shall denote the former by
$\phi^\alpha$. We define therefore the fields  
\be
\phi = \phi^\alpha(x) \tilde t_\alpha\ ,
\label{defmg}
\ee

\noindent their field strengths $\cF = d\phi$ at $p=-1$ as
\be
\cF_I = f^M{}_{\alpha,I}\ \partial_M \phi^\alpha\ ,
\label{fs2}
\ee
and their Bianchi identities at level $p=0$,
\begin{align}
 d\cF= \bigl( \Omega^{IJ} f^M{}_{\alpha,I} f^N{}_{\beta,J} \, \partial_M \partial_N \phi^\beta  \bigr) t^\alpha  = 0 \ ,
\end{align}
up to the section constraint. This field strength is by construction invariant under the gauge transformations $\delta^{\scriptscriptstyle \mathscr{T}}_\Xi \phi = d \Xi$, for a $p=-3$ gauge parameter 
\be
\Xi = \Xi(x)^M P_M\ ,
\label{defgp}
\ee
satisfying the section constraint. More explicitly, this gauge transformation takes the form
\be
\delta^{\scriptscriptstyle \mathscr{T}}_\Xi  \phi^\alpha = D^{\alpha\,M}{}_N \ \partial_M \Xi^N\ ,
\label{gt2}
\ee
and
\bea
\delta^{\mathscr T}_\Xi {\mathcal F}_I &=& f^M{}_{\alpha,I} D^{\alpha\,N}{}_P\, \partial_M \partial_N \Xi^P 
=f^{\Xi}{}_{P,I} \Pi^{MN}{}_{,\Xi} \ \partial_M \partial_N \Xi^P =0\ .
\label{gifs}
\eea
The second equality follows from the Jacobi identity $[\{P^M,P^N\},P_Q]+2[\{P^{(M},P_Q\},P^{N)}]=0$. The superscript ${\mathscr T}$ means that the variation is computed using the commutation relations of the tensor hierarchy algebra ${\mathscr T}$, as opposed to variations without the superscript, which we shall encounter later, corresponding to variations of coset fields that are compensated so that they remain in the coset. By construction the gauge transformations are infinitely reducible, and in the BRST formalism one can interpret the fields at lower degrees $p=-4,\, -5,\,$... as a sequence of ghosts for ghosts for the potentials at degree $p=-2$. Note that in the $\mathfrak{gl}(11)$ decomposition, the gauge invariance at a given horizontal degree $q$ are finitely reducible but in a $E_{11}$ covariant language we have an infinitely reducible gauge invariance.

As we shall see later, only $E_{11}$ is expected to be a symmetry of the full equations of motion, and furthermore only $K(\mf{e}_{11})$ at the linearised level. Therefore it is important to understand the $\mf{e}_{11}$ representation content of the tensor hierarchy algebra. At vertical degree $p=0$, the generators are
\be
t^\alpha = \left( t^{\alpha_0},\,t^{\alpha_1},\,t^{\alpha_2},\,\dots\right) \in  \left( \mathfrak{e}_{11},\,{\mfr}^\ord{0}_1,\,{\mfr}^\ord{0}_2,\, \dots\right)\ , 
\label{level0}
\ee
where the notation means that the total module is a vector space that decomposes into a direct sum of vector spaces (but not $\mf{e}_{11}$ representations) associated with the irreducible highest weight modules labeled by ${\mfr}^\ord{0}_i$.
Here we have the irreducible highest weight representations with Dynkin labels
\bea
\mfr^\ord{0}_1 &=& (0,1,0,0,0,0,0,0,0,0,0)\ ,
\nn\\
\mfr^\ord{0}_2 &=& (0,0,0,0,0,0,0,0,0,1,0)\ ,
\eea
according to the labeling conventions depicted in Figure \ref{fig:e11dynk}. In general we shall use the notation ${\mfr}^{(p)}_{i}$ to denote the representation labeled by $i$, at vertical degree $p$. The ellipses in (\ref{level0}) denote possible irreducible highest weight modules that could arise. Direct inspection of possible irreducible representations at low levels suggests that there may not be any other representation beyond the ones displayed in \eqref{level0}, but this remains to be fully investigated.
 
More precisely, the total module $R_0$ has the following structure. It is known to contain the adjoint representation of ${\mathfrak{e}}_{11}$. Furthermore, factoring out this representation yields a highest weight representation of ${\mathfrak{e}}_{11}$ in the sense that the highest weight state is annihilated by the ${\mathfrak{e}}_{11}$ raising operators, but the resulting weight space need not correspond to a single irreducible representation of ${\mathfrak e}_{11}$. The notation in \eq{level0} indicates that the weight space contains the representation ${\mfr}^\ord{0}_1$. Factoring out this representation, in turn, gives a new highest weight representation of ${\mathfrak e}_{11}$ which contains $\mfr^\ord{0}_2$ and so on. This structure of the module does not mean full reducibility of the
${\mathfrak e}_{11}$ representation $R_0$. Indeed we shall show that $(\mathfrak{e}_{11},\,\bar{\ell}_2)$ is indecomposable while we do not know yet if the remaining components decompose into irreducible higher highest weight modules. The fact that there exists an indecomposable module $(\mathfrak{e}_{11},\bar \ell_{2})$ seems to be related to the fact that the highest weight of $\bar{\ell}_{2}$ is a null root. One can show that the only $\mathfrak{gl}(11)$ irreducible representations in the level decomposition of $R_0$ that do not appear in $\mathfrak{e}_{11}$ itself are associated to the $\mathfrak{gl}(11)$ level decomposition of the highest weight representation $\bar{\ell}_2$ (or its multiples). This implies that one can have a non-trivial mixing of the two representations that cannot be reabsorbed into a redefinition of them.\footnote{For $\mf{e}_{9}$, the vector space replacing $\ell_2$  is one-dimensional, and the corresponding generator is the Virasoro raising operator
\cite{Samtleben:2007an}.} 
Further details can be found in Appendix~\ref{app:THA}. 
The only property that is ensured by the construction of the algebra is that $\mathfrak{e}_{11}$ is a subalgebra, such that we have the commutation relations 
\bea
[t^{\alpha_0}, t^{\beta_0}] &=& C^{\alpha_0\beta_0}{}_{\gamma_0}\, t^{\gamma_0}\ ,
\w2
[t^{\alpha_0}, t^{\beta_i}] &=& \sum_{j\ge 1} D^{\alpha_0\beta_i}{}_{\gamma_j}\, t^{\gamma_j} + T^{\alpha_0\beta_i}{}_{\gamma_0}\, t^{\gamma_0}\ , 
\label{alg0}
\eea
where $D^{\alpha_0\beta_i}{}_{\gamma_j}$ are representation matrices of $\mf{e}_{11}$, which, as a result of the Jacobi identity involving $\{t^{\alpha_0}, t^{\beta_0}, t^{\gamma_j}\}$, satisfy
\be \sum_{k\ge 1}   2 D^{[\alpha_0|\gamma_i}{}_{\eta_k}  D^{|\beta_0]\eta_k}{}_{\delta_j}   = C^{\alpha_0\beta_0}{}_{\eta_0} D^{\eta_0\gamma_i}{}_{\delta_j} \ , 
\ee
whereas $T^{\alpha_0\beta_i}{}_{\gamma_0}$ satisfy
\be
\sum_{j \ge 1} D^{[\alpha_0| \alpha_i}{}_{\beta_j} T^{|\beta_0]\beta_j}{}_{\gamma_0}
=C^{\delta_0[\alpha_0}{}_{\gamma_0} T^{\beta_0]\alpha_i}{}_{\delta_0}
+\frac12 C^{\alpha_0\beta_0}{}_{\delta_0} T^{\delta_0\alpha_i}{}_{\gamma_0}\ .
\ee

The $p=1$ generators decompose similarly as follows,
\be
P^M = \left( P^{M_0},\,P^{M_1},\,\dots\right) \in   \left( {\mfr}^\ord{1}_1,\,{\mfr}^\ord{1}_2,\, \dots\right)\ ,
\label{level1dec}
\ee
and similarly for $P_M$, where
\bea
\mfr^\ord{1}_1 &=& (1,0,0,0,0,0,0,0,0,0,0)\ ,
\nn\\
\mfr^\ord{1}_2 &=& (1,0,0,0,0,0,0,0,0,1,0)\oplus (0,0,0,0,0,0,0,0,0,1) \ .
\label{level1}
\eea
Thus, the $P^M$ and $P_M$ obey the commutation rules
\be
[t^{\alpha_0},P^{M_i}] =\sum_{j \ge 0} D^{\alpha_0M_i}{}_{N_j} P^{N_j}\, , \qquad [t^{\alpha_0}, P_{M_i}] = -\sum_{j\ge 0} D^{\alpha_0 N_j}{}_{M_i} P_{N_j} \ .
\ee
One derives then that the generators at $p=-2$ commute with
the $\mathfrak{e}_{11}$ generators as
\begin{align} 
\label{eq:CRm22}
[ t^{\alpha_0}, \tilde t_{\beta_0}] = - C^{\alpha_0\gamma_0}{}_{\beta_0} \tilde t_{\gamma_0} - \sum_{i\ge 1} T^{\alpha_0 \gamma_i}{}_{\beta_0} \tilde t_{\gamma_i} \ , \qquad [ t^{\alpha_0}, \tilde t_{\beta_i}] = - \sum_{j\ge 1} D^{\alpha_0\gamma_j}{}_{\beta_i} \tilde t_{\gamma_j} \ . 
\end{align}
We will see in the next section that the structure coefficients $T^{\alpha_0\beta_1}{}_{\gamma_0}$ do not vanish, so that the module $R_{-2}$ is not completely reducible. However, the structure coefficients computed in Appendix \ref{app:THA} satisfy 
\be 
T^{\alpha_0\beta_2}{}_{\gamma_0}  = D^{\alpha_0\beta_1}{}_{\gamma_2} = D^{\alpha_0\beta_2}{}_{\gamma_1} = 0 \ , 
\ee
so that the vector space associated to the highest weight $\bar \ell_{10}$ in \eqref{level0} does not mix as an $\mf{e}_{11}$ module with $(\mathfrak{e}_{11},\bar \ell_{2})$. It would be very useful if this extended to all higher highest weight components such that the $\mf{e}_{11}$ module $R_{-2}$ would decompose into the direct sum of an indecomposable module $(\mathfrak{e}_{11},\bar \ell_{2})$ and a (possibly reducible) module including all other components, {\it i.e.}
\be 
T^{\alpha_0\beta_i}{}_{\gamma_0}  = D^{\alpha_0\beta_1}{}_{\gamma_i} = D^{\alpha_0\beta_i}{}_{\gamma_1} = 0 \quad  \forall \; i \ge 2 \ , 
\label{h1}\ee
but this is not necessary for the model based on the tensor hierarchy algebra we are proposing.\footnote{We make the following observations that may be useful for studying the question of decomposability of the tensor hierarchy algebra. One can assign roots to the generators of $\scr T$ and the standard techniques of identifying possible $\mathfrak{gl}(11)$ representations associated with roots gives all the Young tableaux that are listed for example in~\cite[Table 2]{Nicolai:2003fw} or~\cite[App.~B.1]{Kleinschmidt:2003jf}. These give a complete list of possible Young tableaux that can occur at vertical degree $p=0$ and any fixed horizontal degree $q$; the only issue then is to determine what is called the outer multiplicity $\mu$ of a Young tableau. For $\mf{e}_{11}$ this can be done by computer based on the denominator formula; for the tensor hierarchy algebra we unfortunately do not have a similar structure at our disposal, so we have to do it by hand.
As was noted in~\cite{Nicolai:2003fw} the only places where $\mu=0$
occurs is for null roots of $\mf{e}_{11}$ (besides spurious real roots). This observation can be proven by noting that null roots always occur as special elements in the `gradient representations' triggered by the affine subalgebra $\mf{e}_{9}$ and by the fact that one knows the multiplicity of null roots (it is equal to eight). The first null root is the one that corresponds to the potential we call $X_9$. The inclusion of the corresponding
$\mf{e}_{11}$ representation $\bar \ell_2$ increases the outer multiplicity from $\mu=0$ in $\mf{e}_{11}$ to $\mu=1$ in the tensor hierarchy algebra for this and Weyl related null roots. Continuing now to the next additional representations that we add we encounter $\bar \ell_{10}$ on $p=0$. This is not a null root (all dominant null roots are of the form $X_9$, $X_{9,9}$, etc.) and therefore starts out with $\mu>0$ in $\mf{e}_{11}$; this $\mu$ gets even bigger in the tensor hierarchy algebra. By forming suitable linear combinations one should be able to find a lowest weight vector in this larger space that allows to split off a lowest weight representation as a direct summand. An instance of this can be seen in~\eqref{Reducible}.}
The structure coefficients computed in Appendix \ref{app:THA} also satisfy  
\be 
D^{\alpha_0M_0}{}_{N_1} =  D^{\alpha_0M_1}{}_{N_0}  = 0 \ , 
\ee
implying that the vector space associated to $\mfr^\ord{1}_2$ in \eq{level1}
does not mix as an $\mf{e}_{11}$ module with $\mfr^\ord{1}_1$. So once again, it would be very useful if the module $R_1$ decomposed into the direct sum
of the irreducible module $\bar \ell_1$ and a possibly reducible module associated to the remaining highest weights, {\it i.e.} 
\be 
D^{\alpha_0M_0}{}_{N_i} =  D^{\alpha_0M_i}{}_{N_0}  = 0 \quad  \forall\;  i \ge 1 \ . 
\label{h2}
\ee
We shall \textit{assume} this condition even though it is not ruled out that it may not be necessary.  Unlike the \eq{h1}, this condition plays a more important role in the construction of the linearised field equations, as we shall see in the next section. Then, decomposing the potential as $\phi^\alpha = ( \phi^{\alpha_0}, X^{\alpha_i})$, the field strength takes the form
\be
\cF_I = f^{M_0}{}_{\alpha_0,I}\ \partial_{M_0} \phi^{\alpha_0} + f^{M_0}{}_{\alpha_1,I}\ \partial_{M_0} X^{\alpha_1} + \sum_{i\ge 2}f^{M_0}{}_{\alpha_i,I}\ \partial_{M_0} X^{\alpha_i} \ .
\label{fs2E11}
\ee
If \eqref{h1} were to hold as well, one could truncate the system consistently by setting $X^{\alpha_i}=0$ for $i\ge 2$ keeping $E_{11}$ symmetry. 

The $\mf{e}_{11}$ module $R_{-1}$ may also be reducible, in which case we may want to consider only the field strength associated to a minimal indecomposable module. However, this is neither a highest weight nor a lowest weight $\mf{e}_{11}$ module, and there is not much known about the classification of such Kac--Moody algebra modules. 

Given our assumption~\eqref{h2}, there exists a standard non-degenerate bilinear invariant form $M^{M_0N_0}$ on the $\bar \ell_1$ representation that we will use below in the construction of the field equations. We remark that if our assumption~\eqref{h2} was not valid, we would require the existence of a similar non-degenerate invariant bilinear form $M^{MN}$ on all of $R_1$ to construct our theory. In this case, the restriction of $M^{MN}$ to the space indexed by $M_0$ will not agree with the standard bilinear form. As a matter of fact, in our truncation scheme, the difference will not be visible as all the higher level representations mentioned above will be beyond our $\mf{gl}(11)$ level truncation, and only the lowest $\mf{gl}(11)$-level component of $X^{\alpha_1}$ will appear to play an important role.

To define the field equations, one needs eventually to quotient by the right $K(E_{11})$ action to define the theory. Nonetheless, the differential complex described above will serve as a main building block in the construction to be discussed in Section~\ref{sec:THA2}. At this level, $\phi^\alpha$ is still understood as an element of the algebra without constraints, and all quantities are in $\mf{e}_{11}$ representations. This provides a huge simplification, because the construction of the field strength $\cF$ is consistent with the $\mf{gl}(11)$ level (so that the horizontal degree $q$ is preserved unlike in the scheme described in Section \ref{PeterThing} where it is not). In the next subsection we shall exploit this property to present explicit formulas for the transformations and field strengths, and address the problem of defining field equations in the subsequent section.

%%%%%%%%%%%%%%%%%%%%%%%%%%%%%%%%%%%%%%%%%%%%%%%%%%%%%%%%%%%%%%%
\subsection{Explicit formulas for transformations}
%%%%%%%%%%%%%%%%%%%%%%%%%%%%%%%%%%%%%%%%%%%%%%%%%%%%%%%%%%%%%%%

The full structure of the tensor hierarchy algebra $\mathscr{T}$
described above, and defined in Appendix \ref{app:THA}, is not known but 
we can probe it degree by degree both horizontally and vertically. Recall that the horizontal degree $q$
is related to the $\mf{gl}(11)$ level $\ell$ by $q=\ell+\frac32p$ and $\ell$ is the eigenvalue of the Cartan generator $\tfrac13 K^m{}_m$ of $\mf{e}_{11}$. In this section, we shall give the explicit form of the structure coefficients $D^{\alpha\,M_0}{}_{N_0}$, $f^{M_0}{}_{\alpha,I}$ and $\Pi^{M_0N_0}{}_\Xi$, and the explicit transformations of $\phi^\alpha$, $\mathcal{F}_I$ with respect to $\mf{e}_{11}$ in the level truncation we are working with. We recall from the previous section that the `potential' fields $\phi^\alpha$ are associated with vertical degree $p=-2$ in the tensor hierarchy algebra that is dual to  $p=0$. At this stage we do not perform a coset construction, \ie there will be fields associated with \textit{all} generators at $p=-2$. Furthermore, the derivatives, gauge parameters and field strengths are associated with vertical degrees $p=1$, $p=-3$ and $p=-1$, respectively. The assignments of horizontal degrees $q$ within these vertical ones are summarized in the table below.
%Table \ref{tab:pq-degrees}.

\renewcommand{\arraystretch}{1.2}
\begin{table}[ht]
\centering
\begin{tabular}{r|c||c|c|c|c}
&& $q=-3$ & $q=-2$ & $q=-1$ & $q=0$ \\\hline
section constraint & $p=2$ & $L^{n_1\dots n_6,m}$, $L^{n_1\dots n_7}$ & $L^{n_1\dots n_4}$ & $L^n$ &  \\
derivative &  $p=1$ & $\partial^{n_1\ldots n_7,m}$, $\partial^{n_1\ldots n_8}$ &$\partial^{n_1\ldots n_5}$ & $\partial^{n_1n_2}$ & $\partial_m$ \\
field strength &  $p=-1$ & $\cF_{n_1n_2}{}^m$ & $\cF_{n_1\ldots n_4}$ & $\cF_{n_1\ldots n_7}$ & $\cF_{n_1\ldots n_9,m}$, $\cF_{n_1\ldots n_{10}}$ \\
potential &  $p=-2$ & $h^+_n{}^m$ & $A^+_{n_1n_2n_3}$ & $A^+_{n_1\ldots n_6}$ & $h^+_{n_1\ldots n_8,m}$, $X_{n_1\ldots n_9}$\\
gauge parameter &  $p=-3$ & $\xi^m$ & $\lambda_{n_1n_2}$ & $\lambda_{n_1\ldots n_5}$ & $\xi_{n_1\ldots n_7,m}$, $\lambda_{n_1\ldots n_8}$
\end{tabular}
%\caption{\label{tab:pq-degrees} \small\textit{$q$- and $p$-degrees of selected objects in the tensor hierarchy algebra.}}
\end{table}

In order to exhibit the global $E_{11}$ transformations
\be 
\delta \phi^\alpha = - C^{\beta_0\alpha}{}_\gamma u_{\beta_0}  \phi^\gamma \ , 
\ee
it suffices to study the infinitesimal transformations under the level $\ell=q=\pm 1$ generators $E^{n_1n_2n_3}$ and $F_{n_1n_2n_3}$ of $\mf{e}_{11}$ as the higher and lower level transformations can be obtained by iteration/commutation. We denote the parameters $u_{\alpha_0}$ of these transformations by $e_{n_1n_2n_3}$ and $f^{n_1n_2n_3}$, respectively. More precisely, we write the general element at $p=-2$ as\footnote{For the tensor hierarchy algebra at $p=-2$ we write the coordinate associated with the dual of $\mf{gl}(11)$ as $h_n^+{}^m$; it has no particular symmetry properties and so it is akin to the quantity $\varphi_n{}^m$ appearing in the parametrisation of the adjoint of $\mf{e}_{11}$ in~\eqref{eq:GE}.}
\begin{align}
\label{eq:phiexp}
\phi^\alpha \tilde{t}_\alpha &= \ldots + \frac{1}{8!} h_-^{n_1\ldots n_8,m} \tilde{F}_{n_1\ldots n_8,m} + \frac1{6!} A_-^{n_1\ldots n_6}\tilde{F}_{n_1\ldots n_6} + \frac1{3!} A_-^{n_1n_2n_3} \tilde{F}_{n_1n_2n_3} + h^+_n{}^m \tilde{K}^n{}_m \nn\\
&\quad + \frac{1}{3!} A^+_{n_1n_2n_3} \tilde{E}^{n_1n_2n_3} + \frac1{6!} A^+_{n_1\ldots n_6} \tilde{E}^{n_1\ldots n_6} + \frac1{8!} h^+_{n_1\ldots n_8,m} \tilde{E}^{n_1\ldots n_8,m} \nn\\
&\quad + \frac{1}{8!} X_{n_1\ldots n_9} \tilde{E}^{n_1\ldots n_9} + \ldots \,,
\end{align}
where the term in the last line corresponds to the new generator $\tilde{E}^{n_1\ldots n_9}$ with coefficient $X_{n_1\ldots n_9}$ in the tensor hierarchy algebra that is not present in $\mf{e}_{11}$ and that is totally antisymmetric in its nine indices. The transforming rigid $\mf{e}_{11}$ element at level $p=0$ we take as
{\allowdisplaybreaks{\begin{align}
\frac1{3!} f^{n_1n_2n_3} F_{n_1n_2n_3} + \frac1{3!} e_{n_1n_2n_3} E^{n_1n_2n_3} 
\end{align}
and the important new commutator in the tensor hierarchy algebra is
\begin{align}
\label{eq:X9CR}
[E^{n_1n_2n_3} , \tilde{E}^{p_1\ldots p_6}] = -3  \tilde{E}^{n_1n_2n_3p_1\ldots p_6} - 3 \tilde{E}^{p_1\ldots p_6[n_1n_2,n_3]} \,,
\end{align}
whose dual version was given in~\eqref{eq:CXnew}. From the $\mf{e}_{11}$ commutation relations given in Appendix~\ref{app:conv}, and those of the tensor hierarchy algebra given in Appendix~\ref{app:THA}, we derive the following rigid $E_{11}$ transformations at $p=0$ of the `potentials' lying at $p=-2$: 
\begin{subequations}
\label{adjoint}
\begin{align} 
\delta {h}_-^{n_1\cdots n_8,m} &= -56 f^{\lsharp n_1n_2n_3}A_-^{n_4\cdots n_8,m\rsharp} + \cdots \ ,
\\
\delta {A}_-^{n_1\cdots n_6} &= -20 f^{[n_1n_2n_3} A_-^{n_4n_5n_6]} 
+\frac{1}{2} e_{n_7n_8 n_9} h_-^{n_1\cdots n_8,n_9} \ ,
\\
\delta A_-^{n_1n_2n_3} &= \frac{1}{6} e_{p_1p_2p_3} 
  A_-^{n_1n_2n_3p_1p_2p_3} + 3 f^{p[n_1n_2} h^+_p{}^{n_3]} \,,
\\
\delta h^+_n{}^m &=\frac{1}{2}e_{np_1p_2} A^{mp_1p_2}_- - \frac{1}{2} f^{mp_1p_2} A^+_{np_1p_2}\nn\\ &\quad\, 
- \frac{1}{18} \delta_n^m \scal{e_{p_1p_2p_3}  A_-^{p_1p_2p_3} - f^{p_1p_2p_3} A^+_{p_1p_2p_3}}\,, 
\\
\delta A^+_{n_1n_2n_3} &= -\frac{1}{6} f^{p_1p_2p_3} A^+_{n_1n_2n_3p_1p_2p_3} 
- 3 e_{p[n_1n_2} h^+_{n_3]}{}^{p}\,,
\\
\delta A^+_{n_1\cdots n_6} &= 20 e_{[n_1n_2n_3} A^+_{n_4n_5n_6]} 
-\frac{1}{2} f^{n_7n_8n_9} h^+_{n_1\cdots n_8,n_9}  \,,
\\
\delta h^+_{n_1\cdots n_8,m} &= 56 e_{\lsharp n_1n_2n_3} A^+_{n_4\cdots n_8,m\rsharp} + \cdots\,, \\
\label{eq:deltaX}
\delta X_{n_1\cdots n_9} &= - 28 e_{[n_1n_2n_3} A^+_{n_4\cdots n_9]} + \cdots\,.
\end{align}
\end{subequations}

As can be seen in \eqref{eq:phiexp} and in the table above, 
fields with the superscript $+$ or subscript $-$ belong to the part of $R_{-2}$ 
corresponding to the adjoint of $\mf{e}_{11}$ at
$\ell\geq 0$ and $\ell<0$ (that is, $q\geq -3$ and $q < -3$), respectively.
The transformation rules of the latter are obtained from the former ones by raising and lowering all the indices and by interchanging the parameters $e \leftrightarrow -f$.
The fields in the additional part of $R_{-2}$ appear at $\ell \geq 3$ (in particular $X_{n_1\ldots n_9}$ at $\ell=3$)
and have no counterparts at negative levels.
Note that this is the transformation of fields in the whole of $R_{-2}$ (not yet the physical potential associated to the non-linear realisation), so that the
$\mf{gl}(11)$ level $\ell$ is preserved. The parameters $e_{n_1n_2n_3}$ and $f^{n_1n_2n_3}$ have levels $\ell=1$ and $\ell=-1$, respectively, and the fields have $\ell=(N-M)/3$ where $N$ is the number of lower indices and $M$ is the number of upper indices. The ellipses in some of the equations indicate contributions from level $\ell=4$ and $\ell=-4$ fields which are outside the range we are considering.

The most important new ingredient in~\eqref{adjoint} for the tensor hierarchy algebra is the last equation~\eqref{eq:deltaX} that gives the transformation of the new potential $X_{n_1\ldots n_9}$ that is present in the tensor hierarchy algebra but not in ${\mathfrak e}_{11}$. As we can see it transforms back into $\mathfrak{e}_{11}$ under the action of ${\mathfrak e}_{11}$, illustrating the fact that $R_{-2}$ is not the direct sum of the adjoint of ${\mathfrak e}_{11}$ and some other representation of ${\mathfrak e}_{11}$. This crucial fact is necessary to obtain the correct linearised equations of motion in the following section.

The local gauge transformation \eq{gt2}, given by the structure coefficients $D^{\alpha M}{}_N$ shown in Appendix~\ref{app:l1}, more explicitly read as follows:
\begin{subequations}
\label{eq:Xitrm}
\begin{align}
\delta^{\scriptscriptstyle \mathscr{T}}_\Xi {h}_-^{n_1...n_8,m} &= -8\partial^{[n_1\cdots n_7,|m|}\,\xi^{n_8]} 
+\frac{8}{3}  \partial^{m[n_1\cdots n_7}\,\xi^{n_8]} - \frac{8}{3} \partial^{n_1\cdots n_8 }\,\xi^{m} +\cdots \ ,
\\
\delta^{\scriptscriptstyle \mathscr{T}}_\Xi {A}_-^{n_1...n_6} &= -6\partial^{[n_1...n_5}\xi^{n_6]} 
- \partial^{n_1...n_6p,q}\lambda_{pq}
+\partial^{n_1...n_6p_1p_2}\lambda_{p_1p_2} +\cdots \ ,
\\
\delta^{\scriptscriptstyle \mathscr{T}}_\Xi A_-^{n_1n_2n_3} &= 3 \partial^{[n_1n_2} \xi^{n_3]} 
+ \frac{1}{2} \partial^{n_1n_2n_3p_1p_2} \lambda_{p_1p_2} 
+ \frac{1}{4!} \partial^{n_1n_2n_3p_1\cdots p_4,p_5} \lambda_{p_1\cdots p_5} 
\\
&\quad-  \frac{1}{5!} \partial^{n_1n_2n_3p_1\cdots p_5} \lambda_{p_1\cdots p_5} +\cdots \ ,
\nn \\
\delta^{\scriptscriptstyle \mathscr{T}}_\Xi h^+_n{}^m &= \partial_n \xi^m - \partial^{mp} \lambda_{np} 
- \frac{1}{4!} \partial^{mp_1\cdots p_4} \lambda_{np_1\cdots p_4} 
-\frac{1}{6!} \partial^{mp_1\cdots p_6,q} \xi_{np_1\cdots p_6,q} 
\\*
&\quad - \frac{1}{7!} \partial^{p_1\cdots p_7,m} \xi_{p_1\cdots p_7,n}  
 - \frac{8}{7!} \partial^{mp_1\cdots p_7} \lambda_{np_1\cdots p_7} 
\nn\\*
& \hspace{-5mm} + \frac{1}{3} \delta_n^m \left(\frac{1}{2} \partial^{p_1p_2} \lambda_{p_1p_2}
+\frac{2}{5!} \partial^{p_1\cdots p_5} \lambda_{p_1\cdots p_5} 
+ \frac{3}{7!} \partial^{p_1\cdots p_7,q} \xi_{p_1\cdots p_7,q} 
+ \frac{3}{7!} \partial^{p_1\cdots p_8} \lambda_{p_1\cdots p_8}\right) +\cdots \ ,\nn
\\*
\delta^{\scriptscriptstyle \mathscr{T}}_\Xi A^+_{n_1n_2n_3} &= 3 \partial_{[n_1} \lambda_{n_2n_3]} 
+ \frac{1}{2} \partial^{p_1p_2}\lambda_{n_1n_2n_3p_1p_2} 
+ \frac{1}{4!} \partial^{p_1\cdots p_5} \xi_{n_1n_2n_3p_1\cdots p_4,p_5}\nn\\
&\quad - \frac{1}{5!} \partial^{p_1\cdots p_5} \lambda_{n_1n_2n_3p_1\cdots p_5}+\cdots \ ,
\\
\delta^{\scriptscriptstyle \mathscr{T}}_\Xi A^+_{n_1\cdots n_6} &= 6 \partial_{[n_1} \lambda_{n_2\cdots n_6]} 
- \partial^{p_1p_2} \xi_{n_1\cdots n_6p_1,p_2} 
+  \partial^{p_1p_2} \lambda_{n_1\cdots n_6p_1p_2}+\cdots \ ,
\\
\delta^{\scriptscriptstyle \mathscr{T}}_\Xi h^+_{n_1\cdots n_8,m} &= 8 \partial_{[n_1} \xi_{n_2\cdots n_8],m}
+24 \partial_{\lsharp n_1} \lambda_{n_2\cdots n_8 ,m\rsharp}+\cdots \ ,
\\
\delta^{\scriptscriptstyle \mathscr{T}}_\Xi X_{n_1\cdots n_9}  &= 24 \partial_{[n_1} \lambda_{n_2\cdots n_9]}  + \cdots \ .
\end{align}
\end{subequations}
In these gauge transformations one has again a preservation of the horizontal degree $q$.
Note that there is no additional gauge parameter for the potential $X_{n_1\ldots n_9}$ and the transformation of the latter only involve the parameter $\lambda_{n_1\ldots n_8}$ that already enters in the transformation of the dual graviton $h_{n_1\ldots n_8,m}$.  The ellipses denote terms involving derivatives of level $\ell < -\frac{9}{2}$ or gauge parameters of level $\ell > \frac{9}{2}$, that are ignored in our computations. The first new gauge parameters that are in the $\mfr^\ord{1}_2$ module are $\xi_{10,1},\, \lambda_{11}$ and only appear at level $\ell = \frac{11}{2}$.}}

The structure coefficients $f^M{}_{\alpha, I}$, occurring in the definition of the field strengths ${\cal F}^I$ given in \eq{fs2}, are determined by the Bianchi identities, and equivalently by the property that the field strength \eqref{gt2} is gauge invariant modulo the section constraint
\begin{subequations} \label{SCall}
\begin{align}
L^m &= \partial^{mn} \partial_n\ , \label{SC1} \\
L^{n_1n_2n_3n_4} &= 3 \partial^{[n_1n_2} \partial^{n_3n_4]} - \partial^{n_1n_2n_3n_4m}‚\partial_{m} \label{SC2} \ , 
\\ 
L^{n_1n_2n_3n_4n_5n_6,m} &= 15 \partial^{\lsharp n_1n_2} \partial^{n_3n_4n_5n_6,m\rsharp} 
- \partial^{p\lsharp n_1n_2n_3n_4n_5n_6,m\rsharp}‚\partial_{p} \label{SC3} \ , \\ 
L^{n_1n_2n_3n_4n_5n_6n_7} &= 3 \partial^{[n_1n_2} \partial^{n_3n_4n_5n_6n_7]} 
-\frac{3}{7} \partial^{n_1n_2n_3n_4n_5n_6n_7,m}‚\partial_{m} + \partial^{n_1n_2n_3n_4n_5n_6n_7m} \partial_m  \label{SC4} \ ,
\end{align}
\end{subequations}
and transforms to itself with respect to $E_{11}$. Using these constraints and the known $\mf{gl}(11)$ irreducible representations appearing at each horizontal degree $q$ in $R_{-1}$, one computes in this way that 
{\allowdisplaybreaks{
\begin{subequations}
\label{ncfs}
\begin{align}
\cF^{n_1\dots n_8} &= 5 \partial^{[n_1n_2} A_-^{n_3\dots n_8]} + 16 \partial^{[n_1\dots n_5} A_-^{n_6n_7n_8]}- \frac{18}{7} \partial^{[n_1\dots n_7|,q} h^+_q{}^{|n_8]}  - 6 \partial^{q[n_1\dots n_7}   h^+_q{}^{n_8]} +\dots\\*
\cF_m{}^{n_1\dots n_8,p} &= \partial_m h_-^{n_1\dots n_8,p}+ \frac{56}{33} \Bigl( 8 \delta_m^{[n_1} \partial^{|p|n_2} A_-^{n_3\dots n_8]} + 9 \delta_m^{[n_1} \partial^{n_2n_3} A_-^{n_4\dots n_8]p}  -   \delta_m^p \partial^{[n_1n_2}  A_-^{n_3\dots n_8]}
\nn \\*
&\hspace{5mm}+20 \delta_m^{[n_1} \partial^{|p|n_2\dots n_5} A_-^{n_6n_7n_8]}  + 21  \delta_m^{[n_1} \partial^{n_2\dots n_6} A_-^{n_7n_8]p} -  \delta_m^p \partial^{[n_1\dots n_5}A_-^{n_6n_7n_8]}
 \nn \\*
&\hspace{5mm}  -22 \delta_m^{[n_1}\partial^{n_2\dots n_7|q,p|}h^+_q{}^{n_8]} + \delta_m^{[n_1}\partial^{n_2\dots n_7|p,q|}h^+_q{}^{n_8]} - \frac{12}{7} \delta_m^{[n_1}\partial^{n_2\dots n_8],q}h^+_q{}^{p} \nn\\*
&\hspace{5mm} -\frac{3}{7}  \delta_m^{p} \partial^{[n_1\dots n_7|,q}h^+_q{}^{n_8]}
 + 5 \delta_m^{[n_1} \partial^{n_2\dots n_7|pq|}h^+_q{}^{n_8]} + 4 \delta_m^{[n_1} \partial^{n_2\dots n_8]q}h^+_q{}^{p}
 \nn\\*
&\hspace{5mm}+ \delta_m^p \partial^{[n_1\dots n_7|q}h^+_q{}^{n_8]} \Bigr) +8 \partial^{[n_1\dots n_7|,p} h^+_m{}^{|n_8]} + 3 \partial^{\lsharp n_1\dots n_8,} h^+_m{}^{p\rsharp}  +\dots  \ , \\ 
\cF^{n_1n_2n_3n_4,m} &= - 6 \partial^{\lsharp n_1n_2} A_-^{n_3n_4,m\rsharp} 
+ \partial^{q\lsharp n_1n_2n_3n_4,} h^+_q{}^{m\rsharp} 
+ \frac{1}{6} \partial^{p_1p_2p_3\lsharp n_1n_2n_3n_4,m\rsharp} A^+_{p_1p_2p_3} +\dots \,,\\*
\cF_m{}^{n_1\cdots n_6} &= \partial_m A_-^{n_1n_2n_3n_4n_5n_6} 
+6 \partial^{[n_1\cdots n_5} h^+_m{}^{n_6]} + \partial^{n_1\cdots n_6p,q} A^+_{mpq} 
- \partial^{n_1\cdots n_6p_1p_2} A^+_{mp_1p_2}\nn\\*
&\quad +12 \delta_m^{[n_1} \Bigl( \partial^{n_2n_3} A_-^{n_4n_5n_6]} - \partial^{n_2n_3n_4n_5|q} h^+_q{}^{n_6]} 
+ \frac{3}{20} \partial^{n_2\cdots n_6]p_1p_2,q} A^+_{p_1p_2q}\nn\\*
& \hspace{60mm}  - \frac{1}{12}  \partial^{n_2\cdots n_6]p_1p_2p_3} A^+_{p_1p_2p_3}  \Bigr) +\ldots\,,\\
\cF^{n_1,n_2} &= \partial^{q(n_1} h^+_q{}^{n_2)} + \frac{1}{6!} \partial^{p_1p_2p_3p_4p_5p_6(n_1,n_2)} A^+_{p_1p_2p_3p_4p_5p_6}  +\dots\,,\\
 \cF_m{}^{n_1n_2n_3} &= - \partial_m A_-^{n_1n_2n_3} + 3 \partial^{[n_1n_2} h^+_m{}^{n_3]} 
 + \frac{1}{2}\partial^{n_1n_2n_3p_1p_2} A^+_{mp_1p_2}  \nn\\
 &\quad + \frac{1}{4!}    \partial^{n_1n_2n_3p_1p_2p_3p_4,q} A^+_{mp_1p_2p_3p_4q} 
 -  \frac{1}{5!} \partial^{n_1n_2n_3p_1p_2p_3p_4p_5} A^+_{mp_1p_2p_3p_4p_5}
 \nn\\
 &\quad + \frac{3}{2} \delta_m^{[n_1} \Bigl( \partial^{n_2|q} h^+_q{}^{n_3]} 
 - \frac{1}{6} \partial^{n_2n_3]p_1p_2p_3} A^+_{p_1p_2p_3} - \frac{3}{2\cdot 5!}  \partial^{n_2n_3]p_1\dots p_5,q} A^+_{p_1\dots p_5q}\nn\\
& \hspace{40mm}  
+\frac{1}{6!}  \partial^{n_2n_3]p_1\dots p_6} A^+_{p_1\dots p_6} \Bigr) +\dots \,, \\
\cF_{n_1n_2}{}^m &= 2 \partial_{[n_1} h^+_{n_2]}{}^m + \partial^{mp} A^+_{n_1n_2p} 
+ \frac{1}{4!} \partial^{mp_1\cdots p_4} A^+_{n_1n_2 p_1\cdots p_4} \nn\\
&\quad + \frac{1}{6!} \scal{\partial^{mp_1\cdots p_6,q} + \partial^{p_1\cdots p_6 q,m}} h^+_{n_1n_2p_1\cdots p_6,q} 
\nn\\
&\quad+ \frac{1}{3}\delta^m_{[n_1} \Bigl( \partial^{p_1p_2} A^+_{n_2]p_1p_2} 
+ \frac{4}{5!} \partial^{p_1\cdots p_5} A^+_{n_2]p_1\cdots p_5}
\nn\\
& \hspace{20mm} + \frac{6}{7!} \partial^{p_1\cdots p_7,q} h^+_{n_2]p_1\cdots p_7,q} 
+\frac{2}{7!} \partial^{p_1\cdots p_8} h^+_{n_2]p_1\cdots p_7, p_8}\Bigr) 
\nn\\
&\quad - \frac{1}{7!} \partial^{p_1\cdots p_7,m} X_{n_1n_2p_1\cdots p_7}  
+ \frac{3}{7!} \partial^{mp_1\cdots p_7} X_{n_1n_2p_1\cdots p_7}\nn\\
&\hspace{40mm} + \frac{1}{12\cdot 9!}\delta^m_{[n_1} \partial^{p_1\cdots p_8}X_{n_2]p_1\cdots p_8}+\dots \ ,\\
\cF_{n_1n_2n_3n_4} &= 4 \partial_{[n_1} A^+_{n_2n_3n_4]} 
- \frac{1}{2}\partial^{p_1p_2} A^+_{n_1n_2n_3n_4p_1p_2} 
-\frac{1}{4!} \partial^{p_1p_2p_3p_4p_5} h^+_{n_1n_2n_3n_4p_1p_2p_3p_4,p_5} 
\nn\\
&\quad +\frac{1}{5!} \partial^{p_1p_2p_3p_4p_5} X_{n_1n_2n_3n_4p_1p_2p_3p_4p_5}+\dots \ ,
\\
\cF_{n_1\cdots n_7} &= 7 \partial_{[n_1} A^+_{n_2\cdots n_7]} + \partial^{p_1p_2} h^+_{n_1\cdots n_7 p_1,p_2}  
-\frac{1}{2} \partial^{p_1p_2} X_{n_1\cdots n_7p_1p_2}+\dots \ ,
\\
\cF_{n_1\cdots n_9,m} &= 9 \partial_{[n_1} h^+_{n_2\cdots n_9],m} 
+\partial_{\lsharp m,} X_{n_1\cdots n_9\rsharp}+\dots \ ,
\\
\cF_{n_1\cdots n_{10}} &= \partial_{[n_1} X_{n_2\cdots n_{10}]}+\dots \ .
\end{align}
\end{subequations}
where the ellipses denote terms of order $\mathcal{O}(4,4)$, that is, involving either potentials of horizontal degree outside the range $-6\leq q\leq 0$ or derivatives of horizontal degree $q<-3$.
As noted earlier, $q$ is preserved in the expressions for the gauge invariant field strengths. Indeed, reading off the horizontal degrees from Table~\ref{tab:THA}, we note that the field strengths listed above have $q=-6,-5,-5,-4,-4,-3,-2,-1,0,0$, respectively. Note that the list of field strengths displayed above is exhaustive for $-5\le q\le 0$, however, there are other field strengths at $q=-6$ in the reducible
representation $(6,2)+(7,1)$. We do not display these three irreducible components because they do not depend on the dual graviton field $h_{8,1}$, and they will not be relevant in the following. The components in $(7,1)+(8)$ are determined from conditions that will be explained in the next section.}}

It is worth noting that if we restrict the range of the indices to run from $m=0,1,\ldots,7$ in~\eqref{ncfs}, the terms depending on the nine-form potential vanish and the expressions for the symmetry transformations as well as the field strengths discussed above reduce to those one would obtain from the embedding tensor representation of $\mf{e}_8$. The field strength representation can be defined using $\mf{gl}(11)$ tensor calculus and demanding that it is gauge invariant modulo the section constraint and transforms to itself under $\mf{e}_{11}$. It appears that this construction faces an obstruction if one restricts oneself to an ansatz depending on the fields in $\mathfrak{e}_{11}$ only, so that the necessity of introducing a nine-form comes naturally in the construction. So we want to stress that the nine-form does not come only as a consequence of the construction of this representation based on the tensor hierarchy algebra, but is in fact a consequence of the requirement that there exists
such an $\mf{e}_{11}$ representation in which $\partial \phi$ is indeed gauge invariant modulo the section constraint.

Using the definitions \eqref{ncfs}, one computes indeed that the field strengths transform as
\bea \label{E11varyF}
\delta \cF^{n_1\dots n_8} &=&   5\,  f^{q[n_1n_2} \cF_q{}^{n_3\dots n_8]} +\dots\ , \CR
\delta \cF_m{}^{n_1\dots n_8,p} &=& - 56 f^{\lsharp n_1n_2n_3} \cF_m{}^{n_4\dots n_8,p\rsharp} + \frac{56}{33}\Bigl( 8 \delta_m^{[n_1} f^{|qp|n_2}\cF_q{}^{n_3\dots n_8]}   +9 \delta_m^{[n_1} f^{|q|n_2n_3}  \cF_q{}^{n_4\dots n_8]p}\CR
&& \hspace{70mm}  - \delta_m^p f^{q[n_1n_2} \cF_q{}^{n_3\dots n_8]}    \Bigr) + \dots \ ,\CR
\delta \cF^{n_1n_2n_3n_4,m} &=& 6 f^{ p\lsharp n_1n_2} \cF_p{}^{n_3n_4,m\rsharp } 
- 4 f^{[n_1n_2n_3}\cF^{n_4],m} +\cdots \ ,\CR
\delta \cF_{m}{}^{n_1\cdots n_6} &=& 20 f^{[n_1n_2n_3} \cF_{m}{}^{n_4n_5n_6]}  
-12  \delta_m^{[n_1} f^{p|n_2n_3} \cF_p{}^{n_4n_5n_6]}  + \frac12 e_{p_1p_2q} \cF^{n_1\dots n_6p_1p_2,q} + \dots  \ ,
\CR
\delta \cF^{m,n} &=& \frac{1}{2} f^{p_1p_2(m} \cF_{p_1p_2}{}^{n)} 
- \frac{1}{6} e_{p_1p_2p_3} \cF^{p_1p_2p_3(m,n)}\ , \CR
\delta \cF_{m}{}^{n_1n_2n_3} &=& -3 f^{p[n_1n_2} \cF_{mp}{}^{n_3]} 
+ \frac{3}{4} f^{p_1p_2[n_1} \delta_m^{n_2} \cF_{p_1p_2}{}^{n_3]} 
- \frac{1}{6}e_{p_1p_2p_3} \cF_m{}^{n_1n_2n_3p_1p_2p_3}\CR
&&  - e_{mpq} \cF^{n_1n_2n_3p,q} + \frac{3}{8} \delta_m^{[n_1} e_{p_1p_2q} \cF^{n_2n_3]p_1p_2,q} \ ,\CR
\delta \cF_{n_1n_2}{}^m &=& e_{p_1p_2[n_1} \cF_{n_2]}{}^{mp_1p_2} 
- \frac{1}{9} e_{p_1p_2p_3}\delta^m_{[n_1} \cF_{n_2]}{}^{p_1p_2p_3} +  e_{pn_1n_2} \cF^{m,p} 
\CR
&& - \frac{1}{2} f^{mp_1p_2} \cF_{n_1n_2p_1p_2} - \frac{1}{9} f^{p_1p_2p_3} \delta^m_{[n_1} \cF_{n_2]p_1p_2p_3} 
\CR
\delta \cF_{n_1n_2n_3n_4} &=& - 6 e_{p[n_1n_2} \cF_{n_3n_4]}{}^p - \frac{1}{6} f^{p_1p_2p_3} \cF_{n_1n_2n_3n_4p_1p_2p_3} 
\CR
\delta \cF_{n_1\cdots n_7} &=&-35 e_{[n_1n_2n_3} \cF_{n_4\cdots n_7]} 
- \frac{1}{2} f^{p_1p_2q} \cF_{n_1\cdots n_7p_1p_2,q} +\frac{1}{2} f^{p_1p_2q} \cF_{n_1\cdots n_7p_1p_2p_3} 
\CR
\delta \cF_{n_1\cdots n_9,m} &=&-84  e_{\lsharp n_1n_2n_3} \cF_{n_4\cdots n_9,m\rsharp} +\cdots 
\CR
\delta \cF_{n_1\cdots n_{10}} &=&4  e_{[n_1n_2n_3} \cF_{n_4\cdots n_{10}]} +\cdots 
\label{nc}
\eea
with respect to $\mf{e}_{11}$, with the ellipses denoting terms involving field strengths of level $\ell>\frac{3}{2}$ or $\ell<-\frac{7}{2}$, which we do not define in \eqref{ncfs}.

%%%%%%%%%%%%%%%%%%%%%%%%%%%%%%%%%%%%%%%%%%%%%%%%%%%%%%%%%
\section{Field equations from the tensor hierarchy algebra}
\label{sec:THA2}
%%%%%%%%%%%%%%%%%%%%%%%%%%%%%%%%%%%%%%%%%%%%%%%%%%%%%%%%%

In this section we shall derive linearised equations of motion for the potentials. In addition to the standard potential $A$ parametrising the symmetric space
$E_{11}/K(E_{11})$, the theory will involve an additional potential $X$ in the $\mfr^\ord{0}_1$ module (and possibly other potentials completing the $R_{-2}$ module discussed in the preceding section) transforming together in an indecomposable representation of $\mf{e}_{11}$. In this section we will restrict our analysis to the linearised approximation, in which case only the symmetry $K(E_{11})$ is manifest. Extending the indecomposable module discussed in the previous section to a non-linear realisation of $E_{11}$ is beyond the scope of this paper, see, however, Section~\ref{sec:NL}. The structure coefficients of the tensor hierarchy algebra described in the previous section will serve as building blocks for deriving gauge invariant second order differential equations for the fields and an infinite set of first order duality equations, necessary to avoid infinite degeneracy of the physical states.

%%%%%%%%%%%%%%%%%%%%%%%%%%%%%%%%%%%%%%%%%%%%%%%%%%%%%%%%%%%%%%%%%%%%%%%%%%%%%%%%%
\subsection{Twisted selfduality for \texorpdfstring{$E_{11}$}{E11} and field equations}
%%%%%%%%%%%%%%%%%%%%%%%%%%%%%%%%%%%%%%%%%%%%%%%%%%%%%%%%%%%%%%%%%%%%%%%%%%%%%%%%%

To define the field equations we must consider the coset component of the Maurer--Cartan form ${\mathcal{P}}_{M_0}$ in the gauge \eqref{eq:GE} as in Section \ref{NLreal}. We define the projection to the coset component and $K(\mathfrak{e}_{11})$ from the projectors
\be
P_\pm^{\alpha_0}{}_{\beta_0} = \frac12 \left( \delta^{\alpha_0}_{\beta_0} \pm 
\kappa^{\alpha_0\gamma_0} M_{\gamma_0\beta_0}\right)\ ,
\label{pros}
\ee
respectively, which are defined from the $\mf{e}_{11}$ Cartan--Killing form $\kappa^{\alpha_0\beta_0}$ and the $K(\mf{e}_{11})$ invariant bilinear form on $\bar \ell_1$
\be 
M^{M_0N_0} = \sqrt{g} \ \mbox{diag}(g^{mn} , g_{m_1n_1}  g_{m_2n_2} , g_{m_1n_1}  \cdots  g_{m_5n_5},\dots ) \ ,
\label{m1}
\ee
with $M_{\alpha_0 \beta_0} $ related to $M_{M_0 N_0}$ through the relation\footnote{In the case of $GL(d)/SO(d)$, equation~\eqref{m2} is the following statement. Fundamental indices $M$ correspond to standard vector indices $m=1,\ldots,d$ and adjoint indices $\alpha_0=1,\ldots,d^2$ correspond to pairs of fundamental indices ${}^m{}_n$ as on the generators $K^m{}_n$. For a symmetric matrix $M^{mn}$ constructed from the fundamental representation, the corresponding symmetric matrix $M^{m_1}{}_{n_1}{}^{m_2}{}_{n_2}$ in the adjoint is then determined by   $M^{m_1}{}_{n_1}{}^{m_2}{}_{n_2} \delta^{n_2}_n \delta^m_{m_2} = \delta^{m_1}_{n_2} \delta^{m_2}_{n_1} M_{np} M^{mq} ( \delta^{n_2}_{q} \delta^p_{m_1})$ to simply be $M^{m_1m_2} M_{n_1n_2}$. Another way of understanding this equation is to note that the fundamental and its conjugate anti-fundamental representation are related by the Cartan involution.}
\be
M_{\alpha_0 \beta_0} D^{\beta_0\,M_0}{}_{N_0} =  \kappa_{\alpha_0\beta_0} M_{N_0P_0} M^{N_0Q_0} D^{\beta_0\,P_0}{}_{Q_0}\ , 
\label{m2}
\ee
and the inverse of $M^{M_0N_0}$ is denoted by $M_{M_0N_0}$. One can define the Cartan involution such that $g^{mn}= \eta^{mn}$, the $SO(1,10)$ Minkowski metric, but any matrix conjugate to $M_{M_0N_0}$ in $E_{11}$ defines equivalently a $K(E_{11})$ subgroup, and we shall chose $g^{mn}$ to be an arbitrary constant background metric. We want to keep a general constant metric $g_{mn}$ to exhibit in the following that the density factor in $\det g$ will come out correctly. With respect to the projectors~\eqref{pros}, the coset component of the $\mf{e}_{11}$-valued Maurer--Cartan form~\eqref{eq:MCdec} satisfies $P_+ \mathcal{P}_{M_0}=\mathcal{P}_{M_0}$ and $P_-\mathcal{P}_{M_0} = 0$.

In the linearised approximation, the coset component of the Maurer--Cartan form $\mathcal{P}_{M_0}$ is simply the derivative of a Lie algebra element in the coset component:
\begin{align}
\mathcal{P}_{M_0}=\frac{1}{2} \partial_{M_0} \left( A_{\alpha_0} t^{\alpha_0} \right)\,,
\end{align}
where the normalisation is chosen for convenience, and where
\begin{align}
\label{gc2}
P_-^{\beta_0}{}_{\alpha_0} A_{\beta_0} =0
\end{align}
ensures that $A_{\alpha_0} t^{\alpha_0}$ lies in the coset component. This is \textit{not} a gauge condition on $A_{\alpha_0}$; fixing a $K(E_{11})$ gauge determines how the components of $A_{\alpha_0}$ are expressed in terms of the potentials in a gauge-fixed representative of the $E_{11}/K(E_{11})$ coset element. Parametrising the $E_{11}$ coset representative in the parabolic gauge~\eqref{eq:GE} leads to the linearised Maurer--Cartan form
\begin{align}
\label{CosetMC}
g_E^{\; -1} \partial_{M_0} g_{E} &= \partial_{M_0} \Bigl( \scalebox{0.9}{$\varphi_n{}^m K^n{}_m+ \tfrac{1}{3!} A_{n_1n_2n_3} E^{n_1n_2n_3} + \tfrac{1}{6!} A_{n_1\dots n_6} E^{n_1\dots n_6} + \tfrac{1}{8!} h_{n_1\dots n_8,m} E^{n_1\dots n_8,m} + \dots$} \Bigr) \ , \CR
\Rightarrow \quad {\mathcal{P}}_{M_0} &= \partial_{M_0}  \Bigl( \scalebox{0.9}{$ \frac12  h_a{}^b K^a{}_b+ \tfrac{1}{2\cdot 3!} A_{a_1a_2a_3} ( E^{a_1a_2a_3} +F^{a_1a_2a_3}) + \tfrac{1}{2\cdot 6!} A_{a_1\dots a_6} ( E^{a_1\dots a_6}+ F^{a_1\dots a_6})   + \dots $} \Bigr)\ . 
\end{align}
Note that for $\mathfrak{e}_{11}$, the Killing form permits the interpretation of  $\mathcal{P}_{M_0}$ as an element of degree $p=-2$. Doing so we can identify the field $A_{\alpha_0} t^{\alpha_0}$ with the potential $A^{\alpha_0} \tilde t_{\alpha_0}$ at degree $p=-2$ according to the discussion of Section~\ref{sec:tha1}. In terms of~\eqref{eq:phiexp}, one obtains $\mathcal{P}_{M_0}$ by substituting to the components of $\phi^{\alpha_0}$
\begin{align} 
A^+_{n_1n_2n_3} &= \frac12 A_{n_1n_2n_3}  \ , \quad & A^+_{n_1\dots n_6}  &=  \frac12 A_{n_1\dots n_6} \ , \quad & \dots
\nn\\
A_-^{n_1n_2n_3}&= \frac12 g^{n_1p_1} g^{n_2p_2} g^{n_3p_3} A_{p_1p_2p_3} \ , \quad  & A_-^{n_1\dots n_6}&=\frac12  g^{n_1p_1} \dots g^{n_6p_6}A_{p_1\dots p_6} \ , \quad &  \dots 
\label{gff}
\end{align}
and similarly for all higher level fields, and 
\be \label{hSym}
h^+_m{}^n = \frac12 h_m{}^n = \frac12 g_{mp} g^{nq} h_q{}^p \ .
\ee

The gauge transformations for the fields are then obtained from \eqref{eq:Xitrm} by summing the contribution from $A^+_{n_1n_2n_3},\dots $ and their conjugate $g_{n_1p_1} g_{n_2p_2} g_{n_3p_3}  A_-^{p_1p_2p_3} , \dots $, and similarly for $h_n{}^m$ by summing  \eqref{eq:Xitrm}  and its transpose, so that 
\be
\delta_\Xi h  = \delta^{\scriptscriptstyle \mathscr{T}}_\Xi h^+ +\delta^{\scriptscriptstyle \mathscr{T}}_\Xi h^{+}{}^T\ , \quad  \delta_\Xi A  = \delta^{\scriptscriptstyle \mathscr{T}}_\Xi A^+ +\delta^{\scriptscriptstyle \mathscr{T}}_\Xi A^- \ , 
\ee
in agreement with \eqref{eq:GTM}. In order to be consistent with the $K(\mf{e}_{11})$ transformations, the gauge transformation of the field $X$ must also be modified. To do this we observe that $\delta^{\scriptscriptstyle \mathscr{T}}_\Xi A^+$ and $\delta^{\scriptscriptstyle \mathscr{T}}_\Xi A^-$ are obtained from one another by exchanging $\partial_M$ and $\Xi^M$ and by lowering and raising all indices using the background metric $g_{mn}$. We will therefore consider that the gauge transformation of the field $X^{\alpha_i}$ is also modified in the same way, \ie 
\be
\delta_\Xi X =  \delta^{\scriptscriptstyle \mathscr{T}}_\Xi X +\overline{\delta^{\scriptscriptstyle \mathscr{T}}_\Xi X} \ ,  
\label{gtx}
\ee
understanding that $\overline{\delta^{\scriptscriptstyle \mathscr{T}}_\Xi X} $ is obtained from $\delta^{\scriptscriptstyle \mathscr{T}}_\Xi X$ by exchanging the $\partial_M$ and the $\Xi^N$ and raising and lowering the indices with $g_{mn}$. For example, one has 
\be 
\delta_\Xi X_{a_1\cdots a_9}  = 24 \partial_{[a_1} \lambda_{a_2\cdots a_9]} + 24 \partial_{[a_1\cdots a_8} \xi_{a_9]}  + \mathcal{O}(4,4) \; . 
\ee
To define field equations for the $E_{11}/ K(E_{11})$ fields $ h,\, A$ and the additional fields $X^{\alpha_i}$, we need therefore to define in some way the equivalent of the projection to the coset component for the additional fields $X^{\alpha_i}$,  consistently with the gauge transformation \eqref{gtx}. Note that if the assumption of the previous section about the reducibility of these modules were true, one could consistently truncate the coordinate dependence to the one in the $\mfr^\ord{1}_0$ module and the additional fields to $X^{\alpha_1}$ in $\mfr^\ord{0}_1$ only. In this section we assume indeed that one can set the coordinates in $\mfr^\ord{1}_0$ to zero, whereas the second condition is not essential in the following. This simplifying property would nevertheless be very desirable to define a minimal extension of the $E_{11}$ paradigm.

We shall define the physical $K(\mf{e}_{11})$-covariant field strength from the projection to the coset component of the field strength defined from the tensor hierarchy algebra $\cF_I$ as follows
\be
\mathcal{G}_{I}  = {f}^{M_0}{}_{\alpha,I}\, \partial_{M_0}  A^{\alpha} = 
f^{M_0}{}_{\alpha_0,I}\, \partial_{M_0}  A^{\alpha_0} + \sum_{i\ge 1} f^M{}_{\alpha_i,I} \partial_M X^{\alpha_i}  \ , \label{defg2}
\ee
where we define for convenience $A^\alpha = ( A^{\alpha_0} , X^{\alpha_i})$. More schematically, one can obtain all the components of $\cG_{I}$ using \eqref{ncfs} with the above substitutions (\ref{gff}), (\ref{hSym}). Defining the expressions  \eqref{ncfs} as $\mathcal{F}[A^-, h^+,A^+,X]$, one can formally write that 
\be 
\label{eq:FSC}
\mathcal{G}[h,A,X] =  \mathcal{F}[A,h,A,X] \ , 
\ee
where we avoid writing the dependence in the background metric $g_{mn}$ for brevity. The gauge transformation of $A^\alpha$ can be written as
\be
\delta_\Xi A^\alpha = (D^\alpha)^{M_0}{}_{N_0} \ \Bigl( \partial_{M_0} \Xi^{N_0} +M^{N_0P_0} M_{M_0Q_0} \partial_{P_0} \Xi^{Q_0} \Bigr) \ .
\label{gt2Coset}
\ee
The field strength $\mathcal{G}_I$  defined as in \eq{defg2} is \textit{not} gauge invariant, even for gauge parameters satisfying the section constraint. One gets instead 
\be
\delta_\Xi \mathcal{G}_{I} = f^{M_0}{}_{\alpha,I}\  (D^\alpha)^{N_0}{}_{P_0}  M^{P_0Q_0} M_{N_0R_0} \partial_{M_0}  \partial_{Q_0} \Xi^{R_0} \ , \label{vg}
\ee
where we used the gauge invariance of $\mathcal{F}_I$ modulo the section condition \eq{gifs} to simplify the expression. One can check that this gauge variation does not vanish. 
 This is not a contradiction because it is indeed expected that one cannot write gauge invariant first order duality equations, as we already discussed, since the first order duality equations for the metric field are not gauge invariant in ordinary spacetime. We shall see nonetheless that one can define second order field equations that are solved by the solutions to a non-gauge invariant first order constraint, and which turn out to be gauge invariant in a low-level truncation. Assuming that this second order equation is indeed gauge invariant, we find that no higher order field equations are needed in this set-up. In principle, gauge invariant first order duality equations can be written at the expense of introducing additional St\"uckelberg fields as in~\cite{Boulanger:2008nd}.

The great advantage of the above construction is that $\mathcal{G}_I $ is defined in a representation of $\mf{e}_{11}$, and as such preserves the level up to the projection applied to the coset fields. Therefore a component of $\mathcal{G}_I $ of $\mf{gl}(11)$ level $\ell$ admits contributions from level $\ell_d$ derivatives acting on coset potentials of level $|\ell-\ell_d|$  only.  This ensures in particular that at a given level $\ell$, one can only have ordinary derivatives of potentials at level $|\ell+\frac{3}{2}|$. One can therefore consistently consider the restriction of the fields to depend on the eleven space-time coordinates for a given level truncation, without possibly missing contributions from arbitrary high level fields, as it may be the case in the conventional $E_{11}$ paradigm. 

To define the first order duality equation we need an invariant bilinear form on the $R_{-1}$ module. For a finite-dimensional group $G$, the existence of a symplectic form in a $2n$-dimensional representation implies that $G\subset Sp(2n,\mathbb{R})$ and that its maximal compact subgroup $K\subset G$ is a subgroup of $U(n)$ such that there is a $K$ invariant bilinear form in this representation. These building blocks permit to define consistent twisted self-duality equations for $D/2$-form field strengths in dimensions
$D=4$~mod~4, as one finds in supergravity theories in space-time dimension four and eight.

Given that the $R_{-1}$ module admits an $\mf{e}_{11}$ invariant symplectic form $\Omega^{IJ}$, one can understand $E_{11}$ as a symplectic group acting in this representation. Provided that a symmetric non-degenerate bilinear form $M^{IJ}$ exists, one can write a first order duality equation for the field strength $\mathcal{G}_I$:
\be 
M^{IJ} \mathcal{G}_J = \Omega^{IJ} \mathcal{G}_J \ . 
\label{FirstTH} 
\ee
If $R_{-1}$ were irreducible under $\mf{e}_{11}$, it would follow that $M^{IJ}$ existed and was non-degenerate. Independently of the assumption that the module $R_{-1}$  is irreducible, we shall find evidences for the existence of this bilinear form in the low level truncation. It may also be the case that $M^{IJ}$ exists but is not unique; the low level expression we construct in the next section is inspired by the first order formulation of supergravity. The duality equation relates levels $\ell$ and $-\ell$ in the $\mf{gl}(11)$ decomposition of the $R_{-1}$ representation since $\Omega^{IJ}$ has the reflection symmetry discussed in Section~\ref{sec:tha1} and $M^{IJ}$ will be seen below to be diagonal.

The equation \eq{FirstTH} is reminiscent of the twisted selfduality equations appearing in supergravity theories~\cite{Cremmer:1998px}, where $M^{IJ}$ includes all the factors in the metric fields and more generally on the background $E_{11}/K(E_{11})$ coset, whereas $\Omega^{IJ}$ only involves the Levi--Civita tensor. Moreover, $M^{IJ}$ includes the appropriate factor of $\sqrt{g}$ of the background metric, and we shall see that it reproduces the appropriate first order duality equations for the various fields included in our truncation scheme. 

However, the duality equation~\eqref{FirstTH} is not gauge invariant modulo the section constraint. It would be suitable to have a second order field equation that would be gauge invariant, and would be automatically solved by any solution to this first order equation satisfying the section constraint. The tensor hierarchy algebra implies the degree $p=0$ Bianchi identity
\be 
d^2 ( \phi^\alpha  \tilde{t}_\alpha) = \Omega^{IJ} f^{M_0}{}_{\alpha,I} f^{N_0}{}_{\beta,J} \partial_{M_0} \partial_{N_0} \phi^\beta t^\alpha  = 0 \ 
\ee
modulo the section constraint. The field $A^{\alpha_0}$ belongs to the coset component, and the corresponding field equation must therefore also belong to $\mf{p}$. So we define the projected structure coefficients 
\be
 \hat{f}^{M_0}{}_{\alpha_0,I}  \equiv  P_+^{\beta_0}{}_{\alpha_0} {f}^{M_0}{}_{\beta_0,I}  \ , \quad \hat{f}^{M_0}{}_{\alpha_i,I} \equiv   {f}^{M_0}{}_{\alpha_i,I}  \ . 
\label{defhf}
\ee
Because the relevant structure coefficients of 
the tensor hierarchy algebra do not involve the contraction of the indices 
$\alpha,\beta$ in this equation, it is also true that 
\be 
\label{BianchiTHA}
\Omega^{IJ} \hat{f}^{M_0}{}_{\alpha,I} \hat{f}^{N_0}{}_{\beta,J} 
\partial_{M_0} \partial_{N_0} A^\beta = 0 \  ,
\ee
modulo the section constraint. One concludes that any solution to the first order duality equation \eqref{FirstTH} automatically solves the second order differential equation 
\be 
  \hat{f}^{M_0}{}_{\alpha, I} M^{IJ} \partial_{M_0} \mathcal{G}_J  = 0  \ , 
\label{SecondTH0}
\ee
with ${\mathcal G}_I$ and $\hat{f}^{M_0}{}_{\alpha, I}$ defined from \eq{defg2} and \eq{defhf}. Equation \eq{SecondTH0} is very suggestive of the second order field equations one encounters in supergravity, and it turns out to be the equation of motion of a Lagrangian 
\be 
\mathcal{L}^\ord{0} = -  \frac{1}{2} \mathcal{G}_I M^{IJ}  \mathcal{G}_J \ , 
\label{THLagrange0} 
\ee
uniquely determined by the invariant bilinear form $M^{IJ}$. In the following section we shall determine a $K(\mf{e}_{11})$-invariant  bilinear form  ${M}^{IJ}$ that  preserves the $\mf{gl}(11)$ level in our truncation scheme, meaning that the Lagrangian \eqref{THLagrange0} decomposes schematically as $\mathcal{L}^\ord{0} \sim- \sum_\ell |\cG^\ord{\ell-\frac{3}{2}} |^2$. The property that $M^{IJ}$ preserves the level is essential for the consistency of the level truncation scheme.

However, we shall find that the second order equation \eqref{SecondTH0} is not gauge invariant. The lack of gauge invariance seems to be related to the asymmetry of the formalism between the field $A^{\alpha_0}$ that is projected to the coset component and the additional fields $X^{\alpha_i}$ that are not. We will now describe how this problem can be circumvented at the price of introducing another algebraic structure. 

Following this line of thought, we therefore define the spurious field $\bar X_{\alpha_1}$ in the conjugate representation $\bar \ell_2$. For this purpose we define the indecomposable module $R_{-2}^*$, that is obtained from $R_{-2}$ through the action of the Cartan involution. It decomposes into vector spaces as  $R_{-2}^* = \oplus_i \mf{r}_i^\ord{0} = \mf{e}_{11} \oplus \bar \ell_2\oplus \bar \ell_{10} \oplus \dots $, but should not be confused with the conjugate module $R_0$ that is obtained by conjugation and not the Cartan involution. This definition ensures by construction that $R_{-2}^*$ and $R_{-2}$ are identical as $K(\mf{e}_{11})$-modules. The highest $\mf{gl}(11)$ component of $\bar X_{\alpha_1}$ is $\bar X^{n_1\dots n_9}$, it transforms accordingly with respect to $\mf{e}_{11}$ as
\be \label{barXe11}
\delta \bar X^{n_1\cdots n_9}  = 28 f^{[n_1n_2n_3} A_-^{n_4\dots n_9]}  +e_{p_1p_2p_3} ( \dots )  \; . 
\ee
where the terms denoted by ellipses will not be needed at the level truncation we consider below.
We define its gauge transformation as the conjugate transformation $\overline{\delta^{\scriptscriptstyle \mathscr{T}}_\Xi X}$, such that in particular 
\be 
\delta^{\scriptscriptstyle \mathscr{T}}_\Xi \bar X^{n_1\cdots n_9}  = 24 \partial^{[n_1\cdots n_8} \xi^{n_9]}  + \mathcal{O}(4,4) \; .  
\ee
This gauge transformation is by construction consistent with the indecomposable $\mf{e}_{11}$-module structure of $R_{-2}^*$. So just as the coset projection of $A^{\alpha_0}$ is defined such that its gauge transformation is $\delta_\Xi A = \delta^{\scriptscriptstyle \mathscr{T}}_\Xi A^+ +\delta^{\scriptscriptstyle \mathscr{T}}_\Xi A^-$, the gauge transformation of the physical fields $X^{\alpha_i}$ are defined such that $\delta_\Xi X = \delta^{\scriptscriptstyle \mathscr{T}}_\Xi X +\delta^{\scriptscriptstyle \mathscr{T}}_\Xi \bar X$. We will write $\bar \phi_\alpha = (\kappa_{\alpha_0\beta_0}\phi^{\beta_0},\bar X_{\alpha_i}) \in R_{-2}^*$, keeping in mind that $R_{-2}^*$ is not conjugate to $R_{-2}$. In particular, one can write the gauge transformation
\be 
\delta^{\scriptscriptstyle \mathscr{T}}_\Xi \bar \phi_{\alpha} = \bigl(  M_{\alpha\beta}  (D^\beta)^{Q_0}{}_{P_0} \ M^{P_0N_0} M_{Q_0M_0} \bigr)   \partial_{N_0} \Xi^{M_0}  \ , 
\ee
where $M_{\alpha\beta}$ is defined from the condition that the dependence in the background metric $g_{mn}$ drops out in this equation. $M_{\alpha\beta}$ defines the conversion of $\bar \phi_\alpha$ to the physical field $M_{\alpha\beta} A^\beta$, but it {\it does not} define a $K(\mf{e}_{11})$ invariant bilinear form on $R_{-2}$. In the $\mf{gl}(11)$ level decomposition, $M_{\alpha\beta}$ simply lowers all upper indices with the background metric $g_{mn}$ and raises all lower indices with its inverse.

Just like the field strength $\cF_I$ defined from the tensor hierarchy algebra is by construction $\mf{e}_{11}$ covariant and gauge invariant modulo the section constraint, we would like to define a field strength from the  potential $\bar \phi_\alpha $ in a representation of $\mf{e}_{11}$ that would be gauge invariant modulo the section constraint.  Although the tensor hierarchy algebra does not provide such a definition, we shall now argue that one can define such a field strength  $\bar {\cF}_{I_1}$ in the highest weight module $\bar \ell_3$. 

For this purpose we observe the decomposition of the tensor product 
\be 
\bar{\ell}_1 \otimes \bar{\ell}_2 = \overline{(\ell_1+\ell_2)} \oplus \bar{\ell}_3 \oplus \dots \ , 
\ee
into irreducible $\mf{e}_{11}$ representations. The terms on the right-hand side are highest weight representations labelled by their highest weight, for instance, the first term has Dynkin labels $(1, 1, 0, 0, 0, 0, 0, 0, 0, 0, 0)$. The decomposition into highest weight representations allows to define a projector $\Pi_{I_1}{}^{M_0 \alpha_1} $ from $\bar \ell_1 \otimes \bar \ell_2$ to the module $\bar \ell_3$. To define the field strength $\bar {\cF}_{I_1}$  in $\bar \ell_3$ we would need a similar projector from $\bar\ell_1 \otimes R_{-2}^*$ to the module $\bar \ell_3$. The projection to $\bar \ell_3$ is determined by the property that it is a highest weight representation, with a rank eight antisymmetric tensor of level $-\frac{9}{2}$ as its highest level component in the $\mf{gl}(11)$ decomposition. Checking the highest weight condition on an ansatz of $\mf{gl}(11)$ level $- \frac{9}{2}$, \ie one that is annihilated by the action of the $\mf{e}_{11}$ lowering generator ${F}_{p_1p_2p_3}$, one determines the field strength component 
\begin{multline} \label{bF8def}
\bar F^{n_1\cdots n_8}  = \partial_p ( h_-^{n_1\dots n_8,p} + \bar X^{n_1\dots n_8p} ) - 28 \partial^{[n_1n_2} A_-^{n_3\dots n_8]} -56 \partial^{[n_1\dots n_5} A_-^{n_6n_7n_8]} \\ + 8 \partial^{[n_1\dots n_7|,p} h^+_p{}^{n_8]} - 24 \partial^{p[n_1\dots n_7} h^+_p{}^{n_8]} +\mathcal{O}(4,1)  \; , 
\end{multline} 
in the level truncation we consider. One computes moreover that it is gauge invariant modulo the section constraint, 
\be
\delta^{\scriptscriptstyle \mathscr{T}}_\Xi \bar F^{n_1\cdots n_8}   = \mathcal{O}(\partial^\ord{-\frac{3}{2}} \partial^\ord{-\frac{11}{2}} \lambda_2),
\ee
up to derivatives that are beyond our truncation scheme. Terms involving $\xi^m$ drop out identically. Based on these observations, we assume that one can indeed define the field strength $\bar {\cF}_{I_1}$ in $\bar \ell_3$,
\be 
\bar {\cF}_{I_1} \equiv \Pi_{I_1}{}^{M_0 \alpha} \partial_{M_0} \bar \phi_\alpha =  \Pi_{I_1}{}^{M_0 \alpha_0} \partial_{M_0} \phi_{\alpha_0} + \Pi_{I_1}{}^{M_0 \alpha_1} \partial_{M_0} \bar X_{\alpha_1} + \sum_{i\ge2}   \Pi_{I_1}{}^{M_0 \alpha_i} \partial_{M_0} \bar X_{\alpha_i} \ .
\label{barF1} 
\ee
We note that only $\Pi_{I_1}^{M_0\alpha}$ and $\Pi_{I_1}^{M_0\alpha_1}$ are $\mf{e}_{11}$ tensors.\footnote{The   $\Pi_{I_1}{}^{M_0 \alpha_i}$ would vanish for $i\ge 2$, if the structure coefficients \eqref{eq:CRm22} were upper triangular.}
In addition, we assume that $\bar {\cF}_{I_1}$ is gauge invariant modulo the section constraint, \ie
\be\label{GVbF} 
\delta^{\scriptscriptstyle \mathscr{T}}_\Xi  \bar {\cF}_{I_1} =\Pi_{I_1}{}^{M_0 \alpha} M_{\alpha\beta} (D^\beta)^{N_0}{}_{P_0}  M_{N_0R_0} M^{P_0Q_0}  \partial_{M_0} \partial_{Q_0} \Xi^{R_0} = 0\  .
\ee
This is true up to the level we have checked.

Assuming this field strength $\bar{\cF}_{I_1}$ in $\bar \ell_3$ indeed exists and is gauge invariant modulo the section constraint, one can define the $K(\mf{e}_{11})$-covariant physical field strength
\be
\bar{\cG}_{I_1} \equiv \Pi_{I_1}{}^{M_0 \beta} M_{\beta\alpha}   \partial_{M_0}  A^{\alpha}  = \Pi_{I_1\alpha}^{M_0} \partial_{M_0}  A^{\alpha}   = \Pi_{I_1\alpha_0}^{M_0} \partial_{M_0}  A^{\alpha_0}  + \sum_{i\ge1}   \Pi_{I_1 \alpha_i}^{M_0} \partial_{M_0} X^{\alpha_i} \ ,
\label{defg2bar}
\ee
where we defined $\Pi_{I_1\alpha}^{M_0}\equiv \Pi_{I_1}{}^{M_0 \beta} M_{\beta\alpha} $ for convenience. Writing \eqref{bF8def} as $\bar{\mathcal{F}}[A^-, h^+,A^+,\bar X]$, one can formally write that 
\be 
\label{eq:FSCbar}
\bar{\mathcal{G}}[h,A,X] =  \bar{\mathcal{F}}[A,h,A,X] \ .
\ee
As $\cG_I$ defined in~\eqref{defg2}, this field strength $\bar{\cG}_{I_1}$ is not gauge invariant, but its gauge transformation simplifies upon use of \eqref{GVbF} to 
\be
\delta_\Xi \bar{\mathcal{G}}_{I_1} = \Pi_{I_1\alpha}^{M_0}  (D^\alpha)^{N_0}{}_{P_0}  \partial_{M_0}\left( \partial_{N_0} \Xi^{P_0} + M_{N_0R_0} M^{P_0Q_0} \partial_{Q_0} \Xi^{R_0}\right) =   \Pi_{I_1\alpha}^{M_0}  (D^\alpha)^{N_0}{}_{P_0} \partial_{M_0} \partial_{N_0} \Xi^{P_0}  \ .      
\label{vgbar}
\ee
As a highest weight module, $\bar \ell_3$ admits a non-degenerate $K(\mf{e}_{11})$ invariant bilinear form $M^{I_1J_1}$, and one can define the Lagrangian 
\be 
\mathcal{L} =\mathcal{L}^\ord{0}  - \frac12 \bar {\cG}_{I_1} M^{I_1J_1} \bar {\cG}_{J_1} =  - \frac12 \cG_I M^{IJ} \cG_J - \frac12 \bar {\cG}_{I_1} M^{I_1J_1} \bar {\cG}_{J_1}\ , \label{THLagrange}
\ee
that defines the second order equations of motion
\be 
  \mathcal{E}_\alpha  = \hat{f}^{M_0}{}_{\alpha, I} M^{IJ} \partial_{M_0} \mathcal{G}_J  + \widehat{\Pi}_{I_1\alpha}^{M_0} M^{I_1J_1}  \partial_{M_0}  \bar {\cG}_{J_1} \ , 
\label{SecondTH}
\ee
where 
\be \widehat{\Pi}_{I_1\alpha_0}^{M_0} \equiv  P^{\beta_0}{}_{\alpha_0} {\Pi}_{I_1\beta_0}^{M_0}\ , \qquad  \widehat{\Pi}_{I_1\alpha_i}^{M_0} \equiv   {\Pi}_{I_1\alpha_i}^{M_0} \ . \ee
We shall prove in the next section that these second order equations are gauge invariant modulo the section constraint within our level truncation scheme. We therefore conjecture that one can define gauge invariant second order equations to all levels following this construction, or a generalisation thereof involving possibly additional highest weight modules in a similar way. 

These equations \eqref{SecondTH} are automatically solved by the solutions to the first order equations
\begin{subequations}
\begin{align}
\label{dualityBTHA} 
 M^{IJ} \mathcal{G}_J &= \Omega^{IJ} \mathcal{G}_J \ , \\
 \label{gaugeBTHA}
 \bar {\cG}_{I_1} &= 0 \ .
\end{align}
\end{subequations}
It may look rather drastic to set $\bar {\cG}_{I_1}$ to zero. One can interpret $\bar {\cG}_{I_1} = 0$ as a $K(\mf{e}_{11})$-multiplet of gauge-fixing conditions for the field $X^{\alpha_1}$. This is then consistent with the first order duality equations \eqref{dualityBTHA} being not gauge invariant. It might be possible to define gauge invariant first order equation by introducing appropriate St\"{u}ckelberg gauge fields. We expect that within such a formulation, the St\"{u}ckelberg gauge fields would couple these two equations non-trivially. Note that the identification of the correct physical degrees of freedom requires the first order duality equation to be satisfied, which does not derive from the Lagrangian \eqref{THLagrange}. This is similar to the situation one encounters in the democratic formulation of supergravity theories. 

We shall now work out these second order equations within our level truncation scheme, and exhibit that they are indeed gauge invariant modulo the section constraint.

%%%%%%%%%%%%%%%%%%%%%%%%%%%%%%%%%%%%%%%%%%%%%%%%%%%%%%%%%%%%%%%%%%%%
\subsection{Explicit field equations in the level truncation} \label{fieldlevel}
%%%%%%%%%%%%%%%%%%%%%%%%%%%%%%%%%%%%%%%%%%%%%%%%%%%%%%%%%%%%%%%%%%%%

It will be convenient to define the tensors in tangent frame, so we introduce the constant vielbein $e_m{}^a$ associated to the background metric $g_{mn}$ used in the previous section, with determinant $e=\det e_m{}^a$. Since the various field strength components have the same number of indices, we shall use different letters to define them according to their interpretation, as in Section \ref{PeterThing}. At low levels we have 
\bea 
\label{eq:CFS}
\Omega_{a_1\cdots a_9,b} &=&\sqrt{e}\,   
e_{a_1}{}^{n_1} \cdots   e_{a_9}{}^{n_9} e_b{}^m \cG_{n_1\cdots n_9,m} \ , \quad \Omega_{a_1\cdots a_{10}} =\sqrt{e}\,   
e_{a_1}{}^{n_1} \cdots   e_{a_{10}}{}^{n_{10}}  \cG_{n_1\cdots n_{10}} \, , 
\CR
\cG_{a_1\cdots a_7} &=&\sqrt{e}\,   
e_{a_1}{}^{n_1} \cdots   e_{a_7}{}^{n_7} \cG_{n_1\cdots n_7} \; , 
\CR
\cG_{a_1a_2a_3a_4} &=&\sqrt{e}\, 
e_{a_1}{}^{n_1} \cdots   e_{a_4}{}^{n_4} \cG_{n_1n_2n_3n_4} \; , 
\CR
 \Omega_{a_1a_2}{}^b &=&\sqrt{e}\, 
e_{a_1}{}^{n_1}e_{a_2}{}^{n_2} e_m{}^b \cG_{n_1n_2}{}^m \; , 
\CR
\cH_{a}{} ^{b_1b_2b_3} &=&\sqrt{e}\, 
e_{a}{}^{m}  e_{n_1}{}^{b_1}  e_{n_2}{}^{b_2}  e_{n_3}{}^{b_3}  \cG_m{}^{n_1n_2n_3}\; ,  \quad \Theta^{a,b} =\sqrt{e}\,   
 e_m{}^a e_n{}^b   \cG^{m,n}  \, , 
 \CR
\cH_{a}{}^{b_1\cdots b_6} &=&\sqrt{e}\, 
e_{a}{}^{m}  e_{n_1}{}^{b_1}  \cdots e_{n_6}{}^{b_6}  \cG_m{}^{n_1\cdots n_6}\; ,  \quad \Theta^{a_1\dots a_4,b} =\sqrt{e}\,   
 e_{n_1}{}^{a_1} \cdots  e_{n_4}{}^{a_4}  e_m{}^b  \cG^{n_1n_2n_3n_4,m}  \, ,
 \CR
\cH_a{}^{b_1\dots b_8,c} &=&\sqrt{e}\, 
e_{a}{}^{m}  e_{n_1}{}^{b_1}  \cdots e_{n_8}{}^{b_8} e_p{}^c \,  \cG_m{}^{n_1\cdots n_8,p} \, , \quad \mathcal{N}^{a_1\dots a_8} = \sqrt{e}\, 
e_{n_1}{}^{b_1}  \cdots e_{n_8}{}^{b_8}  \,  \bar \cG^{n_1\cdots n_8} \, ,
\eea
where the field strengths $\cG$ are defined in \eq{eq:FSC} with $\cF$ from \eq{ncfs} and $\bar{\cG}$ in~\eqref{defg2bar} with $\bar {\cF}$ from \eqref{barF1}. For example, this gives
\begin{align}
\mathcal{G}_{a_1a_2a_3a_4} &= 4\partial_{[a_1} A_{a_2a_3a_4]} -\frac12 \partial^{b_1b_2} A_{a_1a_2a_3a_4b_1b_2} -\frac1{4!} \partial^{b_1\ldots b_5} h_{a_1a_2a_3a_4b_1\ldots b_4,b_5}\nn\\
&\quad\quad+ \frac1{5!}\partial^{b_1\ldots b_5} X_{a_1a_2a_3a_4b_1\ldots b_5}+\ldots \; ,\nn\\
\mathcal{G}_{a_1\ldots a_7} &=  7 \partial_{[a_1} A_{a_2\ldots a_7]} + \partial^{b_1b_2} h_{a_1\ldots a_7 b_1,b_2} -\frac12 \partial^{b_1b_2} X_{a_1\ldots a_7b_1b_2}+\ldots \, .
\end{align}
Comparing with~\eqref{G4WestCorr} and~\eqref{G7E11}, we see that the lowest order terms coincide, but there are important differences for terms beyond order $\mathcal{O}(2,2)$. In our formulation the additional fields arising from the extension to the tensor hierarchy algebra, in particular the field $X_9$ and its partners in $\ell_2$, allow one to define an $\mf{e}_{11}$ representation for the field strength, so that the $K(\mf{e}_{11})$ representation defining the duality equation is determined. This implies in particular that the field strengths preserve the horizontal degree (modulo the projection of the potentials to the coset component). In~\eqref{G4WestCorr} and~\eqref{G7E11}, there are more terms that are introduced by the requirement of $K(\mf{e}_{11})$ covariance (without $X_9$), that do not preserve the horizontal degree.

The $\mf{gl}(11)$ level of the field strengths in~\eqref{eq:CFS} are determined by their number $N$ of covariant indices and their number $M$ of contravariant indices as $\frac{N-M-11/2}{3}$, so that the action of $E_{11}$ includes the additional factor in the square root of the determinant of the vielbein. Note that the Lagrangian~\eqref{THLagrange} includes therefore the relevant determinant factor for a Lagrange density. The various lines in~\eqref{eq:CFS} correspond to different 
$\mf{gl}(11)$ level components (where the level is related to the horizontal degrees $q$ in the vertical degree $p=-1$ of the tensor hierarchy algebra as $q=\ell-\frac{3}{2}$). The component $\mathcal{N}^{a_1\ldots a_8}$ is added according to its $\mf{gl}(11)$ level. The notation for the various components is in analogy with what happens in double field theory and non-geometric fluxes~\cite{Bergshoeff:2015cba}; such that $\mathcal{G}$ stands for ordinary $p$-form field strengths, $\Omega$ for field strengths associated to the gravitation field or its dual, $\cH$ for field strengths associated to unfolding dualities that involve potentials with at least one column of nine antisymmetrised indices, and $\Theta$ for field strengths associated to non-dynamical dualities that involve potentials with at least one column of ten antisymmetrised indices.

In order to evaluate the Lagrangian \eqref{THLagrange0}, one has to work out the $K(\mf{e}_{11})$ invariant bilinear form $M^{IJ}$ level by level. This can be done using the $\mf{e}_{11}$ transformations \eqref{E11varyF}
restricted to $K(\mf{e}_{11})$ by setting $f^{n_1n_2n_3}=-g^{n_1p_1} g^{n_2p_2} g^{n_3p_3} e_{p_1p_2p_3}$. The result is
\begin{align} 
\label{eq:LLT0}
\mathcal{L}^\ord{0}  &= -\frac{1}{2} \Bigl(   \frac{1}{9!} \Omega_{a_1\cdots a_9,b}  \Omega^{a_1\cdots a_9,b} 
-  \frac{1}{8!} \Omega_{a_1\cdots a_{10}}  \Omega^{a_1\cdots a_{10}}+ \frac{1}{7!} \cG_{a_1\cdots a_7}  
\cG^{a_1\cdots a_7}  + \frac{1}{4!} \cG_{a_1\cdots a_4}  \cG^{a_1\cdots a_4}  
\CR
& \quad +\frac{1}{2} \Omega_{a_1a_2}{}^b \Omega^{a_1a_2}{}_b -\Omega_{ab}{}^b \Omega^{ac}{}_c 
+\frac{4}{6} \cH^{a_4}{}_{[a_1a_2a_3} \cH_{a_4]}{}^{a_1a_2a_3} + \Theta_{a,b} \Theta^{a,b} + \frac{1}{4!} \Theta_{a_1\cdots a_4,b} \Theta^{a_1\cdots a_4,b}
\CR
& \quad + \frac{7}{6!} \cH^{a_7}{}_{[a_1\cdots a_6} \cH_{a_7]}{}^{a_1\cdots a_6} + \frac{9}{8!} \cH^{[a_1}{}_{a_2\dots a_9,b} \cH_{a_1}{}^{a_2\dots a_9],b} - \frac{1}{8!} \cH^{b}{}_{a_1\dots a_8,b}\cH_{c}{}^{a_1\dots a_8,c}+ \dots \Bigr) \ ,   
\end{align}
where the field strengths are ordered with respect to their $\mf{gl}(11)$ level, starting from level $\ell=\frac{3}{2}$ and decreasing down to $\ell=-\frac{9}{2}$. The list of terms is exhaustive up to level $\ell=-\frac{7}{2}$, whereas we have neglected field strengths at level $\ell=-\frac{9}{2}$ that do not depend on the dual graviton field. Although $GL(11)$ representation theory does not distinguish a specific canonical field strength among the linear combination of the level $\ell=-\frac{9}{2}$ field strengths $\cH_1{}^{8,1}$, $\Theta^{8}$ and $\Theta^{7,1}$
\be \cH_{c}{}^{a_1\dots a_8,b} +\alpha \delta_c^{\lsharp b,}   \Theta^{a_1\dots a_8\rsharp }  + \beta  \delta_c^{[a_1} \Theta^{a_2\dots a_8],b}   \ , \ee
$K(\mf{e}_{11})$ invariance determines the Lagrangian to depend on them through the combination
\be \frac{9}{8!} \cH^{[a_1}{}_{a_2\dots a_9,b} \cH_{a_1}{}^{a_2\dots a_9],b} - \frac{1}{8!} \cH^{b}{}_{a_1\dots a_8,b}\cH_{c}{}^{a_1\dots a_8,c} + \frac{1}{8!} \Theta_{a_1\dots a_8} \Theta^{a_1\dots a_8} +  \frac{1}{7!} \Theta_{a_1\dots a_7,b} \Theta^{a_1\dots a_7,b}  \ , \ee 
 in our conventions, which justifies the definition of the field strength $\cH_1{}^{8,1}$ {\it a posteriori}. We refrain from writing out explicitly the additional field strengths $\Theta^{8}$, $\Theta^{7,1}$ and $\Theta^{6,2}$ and their $K(\mf{e}_{11})$ transformations for brevity, because they do not contribute to the field equations described in this paper. 
 
The $K(\mf{e}_{11})$ invariant contribution from the fields $\bar{\cG}_{I_1}$ in our level truncation scheme produces a term quadratic in $\mathcal{N}^{8}$, such that the complete Lagrangian \eqref{THLagrange} becomes
\be  \label{eq:LLT}
\mathcal{L} = \mathcal{L}^\ord{0} - \frac{1}{2} \frac{1}{8!} \mathcal{N}_{a_1\dots a_8} \mathcal{N}^{a_1\dots a_8} + \dots \ .  
\ee
The relative coefficient is compatible with gauge invariance as we shall shortly exhibit. 

Note that upon restricting the fields to depend on the eleven coordinates $x^m$ in $\mathcal{L}$, all the field strengths $\Theta$ drop out, while the contributions from the `gradient' field strengths $\cH$ and $\cN$ become equal to those of the `curl' field strengths $\cG$ and $\Omega$ modulo a total derivative, save for the term containing $\Omega_{2}{}^1$, which is the only one that contributes to the linearised Ricci scalar. After integration by parts the Lagrangian reduces to twice the standard free Lagrangian in the democratic formulation of supergravity, with the correct normalisation 
\begin{multline}
 \frac{1}{2} \mathcal{L} \sim e R(h) - \frac{1}{2} \Bigl( \frac{1}{9!} \Omega_{a_1\cdots a_9,b}  \Omega^{a_1\cdots a_9,b} -  \frac{1}{8!} \Omega_{a_1\cdots a_{10}}  \Omega^{a_1\cdots a_{10}} \Bigr .  \\
 \Bigl . + \frac{1}{7!} \cG_{a_1\cdots a_7}  
\cG^{a_1\cdots a_7}  + \frac{1}{4!} \cG_{a_1\cdots a_4}  \cG^{a_1\cdots a_4}  \Bigr) + \dots 
\end{multline}
We expect this property to extend to all levels, such that each field would get a contribution to its kinetic term from its `curl field' strength and its `gradient' field strength. Though this Lagrangian produces the correct field equations, it is nonetheless formal. Its energy momentum tensor involves, for example, infinitely many copies of the same degrees of freedom through the unfolding mechanism and would therefore require an appropriate regularisation.

Note that the level $\ell-\frac{3}{2}$ field strengths (both $\cG^\ord{\ell-\frac{3}{2}}$ and $\bar{\cG}^\ord{\ell-\frac{3}{2}}$) have the schematic form\footnote{Note that $\partial^\ord{-n-\frac{3}{2}}$ is the horizontal degree $q=-n$ derivative,  $A^\ord{\ell}$ is the degree $q=\ell-3$ potential, and $\cG^\ord{\ell}$ is the degree $q=\ell-\frac{3}{2}$ field strength. The absolute value arises because the coset potentials are identified for positive and negative $\ell$, compare with~\eqref{gff}.} 
\be 
\cG^\ord{\ell-\frac{3}{2}} = \sum_{n\ge 0} \partial^\ord{-n-\frac{3}{2}} A^\ord{|\ell+n|} \ , \label{SchemaG}  
\ee
so only the field strengths of level $\frac{3}{2} \ge \ell \ge -\frac{15}{2}$ have non-trivial contributions up to order $\mathcal{O}(4,4)$. Because the Lagrangian is of the form $\mathcal{L} \sim- \sum_\ell |\cG^\ord{\ell-\frac{3}{2}} |^2- \sum_{n\ge 0}|\bar{\cG}^\ord{-n-\frac{9}{2}} |^2$, the equation of motion for a level $\ell$ field is of the schematic form 
\be 
\mathcal{E}^\ord{\ell} = \sum_{n\ge 0} \partial^\ord{-n-\frac32} \mathcal{G}^\ord{\ell-n-\frac32} +\sum_{n\ge 0} \partial^\ord{-n-\frac32} \mathcal{G}^\ord{-\ell-n-\frac32} \ , 
\label{SchemaE} 
\ee
where the field strengths can be either $\cG$ or $\bar{\cG}$.

To check the gauge invariance of the equations of motion following from this Lagrangian, it will be convenient to introduce a set of spurious fields $L^\alpha$ in the Lie algebra of $K(E_{11})$ and in $\ell_2$, with the gauge transformation 
\be \delta L^{\alpha} = D^{\alpha \, M_0}{}_{N_0} \ ( \partial_{M_0} \Xi^{N_0} -M^{N_0P_0} M_{M_0Q_0} \partial_{P_0} \Xi^{Q_0}) \ee
such that the linear combinations  $A^\pm = \frac{1}{2} ( A \pm L)$ defined as 
\bea
&&\hspace{-5mm}h^\pm_a{}^b = \frac{1}{2} ( h_a{}^b \pm L_a{}^b ) \, , \quad  A^\pm_{a_1a_2a_3} =   \frac{1}{2} ( A_{a_1a_2a_3} \pm L_{a_1a_2a_3} ) \, , \quad  A^\pm_{a_1\dots a_6} =   \frac{1}{2} ( A_{a_1\dots a_6} \pm L_{a_1\dots a_6} ) \, ,\quad \dots \nn\\
&&X^\pm_{a_1\dots a_9} = \frac{1}{2} ( X_{a_1\dots a_9} \pm L_{a_1\dots a_9} ) \ ,\quad \dots 
\eea
transform according to \eqref{eq:Xitrm} for $X^+$ and \eqref{barXe11} for $X^-$. The field strength  \eqref{eq:FSC} is defined by construction using this substitution as $\mathcal{F}_I[h^+,A^+,A^-,X^+]|_{L=0},\bar{\cF}_{I_1}[h^+,A^+,A^-,X^-]|_{L=0}$. In this section we shall prove in the low level truncation that 
\begin{align} 
\label{ConjGaIn} 
\mathcal{E}_\alpha &=\hat{f}^{M_0}{}_{\alpha, I} M^{IJ} \partial_{M_0} \mathcal{G}_J[h,A,X] + \widehat{\Pi}_{I_1 \alpha}^{M_0}  M^{I_1J_1} \partial_{M_0}\bar {\cG}_{J_1}[h,A,X]\\
&= 2 \hat{f}^{M_0}{}_{\alpha, I}  M^{IJ}  \partial_{M_0} \mathcal{F}_J[h^+,A^+,A^-,X^+] +2 \widehat{\Pi}_{I_1 \alpha}^{M_0}  M^{I_1J_1}  \partial_{M_0} \bar{\cF}_{J_1}[h^+,A^+,A^-,X^-]   +\mathcal{O}(4,3) \ , 
\nn
\end{align}
such that the dependence in the spurious fields $L$ drops out automatically. Because the right hand side is linear in the manifestly gauge invariant field strength $\mathcal{F}_I[h^+,A^+,A^-,X^+] $, and $\bar{\cF}[h^+,A^+,A^-,X^-] $  it follows from this equation that the second order field equation is itself gauge invariant in the low level truncation. If the equation~\eqref{ConjGaIn} was valid for all levels, this would show gauge invariance of the second order equations to all levels.

The fact that~\eqref{ConjGaIn} does not depend on $L$ can be understood in the schematic form (\ref{SchemaG},\ref{SchemaE}) as the property that the fields $\phi^\ord{\ell}$ of opposite level $\ell$ always arise with the same tensor structure such that the dependence in $L^\ord{\ell}$ drops out upon using $\phi^\ord{\ell} = \frac{1}{2} ( A^\ord{|\ell|} +\mbox{sign} \, \ell\,  L^\ord{|\ell|})$. Ignoring tensor structures and coefficients the identity to verify is
\begin{align} 
&\quad   \sum_{n\ge 0} \partial^\ord{-n-\frac32} \mathcal{F}^\ord{\ell-n-\frac32} +\sum_{m\ge 0} \partial^\ord{-m-\frac32} \mathcal{F}^\ord{-\ell-m-\frac32} \CR
&=  \sum_{n\ge 0} \partial^\ord{-n-\frac32} \Bigl( \sum_{m\ge 0} \partial^\ord{-m-\frac32} \phi^\ord{\ell+m-n} \Bigr)  +\sum_{m\ge 0} \partial^\ord{-m-\frac32}  \Bigl( \sum_{n\ge 0} \partial^\ord{-n-\frac32} \phi^\ord{-\ell-m+n} \Bigr) \CR
&=  \sum_{m\ge 0} \sum_{n\ge 0}\partial^\ord{-m-\frac32}\partial^\ord{-n-\frac32} (  \phi^\ord{\ell+m-n}+ \phi^\ord{-\ell-m+n} ) \CR
&=  \sum_{m\ge 0} \sum_{n\ge 0}\partial^\ord{-m-\frac32}\partial^\ord{-n-\frac32} A^\ord{|\ell + m-n|}  = \frac{1}{2} \mathcal{E}^\ord{\ell}  \ . 
\end{align}
This scheme allows a consistent level truncation. We shall now exhibit that the cancellation of $L$ is indeed occurring for some relevant examples. One computes from \eq{eq:LLT}, for instance, that the equation of motion for the 3-form potential is
\begin{align} 
\label{E3}
\mathcal{E}_{a_1a_2a_3} &= - \partial^{a_4}  \cG_{a_1a_2a_3a_4} 
+3 \partial_{[a_1a_2} \Omega_{a_3]b}{}^b 
+ 3 \partial_{b[a_1} \Omega_{a_2a_3]}{}^b + \frac{1}{2} \partial_{b_1b_2b_3[a_1a_2}  \cH_{a_3]}{}^{b_1b_2b_3}  \\
& \quad+\frac{1}{2} \partial_{a_1a_2a_3b_1b_2}  \cH_{b_3}{}^{b_1b_2b_3}  - \frac{1}{40} \bigl(  \partial_{[a_1a_2}{}^{b_1\dots b_5,b_6} + \tfrac13 \partial_{[a_1a_2}{}^{b_1\dots b_6}\bigr) \cH_{a_3]b_1\dots b_6} \CR
&\quad + \frac{1}{4!} \bigl( \partial_{a_1a_2a_3b_1\dots b_4,b_5} - \tfrac15 \partial_{a_1a_2a_3b_1\dots b_5} \bigr) \cH_c{}^{b_1\dots b_5c} + \frac{1}{4!} \partial_{a_1a_2a_3b_1\dots b_4,c} \Theta^{b_1\dots b_4,c} \CR
&\quad+ 4 \partial_{[a_1}  \cH^{a_4}{}_{a_2a_3a_4]} + \frac{1}{2} \partial^{b_1b_2}  \cH^{b_3}{}_{a_1a_2a_3b_1b_2b_3} 
+\partial^{b_1b_2}  \Theta_{a_1a_2a_3b_1,b_2} \CR 
&\quad +\frac{1}{4!} 
\partial^{b_1b_2b_3b_4b_5}  \cH^c{}_{a_1a_2a_3b_1\cdots b_4c,b_5}-\frac{1}{5!} 
\partial^{b_1b_2b_3b_4b_5}  ( \cH^c{}_{a_1a_2a_3b_1\cdots b_5,c} - \cN_{a_1a_2a_3b_1\dots b_5}) + \dots  \ ,\nn
\end{align}
where the ellipses stand for terms of the form  $ \partial^\ord{-\frac72} \cG^\ord{-\frac92} $ and  $\partial^\ord{-\frac92} \cG^\ord{-\frac{11}2}$. To check the formula \eqref{ConjGaIn}, we compute the same combination of field strengths using the component expression for $\mathcal{F}_I[h^+,A^+,A^-,X^+]$ and $\bar {\cF}_{I_1} [h^+,A^+,A^-,X^-]$ to exhibit that the dependence 
on $L$ drops out,  {\allowdisplaybreaks{
\begin{align} 
\label{E3Compo}  
&\quad - \partial^{a_4}  \cF^\ord{-\frac{1}{2}}_{a_1a_2a_3a_4} 
+3 \partial_{[a_1a_2} \cF^\ord{-\frac{3}{2}}{}_{a_3]b}{}^b 
+ 3 \partial_{b[a_1} \cF^\ord{-\frac{3}{2}}{}_{a_2a_3]}{}^b + \frac{1}{2} \partial_{b_1b_2b_3[a_1a_2}  \cF^\ord{-\frac{5}{2}}{}_{a_3]}{}^{b_1b_2b_3} 
\nn\\* 
& \quad+\frac{1}{2} \partial_{a_1a_2a_3b_1b_2}  \cF^\ord{-\frac{5}{2}}{}_{b_3}{}^{b_1b_2b_3} - \frac{1}{40} \bigl(  \partial_{[a_1a_2}{}^{b_1\dots b_5,b_6} + \tfrac13 \partial_{[a_1a_2}{}^{b_1\dots b_6}\bigr) \cF^\ord{-\frac72}{}_{a_3]\, b_1\dots b_6} \nn\\*
&\quad + \frac{1}{4!} \bigl( \partial_{a_1a_2a_3b_1\dots b_4,b_5} - \tfrac15 \partial_{a_1a_2a_3b_1\dots b_5} \bigr) \cF^\ord{-\frac72}_{\hspace{4.5mm} c}{}^{b_1\dots b_5c} + \frac{1}{4!} \partial_{a_1a_2a_3b_1\dots b_4,c} \cF^{\ord{-\frac72}b_1\dots b_4,c} \nn\\*
&\quad+ 4 \partial_{[a_1}  \cF^\ord{-\frac{5}{2}}{}^{a_4}{}_{a_2a_3a_4]} + \frac{1}{2} \partial^{b_1b_2}  \cF^\ord{-\frac{7}{2}}{}^{b_3}{}_{a_1a_2a_3b_1b_2b_3} 
+\partial^{b_1b_2}  \cF^\ord{-\frac{7}{2}}{}_{a_1a_2a_3b_1,b_2}  \nn\\*
& \quad+\frac{1}{4!} 
\partial^{b_1b_2b_3b_4b_5}  \cF^\ord{-\frac{9}{2}}{}^c{}_{a_1a_2a_3b_1\cdots b_4c,b_5}-\frac{1}{5!} \partial^{b_1b_2b_3b_4b_5}  ( \cF^\ord{-\frac{9}{2}}{}^c{}_{a_1a_2a_3b_1\cdots b_5,c} - \bar{\cF}_{a_1a_2a_3b_1\dots b_5}) +\dots \nn\\
&= - 4\partial^{a_4}  \partial_{[a_1} A^{}_{a_2a_3a_4]} 
+ \frac{1}{2} \partial^{b_1} \partial^{b_2b_3} A^{}_{a_1a_2a_3b_1b_2b_3} 
+ 12 \partial_{[a_1} \partial_{a_2a_3} h^{{}a_4}{}_{a_4]} + 12 \partial_{b[a_1} \partial_{a_2} h^{{}b}{}_{a_3]} \CR
&\quad + 6 \partial^{b_1b_2} \partial_{[b_1a_1} A^{}_{a_2a_3]b_2} 
- \frac{1}{2} \partial_{b_1b_2a_1a_2a_3} \partial_{b_3} A^{{}b_1b_2b_3} 
- \frac{1}{2} \partial_{b_1b_2b_3[a_1a_2} \partial_{a_3]} A^{{}b_1b_2b_3} 
\nn\\*
&\quad + \frac{1}{4!} \bigl( \partial_{a_1a_2a_3}{}^{b_1\dots b_4,b_5} - \tfrac{1}{5} \partial_{a_1a_2a_3}{}^{b_1\dots b_5} \bigr) \partial^{b_6} A_{b_1\dots b_6} - \frac{1}{40}   \bigl( \partial_{[a_1a_2}{}^{b_1\dots b_5,b_6} + \tfrac{1}{3} \partial_{[a_1a_2}{}^{b_1\dots b_6} \bigr) \partial_{a_3]} A_{b_1\dots b_6} 
\nn\\*
& \quad- \frac{1}{6} \partial_{a_1a_2a_3}{}^{b_1\dots b_4,c}  \bigl( 3 \partial_{b_1b_2} A_{b_3b_4c} + \partial_{c\hspace{0.12mm}b_1} A_{b_2b_3b_4} \bigr) + \frac12 \partial_{a_1a_2a_3b_1b_2} ( \partial^{b_1b_2} h_c{}^c - 4 \partial^{cb_1} h_c{}^{b_2}) 
\nn\\*
&\quad + \frac32 \partial_{b_1b_2b_3[a_1a_2} \partial^{b_1b_2} h_{a_3]}{}^{b_3} + \frac{1}{4!} \partial^{b_1} \partial^{b_2\cdots b_6} h_{a_1a_2a_3b_1\cdots b_5,b_6} - \frac{1}{5!} \partial^{b_1} \partial^{b_2\cdots b_6} X_{a_1a_2a_3b_1\cdots b_6}+\dots 
\end{align}
where we write explicitly the level to avoid confusion between field strengths with the same number of indices.\footnote{For example, %
\begin{align} 
\cF^\ord{\frac{1}{2}}_{a_1\cdots a_7} &=\sqrt{e} \,   
e_{a_1}{}^{n_1} \cdots   e_{a_7}{}^{n_7} \cF_{n_1\cdots n_7} 
\,,&\quad
\cF^\ord{-\frac{1}{2}}_{a_1a_2a_3a_4} &=\sqrt{e} \, 
e_{a_1}{}^{n_1} \cdots   e_{a_4}{}^{n_4} \cF_{n_1n_2n_3n_4} 
\nn\\
\cF^\ord{-\frac{3}{2}}_{a_1a_2}{}^b &=\sqrt{e} \, 
e_{a_1}{}^{n_1}e_{a_2}{}^{n_2} e_m{}^b \cF_{n_1n_2}{}^m 
\,,&\quad
\cF^\ord{-\frac{5}{2}}_{a_1a_2a_3a_4} &=\sqrt{e} \, 
e_{a_1}{}^{m}  e_{n_1 a_2} e_{n_2 a_3}  e_{n_3 a_4}  \cF_m{}^{n_1n_2n_3}\nn
\end{align}}
The ellipses stand for terms of order $\mathcal{O}(4,4)$ beyond the considered level truncation, and the terms of type $\partial^5 (  \partial^2 A_6 + \partial^5 A_3 + \partial^{7;1} h_1{}^1)$ and 
$\partial^{7;1} ( \partial^2 h_{8,1}+ \partial^5 A_6 + \partial^{7;1} A_3 )$, whose dependence in the negative level fields $A^-$ would come from the field strength derivatives $\partial^\ord{-\frac72} \cF^\ord{-\frac92}$ and $ \partial^\ord{-\frac92} \cF^\ord{-\frac{11}2}$ that we have not included in the Lagrangian~\eqref{eq:LLT}. (We recall that the notation $\partial^{7;1}$ includes both the derivatives $\partial^{7,1}$ and $\partial^8$.) The first three terms in the equation~\eqref{E3Compo} reproduce the ones of \eqref{eq:Ans1} that we have obtained within the $E_{11}$ paradigm.}}

One derives similarly from \eq{eq:LLT} the second order equation for the six-form potential 
\begin{align} 
\label{E6}
 \mathcal{E}_{a_1\cdots a_6} &= \partial^{a_7} \cG_{a_1\cdots a_7} 
- 15 \partial_{[a_1a_2} \cG_{a_3a_4a_5a_6]} 
+ 15 \partial_{b[a_1a_2a_3a_4}  \Omega_{a_5a_6]}{}^b 
- 6 \partial_{[a_1a_2a_3a_4a_5}  \Omega_{a_6]b}{}^b 
\\
& - \bigl(  5 \partial_{b_1b_2b_3[a_1\dots a_4,a_5} -\partial_{b_1b_2b_3[a_1\dots a_5} \bigr)  \cH_{a_6]}{}^{b_1b_2b_3} - \bigl(  \partial_{a_1\dots a_6b_1,b_2} -  \partial_{a_1\dots a_6b_1b_2} \bigr)  \cH_c{}^{b_1b_2c} 
\CR
&  + 7 \partial_{[a_1} \cH^{a_7}{}_{a_2\cdots a_7]} + \partial^{b_2b_3} \cH^{b_1}{}_{a_1\cdots a_6b_1b_2,b_3}+ \frac12  \partial^{b_1b_2} ( \cH^{c}{}_{a_1\cdots a_6b_1b_2,c}- \cN_{a_1\dots a_6b_1b_2}) +  \dots  \nn
\end{align}
where the ellipses stand for terms of the form  $ \partial^\ord{-\frac52} \cG^\ord{-\frac92} $ , $ \partial^\ord{-\frac72} \cG^\ord{-\frac{11}2} $ and  $\partial^\ord{-\frac92} \cG^\ord{-\frac{13}2}$. Similarly one checks that this equation can be written in terms of gauge invariant field strengths as
\begin{align}
&\quad \partial^{a_7} \cF^\ord{\frac{1}{2}}_{a_1\cdots a_7} 
- 15 \partial_{[a_1a_2} \cF^\ord{-\frac{1}{2}}_{a_3a_4a_5a_6]} 
+ 15 \partial_{b[a_1a_2a_3a_4}  \cF^\ord{-\frac{3}{2}}{}_{a_5a_6]}{}^b 
- 6 \partial_{[a_1a_2a_3a_4a_5}  \cF^\ord{-\frac{3}{2}}{}_{a_6]b}{}^b \CR
&\quad - \bigl(  5 \partial_{b_1b_2b_3[a_1\dots a_4,a_5} -\partial_{b_1b_2b_3[a_1\dots a_5} \bigr)  \cF^\ord{-\frac52}_{\hspace{2.5mm} a_6]}{}^{b_1b_2b_3} - \bigl(  \partial_{a_1\dots a_6b_1,b_2} -  \partial_{a_1\dots a_6b_1b_2} \bigr)  \cF^\ord{-\frac52}_{\hspace{4.0mm}c}{}^{b_1b_2c} 
\CR
&\quad + 7 \partial_{[a_1} \cF^\ord{-\frac{7}{2}}{}^{a_7}{}_{a_2\cdots a_7]} 
+ \partial^{b_2b_3} \cF^\ord{-\frac{9}{2}}{}^{b_1}{}_{a_1\cdots a_6b_1b_2,b_3}+\frac12 \partial^{b_1b_2} ( \cF^\ord{-\frac{9}{2}}{}^{c}{}_{a_1\cdots a_6b_1b_2,c}- \bar{\cF}_{a_1\cdots a_6b_1b_2}) + \dots \CR
&= 7 \partial^{a_7} \partial_{[a_1} A^{}_{a_2\cdots a_7]} 
- 60 \partial_{[a_1a_2} \partial_{a_3} A^{}_{a_4a_5a_6]} + 60 \partial_{b[a_1\cdots a_4} \partial_{a_5} h^{}_{a_6]}{}^b 
- 42   \partial_{[a_1\cdots a_5} \partial_{a_6} h^{}_{a_7]}{}^{a_7} \CR
&  - \partial_{[a_1} \bigl( 3 \partial_{a_2\dots a_6]}{}^{b_1b_2,b_3} +\partial_{a_2\dots a_6]}{}^{b_1b_2b_3}\bigr) A_{b_1b_2b_3} + \bigl(  \partial_{a_1\dots a_6}{}^{b_1,b_2} -\partial_{a_1\dots a_6}{}^{b_1b_2}\bigr) \partial^{b_3}A_{b_1b_2b_3} \CR
& + \partial^{b_1} \partial^{b_2b_3} h^{}_{a_1\cdots a_6b_1b_2,b_3} 
-\frac{1}{2} \partial^{b_1} \partial^{b_2b_3} X_{a_1\cdots a_6b_1b_2b_3}  +
\dots 
\end{align}
where the ellipses stand for
terms in $\partial^2 (  \partial^2  A_3 + \partial^5 h_1{}^1 + \partial^{7;1} A_3 )$ that get contributions from $\partial^2 \cF^\ord{-\frac92}$ and similarly for $\partial^5 \cF^\ord{-\frac{11}2}$ and $\partial^{7;1} \cF^\ord{-\frac{13}2}$. Note that $X_9$ appears explicitly in this equation, and its gauge variation in $\partial_1 \lambda_8$ is necessary for the equation to be gauge invariant. 
Let us finally give the Einstein equation $\mathcal{R}_a{}^b  - \frac12 \delta_a^b \mathcal{R}_c{}^c= 0 $ through the Ricci tensor 
\begin{align} 
\label{R2}
\mathcal{R}_a{}^b &= 2 \partial^{[b} \Omega_{ac}{}^{c]} 
+ 2\partial_{[a}  \Omega^{bc}{}_{c]} 
+ \frac{1}{2} \partial_{c_1c_2} \cH_a{}^{bc_1c_2} 
- \partial_{ac_1} \cH_{c_2}{}^{bc_1c_2}  
+   \frac{1}{2} \partial^{c_1c_2} \cH^b{}_{ac_1c_2}
\CR
& - \partial^{bc_1} \cH^{c_2}{}_{ac_1c_2} 
+ \frac{1}{3} \delta_a^b \partial_{c_1c_2} \cH_{c_3}{}^{c_1c_2c_3}+ \partial_{ac} \Theta^{b,c}+ \partial^{bc} \Theta_{a,c}  - \frac{1}{4!} \partial_{ac_1\dots c_4} \cH_d{}^{bc_1\dots c_4d} 
\CR
&-\frac{1}{4!} \partial^{bc_1\dots c_4} \cH^d{}_{ac_1\dots c_4d}-\frac{1}{5!} \partial_{c_1\dots c_5} \cH_a{}^{bc_1\dots c_5}-\frac{1}{5!} \partial^{c_1\dots c_5} \cH^b{}_{ac_1\dots c_5}  + \frac{8}{6!} \delta_a^b  \partial_{c_1\dots c_5}  \cH_d{}^{c_1\dots c_5d} 
\CR
& + \frac{1}{4!} \partial_{ac_1\dots c_4} \Theta^{c_1\dots c_4,b}+\frac{1}{4!} \partial^{bc_1\dots c_4} \Theta_{c_1\dots c_4,a}   + \mathcal{O}( \partial^\ord{-\frac92} \cG^\ord{-\frac{9}2} )\ , \qquad \quad 
\end{align}
which can also be written in terms of gauge invariant field strengths as
\begin{align}
& \quad 2 \partial^{[b} \cF^\ord{-\frac{3}{2}}{}_{ac}{}^{c]} 
+ 2\partial_{[a}  \cF^\ord{-\frac{3}{2}}{}^{bc}{}_{c]} 
+ \frac{1}{2} \partial_{c_1c_2} \cF^\ord{-\frac{5}{2}}{}_a{}^{bc_1c_2} 
- \partial_{ac_1} \cF^\ord{-\frac{5}{2}}{}_{c_2}{}^{bc_1c_2}  
+   \frac{1}{2} \partial^{c_1c_2} \cF^\ord{-\frac{5}{2}}{}^b{}_{ac_1c_2}
\CR
&\quad - \partial^{bc_1} \cF^\ord{-\frac{5}{2}}{}^{c_2}{}_{ac_1c_2} 
+ \frac{1}{3} \delta_a^b \partial_{c_1c_2} \cF^\ord{-\frac{5}{2}}{}_{c_3}{}^{c_1c_2c_3}
+ \partial_{ac} \cF^\ord{-\frac{5}{2}}{}^{b,c}+ \partial^{bc} \cF^\ord{-\frac{5}{2}}_{a,c}  -\frac{1}{4!} \partial_{ac_1\dots c_4} \cF^\ord{-\frac72}_{\hspace{4.5mm} d}{}^{bc_1\dots c_4d} 
\CR
&\quad-\frac{1}{4!} \partial^{bc_1\dots c_4} \cF^\ord{-\frac72}{}^d{}_{ac_1\dots c_4d}-\frac{1}{5!} \partial_{c_1\dots c_5} \cF^\ord{-\frac72}_{\hspace{4.5mm}  a}{}^{bc_1\dots c_5}-\frac{1}{5!} \partial^{c_1\dots c_5} \cF^\ord{-\frac72}{}^b{}_{ac_1\dots c_5}  
\CR
&\quad + \frac{8}{6!} \delta_a^b  \partial_{c_1\dots c_5}  \cF^\ord{-\frac72}_{\hspace{4.5mm} d}{}^{c_1\dots c_5d} + \frac{1}{4!} \partial_{ac_1\dots c_4} \cF^{\ord{-\frac72}c_1\dots c_4,b}+\frac{1}{4!} \partial^{bc_1\dots c_4} \cF^\ord{-\frac72}_{c_1\dots c_4,a}   + \mathcal{O}( \partial^\ord{-\frac92} \cF^\ord{-\frac{9}2} ) 
\CR
&= 4 \partial_{[a} \partial^{[b} h^{}_{c]}{}^{c]} + \partial_{ac_1} \partial_{c_2} A^{{}bc_1c_2} 
- \frac{1}{2} \partial_a \partial_{c_1c_2} A^{{}bc_1c_2}+ \partial^{bc_1} \partial^{c_2} A^{}_{ac_1c_2} 
- \frac{1}{2} \partial^b \partial^{c_1c_2} A^{}_{ac_1c_2}
\CR
&  - \frac{1}{3} \delta_a^b \partial^{c_1} \partial^{c_2c_3} A^{}_{c_1c_2c_3} 
+ \partial_{ac_1} \partial^{c_1c_2} h^{}_{c_2}{}^b  +  \partial^{bc_1} \partial_{c_1c_2} h^{}_{a}{}^{c_2} 
- 2 \partial_{ac} \partial^{bd} h^{}_d{}^c -  \partial_{ac} \partial^{bc} h^{}_d{}^d  
\CR
&+\frac12  \partial_{c_1c_2} \partial^{c_1c_2} h^{}_a{}^b + \frac{1}{6} \delta_a^b 
\scal{ 4 \partial_{ce} \partial^{de} h^{}_d{}^c + \partial_{c_1c_2} \partial^{c_1c_2} h^{}_d{}^d}  +\ldots\,,
\end{align}
where the ellipses denote terms that involve at least one derivative of type $\partial^5$ or lower level. The equations of motion of the dual graviton $h_{8,1}$ and $X_9$ are
\begin{align} 
\label{DualGra2nd}
\mathcal{R}_{a_1\cdots a_8,b}  &= \partial^{a_9} \scal{ \Omega_{a_1\cdots a_9,b}  - \Omega_{[a_1\cdots a_8|b|,a_9]} }+8 \partial_{\lsharp b,a_1} \cG_{a_2\cdots a_8\rsharp} - 70   \partial_{\lsharp b,a_1\cdots a_4} \cG_{a_5\cdots a_8\rsharp}  \\
& \hspace{20mm} + 9 \partial_{[a_1} \cH^{a_9}{}_{a_2\dots a_9],b}  - \partial_b \cH^c{}_{a_1\dots a_8,c} + \partial_{\lsharp b,} \cN_{a_1\dots a_8\rsharp} + \dots \CR
\mathcal{R}_{[a_1\cdots a_9]}&= \partial^{b} \scal{ \Omega_{[a_1\cdots a_8|b|,a_9]}  + \Omega_{a_1\cdots a_9b} } +  \partial_{[a_1} \cN_{a_2\dots a_9]}  -4 \partial_{[a_1a_2} \cG_{a_3\cdots a_9]} + 14 \partial_{[a_1\cdots a_5} \cG_{a_6\cdots a_9]}  + \dots \nn
\end{align}
The ellipses stand for terms in  $\partial^\ord{-\frac92} \cG^\ord{-\frac32}$, $\partial^\ord{-\frac52} \cG^\ord{-\frac{11}{2}}$, $\partial^\ord{-\frac72} \cG^\ord{-\frac{13}{2}}$, $\partial^\ord{-\frac92} \cG^\ord{-\frac{15}{2}}$ that are not determined at this order. One checks using the same argument that these equations are indeed gauge invariant modulo the section constraint. Note that in this case the dependence on $X_9$ is very important, and the gauge transformation of $X_9$ into both $\partial_1 \lambda_8$ and $\partial^8 \xi^1$ is required for the gauge invariance of the equations of motion to be satisfied in this level truncation. We stress that the terms in $\mathcal{N}^8$ are crucial for the dependence in $X_9$ to be consistent with gauge invariance. This concludes our computation that the second order equations of motion deriving from the Lagrangian \eqref{eq:LLT} are gauge invariant within the level truncation that we consider.

As explained in the last section, the solutions to the first order duality equations \eqref{FirstTH} solve automatically the second order equations \eqref{SecondTH} modulo the section constraint. We shall now discuss in more detail the equations~\eqref{FirstTH} within the $\mathfrak{gl}(11)$ decomposition. One derives in this case 
\begin{align} 
\label{1stOrder}
\cG_{a_1\cdots a_7} &= \frac{1}{4!} \varepsilon_{a_1\cdots a_7}{}^{b_1\cdots b_4}  \cG_{b_1b_2b_3b_4}\ , \CR
\cG_{a_1\cdots a_4} &= -\frac{1}{7!} \varepsilon_{a_1\cdots a_4}{}^{b_1\cdots b_7} \cG_{b_1\cdots b_7}\ ,  \CR
\Omega_{a_1\cdots a_9,b} &=- \frac{1}{ 2} \varepsilon_{a_1\cdots a_9}{}^{c_1c_2} \eta_{bd} \Scal{\Omega_{c_1c_2}{}^d+\frac15 \delta^d_{c_1} \Omega_{c_2c_3}{}^{c_3}}\ ,  \CR
\Omega_{a_1a_2}{}^b+\frac15 \delta^b_{[a_1} \Omega_{a_2]c}{}^{c}  &= \frac{1}{ 9!} \varepsilon_{a_1a_2}{}^{c_1\cdots c_9} \eta^{bd} \Omega_{c_1\cdots c_9,d}\ ,  \CR
\Omega_{ab}{}^b &=
\frac{1}{9!} \varepsilon_a{}^{b_1\cdots b_{10}} \Omega_{b_1\cdots b_{10}}\ ,
\end{align}
that transform indeed together with respect to $K(\mf{e}_{11})$ as
\begin{align}
&\delta \Scal{\cG_{a_1\cdots a_4} +\frac{1}{7!} \varepsilon_{a_1\cdots a_4}{}^{b_1\cdots b_7}  \cG_{b_1\cdots b_7} } \CR
&\qquad\qquad=\frac{1}{6}  \Lambda^{a_5a_6a_7} \Scal{\cG_{a_1\cdots a_7} - \frac{1}{4!} \varepsilon_{a_1\cdots a_7}{}^{b_1\cdots b_4} \cG_{b_1b_2b_3b_4} } \CR
&\qquad\qquad\quad\, - 6 \Lambda_{b[a_1a_2} \Scal{\Omega_{a_3a_4]}{}^b-\frac{1}{ 9!} \varepsilon_{a_3a_4]}{}^{c_1\cdots c_9} \eta^{bd} \scal{ \Omega_{c_1\cdots c_9,d}-\Omega_{c_1\cdots c_9d}}}
\end{align}
and
\begin{align}
&  \delta \Scal{ \Omega_{a_1a_2}{}^b+\frac15 \delta^b_{[a_1} \Omega_{a_2]c}{}^{c}-\frac{1}{ 9!} \varepsilon_{a_1a_2}{}^{c_1\cdots c_9} \eta^{bd} \Omega_{c_1\cdots c_9,d}}\CR
 &\quad= \frac{1}{2} \Lambda^{bc_1c_2}  \Scal{\cG_{a_1a_2c_1c_2} +\frac{1}{7!} \varepsilon_{a_1a_2c_1c_2}{}^{b_1\cdots b_7}  \cG_{b_1\cdots b_7}  } \CR
 &\quad\quad\, + \frac{1}{10} \delta^b_{[a_1} \Lambda^{c_1c_2c_3} \Scal{\cG_{a_2]c_1c_2c_3} +\frac{1}{7!} \varepsilon_{a_2]c_1c_2c_3}{}^{b_1\cdots b_7}  \cG_{b_1\cdots b_7}  }  \CR
 &\quad\quad\, + \Lambda_{c_1c_2[a_1} \scal{ \cH_{a_2]}{}^{bc_1c_2}  - \tfrac{1}{10} \delta_{a_2]}^b \cH_d{}^{c_1c_2d}} -\tfrac{1}{10} \Lambda_{c_1c_2c_3} \delta_{[a_1}^b \cH_{a_2]}{}^{c_1c_2c_3} + \Lambda_{a_1a_2c} \Theta^{b,c} + \cdots \ .
\end{align} 
The other field strength components $\cH$ are related by duality to fields that we have not yet considered in this level truncation, and we shall discuss them separately below. 

One can then check explicitly the Bianchi identity~\eqref{BianchiTHA} explained in the last section, such that these first order equations defined for the coset component, even if not gauge invariant by themselves, solve the second order equation at the considered truncation level. So that upon using the duality equations, \eqref{E3}, \eqref{E6} and \eqref{R2} vanish automatically up to the section constraint:  
\begin{align}
{\cal E}_{a_1a_2a_3} &=  \frac{1}{8!} \varepsilon_{a_1a_2a_3}{}^{b_1\cdots b_8} \scal{ 8 \partial_{b_1} \cG_{b_2\cdots b_8} - \partial^{b_9c} \Omega_{b_1\cdots b_9,c}+6  \partial^{b_9b_{10}} \Omega_{b_1\cdots b_{10}}} =0\ , \CR
 {\cal E}_{a_1\cdots a_6} \hspace{-0.2mm} &=\hspace{-0.2mm}  \frac{1}{5!} \varepsilon_{a_1\cdots a_6}{}^{b_1\cdots b_5}  \Scal{ 5 \partial_{b_1} \cG_{b_2\cdots b_5} +\frac{1}{2} \partial^{b_6b_7} \cG_{b_1\cdots b_7} +\frac{1}{4!} \partial^{b_6\cdots b_9c} \Omega_{b_1\cdots b_9,c}-\frac{1}{8} \partial^{b_6\cdots b_{10}} \Omega_{b_1\cdots b_{10}}} \!=\!0\ \CR
{\cal R}_{a}{}^b &= - \frac{1}{10!} \varepsilon_a{}^{c_1\cdots c_{10}} \Scal{ 10 \partial_{c_1} \Omega_{c_2\cdots c_{10},}{}^b -9 \partial^b \Omega_{c_1\cdots c_{10} }}+\frac{1}{10!} \varepsilon^{c_1\cdots c_{11}} \delta_a^b \partial_{c_1}  \Omega_{c_2\cdots c_{11}} = 0 \ .
\end{align}
Let us finally consider the equations of motion for the dual graviton, defining for convenience $\mathcal{R}_{a_1\cdots a_8;b}= \mathcal{R}_{a_1\cdots a_8,b}+ \mathcal{R}_{a_1\cdots a_8b}$, that accommodates both $h_{8,1}$ and $X_9$,\footnote{Where we do not write the field strengths of level $\ell = - 9/2$ that are dual to field strength of level $\ell = 9/2$ that are neglected in our truncation scheme.} 
\bea 
\mathcal{R}_{a_1\cdots a_8;b}  &=& \partial^{a_9} \scal{ \Omega_{a_1\cdots a_9,b}  - \Omega_{a_1\cdots a_9b} }+8 \partial_{b[a_1} \cG_{a_2\cdots a_8]} - 12 \partial_{[ba_1} \cG_{a_2\cdots a_8]} \\
&& \hspace{20mm} - 70   \partial_{b[a_1\cdots a_4} \cG_{a_5\cdots a_8]} +84 \partial_{[ba_1\cdots a_4} \cG_{a_5\cdots a_8]}+ \dots \CR
&=& - \frac{1}{6} \varepsilon_{a_1\cdots a_8}{}^{c_1c_2c_3}\eta_{bd}  \biggl( 3  \partial_{c_1} \Omega_{c_2c_3}{}^d  - \partial^{dc_4} \cG_{c_1c_2c_3c_4} + \frac{1}{2} \delta^d_{c_1} \partial^{c_4c_5} \cG_{c_2c_3c_4c_5} \biggr . \CR
&& \hspace{50mm}   \biggl . - \frac{1}{4!}  \partial^{dc_4\cdots c_7} \cG_{c_1\cdots c_7}+\frac{2}{5!}\delta^d_{c_1}   \partial^{c_4\cdots c_8} \cG_{c_2\cdots c_8} +\dots  \biggr) = 0\ , \nonumber \eea
which is indeed automatically solved by the solution to the first order duality equation \eqref{1stOrder} modulo the section constraint.

The gravity first order equation for the graviton and its dual are not gauge invariant, even when restricting the dependence of the fields to the eleven coordinates $x^m$. In ordinary space-time, this problem is resolved by considering the second order duality equations for the linearised Riemann tensor:
\be  \partial^{[b_1} \Omega_{a_1a_2}{}^{b_2]} =  \frac{1}{8!} \varepsilon_{a_1a_2}{}^{c_1\dots c_9} \partial^{[b_1}   \partial_{c_1} h_{c_2\dots c_9,}{}^{b_2]} \ , \label{2ndOrderGradual}\ee
from which one checks that $X_{a_1\dots a_9}$ decouples. This was also observed in the work on dualised gravity at the level of the gauge invariant Riemann tensor rather than the spin connection~\cite{Hull:2000zn,Hull:2001iu}. Generalising the Riemann tensor (rather than the Ricci tensor) in exceptional geometry is known to lead to ambiguities~\cite{Coimbra:2011ky,Hohm:2013vpa} and we do not expect the above equation to be part of a gauge invariant $K(\mf{e}_{11})$ multiplet of well-defined second order duality relations.

By construction \eqref{2ndOrderGradual} implies the standard equation of motion for the dual graviton field 
\be 16  \partial_{[a_1} \partial^{[b} h_{a_2\dots a_8c],}{}^{c]} = 0 \ , \label{DualGra2nd11D}  \ee
however, the second order gauge invariant equation \eqref{DualGra2nd} implies instead 
\be \label{SecondDualGravitonExplicit}
 \partial^{a_9} ( 9 \partial_{[a_1} h_{a_2\cdots a_9],b} + \partial_b X_{a_1\cdots a_9}) =0\ , 
\qquad \partial^{b} ( \partial_{[a_1} h_{a_2\cdots a_9],b} + \partial_{[a_1} X_{a_2\cdots a_9]b} ) = 0  \ . 
\ee
Note that these equations are gauge invariant thanks to the variation of the field $X_{a_1\cdots a_9}$. Using the first order constraint, and the property that $\Omega_{[a_1a_2a_3]}=0$ when the field dependence is restricted to the eleven supergravity coordinates, one obtains the space-time gauge invariant first order constraint 
\be 
\label{CompaStandard}\Omega_{a_1\dots a_8b,}{}^b =  \partial^b \bigl(  h_{a_1\dots a_8,b} + X_{a_1\dots a_8b}\bigr)  + 8 \partial_{[a_1} h_{a_2\dots a_8]b,}{}^b  = 0 \ , 
\ee
which can be used to get back the standard second order field equation \eqref{DualGra2nd11D}. Together with the constraint $\mathcal{N}^8=0$ from \eqref{gaugeBTHA}, this equation imposes the constraint that the curl of the trace of $h_{8,1}$ vanishes as a (partial) gauge-fixing condition. This is consistent with the interpretation of \eqref{gaugeBTHA} as a $K(\mf{e}_{11})$-invariant gauge-fixing condition. Note that this situation, where the constraint is compatible with $K(\mf{e}_{11})$ invariance, is quite different from the problematic case of the gauge-fixing condition encountered in the original formulation of the theory, for which we showed that there was no $K(\mf{e}_{11})$ multiplet of gauge-fixing conditions compatible with the vanishing of the nine-form $X_9$.

%%%%%%%%%%%%%%%%%%%%%%%%%%%%%%%%%%%%%%%%%%%%%%%%%%%%%%%%%%%%%%
\subsection{Unfolding dualities and non-geometric fluxes}
\label{NonGeoFluxes}
%%%%%%%%%%%%%%%%%%%%%%%%%%%%%%%%%%%%%%%%%%%%%%%%%%%%%%%%%%%%%%

We shall now extrapolate these results to higher level. At level $\ell=4$ there are three additional $\mf{e}_{11}$ fields: $A_{9,3}$, $B_{10,1,1}$, $C_{11,1}$ and two additional fields in the $\ell_2$ module: $X_{10,2}$ and $X_{11,1}$ \eqref{level0} (all understood to be in irreducible representations of $\mf{gl}(11)$ according to the displayed symmetrisations). Using the tensor hierarchy algebra one computes the following field strengths,
\bea 
\cH_{n_1\dots n_{10},p_1p_2p_3} &=&  10 \partial_{[n_1} ( A_{n_2\dots n_{10}],p_1p_2p_3} + X_{n_2\dots n_{10}][p_1,p_2p_3]} ) - 12  \partial_{[n_1}  X_{n_2\dots n_{10}][p_1p_2,p_3]}\ , \CR
\Theta_{n_1\dots n_{11},m,n} &=&  11 \partial_{[n_1}  B_{n_2\dots n_{11}],m,n}  + \partial_{(m} ( C_{n_1\dots n_{11},n)}+X_{n_1\dots n_{11},n)}) \ .  
\eea
Note that the indecomposable character of the $\mf{e}_{11}$ representation is such that $X_{11,1}$ is only defined modulo an arbitrary shift in $C_{11,1}$, and we have used this freedom to cancel the contribution of $C_{11,1}$ in $\cH_{10,3}$. One can anticipate using the conservation of the level that there is a duality equation of the form 
\be \label{UnfoldDual} 
\Scal{ \cH_{a}{}^{b_1b_2b_3} -\frac13 \delta^{[b_1}_{a} \cH_{c}{}^{b_2b_3]c}} = \frac{1}{ 10!} \varepsilon_{a}{}^{c_1\cdots c_{10}} \eta^{b_1d_1}\eta^{b_2d_2} \eta^{b_3d_3} \Scal{ \frac{2}{3} \cH_{c_1\cdots c_{10},d_1d_2d_3} + \frac{1}{3} \cH_{c_1\cdots c_{9}[d_1,d_2d_3]c_{10}}}   \ee
such that $A_{9,3}$ is the field dual to the gradient of $A_3$. Properties of the tensor hierarchy algebra suggest that this structure extends to all levels. A potential $A_R$ at $\mf{gl}(11)$ level $n$ for $n\geq 1$ transforming in an irreducible $\mf{gl}(11)$ representation can contribute to a field strength component at level $\ell=-\frac{3}{2} -n$ in $R_{-1}$ obtained by acting on $A_R$ with the usual derivative $\partial_1$ at level $\ell=-\frac32$. At the same time, for each irreducible $\mf{gl}(11)$ representation carried by $A_R$ of $\mathfrak{e}_{11}$ at level $n\ge 1$ there is a $\mf{gl}(11)$ highest weight representation (with outer multiplicity at least one) obtained by tensoring $A_R$ with the nine-form representation at level $n+3$. This is true since one can act with the affine subalgebra $\mf{e}_9$ on any of the generator and adding a block of nine antisymmetric indices corresponds to adding the affine null root at level $\ell=3$. Applying this to the standard fields one generates all possible fields dual to their gradients~\cite{Damour:2002cu,Riccioni:2006az,Boulanger:2015mka}. This is also consistent with the fact that the symplectic form defines duality equations between level $  -\frac{3}{2} -n$ and level $  -\frac{3}{2} +n+3$ field strengths.

One can also anticipate a first order duality equation of the form 
\be \Theta_{a_1\dots a_{11},b,c} =  \varepsilon_{a_1\dots a_{11}}  \eta_{bd} \eta_{ce} \Theta^{d,e} \ . \ee
For a solution to eleven-dimensional supergravity depending only on the coordinates $x^m$ the field strength $\Theta^{1,1}$ vanishes, so that the field strength $\Theta_{11,1,1}$ must vanish as well, or more generally be pure gauge (since the first order duality equations are not gauge invariant). We expect in this way that solving the duality equation for a solution to eleven-dimensional supergravity will impose that all the fields with more than nine antisymmetric indices will all be pure gauge. Such fields should nonetheless contribute non-trivially to non-geometric backgrounds. Let us illustrate this through the example of a Romans mass in type IIA.\footnote{For previous work on massive type IIA in connection with Kac--Moody symmetries see~\cite{Schnakenburg:2002xx,Kleinschmidt:2004dy,Henneaux:2008nr}.} According to \cite{Hohm:2011cp}, the Romans mass can be generated through a linearised metric 
\be h_1{}^{10} = h_{10}{}^1 = m y_{1\hspace{0.12mm}10} \ , \ee
where $y_{1\hspace{0.12mm}10}$ is a component of the level $\frac{5}{2}$ extended coordinate $y_{mn}$, such that 
\be 
\Theta^{a,b} = m  ( \delta^a_{10} \delta^b_{10}-\delta^a_{1}\delta^b_{1} ) \ . 
\ee
In this case one will get a non-trivial $B_{10,1,1}$ field, corroborating the observation that this field should define the ten-form in massive type IIA.\footnote{Note that $\Theta_{11,1,1}$ cannot have an $SO(1,10)$ invariant solution. This is consistent with previous observations that the potential for the Romans mass only appears after breaking the $GL(11)$ symmetry as a particular component of the $B_{10,1,1}$ potential~\cite{Kleinschmidt:2003mf,West:2004st,Henneaux:2008nr}. This non-covariance also arises in attempts to defining an M$9$-brane ancestor of the D$8$-brane coupling to the Romans mass~\cite{Bergshoeff:1996ui,Bergshoeff:1997ak}.} Note that the presence of the additional fields $X_{10,2}$ and $X_{11,1}$ allows one to write gauge invariant second order equations, eliminating the problem of having to consider arbitrarily high order equations for arbitrary high level fields as was proposed in~\cite{Tumanov:2016dxc}.

It is interesting to compare our field strengths with the standard chain of NS fluxes obtained by recursive T-dualities \cite{Shelton:2005cf,Bergshoeff:2015cba}. Considering the reduction on a circle along the $x^{10}$ direction, one can identify the NS fluxes with the field strengths 
\be 
H_{n_1n_2n_3} = \cG_{n_1n_2n_310}\ , \quad f_{n_1n_2}{}^m = \Omega_{n_1n_2}{}^m \ , \quad Q_m{}^{n_1n_2} = \cH_m{}^{n_1n_210} \ , \quad R^{n_1n_2n_3} = \Theta^{n_1n_2n_310,10} \ . 
\ee

To conclude this section, we shall analyse briefly the decomposition of the field strength representation $R_{-1}$ with respect to the branching $\mf{gl}(4)\oplus \mf{e}_{7(7)} \subset \mf{e}_{11}$. Considering the field strengths with all indices along $\mf{sl}(7)$ associated to a generalised torus one identifies 
\be 
\cG_7 , \, \cG_4,\, \Omega_2{}^1,\, \cH_1{}^3 ,\, \Theta^{1,1} ,\, \cH_1{}^6,\, \Theta^{4,1},\, \Theta^{6,2},\, \Theta^{7,1},\, \Theta^{7,4},\, \Theta^{7,7} \in {\bf 912} 
\ee
that reproduces all the components of the embedding tensor representation in four dimensions \cite{deWit:2002vt,Riccioni:2007au,Bergshoeff:2007qi,deWit:2008ta}. The field strengths with one index $\mu$ along $\mathbb{R}^{1,3}$ and all the others along $\mf{sl}(7)$
\be \cG_{\mu 6} ,\, \cG_{\mu 3} ,\, \Omega_{\mu 1}{}^1,\, \cH_\mu{}^{3} ,\, \cH_\mu{}^6 \in \Lambda_1 \otimes \mf{e}_{7(7)}  \ee
reproduce all the components of the conserved $\mf{e}_{7(7)}$ current. The field strengths with two indices  $\mu\nu$ along $\mathbb{R}^{1,3}$ and all the others along $\mf{sl}(7)$
\be 
\Omega_{\mu\nu 7,1},\, \cG_{\mu\nu 5} ,\, \cG_{\mu\nu 2} ,\, \Omega_{\mu\nu}{}^1 \in \Lambda_2 \otimes {\bf 56} 
\ee
reproduce all the components of the Maxwell field strengths. One can then straightforwardly check that the duality equations \eqref{1stOrder} restricted to these field strengths reproduce the twisted self-duality equation satisfied by the Maxwell fields in $\cN=8$ supergravity \cite{Cremmer:1979up} in the linearised approximation. Using moreover the  `reflection symmetry' of the algebra, this implies that the branching of the representation $R_{-1}$ with respect to $\mf{gl}(4)\oplus \mf{e}_{7(7)} \subset \mf{e}_{11}$ includes among infinitely many other representations 
\be 
R_{-1} \cong (\Lambda_0 \otimes {\bf 912} )\oplus (\Lambda_1 \otimes \mf{e}_{7(7)}) \oplus (\Lambda_2 \otimes {\bf 56}) \oplus( \Lambda_3 \otimes  \mf{e}_{7(7)} ) \oplus (\Lambda_4 \otimes {\bf 912})\oplus \dots  \ . 
\ee
One can therefore anticipate that the first order duality equation \eqref{FirstTH}
reproduces the twisted self-duality equation introduced in~\cite{Cremmer:1998px}, including the two-form potentials and the non\hyp{}dynamical 3-form potentials appearing in gauged supergravity \cite{deWit:2002vt}. Considering the potentials up to level $8$, one finds indeed the set of three-form potentials~\cite{Riccioni:2007au,Bergshoeff:2007qi} 
\begin{align} 
&A^\ord{1}_{\mu\nu\sigma},\, A^\ord{2}_{\mu\nu\sigma 3},\, h^\ord{3}_{\mu\nu\sigma 5,1},\, A^\ord{4}_{\mu\nu\sigma 6,3} ,\, B^\ord{4}_{\mu\nu\sigma 7,1,1} ,\, A^\ord{5}_{\mu\nu\sigma 6,6} ,\, B^\ord{5}_{\mu\nu\sigma 7,4,1},\, B^\ord{6}_{\mu\nu\sigma 7,6,2},\nn\\
&\hspace{20mm} B^\ord{6}_{\mu\nu\sigma 7,7,1},\, B^\ord{7}_{\mu\nu\sigma 7,7,4},\, B^\ord{8}_{\mu\nu\sigma 7,7,7}  \in \Lambda_3\otimes {\bf 912} \ , 
\end{align}
whose curl should appear in the four-form field strengths in $\Lambda_4\otimes {\bf 912}$. Note moroever that the non-linear field strength defined from the coset component of the Maurer--Cartan form should naturally inlude couplings allowing for the interpretation of the fluxes in $\Lambda_0 \otimes {\bf 912}$ as non-abelian gauge couplings.

%%%%%%%%%%%%%%%%%%%%%%%%%%%%%%%%
\section{Type IIB}
\label{sec:typeIIB}
%%%%%%%%%%%%%%%%%%%%%%%%%%%%%%%%

The section constraint~\eqref{eq:SC} has two well-known solutions. The first is to consider only the eleven-dimensional coordinates $x^m$ (with $m=0,1,\ldots,10$) and relates the equations above to $D=11$ supergravity. The second is the type IIB solution
where one retains the coordinates $x^\mu$ with $\mu=0,1,\ldots,8$ and the coordinate $y_{9\hspace{0.15mm}10}$ that is interpreted as the T-dual of the ninth spatial direction of $D=10$ type IIA supergravity. It is not hard to check that any fields depending on these ten coordinates satisfy the section constraint~\eqref{eq:SC}. In~\cite{Bossard:2015foa} it was shown that for $E_d$ with $d\leq 8$ these are the only two inequivalent solutions of the section constraint. 

In this section, we will analyse the first and second order field equations that result from our tensor hierarchy algebra analysis from the point of view of the type IIB solution to the section constraint. Type IIB has been discussed in an $E_{11}$ context in~\cite{Schnakenburg:2001he,West:2004st,Tumanov:2014pfa} with correspondence between the two level decompositions given in~\cite{Kleinschmidt:2003mf}. A discussion of the non-linear realisation of $E_{10}$ in a type IIB language was given in~\cite{Kleinschmidt:2004rg} and connections between exceptional field theory and type IIB supergravity can be found for example in~\cite{Coimbra:2011nw,Blair:2013gqa,Hohm:2013vpa,Malek:2015hma,Musaev:2015ces,Berman:2015rcc}. Level decompositions of $\mf{e}_{11}$ for different subgroups $GL(d)\times E_{11-d}$ have been mentioned for example in~\cite{Julia:1997cy,Riccioni:2007au,Bergshoeff:2007qi}.

The type IIB solution of the section constraint means that we only retain the following derivatives:
\begin{align}
\partial_\mu \quad (\mu=0,\ldots,8) \quad \textrm{and} \quad
\partial^{9\hspace{0.14mm} 10} \equiv \partial_9\,,
\end{align}
where $\partial_9$ denotes the derivative in the ninth spatial direction in type IIB supergravity.

%%%%%%%%%%%%%%%%%%%%%%%%%%%%%%%%%%%%%%%%%%
\subsection{Level decomposition}
%%%%%%%%%%%%%%%%%%%%%%%%%%%%%%%%%%%%%%%%%%

We consider the decomposition of $\mf{e}_{11}$ under its $\mf{gl}(10)\oplus \mf{sl}(2)$ subalgebra obtained by deleting node $9$ of its Dynkin diagram shown in Figure~\ref{fig:e11dynk}. $\mf{gl}(10)$ then is further decomposed into $\mf{gl}(9)$ that is common to both type IIB $\mf{gl}(10)$ and M-theory $\mf{gl}(11)$ and corresponds to a further removal of node $11$ from the diagram, while keeping the $\mf{sl}(2)$ associated with node $10$ manifest. The representations are listed in Table~\ref{tab:e11IIB} and are bi-graded where the level $\ell_{\mathrm{IIB}}$ is associated with node $9$ and the Kaluza--Klein level $\ell_{\mathrm{KK}}$ is associated with node $11$ and the reduction of type IIB from $D=10$ to $D=9$.

\renewcommand{\arraystretch}{1.2}
\begin{table}
\centering
\begin{tabular}{|c|c|c|c|}
\hline
Level $\ell_{\mathrm{IIB}}$ & Level $\ell_{\mathrm{KK}}$ & $\mf{sl}(9)\oplus\mf{sl}(2)$ representation & Field\\[1mm]
\hline
0 &
$\begin{matrix}
0\\
0\\
0\\
1
\end{matrix}$
&
$\begin{matrix}
(1,0,0,0,0,0,0,1) (0)\\
(0,0,0,0,0,0,0,0) (0)\\
(0,0,0,0,0,0,0,0) (2)\\
(0,0,0,0,0,0,0,1) (0)
\end{matrix}$
&
$\begin{matrix}
h^{\rm \scriptscriptstyle IIB}_\mu{}^\nu\\
h^{\rm \scriptscriptstyle IIB}_{9}{}^{9}\\
\phi_{i,j}\\
h^{\rm \scriptscriptstyle IIB}_\mu{}^{9}
\end{matrix}$\\
\hline
1 &
$\begin{matrix}
0\\
1
\end{matrix}$&
$\begin{matrix}
(0,0,0,0,0,0,0,1) (1)\\
(0,0,0,0,0,0,1,0) (1)
\end{matrix}$&
$\begin{matrix}
B_{i \mu 9}\\
B_{i \mu\nu}
\end{matrix}$\\
\hline
2 &
$\begin{matrix}
1\\
2
\end{matrix}$&
$\begin{matrix}
(0,0,0,0,0,1,0,0) (0)\\
(0,0,0,0,1,0,0,0) (0)
\end{matrix}$&
$\begin{matrix}
C_{\mu_1\mu_2\mu_3 9}\\
C_{\mu_1\ldots \mu_4}
\end{matrix}$\\
\hline
$3$ & 
$\begin{matrix}
2\\
3
\end{matrix}$&
$\begin{matrix}
(0,0,0,1,0,0,0,0) (1)\\
(0,0,1,0,0,0,0,0) (1)
\end{matrix}$&
$\begin{matrix}
B_{i \mu_1\ldots\mu_5 9}\\
B_{i \mu_1\ldots\mu_6}
\end{matrix}$
\\ %[1mm]
\hline
$4$ & 
$\begin{matrix}
2\\
3\\
3\\
3\\
4\\
4
\end{matrix}$ &
$\begin{matrix}
(0,0,1,0,0,0,0,0) (0)\\
(0,1,0,0,0,0,0,0) (0)\\
(0,1,0,0,0,0,0,0) (2)\\
(0,0,1,0,0,0,0,1) (0)\\
(0,1,0,0,0,0,0,1) (0)\\
(1,0,0,0,0,0,0,0) (2)
\end{matrix}$ &
$\begin{matrix}
h^{\rm \scriptscriptstyle IIB}_{\mu_1\ldots\mu_6 9,9}\\
h^{\rm \scriptscriptstyle IIB}_{\mu_1\ldots\mu_7,9}\\
\phi_{i,j\mu_1\ldots\mu_7 9}\\
h^{\rm \scriptscriptstyle IIB}_{9\mu_1\ldots\mu_6, \mu_7}\\
h^{\rm \scriptscriptstyle IIB}_{\mu_1\ldots \mu_7,\mu_8}\\
\phi_{i,j \mu_1\ldots\mu_8}
\end{matrix}$ \\
\hline    
\end{tabular}
\caption{\label{tab:e11IIB} \small\textit{Level decomposition of $E_{11}$ under its $\mf{gl}(9)\oplus\mf{sl}(2)$ subalgebra described in the text. The $i$-index is a fundamental index of $\mf{sl}(2)$ while the $\mu$-index is a fundamental $\mf{gl}(9)$ index. The index $9$ indicates the ninth spatial direction that is used in the duality to M-theory. The level $\ell_{\mathrm{KK}}$ is identical to the $\mf{gl}(11)$ level $\ell$ used in Table~\ref{tab:e11adj}.}}
\end{table}

The connection to the $\mf{gl}(11)$ decomposition of Table~\ref{tab:e11adj} is that the level $\ell_{\mathrm{KK}}$ corresponds to the level presented there and from this one can immediately read off the connection between the fields in the two theories. For example,
\begin{align}
A_{\mu_1\mu_2\mu_3} = C_{\mu_1\mu_2\mu_3 9}
\end{align}
etc. We note that the decomposition of the fields in $D=11$ also generates terms that are not listed above. For example, there is a component
\begin{align}
h_{\mu_1\ldots \mu_8,i}
\end{align}
of the $D=11$ dual graviton that would arise at level $(\ell_{\mathrm{IIB}}, \ell_{\mathrm{KK}})=(5,3)$ in the table above and that we have truncated away. 

In this section we are using the following index convention. Greek (curved) indices $\mu,\nu,\ldots$ lie in the range $0,1,\ldots,8$ and label the common $\mf{gl}(9)$ of type IIB and M-theory. The tangent space indices of $SO(1,8)$ will be denoted by $\alpha,\beta,\ldots$. We treat the direction $9$ that corresponds to node $11$ of the $E_{11}$ diagram separately. Indices $i,j=1,2$ are fundamental indices of the global $\mf{sl}(2)$ of type IIB (and should be thought of as corresponding to the directions $9$ and $10$ in the M-theory frame). 

We note that the equations that we derived in the previous sections covered at most the generators in the algebra up to level $\ell_{\mathrm{KK}}=3$. Inspecting Table~\ref{tab:e11IIB} we see that this does not cover all possible components of some of the `physical fields' of type IIB theory, that include all the fields of the type IIB supergravity and their duals, including the dual graviton. For example, the component $h^{\rm \scriptscriptstyle IIB}_{\mu_1\ldots \mu_7,\nu}$ of the type IIB dual graviton occurs at level $\ell_{\mathrm{KK}}=4$. 

%%%%%%%%%%%%%%%%%%%%%%%%%%%%%%%%%%%%%%%
\subsection{First order field equations}
%%%%%%%%%%%%%%%%%%%%%%%%%%%%%%%%%%%%%%%

We begin by studying the first order equations that were given in~\eqref{FirstTH} and~\eqref{1stOrder}.

From the decomposition tables one can deduce (up to numerical factors) the following identification of type IIB potentials with potentials of the M-theory $\mf{gl}(11)$ decomposition:
\begin{subequations}
\label{eq:IIBdict}
\begin{align} 
C_{\mu_1\mu_2\mu_39} &= A_{\mu_1\mu_2\mu_3}\,, & C_{\mu_1\mu_2\mu_3\mu_4} &= -A_{\mu_1\mu_2\mu_3\mu_49\hspace{0.12mm} 10}  \,,\\
B_{\mu9i} &=  \varepsilon_{ji} h_\mu{}^j \,, & B_{\mu_1\mu_2i} &= A_{\mu_1\mu_2i}\,, \\
B_{\mu_1\cdots \mu_59i} &=A_{\mu_1\cdots \mu_5i} \,, &B_{\mu_1\cdots \mu_6i} &=-h_{\mu_1\cdots \mu_69\hspace{0.12mm} 10,i}  \,,\\
h^{\rm \scriptscriptstyle IIB}_\mu{}^{9} &= A_{\mu 9\hspace{0.12mm} 10} \,,& h^{\rm \scriptscriptstyle IIB}_\mu{}^{\nu} &= h_\mu{}^\nu + \frac{1}{4} \delta_\mu^\nu h_i{}^i\,,\\
h^{\rm \scriptscriptstyle IIB}_{9}{}^{9} &= - \frac{3}{4} h_i{}^i  \,,& \phi_i{}^j &= h_i{}^j - \frac{1}{2} \delta_i^j h_k{}^k  \,,
\end{align}
with  $\partial^{9\hspace{0.12mm} 10}\equiv \partial_9$ and 
\begin{align}
 h^{\rm \scriptscriptstyle IIB}_{\mu_1\cdots \mu_69,9} = A_{\mu_1\cdots \mu_6} \ , \quad h^{\rm \scriptscriptstyle IIB}_{\mu_1\cdots \mu_6\mu_7,9} =  \varepsilon^{ij} h_{\mu_1\cdots \mu_7i,j} \ , \quad h^{\rm \scriptscriptstyle IIB}_{\mu_1\cdots \mu_69,\nu} = -h_{\mu_1\cdots \mu_6 9\hspace{0.12mm} 10,\nu}
\end{align}
as well as 
\begin{align}
X^{\rm \scriptscriptstyle IIB}_{\alpha_1\cdots \alpha_7 9} =  X_{\alpha_1\cdots \alpha_7 9\hspace{0.12mm} 10} 
\end{align}
\end{subequations}
for the additional field arising in the tensor hierarchy algebra. We have fixed the numerical factors in such a way that the subsequent equations become canonical. The notation above introduces a superscript ${}^{\rm \scriptstyle IIB}$ for the metric $h$ and its dual and for the trace component $X$ that is introduced by the tensor hierarchy algebra. For the gauge potentials in type IIB we have employed the more standard notation $B_{\mu\nu i}$ for the doublet of two-forms (and their duals) as well as $C_{\mu_1\ldots\mu_4}$ for the four-form. 

The reduction of the field strengths~\eqref{eq:FSC} can then be computed where we retain only the ten derivatives $\partial_\mu$ and $\partial_9$ as dictated by the type IIB solution of the section constraint. Using the mapping~\eqref{eq:IIBdict} one computes the reduction of $\cG_{m_1\ldots m_4}$ as
\begin{align}
\label{eq:F4IIB}
\cG_{\alpha_1\ldots \alpha_4}&=  4 \partial_{[\alpha_1} C_{\alpha_2\alpha_3\alpha_4]9} + \partial_9 C_{\alpha_1\alpha_2\alpha_3\alpha_4} 
\CR
&= 5 \partial_{[\alpha_1} C_{\alpha_2\alpha_3\alpha_4 9]} \ ,
\CR
\cG_{\alpha_1\alpha_2\alpha_3 i}&=  3 \partial_{[\alpha_1} B_{\alpha_2\alpha_3]i} \ ,
\CR
\cG_{\alpha_1\alpha_29\hspace{0.12mm} 10}&= 2 \partial_{[\alpha_1}^{\phantom{9}} h^{\rm \scriptscriptstyle IIB}_{\alpha_2]}{}^{9} \ .
\end{align}
We have converted the field strength into tangent space indices. For the seven-form field strength the tensor hierarchy algebra construction gives
\begin{align}
\label{eq:F7IIB}
 \cG_{\alpha_1\dots \alpha_7} &= 7 \partial_{[\alpha_1} h^{\rm \scriptscriptstyle IIB}_{\alpha_2\dots \alpha_7]9,9} -\partial_{9} h^{\rm \scriptscriptstyle IIB}_{\alpha_1\dots \alpha_7,9} -\partial_9  X^{\rm \scriptscriptstyle IIB}_{\alpha_1\cdots \alpha_7 9}\CR
&= 8\partial_{[\alpha_1} h^{\rm \scriptscriptstyle IIB}_{\alpha_2\dots \alpha_79],9} -\partial_9  X^{\rm \scriptscriptstyle IIB}_{\alpha_1\cdots \alpha_7 9}\,,\CR
\cG_{\alpha_1\dots \alpha_6i} &= 6 \partial_{[\alpha_1} B_{\alpha_2\dots \alpha_6]9i} +\partial_{9} B_{\alpha_1\dots \alpha_6i} \CR
  &= 7\partial_{[\alpha_1} B_{\alpha_2\dots \alpha_69]i}\,,\CR
\cG_{\alpha_1\dots \alpha_59\hspace{0.12mm} 10} &=- 5 \partial_{[\alpha_1} C_{\alpha_2\dots \alpha_5]}\,.
\end{align}
Let us finally consider the level $\ell_{\mathrm{IIB}}=5$ field in the type IIB decomposition associated to a gradient of the $B$-field:
\be 
h_{\alpha_1\dots \alpha_7 \hspace{0.12mm} i,\beta} =  B_{\alpha_1\dots \alpha_79,\beta9 \hspace{0.12mm}i} \, , \quad X_{\alpha_1\dots \alpha_8 \hspace{0.12mm}i} = X^{\rm \scriptscriptstyle IIB}_{\alpha_1\dots \alpha_8 9,9 \hspace{0.12mm}i}\ . 
\ee
The duality equation for the dual graviton in \eqref{1stOrder} gives in this decomposition 
\be 
\partial_\alpha B_{\beta 9\hspace{0.12mm} i} = \frac{1}{8!} \varepsilon_\alpha{}^{\gamma_1\dots \gamma_8 9}\bigl(  8 \partial_{\gamma_1} B_{\gamma_2\dots \gamma_8 9,\beta 9\hspace{0.12mm} i} + \partial_\beta X^{\rm \scriptscriptstyle IIB}_{\gamma_1\dots \gamma_8 9,9\hspace{0.12mm} i}  \bigr) \ .
\ee 
Note that $\Omega_{\alpha_1\dots \alpha_8 i , \beta}$ also includes terms in $\partial^{9\hspace{0.12mm} 10} A_{\alpha_1\dots \alpha_8 i , \beta 9\hspace{0.12mm} 10}$ and $\partial^{9\hspace{0.12mm} 10} X_{\alpha_1\dots \alpha_8 \beta i , 9\hspace{0.12mm} 10}$ that must restore $SO(1,9)\subset K(E_{11})$ covariance, since the field strengths we have defined belong by construction in the module $R_{-1}$. We conclude therefore that the type IIB equation should take the form 
\be \label{UnfoldBIIB}
\partial_a B_{b_1b_2 \hspace{0.12mm} i} = \frac{1}{9!} \varepsilon_a{}^{c_1\dots c_9} \bigl( 9 \partial_{c_1} B_{c_2\dots c_9,b_1b_2\hspace{0.12mm} i} + 2 \partial_{[b_1|} X^{\rm \scriptscriptstyle IIB}_{c_1\dots c_9,|b_2] \hspace{0.12mm}i} \bigr)    \ ,
\ee 
where the indices $a,b_1,b_2,$ ... and $c_1, c_2$... run from $0$ to $9$ of $SO(1,9)$. This equation is indeed the expected unfolding duality equation, as we were anticipating in \eqref{UnfoldDual}, such that $B_{8,2\hspace{0.12mm} i}$ is the field dual to the gradient of the field $B_{2\hspace{0.12mm} i}$, and the field $X^{\rm \scriptscriptstyle IIB}_{9,1\hspace{0.12mm} i}$ is necessary for the divergence of $B_{2\hspace{0.12mm} i}$ to do not vanish. Similarly as $X_9$ in \eqref{2ndOrderGradual}, the dependence in $X^{\rm \scriptscriptstyle IIB}_{9,1\hspace{0.12mm} i} $ drops out in the second order unfolding duality equation 
\be 3 \partial_a  \partial_{[b_1} B_{b_2b_3] \hspace{0.12mm} i} =  \frac{3}{8!} \varepsilon_{ac_1\dots c_9}  \partial^{c_1} \partial_{[b_1} B^{c_2\dots c_9,}{}_{b_1b_2]\hspace{0.12mm} i}  \ . \ee
This corroborates the proposal that the $(d-2)$-form fields satisfy to first order duality equations realizing the unfolding mechanism \cite{Boulanger:2015mka}. The field $X^{\rm \scriptscriptstyle IIB}_{a_1\dots a_9,b \hspace{0.12mm}i}$ arises naturally as the general allowed total derivative  when integrating the second order duality equation to the first order constraint \eqref{UnfoldBIIB} \cite{Boulanger:2015mka}.

The constraint $\bar {\cG}_{I_1}= 0$ of equation~\eqref{gaugeBTHA} also gives Lorentz invariant gauge-fixing constraints for the field $X^{\alpha_1}$ 
\bea
\mathcal{N}_{\alpha_1\dots \alpha_6 9\hspace{0.12mm}10} &=& - \partial^\beta h^{\rm \scriptscriptstyle IIB}_{\alpha_1\cdots \alpha_6 9,\beta}- \partial^9  h^{\rm \scriptscriptstyle IIB}_{\alpha_1\cdots \alpha_69,9}- \partial^\beta  X^{\rm \scriptscriptstyle IIB}_{\alpha_1\cdots \alpha_6\beta \hspace{0.12mm} 9} = 0  \ , \CR
\mathcal{N}_{\alpha_1\dots \alpha_7\hspace{0.12mm} i} &=& \partial^\beta ( B_{\alpha_1\dots \alpha_79,\beta 9 \hspace{0.12mm} i} +  X^{\rm \scriptscriptstyle IIB}_{\alpha_1\cdots \alpha_7 9  \beta,9 \hspace{0.12mm} i} ) = 0   \ . 
\eea

Let us now carry out the same analysis in the original $E_{11}$ paradigm, using the definitions~\eqref{G4WestCorr}.  One finds the same decomposition of the four-form
\begin{align}
\mathcal{G}^\West_{\alpha_1\ldots \alpha_4}&=  4 \partial_{[\alpha_1} C_{\alpha_2\alpha_3\alpha_4]9} + \partial_9 C_{\alpha_1\alpha_2\alpha_3\alpha_4} \CR
&= 5 \partial_{[\alpha_1} C_{\alpha_2\alpha_3\alpha_4 9]} \,,\CR
\mathcal{G}^\West_{\alpha_1\alpha_2\alpha_3 i}&=  3 \partial_{[\alpha_1} B_{\alpha_2\alpha_3]i} \,,\CR
\mathcal{G}^\West_{\alpha_1\alpha_29\hspace{0.12mm} 10}&= 2 \partial_{[\alpha_1}^{\phantom{9}} h^{\rm \scriptscriptstyle IIB}_{\alpha_2]}{}^{9} \,,
\end{align}
as in~\eqref{eq:F4IIB}. We have added an additional superscript ${}^\West$ for the $E_{11}$ quantities in order to distinguish them from the field strengths defined using the tensor hierarchy algebra. However, for the type IIB version of the seven-form field strength~\eqref{G7E11} one obtains instead 
\begin{align} \label{WestIIB} 
\mathcal{G}^\West_{\alpha_1\dots \alpha_7} &= 7 \partial_{[\alpha_1} h^{\rm \scriptscriptstyle IIB}_{\alpha_2\dots \alpha_7]9,9} \,,\CR
\mathcal{G}^\West_{\alpha_1\dots \alpha_6i} &= 6 \partial_{[\alpha_1} B_{\alpha_2\dots \alpha_6]9i} +\partial_{9} B_{\alpha_1\dots \alpha_6i} + \varepsilon_{i}{}^j \partial_9 h_{\alpha_1\dots \alpha_6j\beta,}{}^\beta \CR
&= 7\partial_{[\alpha_1} B_{\alpha_2\ldots \alpha_6 9]i} +\varepsilon_{i}{}^j \partial_9 h_{\alpha_1\dots \alpha_6j\beta,}{}^\beta \,, \CR
\mathcal{G}^\West_{\alpha_1\dots \alpha_59\hspace{0.12mm} 10} &=- 5 \partial_{[\alpha_1} C_{\alpha_2\dots \alpha_5]}\,.
\end{align}
These expressions clearly differ from the ones in~\eqref{eq:F7IIB}. Looking at the type IIB reduction of the duality equations~\eqref{1stOrder} for the tensor hierarchy algebra (or the identical in this truncation~\eqref{eq:E11S1}) one sees that only the tensor hierarchy field strengths~\eqref{eq:F4IIB} and~\eqref{eq:F7IIB} give the correct duality relations for type IIB gravity. Without the inclusion of $X^{\rm \scriptscriptstyle IIB}_{\alpha_1\cdots \alpha_7 9}$ in~\eqref{eq:F7IIB} the duality equation for the dual graviton is not Lorentz invariant, since the field strength $\mathcal{G}^\West_{\alpha_1\dots \alpha_7}$ is a $(7,1,1)$ tensor, instead of an $(8,1)$ tensor of $SO(1,9)$. One gets also an extra contribution to the 7-form field strength in \eqref{WestIIB} involving
\be 
\partial_9 h_{\alpha_1\dots \alpha_6\hspace{0.12mm} j\beta,}{}^\beta= \partial_9 B_{\alpha_1\dots a_6\beta9,}{}^{\beta 9}{}_{i}  \ . 
\ee
If one assumes that the dependence in the field $A_{\alpha_1\dots \alpha_8 i , \beta 9\hspace{0.12mm} 10}$ with the correspondence 
\be 
 B_{\alpha_1\dots \alpha_8\hspace{0.1mm} 9,\beta \hspace{0.1mm} 9 \hspace{0.12mm} i} = A_{\alpha_1\dots \alpha_8 i , \beta\hspace{0.1mm}  9\hspace{0.12mm} 10} \ , 
\ee
restores $SO(1,9)$ invariance, one concludes that the type IIB seven-form field strength should be 
\be 
\cG^{\rm \scriptscriptstyle IIB}_{a_1\dots a_7\, i} = 7 \partial_{[a_1} \bigl(  B_{a_2\dots a_7]\, i} + \tfrac12 B_{a_2\dots a_7]c_1c_2,}{}^{c_1c_2}{}_i\bigr)   \ ,
\ee
such that an appropriate identification of the fields would require a non-trivial change of variables. However, this does not explain the lack of Lorentz invariance of the dual graviton equation, and one may expect to encounter an obstruction in trying to extend the construction of the $K(\mf{e}_{11})$-multiplet of first order equations as in Section \ref{PeterThing} to the next level.

%%%%%%%%%%%%%%%%%%%%%%%%%%%%%%%%%%%%%%%%%%%%%%
\subsection{Second order field equations}
%%%%%%%%%%%%%%%%%%%%%%%%%%%%%%%%%%%%%%%%%%%%%%

We now turn to the type IIB frame analysis of the second order field equations~\eqref{SecondTH} as derived from the tensor hierarchy algebra. The various components of the equations decompose as
\begin{align}
\mathcal{E}_{\alpha_1\alpha_2\alpha_3} &= - \partial^{\alpha_3} 
\left( 4 \partial_{[\alpha_1} C_{\alpha_2\cdots \alpha_4]9} + \partial_9 C_{\alpha_1\cdots \alpha_4}\right)
\,,\CR
\mathcal{E}_{\alpha_1\alpha_2 i} &= 3 \partial^{\alpha_3} \partial_{[\alpha_1} B_{\alpha_2\alpha_3] i} 
+ \partial^9 \scal{ \partial_9 B_{\alpha_1\alpha_2 i} + 2 \partial_{[\alpha_1} B_{\alpha_2]9i}} 
\,,\CR
\mathcal{E}_{\alpha_1\cdots \alpha_5 i} &= - \partial^{\alpha_6}
\scal{ 6 \partial_{[\alpha_1} B_{\alpha_2\cdots \alpha_6]9 i} + \partial_9  B_{\alpha_1\cdots \alpha_6 i} } 
\,,\CR
\mathcal{E}_{\alpha_1\cdots \alpha_4 9\,10} &= -  5 \partial^{\alpha_5}  
\partial_{[\alpha_1} C_{\alpha_2\cdots \alpha_5]} -  \partial^9 
\scal{ 4 \partial_{[\alpha_1} C_{\alpha_2\cdots \alpha_4]9} + \partial_9 C_{\alpha_1\cdots \alpha_4}}  
\,,\CR
\mathcal{R}_\alpha{}^i &= - \varepsilon^{ij} \partial^\beta\scal{ \partial_9 B_{\alpha\beta j} 
+ 2 \partial_{[\alpha} B_{\beta]9j}} 
\,,\CR
\mathcal{E}_{\alpha9\,10} &=   \mathcal{R}^{\rm \scriptscriptstyle IIB}{}_\alpha{}^9 
\,,\CR
\mathcal{R}_i{}^j &= \Box \phi_i{}^j - \frac{2}{3} \delta_i^j \mathcal{R}^{\rm \scriptscriptstyle IIB}{}_9{}^9
\,,\CR
\mathcal{R}_\alpha{}^\beta &= \mathcal{R}^{\rm \scriptscriptstyle IIB}{}_\alpha{}^\beta
+\frac{1}{3} \delta_\alpha^\beta \mathcal{R}^{\rm \scriptscriptstyle IIB}{}_9{}^9\,.
\end{align}
The full 6-form equation $\mathcal{E}_{\alpha_1\ldots \alpha_6}$ requires more care because we miss some components of the dual graviton in type IIB that would contribute starting from level $\ell_{\mathrm{KK}}=4$ that has not been derived. By evaluating the derived contributions one obtains
\be 
\mathcal{E}_{\alpha_1\cdots \alpha_6} = 8 \partial_{[9}  \partial^{\beta} 
h^{\rm \scriptscriptstyle IIB}_{\alpha_1\cdots \alpha_6 \beta],9}
- \partial_{9}  \partial^{\alpha_7} X^{\rm \scriptscriptstyle IIB}_{\alpha_1\cdots \alpha_7 9}\,.
\ee
This is not the standard form of the type IIB dual graviton equation\footnote{This would be $16 \partial_{[9}  \partial^{[9} h^{\rm \scriptscriptstyle IIB}_{\alpha_1\cdots \alpha_6 \beta],}{}^{\beta]} = 0.$}
but it coincides nonetheless with the 11-dimensional supergravity equations \eqref{SecondDualGravitonExplicit}. It can be reduced, analogously to the discussion at the end of Section~\ref{fieldlevel}, to the standard equation upon use of the first order duality equation.

%%%%%%%%%%%%%%%%%%%%%%%%%%%%%%%%%%%%%%%%%%%%%%%%%%
\section{Comments on nonlinear dynamics}
%%%%%%%%%%%%%%%%%%%%%%%%%%%%%%%%%%%%%%%%%%%%%%%%%%%
\label{sec:NL}

In this paper we have put forward a proposal to extend the $E_{11}$ paradigm that solves some of the problems of the original formulation that we have exposed. However, this proposal is only defined in the linearised approximation and it is natural to ask if it can be generalised to describe the complete non-linear dynamics. The first difficulty is to define a non-linear realisation that would reproduce the same indecomposable representation of $K(E_{11})$ in the linearised approximation. Because the field $A^\alpha$ is naturally valued in 
$R_{-2}$ rather than $R_0$,
there is no obvious way to define the non-linear realisation from a coset construction $T_{11}/K(E_{11})$ where $T_{11}$ would be a group associated to the Lie algebra $\mathfrak{t}_{11}$, which is the $p=0$ part of the tensor hierarchy algebra. 

Starting from the $E_{11}/K(E_{11})$ non-linear realisation, it seems natural to start with the $E_{11}$ covariant quantity 
\be 
\mathcal{J}_{M_0}{}^{\alpha_0} t_{\alpha_0} \equiv g_E \mathcal{P}_{M_0} g_E^{\; -1} \ , 
\ee
and to write the nonlinear field strength as 
\be\label{NonLinearFieldStrength}   \mathcal{G}_I =2 f^{M_0}{}_{\alpha_0, I} \mathcal{J}_{M_0}{}^{\alpha_0} + \sum_{i\ge 1} f^{M_0}{}_{\alpha_i,I} \nabla_{M_0} X^{\alpha_i} +\mathcal{O}(X^2) \ ,  
\ee
where $\mathcal{J}_{M_0}{}^{\alpha_0} $ is the standard $K(E_{11})$ invariant current defined above, $\nabla_{M_0}$ is an appropriately defined $K(\mf{e}_{11})$ covariant derivative, and the last term stands for some possible non-linear terms in the additional fields $X^{\alpha_i}$ and their derivative. The connection part of $\nabla_{M_0}$ is not fully determined by the theory.

Since the representation $\mfr^\ord{0}_1 $ of $X^{\alpha_i}$ is in the antisymmetric tensor product of two copies of $\mfr^\ord{1}_1 $, one expects that the covariant derivative $\nabla_{M_0} X^{\alpha_1}$ should be uniquely determined by consistency from the covariant derivative relevant to define the gauge transformations at the non-linear level $\nabla_{M_0} \Xi^{N_0}$ as in \eqref{gaugeNL}.  Note, however, that the definition of the latter is already lacking in the original $E_{11}$ paradigm. This problem is due to the fact that there is no unique torsion free connection in exceptional geometry~\cite{Coimbra:2011nw,Coimbra:2011ky,Aldazabal:2013mya,Cederwall:2013naa,Cederwall:2015ica}.

Defining these equations precisely is beyond the scope of this paper, but we would like to discuss this proposal at low level to see if it has any chance to work in the first place. Assuming that the field strength \eqref{NonLinearFieldStrength} can indeed be defined such as to provide a non-linear realisation of $E_{11}$, one may wonder if the Lagrange density \eqref{THLagrange} gives the correct field equations at low levels. 

At low level one can forget about the fields $X^{\alpha_i}$ and write the field strength $\mathcal{G}_I$ in terms of the $E_{11}$ left-invariant momenta $\mathcal{P}_M$ as
\begin{align}
 \mathcal{H}_a{}^{b_1b_2b_3} &= \scalebox{0.9}{$ \sqrt{e} \, e_{a}{}^{m} e_{n_1}{}^{b_1} e_{n_2}{}^{b_2}e_{n_3}{}^{b_3} $}\Scal{\scriptstyle - ( \partial_m+ A_{p_1p_2m} \partial^{p_1p_2}) A^{n_1n_2n_3} + 3 g^{p[n_1}  \partial^{n_2n_3]} g_{mp} + \frac{3}{2} g^{p[n_1}\delta_m^{n_2}  \partial^{n_3]q} g_{pq} } \CR
\Omega_{a_1a_2}{}^b &=\scalebox{0.9}{$ \sqrt{e}\,  e_{a_1}{}^{n_1}  e_{a_2}{}^{n_2} e_m{}^b $}\Scal{2g^{mp} (\scriptstyle  \partial_{[n_1} + A_{p_1p_2[n_1} \partial^{p_1p_2})g_{n_2]p} + \partial^{mp} A_{n_1n_2p} + \frac{1}{3}\delta^m_{[n_1} \partial^{p_1p_2} A_{n_2]p_1p_2} } \\
\mathcal{G}_{a_1a_2a_3a_4} &= \scalebox{0.9}{$  \sqrt{e} \, e_{a_1}{}^{n_1} \cdots e_{a_7}{}^{n_7} $} \Scal{ \scriptstyle  4  ( \partial_{[n_1} + A_{p_1p_2[n_1} \partial^{p_1p_2})A_{n_2n_3n_4]} - \frac{1}{2}  \partial^{p_1p_2} A_{n_1\dots n_4p_1p_2} +5 A_{[n_1n_2n_3 }   \partial^{p_1p_2}A_{n_3p_1p_2] }}   \CR
\mathcal{G}_{a_1\cdots a_7} &= \scalebox{0.9}{$  \sqrt{e} \, e_{a_1}{}^{n_1} \cdots e_{a_7}{}^{n_7} $}\Scal{\scriptstyle 7( \partial_{[n_1} + A_{p_1p_2[n_1} \partial^{p_1p_2}) A_{n_2\cdots n_7]} +70 A_{[n_1n_2n_3}( \partial_{n_4} +  A_{p_1p_2|n_4} \partial^{p_1p_2})  A_{n_5n_6n_7]}} \nn 
\end{align}
to see if \eqref{eq:LLT} would then reproduce the correct Einstein--Hilbert action coupled to the three-form potential, when the fields are assumed to only depend on the eleven coordinates $x^m$. Here the vielbein and the metric are understood to be the dynamical fields, and we shall neglect all derivatives but $\partial_m$. After some manipulations, one can write the Einstein--Hilbert Lagrange density in terms of $\Omega_{mn}{}^p$ as follows 
\begin{multline}4 \sqrt{-g} R = - 2 \partial_{m} \scal{ \sqrt{-g} g^{mn} \Omega_{np}{}^p} -  \sqrt{-g} \Scal{ \frac{1}{2} g^{n_1p_1}   g^{n_2p_2} g_{mq} \Omega_{n_1n_2}{}^m \Omega_{p_1p_2}{}^q - g^{mn} \Omega_{mp}{}^p \Omega_{nq}{}^q } \\
 -  \frac{1}{2} \sqrt{-g} \; g^{mn} g^{q[p} g^{r]s} \partial_p g_{qm} \partial_r g_{sn} \ . \end{multline}
The first term is a total derivative, and the second is precisely the term that \eqref{eq:LLT} reproduces with the substitution $\Omega_{n_1n_2}{}^m = 2 g^{mp} \partial_{[n_1} g_{n_2]p}$.  However, the term in the second line remains, and cannot be written in terms of $\Omega_{n_1n_2}{}^m$ only. 

We see therefore that the correct action cannot be defined in terms of the field strength \eqref{NonLinearFieldStrength} only. One may hope that the extra terms can be understood as some kind of Chern--Simons terms for the field strength \eqref{NonLinearFieldStrength} and its potential, but this is far from obvious. 

Let us now discuss the fate of the twisted self-duality equation \eqref{FirstTH}. The first main difficulty is to be able to describe dual gravity at the non-linear level, so let us try to write the Einstein equations in a suggestive way. The Riemann tensor can be expressed as
\begin{align} 
4R_{mn} &= - \frac{1}{\sqrt{-g}} g_{p_1(m} g_{n)p_2} \partial_q \Scal{ \sqrt{-g} g^{qr_1} g^{p_1r_2} \Omega_{r_1r_2}{}^{p_2}}   - \frac{1}{\sqrt{-g}} g_{q(m} \partial_p \Scal{ \sqrt{-g} g^{pq} \Omega_{n)r}{}^r}\CR
&\quad - 2 g^{pq} g_{rs} \Omega_{np}{}^r \Omega_{nq}{}^s + \Omega_{mp}{}^q \Omega_{nq}{}^p + \Omega_{mp}{}^p \Omega_{nq}{}^q \\
&\quad+ \frac{1}{2} g^{pq} \scal{ \Omega_{(m|p}{}^r +  \delta^r_{(m|} \Omega_{ps}{}^s-  \delta^r_{p} \Omega_{(m|s}{}^s} \partial_r g_{n)q}
- \frac{1}{4} g^{pq} g^{rs} \Scal{ \partial_{(m} g_{n)p} \partial_r g_{sq} - \partial_r g_{p(m} \partial_{n)} g_{sq}  }\,. \nn
\end{align}
Let us try to use these equations to define a non-linear version of the gravity duality equation. For this purpose we define the dual graviton field strength
\be Y_{n_1\dots n_9;m} = - \frac{1}{2 \sqrt{-g}} g_{n_1p_2}\dots g_{n_9p_9}  g_{mq} \varepsilon^{n_1\dots n_{9} r_1r_2} \scal{ \Omega_{r_1r_2}{}^q + 2 \delta^q_{r_1} \Omega_{r_2s}{}^s} \ , \ee
The Einstein equation 
\be R_{mn} - \frac{1}{2} g_{mn} R = T_{mn} \ , \ee
can then be expressed as 
\bea \partial_{[n_1} Y_{n_2\dots n_{10}];m} &=& \frac{1}{2} g^{pq} \partial_{m} g_{p[n_1} Y_{n_2\dots n_{10}];q} - \frac{11}{4} \Omega_{[mn_1}{}^q Y_{n_2\dots n_{10}];q}\CR
 && - \frac{1}{10} g_{n_1p_1} \dots g_{n_{10}p_{10}} \varepsilon^{p_1\dots p_{10}q} \Scal{ 4 T_{qm}+ \tfrac{1}{2} g^{tu} g^{rs}  \partial_{[q|} g_{mt} \partial_{|r]} g_{su} } \CR
 && + \frac{1}{40}g_{n_1p_1} \dots g_{n_{10}p_{10}}g_{m p_{11}} \varepsilon^{p_1\dots p_{11}}  g^{tu} g^{q[p} g^{r]s} \partial_p g_{qt} \partial_r g_{su} + g_{m[n_1} ( \dots ) \qquad    \eea
and more specifically for eleven-dimensional supergravity 
\be - \frac{1}{10} g_{n_1p_1} \dots g_{n_{10}p_{10}} \varepsilon^{p_1\dots p_{10}q} T_{qm}
= \frac{21}{4} F_{m[n_1\dots n_6} F_{n_7\dots n_{10}]} - 3 F_{m[n_1n_2n_3} F_{n_4\dots n_{10}]} \ee
and 
\bea &&  \partial_{[n_1} Y_{n_2\dots n_{10}];m} - \frac{1}{2} g^{pq} \partial_{m} g_{p[n_1} Y_{n_2\dots n_{10}];q} -  g_{m[n_1} ( \dots ) \CR
 &=& -\frac{11}{4} \Omega_{[mn_1}{}^q Y_{n_2\dots n_{10}];q}+21 F_{m[n_1\dots n_6} F_{n_7\dots n_{10}]} - 12 F_{m[n_1n_2n_3} F_{n_4\dots n_{10}]}   \label{DualGravitonEqua}\\
 &&  - \frac{1}{40} g_{\scalebox{0.7}{$n_{1}p_{1}$}} \dots g_{\scalebox{0.6}{$n_{10}p_{10}$}} \varepsilon^{p_1\dots p_{10}q} g^{tu} g^{rs}  \Scal{    \partial_{(q} g_{m)t} \partial_{r} g_{su}-  \partial_{r} g_{t(m} \partial_{q)} g_{su} -g_{mq}  g^{np}  \partial_{[p|} g_{nt} \partial_{|r]} g_{su} }\ .  \nn
  \eea
One can interpret the first line as a covariant exterior derivative of the dual graviton field strength (with the dots meaning that we take the traceless component), the second line is a wedge product of field strengths, whereas the last line cannot be rewritten in terms of $\Omega_{n_1n_2}{}^m$. This last line cannot be reproduced by equation \eqref{FirstTH} with an ansatz of the form \eqref{NonLinearFieldStrength}. Even assuming that this component would vanish, this equation does not define an integrable Bianchi identity that would permit to define the dual graviton field, meaning that there is no local solution for $Y_{n_1\dots n_{9},m} $ as a polynomial in the fields $g_{mn},\, g^{mn},\,A_{n_1n_2n_3},\, A_{n_1\dots n_6},\, h_{n_1\dots n_8,m},\, X_{n_1\dots n_9}$ and their derivative consistent with the grading that satisfies \eqref{DualGravitonEqua}. 

It seems therefore that one must modify \eqref{FirstTH}. Following \cite{Boulanger:2008nd}, it is natural to consider a solution to this equation of the form 
\be Y_{n_1\dots n_{9};m} = 9 \partial_{[n_1} \scal{  h_{n_2\dots n_9],m} + X_{n_2\dots n_9]m}}+B_{n_1\dots n_9,m} \ee
where  $B_{9,1}$ is a St\"{u}ckelberg gauge field that allows the restoration of gauge invariance, and $X_9$ is the antisymmetric component of the dual graviton.  In the linearised approximation $Y$ is a total derivative and one can eliminate the St\"{u}ckelberg gauge field to get back linarised dual gravity. 

To incorporate such a St\"{u}ckelberg gauge field in the $E_{11}$ construction one can for example consider an equation of the form 
\be 
M^{IJ} \mathcal{G}_J = \Omega^{IJ} \mathcal{G}_J + \mathcal{B}^I \ , 
\ee
where $\mathcal{B}^I$ would be St\"{u}ckelberg type gauge field in the degree $p=-1$ representation of $\mf{e}_{11}$, or possibly a proper $K(\mf{e}_{11})$ subrepresentation within $R_{-1}$. Considering for example the level $3n+1$ field $A_{9^n,3}$ which gauge invariant field strength is $\mathcal{R}_{10^n,4} = d^{n+1} A_{9^n,3}$ in the unfolding formalism \cite{Boulanger:2015mka}, it is to be expected that a similar analysis will lead to the need for a chain of  St\"{u}ckelberg type gauge fields $B_{10,9^{n-1},3}$, $C^{(0)}_{10^2,9^{n-2},3}$, up to $C^{(n-2)}_{10^n,3}$.  This proposal seems therefore to necessarily lead to an infinite hierarchy of higher order St\"{u}ckelberg gauge fields needed for the integrability of the previous equation, that may write schematically
\begin{align} 
\label{eq:Stueck}
p&=0:& 
\mathcal{C}^{(0)}_\alpha &= \hat{f}^{M_0}{}_{\alpha,I} \nabla_{M_0} \bigl( \Omega^{IJ} \mathcal{G}_J + \mathcal{B}^I \bigr)     \ ,  \CR
p&=1:&
\mathcal{C}^{(1)}_M P^M &=  [ P^{M_0} , t^\alpha ]\nabla_{M_0} \mathcal{C}^{(0)}_{\alpha} +\dots  \ , \CR
p&=2:&
\mathcal{C}^{(2)}_{MN}  P^{MN}  &=  [ P^{M_0} , P^N ]\nabla_{M_0} \mathcal{C}^{(1)}_{N} +\dots \ , \CR
&\,\,\vdots&&\,\,\vdots 
\end{align}
The first equation at degree $p=0$ is the projection to the Bianchi identity and at the non-linear level we expect there to be an infinite sequence of St\"uckelberg fields $\mathcal{C}^{(p)}$ needed for all $p>-1$. The covariant derivative $\nabla$ is the non-linear extension of the differential $d$ that appeared in Section~\ref{sec:THA} and the St\"uckelberg field at degree $p$ is projected to a suitable $K(\mf{e}_{11})$ representation in $R_p$. Thinking of the introduction of these St\"uckelberg fields iteratively by the horizontal degree $q$, the only way this construction could possibly make sense would be if the higher rank St\"{u}ckelberg gauge fields were all associated to highest weight representations of $\mf{e}_{11}$ as  for $p>0$ we only have highest weight representations. Similarly for $p=0$, one would expect that $\mathcal{C}^{(0)}_{\alpha_0}=0, \mathcal{C}^{(0)}_{\alpha_1}=0$, such that only highest weight representations would appear.\footnote{Note that under the reasonable assumption that the degree $p=0$ subalgebra of the tensor hierarchy algebra decomposes into an indecomposable representation $\mf{e}_{11} \oplus \ell_2$ and the remaining module, as is discussed in Section~\ref{sec:tha1}, the latter module $\mfr^\ord{0}_2$ would provide an appropriate candidate for the definition of such a St\"{u}ckelberg field.} Then they would only contribute to the duality equation for high level gauge fields. Along this line of ideas, one may need to use all components of the tensor hierarchy algebras, understanding that level $p\ge -1$ are associated to St\"{u}ckelberg type gauge fields reproducing somehow the tensor hierarchy~\cite{deWit:2008ta} appearing in supergravity for finite-dimensional groups ${E_d}$ with $d\le 8$. For the tensor hierarchy, the representations at vertical degree $p$ support the dynamical $p$-forms of supergravity in $11-d$ space-time dimensions~\cite{Palmkvist:2013vya,Greitz:2013pua}.

%%%%%%%%%%%%%%%%%%%%%%%%%%%
\section{Conclusions}
\label{sec:concl}
%%%%%%%%%%%%%%%%%%%%%%%%%%%

Finding a unified description of all maximal supergravity theories in order to obtain a better handle on the effective description of M-theory at low energy has been a long-standing goal. There are various approaches based on (infinite-dimensional) symmetry algebras~\cite{Julia,Cremmer:1997ct,Cremmer:1998px,West:2001as,Damour:2002cu,Henneaux:2010ys,HenryLabordere:2002dk,Bandos:2016ppv}. In the construction of this article, the starting point was the proposal by West and collaborators that the Lorentzian Kac--Moody algebra $\mf{e}_{11}$ should play a fundamental role~\cite{West:2001as}. We have reviewed some aspects of the $\mf{e}_{11}$ proposal and have highlighted several open questions that we recapitulate.

First, there is no mathematical definition of the  $K(E_{11})$  representation defining the first order equation describing the dynamics of the theory in the sense of~\eqref{eq:tpdec}. Its construction can only be carried out order by order in the $\mf{gl}(11)$ level decomposition starting from the duality equation in $D=11$ supergravity as discussed in Section~\ref{PeterThing}. In this way one cannot be sure that there will not be obstructions at higher level, contradicting the existence of a $K(\mf{e}_{11})$ multiplet of non-trivial first order duality equations. Moreover, $K(\mf{e}_{11})$ symmetry alone does not allow to prove that the ordinary derivative $\partial_m$ of arbitrary high level fields will not appear in the low level components of the first order equation. Since higher level fields cannot be consistently truncated in the theory, this implies that one cannot show in this way that one reproduces consistently the supergravity field equations when restricting to eleven-dimensional space-time. 

 Second, as has been noted in \cite{West:2011mm} and discussed here in Section~\ref{sec:trace},  the first order $\mf{e}_{11}$ duality equation for gravity that relates the spin connection to a suitable derivative of the dual graviton is not entirely correct, as it lacks a required nine-form potential which is not present in the theory. This problem seems to be related to the fact that the Maurer--Cartan form ${\cal V}$ rather than its coset component ${\cal P}$ was used to define the dynamics, such that the first order duality gravity equation does not transform homogeneously under Lorentz transformations. Interpreting the gravity duality equation modulo a local Lorentz transformation does not allow to identify unambiguously the required nine-form potential. However, in this article we extended the computation of the first order duality equation to higher level with the result that the terms transforming inhomogeneously under Lorentz transformations are incompatible with $K(\mf{e}_{11})$. Thus, the first order duality equation should be written in terms of the coset component ${\cal P}$ only. This implies in particular that the relevant object entering the duality equation is not the spin connection, but the object defined in \eqref{Omega21sugra}. The corresponding first order duality equation is then Lorentz invariant in the linearised approximation, and the nine-form potential is indeed missing. For a second order dualisation of linearised gravity, an $(8,1)$ hook field is sufficient as shown in~\cite{Hull:2001iu} and also discussed around~\eqref{2ndOrderGradual}. However, it is not clear whether this second order duality equation can be part of a $K(\mf{e}_{11})$ multiplet of duality equations with non-trivial propagation. 

Third, there is the issue of generalised gauge invariance of these equations. In \cite{Tumanov:2016dxc}, it was observed that higher level fields have gauge invariant field equations of increasing order in the number of derivatives at the linearised level. However, $K(\mf{e}_{11})$ symmetry preserves the number of derivatives, so one cannot define an irreducible $K(\mf{e}_{11})$ multiplet of differential equations of different orders. There is hence no gauge and $K(E_{11}$) invariant system of differential equations if one truncates at some derivative order. To exhibit the $K(\mf{e}_{11})$ symmetry of such a system, one would need to introduce an infinite hierarchy of St\"{u}ckelberg type fields to be able to write down  $K(\mf{e}_{11})$ invariant first order equations that would imply this infinite chain of higher order equations for higher level fields. In this paper we considered the more conservative approach that one should be able to define gauge invariant second order equations as integrability conditions for the (not gauge-invariant) first order  duality equations whose integrability conditions are the field equations, without introducing additional St\"{u}ckelberg type fields. We showed that this requirement implied that the fields must satisfy the section constraint \cite{Hohm:2013vpa,Hohm:2013uia,Hohm:2014fxa,West:2012qm}. The section constraint has so far played only a marginal role in the work on $E_{11}$,  but one conclusion we draw from our analysis is that it will likely be crucial for finding gauge invariant dynamics in any $E_{11}$-related set-up. 

In this paper we proposed a natural extension of the $E_{11}$ paradigm based on the infinite tensor hierarchy algebra $\scr T$ that includes $\mathfrak{e}_{11}$ as a subalgebra. We have exhibited that this allows us to resolve, at least partially, the three open problems summarized above. For some of the points we could provide all level arguments while other aspects rely on assumptions that we could only investigate at low levels in a level decomposition.

We proved that the tensor hierarchy algebra exists. It is a $\mathbb{Z}$-graded superalgebra whose degree $p=0$ subalgebra is a non semi-simple extension of  $\mf{e}_{11}$. We showed that one can define a  degree $p=1$ differential on fields valued in this algebra that depend on the $\ell_1$ module coordinates and satisfy the section constraint. This defines a differential complex for the fields of the theory that gives a group theoretical foundation for the construction of the gauge transformations, field strengths and Bianchi identities. We proved moreover that the tensor hierarchy algebra admits a non-degenerate quadratic Casimir of degree $p=-2$, which defines a non-degenerate symplectic form on the degree $p=-1$ module in which the generalized field strength is defined. The potentials are valued in the degree $p=-2$ module, which is conjugate to the $p=0$ module.

The symplectic form allows us to define a first order duality equation~\eqref{FirstTH}, by requiring that the coset component of the Maurer--Cartan form $\mathcal{P}$ projected to the $E_{11}$ module defined by the degree $p=-1$ component of the tensor hierarchy algebra vanishes on a $K(E_{11})$ invariant subspace. This first order equation is a natural generalisation of the twisted self-duality equation $\star \mathcal{G} = S \mathcal{G}$ introduced in~\cite{Cremmer:1998px}, where the Levi-Civita symbol is replaced by the $E_{11}$ invariant symplectic form $\Omega$, while the metric and scalar factors are recast into the field-dependent $E_{11}$ matrix $M$. Although the field strength $\mathcal{G}$ only includes the $p$-form field strengths in the original twisted self-duality equation, both $M$ and $\mathcal{G}$ involve all the fields of the theory, including the metric $g_{mn}$. It is worth noting that while there is no automorphism of the tensor hierarchy algebra extending the Cartan involution on $\mf{e}_{11}$, an analogue operation defines a Cartan image of the $p=-2$ module of the tensor hierarchy algebra, which plays an important role in the construction of additional first order constraints necessary to reproduce the correct degrees of freedom of eleven-dimensional supergravity. Because these field strengths are both in representations of $\mf{e}_{11}$, the $\mf{gl}(11)$ level is preserved by the equations. Thus one is ensured that the low level equations cannot have contributions from ordinary space-time derivatives of higher level fields. This allows us to interpret safely the equations of motion when truncated to fields defined on the eleven-dimensional space-time, and to compare them consistently with eleven-dimensional supergravity field equations. 

In addition to the fields parametrising $E_{11}/K(E_{11})$, the degree $p=-2$ module includes infinitely many additional fields. This introduces in particular an additional nine-form potential $X_9$, that cannot be set to zero consistently, along with its infinite set of higher level partners defining the $\ell_2$ module. We showed that $X_9$ provides the missing component of the dual graviton field, and that the first order equation discussed above reproduces the correct duality equation for the dual graviton in the linearised approximation. We analysed moreover the same equations in the type IIB frame, and exhibited that the corresponding equations have also a well defined interpretation in type IIB supergravity, when restricting the support of the fields to the corresponding ten-dimensional space-time. In particular, we get a first order duality equation exhibiting the unfolding mechanism advocated in \cite{Boulanger:2015mka}. On the contrary, the first order duality equations defined in the original $E_{11}$ paradigm do not seem to lead to consistent equations in the type IIB frame.

We define moreover a Lagrangian for the second order field equations, which are by construction solved 
by the solutions of the first order duality equations using Bianchi identities. Comparing equations~\eqref{E3},~\eqref{E6} and~\eqref{R2} with the second order equations of~\cite{Tumanov:2016abm} in the linearised approximation, one finds that they agree at lowest $\mf{gl}(11)$ levels but differ at higher levels.
We note that the consistency between the first order duality equations and the second order differential equations requires the fields to satisfy the section constraint. In this paper we have exhibited these equations explicitly for the supergravity fields, and checked that they are gauge invariant modulo the section constraint in the level truncation scheme we consider (including all the supergravity fields and the dual graviton). Note that the impossibility of defining gauge invariant second order field equations for higher level fields explained in \cite{Tumanov:2016abm} is overcome in our construction by the presence of additional fields in the $\bar \ell_2$ module. 

The property that the field strengths are constructed from a representation of $\mf{e}_{11}$ is also extremely useful in computing its components at higher level efficiently. Moreover, this makes it possible to prove some statements at all levels. We have been able in this way to exhibit some of the desirable properties for the general theory. The symplectic form and the $GL(11)$ representations appearing in the degree $p=-1$ module corroborate the interpretation of the fields associated to null roots (potentials including nine-forms in their tensor structures) in \cite{Boulanger:2015mka} to realise the unfolding mechanism. We also corroborate the validity of the proposal that potentials including ten-forms in their tensor structure source non-geometrical fluxes, and in particular that the $B_{10,1,1}$ flux can be interpreted as the Romans mass \cite{Kleinschmidt:2003mf,West:2004st,Henneaux:2008nr}. 

Despite this progress, we have made certain assumptions in our tensor hierarchy algebra proposal that require further investigation to be proved rigorously. The tensor hierarchy algebra as presented here introduces $E_{11}$ modules that are strictly bigger than the irreducible $E_{11}$ modules appearing in the original construction. In this paper we have assumed that the degree $p=1$ module of the tensor hierarchy algebra is reducible to the $\bar \ell_1$ irreducible module plus the remaining module. Although we provided indications at low levels that this might be true, we have not been able to prove it. This assumption is very important in order for our proposal to remain a reasonably mild extension of the original $E_{11}$ paradigm, and there would be many new open questions if it was not true. This was discussed in more detail in Section~\ref{sec:tha1}. Also we have not proved the existence or uniqueness of a $K(E_{11})$ non-degenerate symmetric bilinear form $M^{IJ}$ on the degree $p=-1$ module. This non-degenerate bilinear form is essential for the definition of the field equations. Its existence would be guaranteed if the degree $p=-1$ module was either irreducible or decomposable into an irreducible submodule (defining then the relevant field strength representation) and a remaining module. We have nonetheless been able to define this bilinear form in the level truncation scheme we considered in this paper. 

It would also be very desirable to understand the gauge invariance of the second order field equations at all levels. The fact that we have been able to prove gauge invariance up to  the level including the dual graviton is very encouraging, but it does by no means guarantee that gauge invariance will not fail at a higher level. Would it fail, it would be likely that one would need to introduce an additional St\"{u}ckelberg type field in a highest weight $E_{11}$ module to restore gauge invariance.

Even though the tensor hierarchy algebra underlies the construction of our dynamical quantities, the actual symmetry of the linearised equations of motion is $K(\mf{e}_{11})$ as in the original construction. The generalisation of our equations to the non-linear level is expected to exhibit the full $E_{11}$ symmetry. However, there are many open questions regarding the non-linear generalisation of the equations of motion. The first challenge is to define the non-linear realisation such as to incorporate the additional component $\bar \ell_2$, consistently with the indecomposability of the $\mf{e}_{11}$ module $\mf{e}_{11}\oplus \bar \ell_2$. We also exposed in Section~\ref{sec:NL} that the naive non-linear generalisation of our proposal does not lead to consistent first order duality equations for the gravitational field. It is in fact to be expected that gauge invariance of the first order duality equation must be realised in order to define the non-linear extension. 

Analysis of the tensor hierarchy algebra suggests that the introduction of an infinite sequence of St\"uckelberg type fields depicted in~\eqref{eq:Stueck} might be necessary to define the non-linear theory. Since one may need to consider fields in all the components of the tensor hierarchy algebra it would be very interesting if it could play a more predominant role at the non-linear level, beyond the definition of the underlying differential complex. 

For extending our formulation to the non-linear level, one needs to define a $K(\mf{e}_{11})$ covariant derivative $\nabla$, not only for the non-linear gauge transformations, but also for the field strengths of the various fields of the theory, including $X_9$ and the St\"uckelberg type fields discussed above. This connection is not uniquely determined from the non-linear realisation, and its definition is an open problem that remains to be investigated~\cite{West:2014eza}. One expects nonetheless that its definition on the gauge parameters, required to define the non-linear gauge transformations, will determine consistently the covariant derivative of the other fields of the theory. Clarifying these issues could shed some light on the elusive non-linear dualisation of gravity beyond the proposal in~\cite{Boulanger:2008nd}.

In this paper we have discussed the restrictions of the fields to eleven-dimensional supergravity and to ten-dimensional type IIB supergravity. It would be very interesting to analyse other (partial) solutions to the section constraint to understand exceptional field theories in this formalism \cite{Hohm:2013vpa,Hohm:2013uia,Hohm:2014fxa}. An interesting future avenue would be to explore the realisation of gauged supergravity theories in our formalism, building for example on~\cite{Bergshoeff:2007qi,Riccioni:2007ni,Berman:2012uy,
Greitz:2013pua,Howe:2015hpa,Cassani:2016ncu}, or massive type IIA supergravity, building on~\cite{Schnakenburg:2002xx,Henneaux:2008nr,Ciceri:2016dmd}. We note that in~\cite{Bergshoeff:2007vb} the massive Romans theory was analysed and, based on an analysis of the gauge algebra, an extension of the $\mf{e}_{11}$ algebra was proposed. The new generator appearing in this investigation is different from the new generators found in the tensor hierarchy algebra in our work as it sits at a different level in the level decomposition compared to the tensor hierarchy algebra.\footnote{The new generator of~\cite{Bergshoeff:2007vb} would appear more naturally in a free Lie algebra extension of $\mf{e}_{11}$.}

\subsection*{Acknowledgements}

We would like to thank N.~Boulanger, M.~Cederwall, J.~Gomis, W.~Linch, H.~Samtleben, P.~Sundell and P.~West for discussions. GB and AK gratefully acknowledge the warm hospitality of Texas A\&M University where this work was begun, and GB and JP likewise acknowledge the warm hospitality of the Albert Einstein Institute.  GB, JP, CNP and ES also acknowledge the Mitchell Family Foundation for hospitality at the Brinsop Court workshop on strings and cosmology. ES would like to thank Universidad San Sebasti\'an and Pontifica Universidad Cat\'olica de Valparaiso for hospitality. The work of ES is supported in part by NSF grant PHY-1214344 and in part by Conicyt MEC grant PAI 80160107. CNP is supported in part by DOE grant DE-FG02-13ER42020, and JP is supported in part by NSF grant PHY-1214344.

\appendix

\section{Conventions for \texorpdfstring{$E_{11}$}{E11} and its representations}
\label{app:conv}

In this appendix, we give our conventions for the Kac--Moody algebra $\mf{e}_{11}$ with the Dynkin diagram displayed in Figure \ref{fig:e11dynk}, and two of its representations. The first one is the representation $\ell_1$ for which the lowest weight is the negative of the fundamental weight corresponding to node $1$ in this labelling, \ie  with Dynkin labels $(1,0,0,0,0,0,0,0,0,0,0)$ and that corresponds to the representation in which the derivatives transform. The second $\mf{e}_{11}$ representation is the $\ell_{10}$ representation that appears in the section constraint. It has Dynkin labels $(0,0,0,0,0,0,0,0,0,1,0)$. We reiterate that we label the lowest weight representations by \textit{minus} the Dynkin labels of the lowest weight vectors.

\subsection{\texorpdfstring{$E_{11}$}{E11}}

The Kac--Moody algebra $\mf{e}_{11}$ is the Lie algebra generated by Chevalley generators $e_I,f_I,h_I$ (with $I=1,\ldots,11$ labelling the nodes in the Dynkin diagram) modulo the Chevalley relations
\begin{align} \label{chev-rel-1}
[h_I,e_J]&=A_{IJ}e_J\ , & 
[h_I,f_J]&=-A_{IJ}f_J\ , & [e_I,f_J]&=\delta_{IJ}h_J\ ,
\end{align}
and the Serre relations 
\begin{align} \label{serre-rel1}
(\text{ad } e_I)^{1-A_{IJ}} (e_J) = (\text{ad } f_I)^{1-A_{IJ}} (f_J) &=0\ ,
\end{align}
where $A_{IJ}$ is the Cartan matrix given by the Dynkin diagram in Figure \ref{fig:e11dynk}.
If two different nodes $I$ and $J$ are connected with a line, then $A_{IJ}=A_{JI}=-1$, otherwise $A_{IJ}=A_{JI}=0$. On the diagonal we have $A_{II}=2$ (no summation).

In $\mf{e}_{11}$ covariant expressions we use the indices $\alpha, \beta, \ldots$ for the adjoint representation, and $M,N,\ldots$ for $\ell_1$, with corresponding basis elements $t^\alpha$ and $P_M$. However, in the application to eleven-dimensional supergravity it is more  convenient to describe  the structure of $\mf{e}_{11}$ and of $\ell_1$ in terms of $\mf{gl}(11)$ level decompositions, where the $\mf{gl}(11)$ subalgebra is obtained by removing node $11$ from the Dynkin diagram.  Any representation of $\mf{e}_{11}$ then decomposes into a direct sum of $\mf{gl}(11)$ representations which can be assigned integer levels $\ell$. For the adjoint, the $\mf{gl}(11)$ representation at level $-\ell$ is the conjugate of the 
representation at level $\ell$, reflecting the structure of positive and negative roots. 

The decompositions of $\mf{e}_{11}$ and $\ell_1$ into $\mf{gl}(11)$ representations for low levels are given in Table~\ref{tab:e11adj} \cite{West:2001as,West:2002jj,Nicolai:2003fw} and Table~\ref{tab:e11l1} \cite{West:2003fc,Kleinschmidt:2003jf}, respectively, together with
our notation for the corresponding potential fields, coordinates and parameters. In the adjoint representation, we denote the generators at the first three 
positive levels by
$E_{n_1n_2n_3}$, $E_{n_1\cdots n_6}$, 
$E_{n_1\cdots n_8,m}$,
and those at the first three
negative levels by $F_{n_1n_2n_3}$, $F_{n_1\cdots n_6}$, 
$F_{n_1\cdots n_8,m}$.  Our convention is such that 
\begin{align}
E^{n_1n_2n_3} - \eta^{n_1p_1}  \eta^{n_2p_2}  \eta^{n_3p_3} F_{p_1p_2p_3}
\end{align}
belongs to the `compact' subalgebra $K(\mf{e}_{11})$. The metric appearing here is the invariant metric of 
$\mf{so}(1,10)$ and therefore the subalgebra is a Wick-rotated form of the standard maximal compact 
subalgebra obtained by the Cartan involution. The involution is sometimes called temporal involution 
and discussed in for example~\cite{Englert:2003py}. The generators at level $-\ell$ are 
then defined with the opposite sign compared to the those at level $\ell\geq2$
\begin{align} \label{112}
E^{n_1 \cdots n_6} &\equiv [E^{n_1n_2n_3},E^{n_4 n_5 n_6}], &
F_{n_1 \cdots n_6} &\equiv -[F_{n_1n_2n_3},F_{n_4 n_5 n_6}],
\end{align}
\begin{align}
E^{n_1\cdots n_8,m}
&\equiv \tfrac83 [ E^{[n_1n_2n_3},E^{n_4\cdots n_8]m}] =\tfrac43 [ E^{[n_1\cdots n_6},E^{n_7 n_8]m}]\ ,
\nn\\
F_{n_1\cdots n_8,m}
&\equiv -\tfrac83 [ F_{[n_1n_2n_3},F_{n_4\cdots n_8]m}] =-\tfrac43 [ F_{[n_1\cdots n_6},F_{n_7 n_8]m}]\ . \label{level3}
\end{align}
The last equations can be inverted to
\begin{align}
[E^{n_1n_2n_3},E^{p_1\cdots p_6}]&=6\, E^{n_1 n_2 n_3[p_1 \cdots p_5,p_6]}=-3\,E^{p_1\cdots p_5[n_1 n_2,n_3]}\ , 
\nn\\
[F_{n_1n_2n_3},F_{p_1\cdots p_6}]&=-6\, F_{n_1 n_2 n_3[p_1 \cdots p_5,p_6]}=3\,F_{p_1\cdots p_5[n_1 n_2,n_3]}\ . \label{level3inv}
\end{align}

For the $\mf{gl}(11)$ representations we employ the following notation. Every tensor displayed 
is either an irreducible representation (if it has only upper or only lower indices) or the full tensor product of two irreducible representations (if it has both upper and lower indices). In the irreducible case it thus corresponds 
to a fixed Young tableau where each box corresponds to an index
$m,n,\ldots=0,1,\ldots,10$. Indices in the same column are antisymmetric, 
usually written with the same letter, and different columns are separated by a comma.
For example, the generator appearing at level $\ell=3$ in $\mf{e}_{11}$ satisfies the 
symmetry and irreducibility constraints
\begin{align} \label{hook}
E^{[n_1\cdots n_8], m} &= E^{n_1\cdots n_8,m}\ , &
E^{[n_1\cdots n_8, m]}&=0\ .
\end{align}
Antisymmetrisations occur always with strength one, and the first equation above just reflects 
the convention that indices in one column are automatically antisymmetric by the Young symmetries. 
The second equation is the Young irreducibility constraint. 
Occasionally, we use \raisebox{.4ex}{$\lsharp\,\rsharp$} to denote projection on an irreducible representation of this type, for which the Young tableau is a hook with two columns, only one box in one column, and an arbitrary number of boxes in the other. With this notation, both conditions (\ref{hook}) can thus be expressed together as
\begin{align}
E^{n_1\cdots n_8,m} = E^{\lsharp n_1\cdots n_8,m \rsharp}\ .
\end{align}
We define the projector on the $(k,1)$ hook symmetry structure in general as
\begin{align}
\label{hookproj}
T^{\lsharp  n_1\cdots n_k,m\rsharp} \equiv T^{[n_1\cdots n_k]m}-T^{[n_1\cdots n_km]}\ .
\end{align}
for a general tensor $T^{n_1\ldots n_km}$ with $k+1$ indices without any particular symmetrisation.\footnote{When a tensor already includes a comma, as in the case above, one understands that the comma is at the same place before and after the projection so that \eg $E^{n_1n_2\lsharp n_3\cdots n_8,m \rsharp} = E^{n_1n_2[n_3\cdots n_8],m}-E^{n_1n_2[n_3\cdots n_8,m]}$.}

Any tensor density with both upper and lower indices transforms
in the full reducible tensor product of the two irreducible representations, \textit{i.e.} it contains traces.
For example, at $\ell=0$ in the adjoint of $\mf{e}_{11}$ we let $K^m{}_n$ denote the generators in the
adjoint of
$\mf{gl}(11)$ which is reducible and decomposes into a direct sum
of two irreducible representations:
the traceless part $\mathfrak{sl}(11)$ and the trace $K\equiv K^m{}_m$.

Sometimes we use a shorthand notation for the $\mathfrak{gl}(11)$ tensor densities
where
the sub- and superscripts denote numbers of (lower and upper, respectively) antisymmetric indices in
the blocks corresponding to columns in the Young tableau
separated by commas.
For example, the generators of $\mf{e}_{11}$ above at levels $\ell=3$ and
$\ell=0$ are then denoted by $E^{8,1}$ and $K^1{}_1$, respectively.

In the adjoint 
representation, the generators are true tensors of $\mf{gl}(11)$ and transform as, for example
\begin{align} \label{KonEandF}
\lb K^m{}_n, E^{p_1p_2p_3} \rb &= 3\delta_n^{[p_1} E^{p_2p_3]m}\ , & 
\lb K^m{}_n, F_{p_1p_2p_3} \rb &= -3\delta^n_{[p_1} F_{p_2p_3]m}\ .
\end{align}
Therefore, the action of $K$ counts the number of upper minus the number of lower indices, 
which is three times the level $\ell$.
By contrast, in the lowest weight representation $\ell_1$ the generators are \emph{not} true tensors of 
$\mf{gl}(11)$ but rather tensor densities of non-trivial weight. Here, the eigenvalue of $K$ is 
the number of upper minus the number of lower indices \emph{plus} $\tfrac{11}2$.

The commutation relations of $\mf{gl}(11)$ are
\begin{align}
\lb K^m{}_n, K^p{}_q \rb = \delta^p_n K^m{}_q - \delta^m_q K^p{}_n\ ,
\end{align}
and those of type $[E,F]$ up to level $\ell=\pm 3$ in $\mf{e}_{11}$ are
\begin{align} \label{EwithF}
[E^{n_1n_2n_3},F_{p_1p_2p_3}]=18\,\de^{[n_1n_2}_{[p_1p_2}\,K^{n_3]}_{p_3]}-2\,\de^{n_1n_2n_3}_{p_1p_2p_3}\,K\ , 
\end{align}
\begin{align}
[E^{n_1\cdots n_6},F_{p_1\cdots p_6}]=480\,(9\,\de^{[n_1\cdots n_5}_{[p_1\cdots p_5}\,K^{n_6]}{}_{p_6]}
-\de^{n_1\cdots n_6}_{p_1\cdots p_6}\,K)\ , %-L)),  %-L)) ,
\end{align}
\begin{align}
[E^{n_1\cdots n_8,m},F_{q_1\cdots q_8,p}]=
35840\,(&-\delta^{n_1\cdots n_8}_{q_1\cdots q_8}K^m{}_p
+\delta^{m[n_1\cdots n_7}_{q_1\cdots q_8}K^{n_8]}{}_p\nn\\
&+\delta^{n_1\cdots n_8}_{p[q_1\cdots q_7}K^{m}{}_{q_8]}
-\delta^{m[n_1\cdots n_7}_{p[q_1 \cdots q_7}K^{n_8]}{}_{q_8]}\nn\\
&+\tfrac23 \delta^{n_1\cdots n_8}_{q_1\cdots q_8}\delta^m_p K
-\tfrac23 \delta^m_{[q_1}\delta^{[n_1\cdots n_7}_{q_2 \cdots q_8]}\delta^{n_8]}_p K)\ ,
\end{align}
\begin{align} \label{123}
[F_{n_1n_2n_3},E^{p_1\cdots p_6}]&=120 \,\de^{[p_1p_2p_3}_{n_1n_2n_3}\,E^{p_4p_5p_6]}\ ,\nn\\
[E^{n_1n_2n_3},F_{p_1\cdots p_6}]&= -120 \,\de^{n_1n_2n_3}_{[p_1p_2p_3}\,F_{p_4p_5p_6]}\ ,
\end{align}
\begin{align}
[F_{n_1n_2n_3},E^{p_1\cdots p_8,q}]&=-112\,(\de^{q[p_1p_2}_{n_1n_2n_3}\,E^{p_3\cdots p_8]}
-\de^{[p_1p_2p_3}_{n_1n_2n_3}\,E^{p_4\cdots p_8]q})\ ,\nn\\
[E^{n_1n_2n_3},F_{p_1\cdots p_8,q}]&=112\,(\de^{n_1n_2n_3}_{q[p_1p_2}\,F_{p_3\cdots p_8]}
-\de^{n_1n_2n_3}_{[p_1p_2p_3}\,F_{p_4\cdots p_8]q})\ , \label{132}
\end{align}
\begin{align}
[F_{n_1\cdots n_6},E^{q_1\cdots q_8,p}]&=
-13440\,(\delta_{n_1\cdots n_6}^{[q_1\cdots q_6}\,E^{q_7q_8]p}
-\delta_{n_1\cdots n_6}^{p[q_1\cdots q_5}\,E^{q_6q_7q_8]})\ ,\nn\\
[E^{n_1\cdots n_6},F_{q_1\cdots q_8,p}]&=
13440\,(\delta^{n_1\cdots n_6}_{[q_1\cdots q_6}\,F_{q_7q_8]p}
-\delta^{n_1\cdots n_6}_{p[q_1\cdots q_5}\,F_{q_6q_7q_8]})\ .
\end{align}

The Chevalley generators can be expressed in terms of the basis elements above in the following 
way ($i=1,\ldots, 10$):
\begin{align}
e_i &= K^i{}_{i+1}\ , & f_i &= K^{i+1}{}_{i}\ , & h_i &= K^i{}_i-K^{i+1}{}_{i+1}\ ,
\end{align}
\begin{align}
e_{11}&=E^{9\,10\,11}, & f_{11}&=F_{9\,10\,11}\ , &
h_{11}&=K^9{}_9+K^{10}{}_{10}+K^{11}{}_{11}-\tfrac13 K\ .
\end{align}

The Kac--Moody algebra $\mf{e}_{11}$ admits a symmetric invariant bilinear form (`Killing form')~\cite{Kac2} that we denote by $(t^\alpha,t^\beta)=\kappa^{\alpha\beta}$. It is given by
\begin{align}
(h_I,h_J) &= A_{IJ}\ ,
&
(e_I,f_J) &=\de_{IJ}\ ,
\end{align}
for the Chevalley generators, which gives
\begin{align}
(K^m{}_n,K^p{}_q) &=\delta^p_n \de^m{}_q -\tfrac12 \delta^m_n \de^p{}_q\ ,
\\
(E^{n_1n_2n_3},F_{p_1p_2p_3}) &= 3!\, \de^{n_1n_2n_3}_{p_1p_2p_3}\ ,
\\
(E^{n_1\cdots n_6},F_{p_1\cdots p_6}) &= 6!\, \de^{n_1\cdots n_6}_{p_1\cdots p_6}\ ,
\\
(E^{n_1\cdots n_8|m},F_{q_1\cdots q_8|p}) &=
\tfrac89\cdot8!\,(
-\de^{m}_{[q_1}\de^{[n_1\cdots n_7}_{q_2\cdots q_8]}\de^{n_8]}_p+\de^{n_1\cdots n_8}_{q_1\cdots q_8}\de^m_p)\ .
\end{align}

\subsection{The \texorpdfstring{$\ell_1$}{l1} representation}   
\label{app:l1}

Our notation for the low-lying generators of the lowest weight $\ell_1$ representation was given in Table~\ref{tab:e11l1}. They are
\begin{align}
P_m,\, Z^{n_1n_2},\, Z^{n_1\ldots n_5},\, P^{n_1\ldots n_7,m} ,\, P^{n_1\ldots n_8},\,\ldots\ .
\end{align}
In the semidirect sum of $\mf{e}_{11}$ and $\ell_1$, the basis elements of $\mf{e}_{11}$ act in the following way at low levels:
\begin{align}
[F_{n_1\cdots n_8,m},P_p]&=0\ ,\nn\\
[F_{n_1\cdots n_6},P_{m}]&=0\ ,\nn\\
[F_{n_1n_2n_3},P_m]&=0\ ,\nn\\
[K^m{}_n, P_p]&=-\de^m_p P_n+\tfrac12\de^m_n P_p\ ,\nn\\
[E^{n_1n_2n_3},P_m]&=-3 \,\de^{[n_1}_m\, Z^{n_2n_3]}\ , \nn\\
[E^{n_1\cdots n_6},P_m]&=-6 \,\de^{[n_1}_m\, Z^{n_2\cdots n_6]}\ ,\nn\\
[E^{n_1\cdots n_8,m},P_p]&=\tfrac83 \de^m_p P^{n_1\cdots n_8}
-\tfrac83 \de^{[n_1}_pP^{n_2\cdots n_8]m}\nn\\&\quad\,-8\de^{[n_1}_pP^{n_2\cdots n_8],m}\ ,
\end{align}
\begin{align}
[F_{n_1\cdots n_8,m},Z^{pq}]&=0\ ,\nn\\
[F_{n_1\cdots n_6},Z^{pq}]&=0\ ,\nn\\
[F_{n_1n_2n_3},Z^{pq}]&=-6\,\de^{pq}_{[n_1n_2}\,P_{n_3]}\ ,\nn\\
[K^m{}_n,Z^{pq}]&=-2 \,\de^{[p}_{n}\,Z^{q]m}+\tfrac12\de^m_n Z^{pq}\ ,\nn\\
[E^{n_1n_2n_3},Z^{pq}]&=-Z^{n_1n_2n_3pq}\ ,\nn\\  
[E^{n_1\cdots n_6},Z^{pq}]&=-2\,P^{n_1\cdots n_6pq}-6\, P^{pq[n_1\cdots n_5,n_6]}\nn\\
&=-2\,P^{n_1\cdots n_6pq}+2\, P^{n_1\cdots n_6[p,q]}\ ,
\end{align}
\begin{align}
[F_{n_1\cdots n_8,m},Z^{p_1 \cdots p_5}]&=0\ ,\nn\\
[F_{n_1\cdots n_6},Z^{p_1\cdots p_5}]&= 720 \,\de^{p_1\cdots p_5}_{[n_1\cdots n_5}\,P_{n_6]}\ ,\nn\\
[F_{n_1n_2n_3},Z^{p_1 \cdots p_5}]&=-60 \,\de^{[p_1p_2p_3}_{n_1n_2n_3}\,Z^{p_4p_5]}\ ,\nn\\
[K^m{}_n,Z^{p_1 \cdots p_5}]&=5 \,\de^{[p_1}_{n}\,Z^{p_2p_3p_4p_5]m}+\tfrac12\de^m_n Z^{p_1\cdots p_5}\ ,\nn\\
[E^{n_1n_2n_3},Z^{p_1 \cdots p_5}]&=P^{n_1n_2n_3p_1 \cdots p_5}-5\, 
P^{n_1n_2n_3[p_1p_2p_3p_4,p_5]}\nn\\&=P^{n_1n_2n_3p_1\cdots p_5}
-3\, P^{p_1\cdots p_5[n_1n_2,n_3]}\ ,
\end{align}
\begin{align}
[F_{n_1\cdots n_8,m},P^{p_1\cdots p_8}]&= \tfrac{8!}3(\de^{p_1\cdots p_8}_{n_1\cdots n_8}P_{m}
-\de^{p_1\cdots p_8}_{m[n_1\cdots n_7}P_{n_8]})\ ,\nn\\
[F_{n_1\cdots n_6},P^{p_1\cdots p_8}]&=-7!\,\de^{[p_1 \cdots p_6}_{n_1\cdots n_6}Z^{p_7 p_8]},\nn\\
[F_{n_1n_2n_3},P^{p_1\cdots p_8}]&=  42\, \de^{[p_1p_2p_3}_{n_1n_2n_3}Z^{p_4\cdots p_8]}\ , \nn\\
[K^m{}_n,P^{p_1\cdots p_8}]&=8\,\de^{[p_1}_n P^{|m|p_2\cdots p_8]}+\tfrac12\de^m_nP^{p_1\cdots p_8}\ ,
\end{align}
\begin{align}
[F_{n_1\cdots n_8,m},P^{q_1\cdots q_7,p}]&= 7\cdot7!\,(\de^p_{[n_1}\de^{[q_1}_{|m|}
\de^{q_2\cdots q_7]}_{n_2\cdots n_7}P_{n_8]}+\de^p_m\de^{q_1\cdots q_7}_{[n_1\cdots n_7}P_{n_8]})\ ,\nn\\
[F_{n_1\cdots n_6},P^{q_1\cdots q_7,p}]&=3780\,(\de^{[q_1\cdots q_6}_{n_1\cdots n_6}Z^{q_7]p}
+\de^{p[q_1\cdots q_5}_{n_1\cdots n_6}Z^{q_6q_7]})\ ,\nn\\
[F_{n_1n_2n_3},P^{q_1\cdots q_7,p}]&=-\tfrac{315}4 (\de^{[q_1q_2q_3}_{n_1n_2n_3}Z^{q_4\cdots q_7]p}+
\de^{p[q_1q_2}_{n_1n_2n_3}Z^{q_3\cdots q_7]})\ ,\nn\\
[K^m{}_n,P^{q_1\cdots q_7,p}]&=7\de^{[q_1}_n P^{q_2\cdots q_7] m,p}+\de^p_n P^{q_1\cdots q_7,m}
+\tfrac12\de^m_n P^{q_1\cdots q_7,p}\ .
\end{align}

%%%%%%%%%%%%%%%%%%%%%%%%%%%%%%%%%%%%%%%%%%%%%%%%%%%%%%%%%%%%%%%%%%%%%%%%%%%%%%%%%%%%%%%%
\subsection{The section constraint representation \texorpdfstring{$\ell_{10}$}{l10}}
\label{app:SCmult}
%%%%%%%%%%%%%%%%%%%%%%%%%%%%%%%%%%%%%%%%%%%%%%%%%%%%%%%%%%%%%%%%%%%%%%%%%%%%%%%%%%%%%%%%

In Table~\ref{tab:e11l10}, we list the low-lying generators of the lowest weight representation $\ell_{10}$ of $\mf{e}_{11}$ in a $\mf{gl}(11)$ decomposition. The representation $\ell_{10}$ arises in the symmetric tensor product of two $\ell_1$ representations. Writing things dually one can think of the various components in the following way
\begin{subequations}
\label{eq:C1}
\begin{align}
L^m &= \partial^{mn} \partial_n \,,\\
L^{n_1n_2n_3n_4} &= 3 \partial^{[n_1n_2} \partial^{n_3n_4]} - \partial^{n_1n_2n_3n_4m} \partial_{m}  
\,,\\
L^{n_1n_2n_3n_4n_5n_6,m} &= \frac{30}{7}\left(\partial^{[n_1n_2} \partial^{n_3n_4n_5n_6]m} -\partial^{m[n_1}\partial^{n_2n_3n_4n_5 n_6]}\right)\nn\\ 
&\quad  - \frac{6}{7}
\left(\partial^{p n_1n_2n_3n_4n_5n_6,m} \partial_{p} -\partial^{pm[n_1n_2n_3n_4 n_5,n_6]} \partial_p\right)\,,\\
L^{n_1n_2n_3n_4n_5n_6n_7} &= 3 \partial^{[n_1n_2} \partial^{n_3n_4n_5n_6n_7]} 
-\frac{3}{7} \partial^{n_1n_2n_3n_4n_5n_6n_7,m} \partial_{m} + \partial^{n_1n_2n_3n_4n_5n_6n_7m} \partial_m  \,.
\end{align}
\end{subequations}
These constraints can be generated using the action of $\mf{e}_{11}$ on the lowest weight vector $L^m$. 

\renewcommand{\arraystretch}{1.5}
\begin{table}[t!]
\centering
\begin{tabular}{|c|c|c|c|}
\hline
Level $\ell$ & $q=\ell-3$ & $\mf{sl}(11)$ representation & Generator structure\\[1mm]
\hline
$4$ & $1$ &
   $(0,0,0,0,0,0,0,0,0,1)$ & $L^m$ \\ %[1mm]
\hline
$5$ & $2$ &
   $(0,0,0,0,0,0,1,0,0,0)$ & $L^{n_1\ldots n_4}$ \\ %[1mm]
\hline
$6$ & $3$ &$\begin{matrix} (0,0,0,1,0,0,0,0,0,0) \\ (0,0,0,0,1,0,0,0,0,1) \end{matrix}$
     & $\begin{matrix} L^{n_1\ldots n_7} \\ L^{n_1\ldots n_6,m} \end{matrix}$\\
\hline
$7$ & $4$ & $\begin{matrix}
         (1,0,0,0,0,0,0,0,0,0)\\ (0,1,0,0,0,0,0,0,0,1)\\  (0,1,0,0,0,0,0,0,0,1)\\   (0,0,1,0,0,0,0,0,1,0) \\ (0,0,1,0,0,0,0,0,0,2) \\  (0,0,0,1,0,0,0,1,0,0)\end{matrix}$ 
       &$\begin{matrix} L^{n_1\ldots n_{10}} \\ L^{n_1\ldots n_{9},m} \\\tilde{L}^{n_1\ldots n_{9},m} \\ L^{n_1\ldots n_{8},m_1m_2} \\ L^{n_1\ldots n_{8},m,p} \\ L^{n_1\ldots n_{7},m_1m_2m_3}\end{matrix}$\\ \hline%[1mm]       
\end{tabular}
\caption{\label{tab:e11l10} \small\textit{Level decomposition of the $\ell_{10}$ 
representation of $E_{11}$ under $\mf{gl}(11)$. This is a lowest weight representation 
and therefore the top entry is annihilated by all lowering generators. The name of the 
corresponding tensor structure reflects its role in the section constraint~\eqref{eq:SC}. 
At level $\ell=7$ we have for the first time a degeneracy in the tensor type, indicated 
by two letters $L$ and $\tilde{L}$. The degree here is related to $\ell$ by $q=\ell-3$.}}
\end{table}

We stress that $\ell_{10}$ is only the beginning of the full section constraint. According to~\eqref{eq:SC} there will be more $\mf{e}_{11}$ lowest weight representations that constitute the full section constraint. Continuing the symmetric tensor product to the next term for $\mf{e}_{11}$ gives
\begin{align}
(\ell_1 \otimes \ell_1)_{\textrm{sym}} = (2\ell_1) \oplus \lb \ell_{10} \oplus (\ell_2+\ell_{10})\oplus\ldots \rb\,.
\end{align}
The lowest weight representation $\ell_2+\ell_{10}$ starts contributing from $\mf{gl}(11)$ level $\ell=7$; at that level it contains only the $\mf{sl}(11)$ representation $(0,1,0,0,0,0,0,0,0,1)$ that is also contained in the $\ell_{10}$ representation as is visible from Table~\ref{tab:e11l10} such that this $\mf{gl}(11)$ tensor structure appears in total three times in the section constraint. The third section constraint of type $(9,1)$ that belongs to $\ell_2+\ell_{10}$ does not have any contribution up to the derivative order we are considering here.

%%%%%%%%%%%%%%%%%%%%%%%%%%%%%%%%%%%%%%%%%%%%%%%%%%%%%%%%%%%%%%%%%%%%%%%%%%%
\section{Construction of the tensor hierarchy algebra}
\label{app:THA}
%%%%%%%%%%%%%%%%%%%%%%%%%%%%%%%%%%%%%%%%%%%%%%%%%%%%%%%%%%%%%%%%%%%%%%%%%%%

In this appendix, we present a proof of the existence of the tensor hierarchy algebra  based on the formalism of local Lie (super)algebras as developed by Kac~\cite{Kac2}. We shall give two different characterisations of the tensor hierarchy algebra; one direct algebraic construction using {(anti-)} commutation relations and one dual characterisation using the BRST formalism. We will also demonstrate the existence of an involution that is used in the first order duality relations.

\subsection{Local Lie algebra constructions}
\label{sec:LLA}

As in both formulations we will make use of Kac's construction based on local Lie (super)algebras, we briefly recall the basic statements from~\cite{Kac2}.

A local Lie superalgebra is a direct sum $\scr T_{-1} \oplus \scr T_{0} \oplus \scr T_1$ of three vector spaces together with a 
bilinear bracket
\begin{align}
\scr T_{-1} \times \scr T_1 &\to \scr T_0, & \scr T_{0} \times \scr T_1 &\to \scr T_1, &\scr T_{0} \times \scr T_{-1} &\to \scr T_{-1},& (x,y) \mapsto [x,y] \label{localbracket}
\end{align}
such that $[x,y]=-(-1)^{|x||y|}[y,x]$ for any 
two homogeneous elements
$x,y \in \scr T_{-1} \oplus \scr T_{0} \oplus \scr T_1$, and the Jacobi identity
\begin{align}
[x,[y,z]]=[[x,y],z]-(-1)^{|x||y|}[y[x,z]]
\end{align}
is satisfied whenever the brackets in this identity are defined.

As shown in~\cite[Prop.~1.2.2]{Kac2}, any local Lie superalgebra can be extended to a
unique minimal $\mathbb{Z}$-graded Lie superalgebra $\scr T=\oplus_{k\in\mathbb{Z}} \scr T_k$,
constructed
in two steps.
First, 
modulo the relations given by (\ref{localbracket}),
the local Lie superalgebra $\scr T_{-1} \oplus \scr T_0 \oplus \scr T_1$
generates a maximal Lie superalgebra
$\tilde{\scr T} = \bigoplus_{k \in \mathbb{Z}} \tilde{\scr T_k}$,
where $\tilde{\scr T_k}={\scr T_k}$ for $k=0,\pm1$, and the subalgebras
$\tilde{\scr T}_\pm = \bigoplus_{k < 0} \tilde{\scr T_{\pm k}}$ 
are freely generated by $\scr T_1$ and $\scr T_{-1}$, respectively.
Among the graded ideals $D$ of $\tilde{\scr T}$ (which means that $D$ is
a direct sum of subspaces $D \cap \tilde{\scr T_k}$ for all integers $k$) intersecting the local part
$\scr T_{-1} \oplus \scr T_{0} \oplus \scr T_{1}$
trivially, there is a maximal one.
In the second step we factor out this maximal ideal $D$ from $\tilde{\scr T}$
and set $\scr T=\tilde{\scr T} / D$.\footnote{In the context of standard Kac--Moody algebras, the local Lie algebra corresponds to the simple Chevalley generators and relations, and the maximal ideal corresponds to the Serre relations.}
This minimal Lie superalgebra $\scr T$ will be the tensor hierarchy algebra in our case. (There also exist other, non-minimal, Lie superalgebras that can be constructed from a local Lie superalgebra but they will play no role in our analysis.) Using Proposition 1.5 in
\cite{Kac}, one can show that any ideal $D$ of $\tilde{\scr T}$ is in fact
graded in our case,
and thus the tensor hierarchy algebra $\scr T$ that we define is simple.

%%%%%%%%%%%%%%%%%%%%%%%%%%%%%%%%%%%%%%%%%%%%%%%%%%%%%
\subsection{Direct algebraic characterisation}
%%%%%%%%%%%%%%%%%%%%%%%%%%%%%%%%%%%%%%%%%%%%%%%%%%%%%

The first characterisation of the tensor hierarchy algebra is a direct application of the Kac construction. 

\subsubsection{Definition of the local Lie superalgebra}

In our case, the local Lie superalgebra $\scr T_{-1} \oplus \scr T_{0} \oplus \scr T_1$ is defined as the tensor product of two $\mathbb{Z}$-graded vector spaces $\Lambda$ and $U$.

The vector space $\Lambda$ is the exterior (Grassmann) algebra of a $d$-dimensional vector space, and is thus (as an algebra)
generated by $d$ elements $\theta_m$
with an associative product such that $\theta_m \theta_n = - \theta_n \theta_m$. As a $\mathbb{Z}$-graded algebra,
$\Lambda$ can be decomposed into a direct sum
\begin{align}
\Lambda = \Lambda_0 \oplus \Lambda_1 \oplus \cdots \oplus \Lambda_d
\end{align}
of subspaces such that $\Lambda_i\Lambda_j = \Lambda_{i+j}$, where, for any    
$k=0,1,\ldots$,
the set of all monomials
\begin{align}
\theta_{n_1\cdots n_k} \equiv \theta_{n_1} \theta_{n_2} \cdots \theta_{n_k}\quad\quad(1\leq n_1<n_2<\ldots<n_k\leq d)
\end{align}
is a basis of the subspace $\Lambda_k$.
We write this as $\Lambda_k=\langle \theta_{n_1\cdots n_k} \rangle$.
As a $\mathbb{Z}_2$-graded algebra, $\Lambda$ decomposes into a direct sum $\Lambda = \Lambda_{(0)} \oplus \Lambda_{(1)}$ 
where $\Lambda_k \subseteq \Lambda_{(0)}$ if $k$ is even and $\Lambda_k \subseteq \Lambda_{(1)}$ if $k$ is odd.
For any $m=1,2,\ldots, d$ we define the interior product $\iota^m$ on $\Lambda$ as the linear map $\iota^m : \Lambda_k \to \Lambda_{k-1}$
given by
\begin{align}
\iota^m \theta_{n_1\cdots n_k} = k \delta_{[n_1}^m \theta_{n_2\cdots n_k]}.
\end{align}

The $\mathbb{Z}$-graded vector space $U=U_{(0)}\oplus U_{(1)}$ is spanned by $\gl(d)$ tensors 
$E^{n_1n_2n_3}$, $E^m$, $E^{m,n}$, $P^m$, $F$
(where, following our conventions, $E^{n_1n_2n_3}=E^{[n_1n_2n_3]}$ and $E^{m,n}=E^{n,m}$)
such that
\begin{align}
U_{(0)} &= \langle E^{n_1n_2n_3} \rangle \oplus \langle E^m \rangle, & U_{(1)} &= \langle E^{m,n} \rangle \oplus \langle P^m \rangle
\oplus \langle F \rangle.
\end{align}
We then decompose the tensor product $\Lambda \otimes U$ into
a direct sum $\scr T_{-1} \oplus \scr T_0 \oplus \scr T_1$ such that
\begin{align}
\label{eq:LLA}
\scr T_{-1}&= \Lambda \otimes \langle F \rangle, & 
\scr T_{0} &= \Lambda \otimes \langle P^m \rangle, &
\scr T_{1}
&= \Lambda \otimes \big(\langle E^{n_1n_2n_3} \rangle \oplus \langle E^{m,n} \rangle \oplus \langle E^m \rangle \big).
\end{align}
The $\mathbb{Z}_2$-degree of an element $au \in \Lambda \otimes U$, where $a \in \Lambda$ and $u \in U$, is given by the product of the 
$\mathbb{Z}_2$-degrees of $a$ and $u$. We write this as $|au|=|a||u|$.

\allowdisplaybreaks{
The bracket on $\scr T_{-1} \oplus \scr T_0 \oplus \scr T_1$ is defined by the following commutation relations,
\begin{subequations}
\label{localcommrel}
\begin{align}
[aE^{n_1n_2n_3},bF]
&=\tfrac3{16}(\iota^{[n_1}\iota^{n_2}a)bP^{n_3]} 
+\tfrac32(-1)^{{|a|}}(\iota^{[n_1}a)(\iota^{n_2}b)P^{n_3]}
+3a(\iota^{[n_1}\iota^{n_2}b)P^{n_3]}, \\
[aE^{m,n},bF]&=a(\iota^{(m} b) P^{n)} + \tfrac14(-1)^{|a|}(\iota^{(m}a)b P^{n)}, \\
[aE^m,bF]&=abP^m,\\
[aP^m, bF]&=a(\iota^m b)F+\tfrac13(-1)^{|a|}(\iota^m a)bF,\\
[aP^m,bP^n]&=a(\iota^m b)P^n +(-1)^{|a|}(\iota^n a)b P^m,\\
[aP^m,bE^{n_1n_2n_3}]&=a(\iota^m b)E^{n_1n_2n_3}+3(-1)^{|a|}(\iota^{[n_1}a)bE^{n_2n_3]m}-\tfrac13(-1)^{|a|}(\iota^m a)bE^{n_1n_2n_3}\\
&\quad\,-3(-1)^{|b|}(\iota^{[n_1}\iota^{n_2}a)bE^{n_3],m}
+\tfrac3{16}(-1)^{|a|}(\iota^{[n_1}\iota^{n_2}\iota^{n_3]}a)bE^m\nn\\
&\quad\,-\tfrac3{16}(-1)^{|a|}(\iota^{m}\iota^{[n_1}\iota^{n_2}a)bE^{n_3]}
+\tfrac9{16}(\iota^{[n_1}\iota^{n_2}a)(\iota^{|m|}b)E^{n_3]},\\
[aP^m,bE^{n,p}]&=a(\iota^m b)E^{n,p}+2(-1)^{|a|}(\iota^{(n} a)b E^{p),m}-\tfrac13(-1)^{|a|} (\iota^m a)b E^{n,p},\nn\\
[aP^m,bE^{n}]&=a(\iota^m b)E^n +(-1)^{|a|}(\iota^n a)b E^m-\tfrac13(-1)^{|a|}(\iota^m a)b E^n. 
\end{align}
\end{subequations}
One can verify that all Jacobi identities are satisfied and thus $\scr T_{-1} \oplus \scr T_0 \oplus \scr T_1$ provides a starting point for the local Lie superalgebra construction.}

The reason for starting with this particular local Lie superalgebra comes from supergravity and its relation to $\mf{e}_{11}$. This connection will become more apparent below when we list some of the further generators of $\scr T$ in $\mf{gl}(11)$ form. 
The tensor hierarchy algebra $\scr T$ associated to $\mf{e}_d$ for $4 \leq d \leq 8$ was defined in~\cite{Palmkvist:2013vya}.
The construction in this appendix is a different $\mathfrak{gl}(d)$ 
covariant definition and has the advantage of also being applicable also to the case
$d\geq 9$. For $d=11$ we obtain the tensor hierarchy algebra $\scr T$ considered in this paper. 

\subsubsection{The tensor hierarchy algebra}

The tensor hierarchy algebra $\scr T$ is now defined as the minimal Lie superalgebra
with the local part above,
and can be constructed from this local part following the steps in Section~\ref{sec:LLA}.
It then comes with 
a $\mathbb{Z}$-grading
$\scr T=\oplus_{k\in\mathbb{Z}} \scr T_k$,
where we
for any $x \in \scr T_k$ set $q(x)=k$. This $\mathbb{Z}$-grading is not consistent:
$x$ does not necessarily 
have the same $\mathbb{Z}_2$-degree as the integer $q(x)$.
However, $\scr T$ can be equipped with a different $\mathbb{Z}$-grading that is consistent.
We denote the $\mathbb{Z}$-degree of a homogeneous element $x$ with respect to this consistent $\mathbb{Z}$-grading by $p(x)$. For the local part $\Lambda \otimes U$
it is given by $p(au)=p(a)+p(u)$ where $p(a)$ refers to the $\mathbb{Z}$-grading of
$\Lambda$ above, $p(a)=k$ if $a \in \Lambda_k$, and $p(u)$ is given by the assignments
\begin{align}
p(F)&=3, & p(P^m)&=1, & p(E^{n_1n_2n_3})&=0, & p(E^{m,n})&=-1, & p(E^m)&=-2.
\end{align}

As in Section~\ref{sec:tha1}, we refer to $p$ and $q$ as vertical and horizontal degrees, respectively. As will be shown below, the subalgebra at $(p,q)=(0,0)$ is $\mf{gl}(d)$,
and the $\mf{gl}(d)$ level is given by
\begin{align}
\ell=q+\frac{3}{9-d}p.
\end{align}

We can now probe the tensor hierarchy algebra degree by degree both vertically and horizontally.
It then follows
that the subspace $\scr T_{-2}$ is the tensor product of $\Lambda$
and a one-dimensional vector space spanned by an element $G$.
We choose a normalization of it such that
\begin{align} \label{FFG}
[aF,bF]=(-1)^{|a|} (ab)G.
\end{align}
\newpage
\noindent
The commutation relations of the form
$[\scr T_{1}, \scr T_{-2}]=\scr T_{-1}$ are then given by 
\begin{align}
[aE^{n_1n_2n_3},bG]&=\tfrac18(-1)^{|a|}(\iota^{n_1}\iota^{n_2}\iota^{n_3}a)bF
+\tfrac34(\iota^{[n_1}\iota^{n_2}a)(\iota^{n_3]}b)F\nn\\
&\quad\,+\tfrac32(-1)^{|a|}(\iota^{[n_1}a)(\iota^{n_2}\iota^{n_3]}b)F
+a(\iota^{n_1}\iota^{n_2}\iota^{n_3}b)F,\nn\\
[aE^{m,n},bG]&=0,\nn\\
[aE^m,bG]&=\tfrac23(-1)^{|a|}(\iota^m a)bF+\tfrac43 a (\iota^m b) F,
\end{align}
and those of the form $[\scr T_{0}, \scr T_{-2}]=\scr T_{-2}$ by
\begin{align}
[aP_m,bG]&= -(-1)^{|a|}a(\iota^m b )G -\tfrac23(\iota^m a )b G.
\end{align}
Continuing to $q=-3$ we find that $\scr T_{-3}$ is the tensor product of $\Lambda$ and a $d$-dimensional vector space spanned by an element $H_m$, such that
\begin{align} \label{FGH}
[aF,bG]=\tfrac23 (\iota^m a)b H_m -\tfrac13(-1)^{|a|}a(\iota^m b)H_m.
\end{align}
The commutation relations of the form $[\scr T_1, \scr T_{-3}] = \scr T_{-2}$
are then given by
\begin{subequations}
\begin{align}
\label{eq:CRm31}
[aE^{n_1n_2n_3},bH_m]&=-\tfrac{27}{16}\de^{[n_1}_m(\iota^{n_2}\iota^{n_3]}a)bG\nn\\
&\quad\,-\tfrac92 \de^{[n_1}_m (\iota^{n_2} a)(\iota^{n_3]} b)G-3(-1)^{|a|}\de^{[n_1}_ma(\iota^{n_2}\iota^{n_3]}b)G,\\
[aE^{p,q},bH_m]&=-(-1)^{|a|}\de^{(p}_m(\iota^{q)} a)bG -\tfrac34 \de^{(p}_m a (\iota^{q)}b)G,\\
[aE^{n},bH_m]&=-(-1)^{|a|}\de^{n}_m ab G,
\end{align}
\end{subequations}
and those of the form $[\scr T_0, \scr T_{-3}] = \scr T_{-3}$ by
\begin{align}
[aP^m,bH_n]=-(-1)^{|a|}a(\iota^m b)H_n -(\iota^m a)b H_n +\de^m_n(\iota^p a)b H_p.
\end{align}
At the first positive horizontal degrees beyond $q=1$, the structure of the tensor hierarchy algebra is more complicated. We will not describe it in detail here, but refer to Table~\ref{tab:THA3} where some of the generators at $q=2,3$ are given, together with those described here for $-3 \leq q \leq 1$. See also Table~\ref{tab:THA}, where other symbols are used for the $\mf{gl}(11)$ tensor densities, and some of them have been dualised using the $\mf{sl}(11)$ invariant epsilon tensor (making Table~\ref{tab:THA} valid only for $d=11$,
whereas Table~\ref{tab:THA3} is valid for any $d$).

We identify two important subalgebras of $\scr T$. First, by restricting to horizontal degree $q=0$ but allowing for arbitrary vertical degrees we find the extension of $\mf{gl}(11)$ to the Cartan superalgebra $W(d)$,
which is the derivation superalgebra of $\Lambda$ \cite{Kac2}. Second, the subalgebra generated by $E^{n_1n_2n_3}$ and
$\theta_{n_1n_2n_3}F$ at $p=0$ is $\mathfrak{e}_{d}$. To see this we set
\begin{align}
K^m{}_n &\equiv -\theta_n P^m - \frac1{9-d}\de_n^m \theta_p P^p\ ,
& F_{n_1n_2n_3}&\equiv\theta_{n_1n_2n_3}F\ . &
\end{align}
The commutation relations of $E^{n_1n_2n_3}$, $F_{n_1n_2n_3}$ and $K^m{}_n$ are exactly those of $\mathfrak{e}_d$ in $\mf{gl}(d)$ decomposition and the Lie algebra they generate is by construction $\mathfrak{e}_d$. It is contained in the subalgebra $\mf{t}_d$ of $\scr T$ 
consisting of all elements with $p=0$. However, for $d\geq 9$ this subalgebra contains also additional generators, in particular $H_9$ at $(p,q)=(0,-3)$, which can be seen in Table~\ref{tab:THA3}. This $H_9$ plays an important role in the low level considerations in the body of the paper as it is related to the new field $X_9$ that carries the dual of the trace of the spin connection.

To see how the additional generators appear, we continue along $p=0$ and set
\begin{align}
G_{n_1\cdots n_6}&\equiv \theta_{n_1\cdots n_6}G\ , &
H_{n_1\cdots n_8;m}&\equiv \theta_{n_1\cdots n_8}H_m
\end{align}
at $q=-2$ and $q=-3$ respectively.
The generator $G_{n_1\cdots n_6}$ corresponds to
$F_{n_1 \cdots n_6}$ in
Appendix~\ref{app:conv}\footnote{The $\mathfrak{e}_{d}$
generator $F_{n_1 \cdots n_6}$ in Appendix~\ref{app:conv}
should not be confused with the 
generator
$\theta_{n_1\cdots n_6} F$
appearing in the extension of $\mathfrak{e}_d$ to $\scr T$ that we consider here
at $(p,q)=(-3,-2)$.
This is the reason
why we
use different letters for different $q<0$ in the present appendix.}, as can be seen by comparing (\ref{FFG}), for $a,b \in \Lambda_3$,
with (\ref{112}).
The generator $H_{n_1\cdots n_8;m}$ transforms under $\mathfrak{gl}(11)$ in the full tensor product of the two representations corresponding to the blocks of indices on the two sides of the semicolon, and can in the usual way be decomposed into irreducible parts as
\begin{align} \label{H-irreps}
H_{n_1\cdots n_8m}&=H_{[n_1\cdots n_8;m]}, &
H_{n_1\cdots n_8,m}=H_{n_1\cdots n_8;m} - H_{n_1\cdots n_8m}.
\end{align}
In the case where $a \in \Lambda_3$ and $b \in \Lambda_6$ in (\ref{FGH}),
the fully antisymmetric
part drops out of the right hand side, and the equation (\ref{FGH})
reduces to the second row of
(\ref{123}) (with $H_{n_1\cdots n_8,m}$ replacing $F_{n_1\cdots n_8,m}$ according to the different notations used here and in Appendix~\ref{app:conv}).
However, when we consider the full tensor hierarchy algebra $\scr T$ we can
take for example $a \in \Lambda_2$ and $b \in \Lambda_7$, writing
\begin{align}
F_{n_1n_2}&\equiv\theta_{n_1n_2}F, & G_{n_1\cdots n_7}&\equiv\theta_{n_1\cdots n_7}G,
\end{align}
and then (\ref{FGH}) gives
\begin{align}
[F_{n_1n_2},G_{p_1\cdots p_7}]&= -H_{n_1n_2p_1\cdots p_7}+2H_{p_1\cdots p_7 [n_1,n_2]}\nn\\
&=-H_{n_1n_2p_1\cdots p_7}-7H_{n_1 n_2 [p_1\cdots p_6, p_7]}\ ,
\end{align}
where now also the fully antisymmetric part $H_9$ is present on the right hand side.

{}From the irreducible pieces in $(p,q)=(0,-3)$ given in~\eqref{H-irreps} and the commutator in~\eqref{eq:CRm31} we can deduce the following commutator in the $p=0$ subalgebra $\mf{t}_{d}$:
\begin{align}
\label{eq:CXnew}
\lb E^{n_1n_2n_3} , H_{p_1\ldots p_9} \rb = 168 \, \delta^{n_1n_2n_3}_{[p_1p_2p_3} G_{p_4\ldots p_9]}^{\phantom{n_1}}\,.
\end{align}
This commutator (when dualised to $p=-2$ as will be argued below) is the reason for~\eqref{eq:deltaX} that is used crucially for the gauge invariance discussion of the tensor hierarchy algebra structures. The relation above demonstrates that within the tensor hierarchy algebra the coefficient $T^{\alpha_0\beta_1}{}_{\gamma_0}$ in~\eqref{eq:CRm22} does not vanish.

Since $\mathfrak{e}_d$ is contained in the subalgebra $\mf{t}_d\subset\scr T$ at $p=0$,
the subspace of $\scr T$ at any vertical degree $p$ is a representation $R_p$ of $\mathfrak{e}_d$.
As we will see below, $R_p$ is the conjugate of $R_{9-d-p}$ for any $p$.
In the case $d=11$, this means that the adjoint of $\mathfrak{e}_{11}$ can be obtained
from $R_{-2}$ by factoring out additional generators, in particular the trace part of
$P_3{}^1$, which is dual to the additional generator $H_9$ in $R_0$. To make this more clear, set
\begin{align}
\widetilde{E}^{n_1n_2n_3}&=
\tfrac1{8!}\varepsilon^{n_1n_2n_3p_1\cdots p_8}G_{p_1\cdots p_8}\ ,\nn\\
\widetilde{E}^{n_1\cdots n_6}&=\tfrac1{5!}\varepsilon^{n_1\cdots n_6p_1\cdots p_5}F_{p_1\cdots p_5}\ ,\nn\\
\widetilde{E}^{n_1\cdots n_8;m}&=\tfrac1{3!}\varepsilon^{n_1\cdots n_8p_1p_2p_3}
P_{p_1 p_2 p_3}{}^m\ ,
\end{align}
in accordance with the notation in Table \ref{tab:THA}. In the same way as in
(\ref{H-irreps}), the generator $\widetilde{E}^{n_1\cdots n_8;m}$ can be decomposed into
the irreducible parts $\widetilde{E}^{n_1\cdots n_8m}$ and
$\widetilde{E}^{n_1\cdots n_8,m}$. We now get, for example, the relations
\begin{align}
[E^{n_1n_2n_3},\widetilde{E}^{p_1p_2p_3}]&=\widetilde{E}^{n_1n_2n_3p_1p_2p_3}\ ,\nn\\
[F_{n_1n_2n_3},\widetilde{E}^{p_1\cdots p_6}]&=120\,\delta^{[p_1p_2p_3}_{n_1n_2n_3}\,
\widetilde{E}^{p_3p_4p_6]}\ ,
\end{align}
which can be compared to (\ref{112}) and (\ref{123}). However, when we act with
$E^3$ on $\widetilde E^6$ we see that this is not the adjoint representation of $\mathfrak{e}_{11}$,
since, compared to (\ref{level3inv}), we get an additional term containing the fully antisymmetric generator $\widetilde E^9$, 
\begin{align}
[E^{n_1n_2n_3},\widetilde{E}^{p_1\cdots p_6}]&=-3\widetilde{E}^{p_1\cdots p_6[n_1n_2;n_3]}
=-3\widetilde{E}^{p_1\cdots p_6[n_1n_2,n_3]}-3\widetilde{E}^{p_1\cdots p_6n_1n_2n_3}\ .
\end{align}
When we act on $\widetilde E^{8,1}$ and $\widetilde E^9$ with $F_3$ we find
\begin{align}
[F_{n_1n_2n_3},\widetilde{E}^{p_1\cdots p_8,q}]= -112\,(\de^{q[p_1p_2}_{n_1n_2n_3}\,\widetilde{E}^{p_3\cdots p_8]}
-\de^{[p_1p_2p_3}_{n_1n_2n_3}\,\widetilde{E}^{p_4\cdots p_8]q})
\end{align}
in accordance with (\ref{132}), and
\begin{align}
[F_{n_1n_2n_3},\widetilde{E}^{p_1\cdots p_9}]=0\ ,
\end{align}
which means that $\widetilde{E}^9$
can be set to zero consistently as a generator in the $\mathfrak{e}_{11}$ representation $R_{-2}$
(but not as a generator in the full Lie superalgebra $\scr T$).

%%%%%%%%%%%%%%%%%%%%%%%%%%%%%%%%%%%%%%%%%%%%%%%%%%%%%%%%%%%%%%%%%%%%%%%%%%%%%%%%%%%%%%%%%%%
\begin{table}[ht]
\setlength{\arraycolsep}{3.5pt}
\renewcommand{\arraystretch}{1.5}
\begin{align*}
\begin{array}{r|c|cc|c|c|c|ccc|ccc|cc|c}
p&\cdots&
\multicolumn{2}{c|}{q=-3}&q=-2 & q=-1 & {q=0} &  \multicolumn{3}{c|}{q=1} & \multicolumn{3}{c|}{q=2}& \multicolumn{2}{c|}{q=3}&
\cdots \\
\hline
\vdots &\ddots
&\vdots&\vdots&\vdots&&  &&&&&&&&&\\ 
3
&\cdots& H_6& H_{5,1}& G_3& F_0&&&&&&&   & &&\\ 
2
&\cdots&H_7& H_{6,1}& G_4& F_1&&&&&&  & &&&\\ 
1
&\cdots&H_8& H_{7,1}& G_5& F_2 & {P_0{}^1} &&& && & &  &&\\ 
0 &\cdots&H_9 & H_{8,1}& G_6& F_3 & P_1{}^1 & E_0{}^3 &  &&
E_0{}^6 & &  & E_0{}^{8,1}&&\cdots\\ 
-1 &\cdots
&H_{10}& H_{9,1}& G_7
& F_4 & P_2{}^1 &
E_1{}^3 & E_0{}^{1,1}&& E_1{}^6 & E_0{}^{4,1}
&  &E_1{}^{8,1} &\cdots&\cdots\\ 
-2 &\cdots
&H_{11} & H_{10,1}    & G_8& F_5 & P_3{}^1 & 
E_2{}^3 & E_1{}^{1,1}  
& E_0{}^1
& E_2{}^6&E_1{}^{4,1}&\cdots&E_2{}^{8,1}&\cdots&\cdots\\ 
-3 
&\cdots& H_{12} & H_{11,1}  & G_9 & F_6 & P_4{}^1 &  E_3{}^3  &E_2{}^{1,1}  & 
E_1{}^1 
& E_3{}^6
& E_2{}^{4,1}&\cdots&E_3{}^{8,1}&\cdots&\cdots\\ 
\vdots &\iddots&\vdots&\vdots &\vdots &\vdots &\vdots &\vdots &\vdots&\vdots&\vdots   & \vdots&\ddots&\vdots&\ddots&\ddots
\end{array}
\end{align*}
\caption{\it Part of the tensor hierarchy algebra $\mathscr T$ for a general $d$, decomposed under $\mathfrak{gl}(d)$.}\label{tab:THA3}
\end{table}
%%%%%%%%%%%%%%%%%%%%%%%%%%%%%%%%%%%%%%%%%%%%%%%%%%%%%%%%%%%%%%%%%%%%%%%%%%%%%%%%%%%%%%%%%%%

\subsubsection{Existence of an invariant bilinear form}

We will now prove the existence of a non-degenerate
supersymmetric and invariant bilinear form $\jp( x, y )$ on $\scr T$. 
Here supersymmetry (following the mathematics terminology) means $\mathbb{Z}_2$-graded symmetry, that is
$\jp( x , y ) = (-1)^{|x||y|} \jp( y , x )$.
Invariance means 
\begin{align} \label{inv-rel}
\jp( [x,y],z ) = \jp( x , [y,z] )
\end{align}
for all elements $x,y,z$ regardless of their $\mathbb{Z}_2$-degrees.
Our proof follows to a large extent the proof of Proposition 7 in \cite{Kac1}.
The bilinear form that we will define
has the properties $\jp(\scr T_{i},\scr T_{j})=0$ unless $i+j=-3$
and $\Omega(R_i,R_j)=0$ unless $i+j=9-d$. Thus it gives
a symplectic form on $R_{-1}$ in the case $d=11$.

We say that a bilinear form $\jp$ defined on some subspace of $\scr T$ is invariant with respect to some subspace $\scr U$ of $\scr T$
if (\ref{inv-rel}) holds for all $x,y,z$
such that both sides of (\ref{inv-rel}) are defined and $y \in \scr U$.

For $s\geq 3$, suppose that $\jp^{(s-1)}$ is a bilinear form on the subspace
$\scr T_{-s-1} \oplus \cdots \oplus \scr T_{s-2}$ of $\scr T$
which is supersymmetric and 
invariant with respect to all $\scr T_k$ with $k\neq 0$, or equivalently, with respect to $\scr T_{\pm 1}$.
Let $\jp^{(s)}$ be an extension of $\jp^{(s-1)}$ to
$\scr T_{-2-s} \oplus \cdots \oplus \scr T_{s-1}$ defined in the following way. First, set
$\jp(\scr T_{i},\scr T_{j})=0$ if one of the integers $i$ and $j$ is equal to $(s-1)$ or $(-2-s)$ and $i+j \neq -3$. Then, for
$w \in \scr T_{s-1}$ and $z \in \scr T_{-s-2}$, write $w$ and $z$ as sums of terms $[u,v]$ and $[x,y]$, respectively,
where
\begin{align}
u,v &\in \scr T_{1} \oplus \cdots \oplus \scr T_{s-2}, & 
x,y &\in \scr T_{-s-1} \oplus \cdots \oplus \scr T_{-1}.
\end{align}
We can without loss of generality assume that there is only one term in each of these sums, and write $w=[u,v]$ and $z=[x,y]$.
We then define $\jp^{(s)}(w,z)=(-1)^{wz}\jp^{(s)}(z,w)$ by
\begin{align} \label{omega-def}
\jp^{(s)}(w,z) = \jp^{(s)}([u,v],[x,y]) \equiv \jp^{(s-1)}([[u,v],x],y).
\end{align}
Using the supersymmetry and invariance of $\jp^{(s-1)}$ with respect to $\scr T_{\pm 1}$,
and the Jacobi identity, we then get 
\begin{align} \label{omega-inv}
\jp^{(s-1)}([[u,v],x],y)
&=\jp^{(s-1)}([u,[v,x]],y)-(-1)^{uv}\jp^{(s-1)}([v,[u,x]],y)\nn\\
&=-(-1)^{u(v+x)}\jp^{(s-1)}([v,x],[u,y])+(-1)^{vx}\jp^{(s-1)}([u,x],[v,y])\nn\\
&=(-1)^{u(v+x+y)}\jp^{(s-1)}([[v,x],y],u)+(-1)^{vx}\jp^{(s-1)}(u,[x,[v,y]])\nn\\
&=\jp^{(s-1)}(u,[[v,x],y])+(-1)^{vx}\jp^{(s-1)}(u,[x,[v,y]])\nn\\
&=\jp^{(s-1)}(u,[v,[x,y]]).
\end{align}
Thus $\jp^{(s)}$ is well defined and invariant with respect to all $\scr T_k$ with $k\neq 0$, or equivalently, with respect to $\scr T_{\pm 1}$.

We define a linear (volume) form on $\Lambda$ by
\begin{align}
V(\theta^{n_1\cdots n_{p}}) = \varepsilon^{n_1\cdots n_{d}}
\end{align}
if $p=d$, and $V(\theta^{n_1\cdots n_{p}})=0$ otherwise.
Then the bilinear form $\Omega^{(0)}$ on $\scr T_{-2} \oplus \scr T_{-1}$ defined by
\begin{align}
\Omega^{(0)}(aF,bG)=(-1)^{(|a|+1)|b|}\Omega^{(0)}(bG,aF)&=V(ab),\nn\\
\Omega^{(0)}(aF,aF)=\Omega^{(0)}(aG,aG)&=0
\end{align}
is invariant with respect to $\scr T_{-1} \oplus \scr T_{0} \oplus \scr T_{1}$.
We then define $\Omega^{(1)}$ by (\ref{omega-def}) with
$u \in \scr T_{-1}$, $v \in \scr T_{1}$, $x \in \scr T_{-1}$ and $y \in \scr T_{-2}$.
Explicitly we get
\begin{align}
\Omega^{(1)}(aP^m,bH_n)=
V(ab)\delta^m_n.
\end{align}
By the invariance of $\Omega^{(0)}$
with respect to $\scr T_{-1} \oplus \scr T_{0} \oplus \scr T_{1}$ and a calculation similar to (\ref{omega-inv}) it then follows
that $\Omega^{(1)}$
is well defined and invariant with respect to $\scr T_{\pm 1}$. Finally we
define $\jp^{(2)}$ on $\scr T_{-4} \oplus \cdots \oplus \scr T_{1}$ again 
by (\ref{omega-def}) 
for $u \in \scr T_0$, $v \in \scr T_{1}$ and
$x,y \in \scr T_{-4} \oplus \cdots \oplus \scr T_{-1}$.
By the same calculation (\ref{omega-inv}) it follows that $\jp^{(2)}$ is well defined and invariant with respect to $\scr T_{\pm 1}$.
We can then recursively extend the bilinear forms $\jp^{(s)}$ and define a bilinear form $\jp$ on the whole of $\scr T$ which is supersymmetric and
invariant with respect to $\scr T_{k}$ for $k \neq 0$. It then follows that $\jp$ is invariant also with respect to $\scr T_0$. The non-degeneracy of the bilinear form
$\jp$ follows from its invariance and the fact that $\scr T$ is a simple Lie superalgebra.

\subsection{BRST form of the tensor hierarchy algebra}

The BRST formalism we shall now use give an equivalent definition of  the tensor hierarchy algebra $\mathscr{T}$ corresponds to defining a nilpotent differential $\delta$ transforming the parameters of the algebra (rather than working with the generators). An important point is that the parameters are `ghosts', meaning that their $\mathbb{Z}_2$ Grassmann degree is shifted: Grassmann even generators of the algebra are associated with Grassmann odd parameters whereas Grassmann odd generators are associated with Grassmann even parameters.  In this way, the nilpotency $\delta^2=0$ of the differential
\be 
\delta c^A = \frac{1}{2} C^A{}_{BC} c^B c^C \qquad \longleftrightarrow \qquad ( c^A T_A )^2 = \delta c^A T_A \  , 
\ee
is equivalent to super-Jacobi identity on $C^A{}_{BC}$. Here, $c^A$ denotes a generic parameter associated with a generator $T_A$ and $C^A{}_{BC}$ are the structure constants of the algebra. This way of writing the transformations, \ie commutators, makes some of the calculations simpler. 

The parameters we are using in this section are related to the superform generators of the last section through
\begin{multline} 
c^A T_A =\dots +  ( \upsilon^m , H_m ) +  (\omega , G) +   ( S , F) + ( V_m , P^m ) \\ 
+ \tfrac{1}{3!} ( \psi_{n_1n_2n_3} , E^{n_1n_2n_3})  +  \tfrac{1}{2} ( T_{m,n} , E^{m,n})+ ( \lambda_m , E^m) + \dots \ , 
\end{multline}
where $(\cdot,\cdot)$ is understood as the standard pairing for generalised forms using the top form in the exterior algebra.

\subsubsection{Local Lie superalgebra and tensor hierarchy algebra}

We now rephrase the definition of the tensor hierarchy algebra in the BRST formalism starting from the local algebra. As $\mathscr{T}_0$ we will take the $W(d)$ superalgebra of super diffeomorphisms defined by Kac in~\cite{Kac2}. It can be parametrised by a Grassmann odd vector-valued extended form in $d$ dimensions, which we will defined as Grassmann even according to the BRST formalism. This means that we have a parameter $V_m$ that lies in the tensor product of the vector representation of $GL(d)$ with the exterior algebra $\Lambda$ in $d$ dimensions. 
The $W(d)$ algebra can be written as
\be 
\label{eq:deltaV}
\delta V_m = V_n \iota^n V_m \ , 
\ee
where $\iota^m$ is the contraction operator whereas the forms are multiplied through the wedge product.  It is easy to check that this transformation is nilpotent. However, this is not the complete transformation in the local algebra as one has to include contributions from $\mathscr{T}_{\pm 1}$. These will be displayed below. The decomposition of $V_m$ in form degree is dual to the column $q=0$ of table~\ref{tab:THA3}. 

The remaining elements of $\scr T_{\pm 1}$ can be written in terms of a Grassmann-even scalar-valued form $S$ (for $q=-1$) and a Grassmann-odd rank-three generator $\psi_{n_1n_2n_3}$, a Grassmann-odd rank-one generator $\lambda_m$ and Grassmann-even symmetric two-form generator $T_{m,n}$ for $q=+1$. These are all forms valued in the exterior algebra $\Lambda$ in $d$ dimensions and correspond to the generators also listed in~\eqref{eq:LLA}.

In BRST form the transformations in the local Lie superalgebra (cf.~\eqref{localcommrel}) take the form
\begin{subequations}
\begin{align}
\delta S &= V_n \iota^n S + \frac{w}{3} \iota^n V_n S\,,\\
\delta V_m &= V_n \iota^n V_m + \psi_{mp_1p_2} \iota^{p_1} \iota^{p_2} S 
+ \left( \frac{1}{2} \iota^p \psi_{mnp}  + T_{m,n}\right)\iota^nS \nn\\
&\hspace{25mm}+ \frac{1}{16} \left(\iota^{p_1} \iota^{p_2} \psi_{mp_1p_2} + 4\iota^n T_{m,n} +  \lambda_m \right)S \,,\\
\delta \psi_{n_1n_2n_3} &= V_p \iota^p  \psi_{n_1n_2n_3}  + 3 \iota^p V_{[n_1}  \psi_{n_2n_3]p} 
- \frac{w}{3} \iota^p V_p  \psi_{n_1n_2n_3} \,,\\
\delta \lambda_m &= V_p \iota^p  \lambda_m +  \iota^p V_{m}  \lambda_p - \frac{w}{3} \iota^p V_p \lambda_{m}  
+ 3\iota^{p_1} \iota^{p_2} V_{q} \iota^q \psi_{mp_1p_2}\nn\\
& \hspace{25mm} +\iota^{p_1} \iota^{p_2} \iota^{p_3} V_{m} \psi_{p_1p_2p_3}
-\iota^{p_1} \iota^{p_2} \iota^{q} V_{q} \psi_{mp_1p_2} \,,\\
\delta T_{m,n} &= V_p \iota^p  T_{m,n}  + 2 \iota^p V_{(m}  T_{n),p} - \frac{w}{3} \iota^p V_p T_{m,n}  
+ \iota^{p_1} \iota^{p_2} V_{(m} \psi_{n)p_1p_2}\,. 
\end{align}
\end{subequations}
One can check that the transformation $\delta$ becomes nilpotent with these rules and there is no redefinition that would allow to remove some of the generators. Therefore the above is an equivalent presentation of the local Lie superalgebra that can be used as a starting point for Kac' construction. The algebra defined in this way is dual to $\scr T$ and we shall now list some of its other generators.

It follows for example that the level $q=-2$ component is  parametrised by a Grassmann 
odd generalised form $\omega$, the level $q=-3$ by a Grassmann odd  co-vector generalised form $\upsilon_n$, and 
the level $q=-4$ by a Grassmann even 3-form, an even 1-form, and a Grassmann odd symmetric tensor. 

One can compute the extension of the BRST transformations (=commutation relations) to these levels. One gets the following identities 
\bea 
\delta S &=& V_n \iota^n S + \frac{1}{3} \iota^n V_n S
-\frac{1}{6} \psi_{n_1n_2n_3} \iota^{n_1}\iota^{n_2}\iota^{n_3}\omega 
+\frac{1}{4} \iota^{n_3}\psi_{n_1n_2n_3} \iota^{n_1}\iota^{n_2}\omega 
- \frac{1}{8} \iota^{n_2}\iota^{n_3}\psi_{n_1n_2n_3} \iota^{n_1} \omega
\CR
&& +\frac{1}{48}\iota^{n_1}\iota^{n_2}\iota^{n_3} \psi_{n_1n_2n_3} \omega 
-\frac{1}{24} \lambda_n \iota^n \omega + \frac{1}{48} \iota^n \lambda_n \omega + \dots 
\CR
\delta \omega &=& V_p \iota^p \omega + \frac{2}{3} \iota^p V_p \omega + S^2 
- \frac{1}{3} \psi_{n_1n_2n_3} \iota^{n_1} \iota^{n_2} \upsilon^{n_3} 
+\left( \tfrac{1}{2} \iota^p \psi_{mnp}+\tfrac{1}{3} T_{m,n}\right) \iota^m \upsilon^n   
\CR
&& \qquad -\frac{1}{16}  \left(3  \iota^{p_1} \iota^{p_2} \psi_{mp_1p_2} 
-4 \iota^n T_{m,n} + \tfrac{1}{3} \lambda_m\right) \upsilon^m + \dots 
\CR
\delta \upsilon^m &=&V_p \iota^p \upsilon^m - \iota^m V_p \upsilon^p + \iota^p V_p \upsilon^m 
+ S \iota^m \omega - 2 \iota^m S \omega  + \dots \,.
\eea 

We have focussed on these $q$-levels as they are dual to the local algebra by an involution that exchanges level $q$ with level $-3\!-\!q$. This is the invariant already encountered above in the direct formulation.

\subsubsection{Involution and symplectic invariant}

We shall now show that there is an involution relating the $W(d)$ representation on level $q$ of the algebra to the conjugate 
$W(d)$ representation of the level $-3\!-\!q$ component, and which is obtained by the use the Hodge-star 
operator on the generalised form. Here, level $q$ follows from Kac' construction and is displayed in table~\ref{tab:THA3}.

Denoting a general field of the tensor hierarchy algebra $\mathscr{T}$ by 
\be
c=(\ldots; \upsilon^n; \omega; \psi_{n_1n_2n_3}, T_{m,n}, \lambda_m; V_m; S; \ldots)
\ee
we define an antisymmetric  bilinear form on $\mathscr{T}$ in components as the top-form component of 
\be 
\Omega(c_1, c_2) = \left( 3 S_1 \omega_2 + V_{1\,n} \upsilon_{2}{}^{n} + \dots - 3 S_2 \omega_1 
- V_{2\,n} \upsilon_{1}{}^{n} + \dots\right)_{\rm top} \,.
\ee
One can check using the above transformations that this is invariant to the level given. We shall now show, starting from the local algebra, that an extension of the invariant $\Omega(x_1, x_2)$ to all levels exists.

We need to check that this bilinear form satisfies 
\be
\jp( [x,y],z ) = \jp( x , [y,z] )
\ee
for the three cases in which: all $x$, $y$ and $z$ are level $q=-1$, when $x$ is level $-2$, $y$ level $0$ and $z$ level $-1$, and when $x$ and $z$ are level $-2$ and $y$ level $1$ and $x$, $y$ degree zero and $z$ degree $-3$. The first case trivially follows from the associativity of the wedge product 
\be  \left( ( S_1  S_2 ) S_3 \right)_{\rm top} =   \left( S_1 ( S_2  S_3)  \right)_{\rm top} \; . \ee
The second follows using integration by part, {\it i.e.} the property that the top form of a total contraction vanishes 
\be  \left( ( V_n \iota^n \omega + \tfrac23 \iota^n V_n \omega) S \right)_{\rm top} =   \left( \omega ( V_n \iota^n S + \tfrac13 \iota^n V_n S)  \right)_{\rm top} \; . \ee
The last one is obtained in the same way with few more steps as
\def\na{{n_{\scalebox{0.5}{$1$}}}}
\def\nb{{n_{\scalebox{0.5}{$2$}}}}
\def\nc{{n_{\scalebox{0.5}{$3$}}}}
\begin{align}
 &\quad\bigg( \Big( -\tfrac{1}{6} \psi_{\na\nb\nc} \iota^{\na}\iota^{\nb}\iota^{\nc}\omega_1 
+\tfrac{1}{4} \iota^{\nc}\psi_{\na\nb\nc} \iota^{\na}\iota^{\nb}\omega_1 \nn\\
&\hspace{20mm} - \tfrac{1}{8} \iota^{\nb}\iota^{\nc}\psi_{\na\nb\nc} \iota^{\na} \omega_1+\tfrac{1}{48}\iota^{\na}\iota^{\nb}\iota^{\nc} \psi_{\na\nb\nc} \omega_1  \Big)\omega_2\bigg)_{\rm top} \nn\\
&=-\bigg(\omega_1 \Big( -\tfrac{1}{6} \psi_{\na\nb\nc} \iota^{\na}\iota^{\nb}\iota^{\nc}\omega_2 
+\tfrac{1}{4} \iota^{\nc}\psi_{\na\nb\nc} \iota^{\na}\iota^{\nb}\omega_2\nn\\
&\hspace{20mm}- \tfrac{1}{8} \iota^{\nb}\iota^{\nc}\psi_{\na\nb\nc} \iota^{\na} \omega_2+\tfrac{1}{48}\iota^{\na}\iota^{\nb}\iota^{\nc} \psi_{\na\nb\nc} \omega_2 \Big) \bigg)_{\rm top} \nn
\end{align}
and
\be \Scal{ \scal{  -\tfrac{1}{24} \lambda_n \iota^n \omega_1 + \tfrac{1}{48} \iota^n \lambda_n \omega_1} \omega_2 }_{\rm top} =  - \Scal{\omega_1 \scal{  -\tfrac{1}{24} \lambda_n \iota^n \omega_2 + \tfrac{1}{48} \iota^n \lambda_n \omega_2} }_{\rm top}  \; , \ee
where the minus one comes from the fact that we have reverse the Grassmann degree of the generator to define the Cartan differential. 

The last case follows by the manifest invariance with respect to the zero level symmetry
\be  \left( ( V_{1\,m} \iota^m V_{2\,n} +\iota^m V_{1\,n}  V_{2\,m}) \upsilon^n \right)_{\rm top} =  - \left( V_{1\,m} ( V_{2\,n} \iota^n   \upsilon^m + \iota^n V_{2\,n} \upsilon^m-\iota^m V_{2\,n} \upsilon^n  ) \right)_{\rm top}\ .  \ee
These identities show that there is an antisymmetric invariant bilinear form on $\mathscr{T}$ that pairs level $q$ with level $-3-q$. The projection on the top component in form-degree also relates level $p$ to level $-2-p$.

\subsection{Remarks on the relation between \texorpdfstring{$E_{11}$}{E11} and the tensor hierarchy algebra}

The tensor hierarchy algebra contains $E_{d}$ as a subalgebra and we will now fix $d=11$ for concreteness. We know that level $p=0$ (see Table~\ref{tab:THA3}) contains $E_{11}$ but also additional new generators beyond the Kac--Moody structure. As we had already seen we crucially obtain one new generator $X_{n_1\ldots n_9}$ associated with $q=-3$. This is the beginning of an $E_{11}$ multiplet associated with the highest weight representation $\ell_2$ of $E_{11}$. We shall now probe whether there are additional $E_{11}$ multiplets contained in $p=0$ and what their reducibility structure is.

First we  decompose the forms as 
\bea 
S &=& M + \theta_n L^n + \frac{1}{2} \theta_{n_1n_2} \Lambda^{n_1n_2} 
+  \frac{1}{6} \theta_{n_1n_2n_3} e^{n_1n_2n_3} 
\CR
&& \qquad +   \frac{1}{4!} \theta_{n_1\cdots n_4} F^{n_1\cdots n_4} 
+   \frac{1}{5!} \theta_{n_1 \cdots n_5} B^{n_1\cdots n_5}+ \dots
\CR
\psi_{n_1n_2n_3}  &=& f_{n_1n_2n_3} + \theta_m\left(  F^m{}_{n_1n_2n_3}-3 \delta^m_{[n_1} F^p{}_{n_2n_3]p}\right)
+\frac{1}{2} \theta_{n_1n_2}B^{n_1n_2}{}_{n_1n_2n_3}  +\dots 
\CR
T_{m,n}  &=& F_{m,n} + \theta_p B^p{}_{m,n}   +\dots 
\CR
\lambda_{m}  &=& B_m+\dots 
\eea
Here, we have used the fields already encountered to parametrise the algebra, although we shall 
see that they parametrise in fact an element of the co-algebra. In this notation the B-fields parametrise 
the Bianchi identity, and one finds for instance three independent Bianchi identities 
\bea 
B_{n_1n_2}{}^{p_1p_2p_3} &=& 2 \partial_{[n_1} F_{n_2]}{}^{p_1p_2p_3} 
- 3 \partial^{[p_1p_2} F_{n_1n_2}{}^{p_3]} + 3 \delta_{[n_1}^{[p_1} \partial^{p_2|q} F_{n_2]q}{}^{p_3]} 
- \frac{1}{2} \partial^{p_1p_2p_3q_1q_2} F_{n_1n_2q_1q_2} 
\CR
&& \qquad - \frac{1}{2} \delta_{[n_1}^{[p_1} \partial^{p_2p_3]q_1q_2q_3} F_{n_2]q_1q_2q_3} 
+ \frac{1}{4!} \delta_{n_1n_2}^{[p_1p_2} \partial^{p_3]q_1\dots q_4} F_{q_1\dots q_4} +\dots 
\CR
B_m{}^{n_1,n_2} & =& \partial_m F^{n_1,n_2} - \partial^{p(n_1} F_{mp}{}^{n_2)} 
+ \frac{1}{72} \delta_m^{(n_1} \partial^{n_2)p_1\dots p_4} F_{p_1\dots p_4} + \dots\CR
B^m & =& \frac{1}{2}  \partial^{n_1n_2} F_{n_1n_2}{}^{m} 
+ \frac{1}{36}  \partial^{mn_1\dots n_4} F_{n_1\dots n_4} + \dots 
\eea 
where one sees that $B_m$ is indeed not a linear combination of the trace of the two others. The Bianchi identities are dual to the entries listed as $(p,q)=(-2,1)$ in Table~\ref{tab:THA3}. This entry is dual to $(p,q)=(0,-4)$ under the involution of the preceding section. The fact that there are three vectors in the representation implies that something for the $E_{11}$ representations needed to extend $E_{11}$ in the tensor hierarchy algebra. We already know that one needs $\ell_2$ (which triggers the important field $X_{n_1\ldots n_9}$ discussed at length in this paper). However, the adjoint of $E_{11}$ and $\ell_2$ together contain only two vectors for $q=-4$ and therefore we deduce that $\mathscr{T}$ also contains the highest weight representation $\ell_{10}$ that starts at $q=-4$. We expect that there is an infinite number of $E_{11}$ highest weight representation needed to extend $E_{11}$ to the level $p=0$ of the tensor hierarchy algebra $\mathscr{T}$.

We note however a difference between the representation $\ell_2$ extending the adjoint representation and the others. Indeed, taking the variation with the zero form component of $\psi_{n_1n_2n_3}$ only, which is associated to the action of the raising generator $E^{n_1n_2n_3}$,
one derives that 
\be 
E^{n_1n_2n_3} \iota^p V_p = - \iota^{n_1} \iota^{n_2} \iota^{n_3} S \ ,
\ee
whereas
\be 
E^{n_1n_2n_3} \bigl( \lambda_m + \tfrac45 \iota^p T_{m,p} - \tfrac35 \iota^{p_1} \iota^{p_2} \psi_{mp_1p_2}\bigr)  = 0 \ . 
\label{Reducible}  
\ee
It follows that the trace $d-9$ form component of $\iota^p V_p$ defines a primitive vector with respect to $E_{d}$ that varies to the adjoint representation, whereas $\lambda_m + \tfrac45 \iota^p T_{m,p} - \tfrac35 \iota^{p_1} \iota^{p_2} \psi_{mp_1p_2}$ does not vary to the corresponding representation. For $E_{11}$, this component includes the lowest weight vector of $\ell_{10}$ in $\scr T_{-2}$ and the lowest weight vectors of  $\ell_1+\ell_{10}\oplus \ell_{11}$ in $\scr T_{-3}$. This implies that the corresponding representations do not decompose as a direct sum in the first case, but do for the second. That this extends to all other higher representations was used as an assumption in Section~\ref{sec:THA}.

We use the convention that the forms are written as 
\be 
V_m = \xi_m + \theta_n V_m{}^n + \frac{1}{2} \theta_{n_1n_2} V_m{}^{n_1n_2}  +  \frac{1}{6} \theta_{n_1n_3}  V_m{}^{n_1n_2n_3}  + \dots 
\ee 
and we define the irreducible $\mf{gl}(11)$ components with a hat, as
\bea 
{V}_m{}^{n_1n_2n_3} &=& \hat{V}_m{}^{n_1n_2n_3}+\frac{1}{3} \delta_m^{[n_1} \hat{V}^{n_2n_3]} \ , \qquad V_q{}^{n_1n_2 q} = \hat{V}^{n_1n_2} \; ,  \CR
\psi_{p_1p_2p_3}{}^{n_1n_2} &=& \hat{\psi}_{p_1p_2p_3}{}^{n_1n_2} - \frac34 \delta^{[n_1}_{[p_1} \hat{\psi}_{p_2p_3]}{}^{n_2]} + \frac{1}{15} \delta^{n_1n_2}_{[p_1p_2} \hat{\psi}_{p_3]} \ , \CR
\psi_{p_1p_2q}{}^{nq} &=& \hat{\psi}_{p_1p_2}{}^{n}- \frac{1}{5} \delta^n_{[p_1} \hat{\psi}_{p_2]}  \ , \qquad {\psi}_{pq_1q_2}{}^{q_1q_2} = \hat{\psi}_p \ , \CR
T_{m,p}{}^n &=& \hat{T}_{m,p}+ \frac{1}{6} \delta^n_{(m} \hat{T}_{p)} \ ,\qquad T_{m,p}{}^p = \hat{T}_m \ . 
\eea 
Using these definitions one computes that 
\bea 
\delta \hat{V}^{n_1n_2} &=& - \hat{\psi}_{p_1p_2p_3} f^{p_1p_2p_3} +\frac{9}{4} \hat{\psi}_{p_1p_2}{}^{[n_1} f^{n_2]p_1p_2} + \frac{1}{16} ( \tfrac13 \hat{\psi}_{p} + \lambda_p  - 4 (\tfrac15 \hat{\psi}_{p} +3 \hat{T}_{p}) ) f^{n_1n_2p}  
\CR
 \delta \hat{V}_m{}^{n_1n_2n_3} &=& - 3 \hat{\psi}_{mp_1p_2}{}^{[n_1n_2} f^{n_3]p_1p_2} + \frac{1}{3} \delta_m^{[n_1} \hat{\psi}_{p_1p_2p_3}{}^{n_2n_3]} f^{p_1p_2p_3} + \frac{1}{16} ( \tfrac13 \hat{\psi}_{m} + \lambda_m) f^{n_1n_2n_3} 
\CR
&& \hspace{10mm} - 3 \hat{T}_{m,q}{}^{[n_1} f^{n_2n_3]q} - \frac{1}{48}  (  \tfrac13 \hat{\psi}_{p}+ \lambda_p)\delta_m^{[n_1} f^{n_2n_3]p} 
\label{deltaV1,3}  
\eea
and consistently with the property that $  \hat{\psi}_{m} +3 \lambda_m$ corresponds to the element of $\mathfrak{e}_{11}$ associated to the field $C_{11,1}$ that 
\be 
\delta (  \hat{\psi}_{m} + 3 \lambda_m) = -2 \hat{V}_m{}^{n_1n_2n_3} e_{n_1n_2n_3} \ . 
\ee
Then the field $\hat{\psi}_m + 15 \hat{T}_m$ corresponds to the field $X_{11,1}$ in $\ell_2$ plus an arbitrary multiple of $C_{11,1}$ that is not uniquely fixed by the representation (beacuse of its indecomposable character). The field $3 \hat{\psi}_m + 4 \hat{T}_m + 5 \lambda_m$ corresponds instead to the field $Y_{11,1}$ defining the highest weight component of the $\ell_{10}$ module. The $E_{11}$ transformations \eqref{Reducible} and \eqref{deltaV1,3}  imply that there is no mixing between $\mathfrak{e}_{11} \oplus \ell_2$ and $\ell_{10}$. 

Using a similar decomposition in irreducible components 
\bea 
\psi_{p_1p_2p_3}{}^{n_1n_2n_3} &=&  \hat{\psi}_{p_1p_2p_3}{}^{n_1n_2n_3} + \frac{9}{7} \delta_{[p_1}^{[n_1} \hat{\psi}_{p_2p_3]}{}^{n_2n_3]} + \frac{1}{4} \delta_{[p_1p_2}^{[n_1n_2} \hat{\psi}_{p_3]}{}_{n_3]} + \frac{1}{165} \delta_{p_1p_2p_3}^{n_1n_2n_3} \hat{\psi} \ , 
\CR
T_{m,p}{}^{n_1n_2} &=& \hat{T}_{m,p}{}^{n_1n_2} - \frac{4}{11} \delta_{(m}^{[n_1} \hat{T}_{p)}{}^{n_2]} \ , \qquad \lambda_m{}^n = \hat{\lambda}_m{}^n + \frac{1}{11} \delta_m^n \hat{\lambda} \ , 
\eea
one computes that 
\bea 
\label{deltaV1,4} \delta V_m{}^{n_1n_2n_3n_4} &=& - 4 \hat{\psi}_{mp_1p_2}{}^{[n_1n_2n_3}f^{n_4]p_1p_2} + \frac{3}{7}( 5 \hat{\psi}_{mp}{}^{[n_1n_2} f^{n_3n_4]p} -  \delta_m^{[n_1} \hat{\psi}_{p_1p_2}{}^{n_2n_3} f^{n_4]p_1p_2})  \CR
&& \hspace{5mm} + 6 \hat{T}_{m,p}{}^{[n_1n_2} f^{n_3n_4]p}   - \frac{1}{11} (\hat{T}_m{}^{[n_1} f^{n_2n_3n_4]} - 12 \delta_m^{[n_1}\hat{T}_p{}^{n_2} f^{n_3n_4]p}) \CR
& & \hspace{10mm} + \frac{1}{12} ( \hat{\psi}_m{}_{[n_1} + \tfrac{1}{11} \delta_m^{[n_1}  \hat{\psi} + \lambda_m{}^{[n_1}  ) f^{n_2n_3n_4]}  \ . 
\eea
In this case one finds that the corresponding components of $\iota^{p_1} \iota^{p_2} \psi_{mp_1p_2}  - 3 \lambda_m$ and $\iota^{p} T_{m,p}$ belong to $\ell_1$, whereas the corresponding components of $ 5\lambda_m + 4 \iota^p T_{m,p} - 3 \iota^{p_1} \iota^{p_2} \psi_{mp_1p_2}$ define the highest weight vectors of  $\ell_1+\ell_{10} $ and $\ell_{11}$. The commutation relation following from \eqref{Reducible} and \eqref{deltaV1,4} are such that there is no mixing between $\ell_1$ and  $\ell_1+\ell_{10} \oplus \ell_{11}$. 

\begin{raggedright}

\end{raggedright}

\end{document}